# Generalization and the Rise of System-level Creativity in Science

Hongbo Fang[†], James Evans[†‡]
[†]University of Chicago, [‡]Santa Fe Institute

**Scientific progress is widely understood as combinatorial, yet existing measures focus on the inputs a paper draws together, saying little about how knowledge is supplied for downstream reuse. Analyzing citation networks from tens of millions of publications in OpenAlex and the Web of Science, we decompose scientific contributions into three functional types, foundations, extensions, and generalizations, distinguishable by the roles they play in future science. Foundational and extensional works, which build and elaborate ideas within disciplines, dominated the post-war decades but declined steadily after the 1990s, while generalizations that catalyze innovation across many contexts rose sharply. Challenging the narrative of declining innovation inferred from the disruption index, our findings suggest the locus of innovation has shifted from within disciplines to between them, a transition the Internet and large language models may have facilitated. Our indices disentangle innovation modes conflated by existing measures, enabling reinterpretation of landmark findings of 'disruptive' science.**

Maintaining a healthy innovation system requires a clear understanding of how scientific progress unfolds. This challenge is central not only for advancing knowledge but also for guiding research funding, institutional design, and innovation policy[1,2]. A long tradition in innovation theory, from Schumpeter's "new combinations" through recombinant growth[3], recombinant search[4,5], and theories of the adjacent possible [6], recognizes that breakthroughs frequently arise from the novel integration of existing components rather than isolated acts of discovery[7]. A growing empirical literature has developed measures to capture recombination in science: bridges and jumps across the network of knowledge[8]; atypical combinations of referenced journals[9], novel pairings of distant knowledge elements[10], and surprising conjunctions of both content and context[11,12]. Related work in scientometrics and informetrics has likewise characterized the concentration, dispersion, and interdisciplinarity of scientific attention[13–16]. These measures have advanced our understanding of science's demand for diverse knowledge to innovate. But how is knowledge supplied to facilitate the knowledge integration and recombinations?

We answer this question by shifting the unit of analysis from the composition of a paper's inputs to the role it plays in subsequent research. To this end, we introduce three complementary measures, the foundation (F), extension (E), and generalization (G) indices, that evaluate the role that a focal paper plays in each of its downstream citations. Specifically, if a focal paper is cited together with more of its citing papers (i.e., the papers that cite the focal paper) than its referenced papers (i.e., the papers cited by the focal paper), we classify the citation as a foundational citation (bold red arrow in Fig. 1a), reflecting the role of the focal paper as a foundation upon which the citing work builds. Conversely, if the focal paper is cited together with more of its referenced papers than its citing papers, we classify it as an extensional citation (bold orange arrow in Fig. 1a), reflecting the role of the focal paper in extending existing paradigms. Finally, if the focal paper is cited in isolation, without any of its referenced or citing papers being co-cited, we classify it as a generalized citation (bold blue arrow in Fig. 1a), reflecting the role of the focal paper as a modular knowledge component imported from, often, a distant domain. The foundation, extension, and generalization indices of the focal paper are then calculated as the proportions of its citations classified as foundational, extensional, and generalizational, respectively.

At a high level, foundational and extensional contributions represent two well-established categories of scientific advances: field-defining milestones and subsequent works that incrementally extend these milestones. Two properties characterize generalizing work and motivate our interpretation of it. The first is modularity: the capacity of a contribution to be used independently of the intellectual context in which it originated (Fig. 1d). The second is general purpose: the capacity to catalyze advance across a wide range of domains without exerting dominant influence within any single one (Fig. 1e). Throughout the paper we validate both properties directly rather than inferring them from citation structure alone. We do this with externally annotated citation purposes, the placement and frequency of citations within full texts, and the breadth and concentration of downstream influence across research domains and countries.

The intuition underlying the three indices is that a paper cited as a foundation for new work is likely to be discussed alongside many of the studies that extend it, for example, to position the proposed contribution with respect to existing approaches. As a result, foundational papers tend to be co-cited intensively with subsequent extensions. In contrast, when a paper is cited solely as a self-contained knowledge component, particularly when imported from a distant domain, the broader knowledge context in which the cited work is situated, including the literature that cites it or is cited by it, is either not pertinent to the new work and so remains omitted from its references or is distant from the new work and so remains unknown to the authors. We illustrate this distinction using the paper that introduced the modern AI transformer, *Attention Is All You Need*[17], hereafter the *Attention* paper, as the citing paper. Under our framework, the citation from *Attention* to the *Long Short-Term Memory* (LSTM) paper[18] is classified as a foundational citation (Fig. 1b). This citation appears in the Introduction, where the *Attention* paper discusses the LSTM model alongside many of its extensions to contextualize its contribution (Fig. 1c). In contrast, the citation to the *Layer Normalization* paper[19] is classified as a generalized citation (Fig. 1b). It is incorporated as a modular component of the transformer architecture and mentioned only briefly in the "Encoder and Decoder Stacks" subsection, without discussion of the broader body of research surrounding the method (Fig. 1c). Given the conceptual rationale underlying the three indices, their operationalization may vary slightly provided that it remains consistent with the intended rationale. For example, the paper *Dropout: A Simple Way to Prevent Neural Networks from Overfitting*[20] (Fig. 1b) may be classified as a generalizing reference under a slightly less restrictive definition (see SI Supplementary Methods Section 2).

The concept underlying foundation, extension and generalization can be traced back to Herbert Simon, who characterized complex systems as "nearly decomposable", and so strongly coupled within modules, weakly coupled between them[21]. In Simon's terms, foundations and extensions characterize progress within modules of the system, while generalizations characterize the creation of interfaces between them. The decomposition can be applied to any system in which credit or influence can be traced through networks of claimed or inferred connection[22]. As we will demonstrate, through much of the postwar era, science approximated this structure. Disciplines functioned as semi-autonomous modules with dense internal citation networks, shared methods, and distinct foundational literatures. Recent scholarship has suggested that contemporary science may operate under conditions of growing connectivity[23,24]. These new indices are directly designed to characterize and track this transition from forging and extending disciplinary foundations to compressing and transmitting useful generalizations across science[25–28]. We further examine the role of information technology in accelerating this process. Recombinant measures capture whether a paper's inputs are drawn from distant domains. Our approach inverts this focus to identify the role research outputs play in downstream research, to build up fields or transgress them over time.

Our proposed indices, especially the foundation and generalization indices, are closely connected to established measures of scientific innovation. In the Supplementary Information (Section 20 in the Supplementary Results), we compare our indices with their closest counterpart, the disruption index[22,29], conceptually and empirically, and show that the foundation and generalization indices provide an effective decomposition of the disruption index. The foundation index captures the type of "disruption" that leads to paradigm shift and substantial change in research trajectories, while the generalization index captures the type of "disruption" that acts as a generalized catalyst of scientific progress, tracing work that may not itself be the locus of field-level breakthrough, but has ambient impact that facilitates innovation across many domains (Fig. 1d,e). We present examples of papers that score high on the disruption index yet range from highly foundational to highly general (Fig. 2a).

Using these measures at scale, we find that, since the late twentieth century, scientific production has shifted away from both foundational and extensional contributions toward greater knowledge compression and generalization. This shift is robust to null models that preserve the temporal growth of publications, references, and citations (Fig. 3b) and is closely associated with a marked change in the composition of scientific output toward reusable research products, including reviews, software, datasets, and tools (Fig. 4a). Together, these patterns point to a reorientation in the locus of scientific innovation: from advancing knowledge primarily within individual fields toward synthesizing and recombining knowledge across the scientific system as a whole (Fig. 3).

Our findings suggest that contemporary scientific innovation increasingly arises not only from breakthroughs within individual fields, but also from the integration and recombination of modular, compressed knowledge across disciplinary boundaries. Advances originating in one domain can be adopted, generalized, and recombined across many others, thereby contributing collectively to the expansion of the scientific frontier. We refer to this emerging mode of scientific progress as **"system innovation,"** in which innovation increasingly operates at the level of the broader knowledge system rather than within disciplinary domains alone. We conclude by discussing the implications of these findings for how scientific progress is understood, measured, and managed.

## Results

We begin by presenting a series of analyses that characterize the substantive meaning of our proposed indices and validate them against independent evidence of how cited work is used in subsequent research. A comparison between our metrics and their closest counterpart in measurement approach, the disruption index, yields a key insight into the nature of scientific breakthroughs (Fig. S26). The generalization index exhibits the strongest association with disruption, showing a positive and monotonic relationship ($r = 0.37$, $p < 2 \times 10^{-16}$, $df = 23{,}448{,}429$). This is followed by the foundation index, which manifests a U-shaped association with disruption ($r = 0.05$, $p < 2 \times 10^{-16}$, $df = 23{,}448{,}429$). The extension index, by contrast, is negatively correlated with disruption ($r = -0.37$, $p < 2 \times 10^{-16}$, $df = 23{,}448{,}429$). At the citation level, 73.96% of all disruptive citations (a paper is cited without its references being co-cited) are essentially generalizational (a paper is cited without either its references or the papers citing it being co-cited ).

These results suggest that a substantial portion of what existing metrics classify as "disruptive" may, in fact, reflect processes of generalization. Rather than displacing existing knowledge within a given field and disrupting the ongoing research trajectory, generalizations synthesize and compress knowledge into modular forms that enable it to diffuse and recombine with concepts in distant domains that would never have been at risk of citing the prior work. The conceptual intuition underlying the disruption index remains valuable, and our measurement approach factorizes it to separate foundational and extensional work from the process of generalization that has emerged as the most important change in the structure of science across recent decades. In the Supplementary Information (Supplementary Results Section 20), we develop this decomposition formally, showing that as the downstream use of a paper shifts from foundational to generalization, its citations pass through the configuration that disruption index scores most highly such that disruption occupies an intermediate position along the foundation–generalization continuum, with stronger affinity to generalization. This reinterpretation has important implications for prior empirical results derived from the disruption index, several of which we examine below.

Next, we examine the relationship between a work's classification as a foundation, extension, or generalization and two measures of knowledge distance: the embedding distance among a paper's references and the semantic diversity within the paper itself. These measures are constructed using continuous machine-learned representations of words and references, derived from their co-presence across 23 million scientific publications in the OpenAlex dataset (Methods). This analysis connects naturally to prior measures of atypicality[9] and context surprise[11], which quantify the extent to which a paper draws on knowledge from distant domains.

We find that papers scoring highly on the F, E, and G indices correspond to distinct strategies of knowledge production (Fig. 2b). Foundational papers cite references that are, on average, 0.01 standard deviations more distant than the overall mean ($t = 4.2$, $p = 3.3 \times 10^{-5}$, $df = 194{,}286$; two-sided Welch's $t$-test), and combine title tokens 0.06 standard deviations more semantically distant than average ($t = 22.7$, $p < 2 \times 10^{-16}$, $df = 199{,}583$). Extension papers, by contrast, cite references 0.35 standard deviations closer than average ($t = 136.7$, $p < 2 \times 10^{-16}$, $df = 202{,}274$), while employing title tokens that are more semantically dispersed, 0.05 standard deviations above the mean ($t = 20.4$, $p < 2 \times 10^{-16}$, $df = 197{,}154$). Generalization papers exhibit a distinctive dual pattern: they cite references that are 0.05 standard deviations more distant than average ($t = 16.4$, $p < 2 \times 10^{-16}$, $df = 177{,}211$), while compressing semantic distance in their titles, with token distances 0.29 standard deviations below the mean ($t = 100.27$, $p < 2 \times 10^{-16}$, $df = 170{,}993$). These patterns are robust to alternative metrics of semantic distance that incorporate full abstracts, and to metrics of reference distance based on specific papers rather than venues (Fig. S16). Notably, the correlations between atypicality or surprise and the F, E, and G indices are modest, particularly for the F index, indicating that atypicality alone cannot account for the dimensions captured by our framework. Atypicality and surprise measure how a paper is constructed, whereas the F, E, and G indices measure how a paper is subsequently used (Supplementary Information, Figs. S15–S16, Tables S6–S7).

These results suggest distinct underlying mechanisms. Generalizing works tend to draw upon ideas from distant domains while expressing them in familiar conceptual and linguistic frameworks, consistent with a process of knowledge compression. Extensional works build upon closely related prior knowledge but expand it through increased semantic complexity, potentially introducing new representations or conceptual frameworks. Foundational works occupy an intermediate position, combining moderately distant references with moderately novel semantic expression, the only category atypical in both citation and semantic spaces, although the effect size is modest (Fig. 2b; Extended Data Fig. 1).

Examining the linguistic signatures of papers with high F, E, and G indices further corroborates their distinct characteristics (Fig. 2d; Extended Data Fig. 2). Papers with high generalization scores employ language associated with review articles that explicitly synthesize and compress knowledge from disparate domains, or draw upon terms such as “tools,” “devices,” and “software,” designed to apply and diffuse across diverse contexts. In contrast, high-foundation papers tend to feature terms that signal the initiation of new research trajectories, including “discovery,” “observation,” and “dataset.” High-extension papers are characterized by frequent use of terms such as “theory,” “metric,” and “hypothesis,” reflecting analytical refinement, formalization, and the consolidation of knowledge.

**Validation against citation purpose, placement, and diffusion**

The most consequential assumptions of our framework are that a paper cited in relative isolation indicates the likelihood of modular reuse and generalized catalysis. We undertake four direct tests of these assumptions. First, we link citations of each structural type to two external datasets with citation-purpose annotations derived from the textual context surrounding citations: the multi-domain ACT2 dataset[30] and the natural-language-processing dataset of Jurgens et al.[31]. Generalizing citations exhibit the highest probability of serving a “uses” purpose (i.e., the citing paper employs the cited work as a component), whereas foundational and extensional citations are more frequently deployed to motivate the study, extend prior work, or compare against existing methods (Fig. 2f; Figs. S17–S18). Subsequent studies primarily use generalizing work without attempting to extend, compare against, or challenge it.

Second, we analyze the full text of approximately 3.2 million papers from Semantic Scholar to identify the frequency and contextual placement of citations. Generalizing citations are mentioned, on average, 1.31 times within the main text of citing papers, significantly less than foundational citations (1.59 times; $t = 441.5$, $p < 2 \times 10^{-16}$, $df = 15{,}625{,}490$, two-sided Welch’s $t$-test) and extensional citations (1.82 times; $t = 691.2$, $p < 2 \times 10^{-16}$, $df = 14{,}160{,}213$), suggesting that generalizing works, despite broad applicability, rarely occupy a central role in the intellectual structure of the citing paper (Fig. 2g). In terms of section-wise placement, generalizing citations are concentrated in discussion-related sections (7.6% in “Results and Discussion,” versus 5.9% for extensional and 5.8% for foundational citations; and 32.4% in “Discussion”), with substantial representation in “Methods” (12.1% versus 10.0% for extensional citations; $t = 136.8$, $p < 2 \times 10^{-16}$, $df = 13{,}437{,}973$), where prior work is adopted as established knowledge. Foundational citations are most frequent in the “Introduction” (45.7%) and “Methods” (13.4%), reflecting their role in framing research questions and grounding methodological design, while extensional citations appear disproportionately in results-oriented sections, where prior work is evaluated, extended, or compared (full statistics in Fig. 2g; robustness in Extended Data Fig. 3 and Figs. S59-S60).

Third, we characterize the general-purpose property directly by measuring the breadth and concentration of downstream influence (Fig. 2c). Controlling for publication year, domain, and citation count, top-decile generalizing papers are cited from a substantially broader range of OpenAlex topics than foundational or extensional papers, while the share of their citations contributed by their single largest topic is markedly lower, indicating broad influence without dominant influence in any one domain. The same pattern holds geographically: generalizing work diffuses more broadly across national boundaries, whereas foundational work exhibits the strongest tendency to be cited by researchers within the same country (Fig. 2c, right). Distance-based analyses reinforce this picture: when papers are divided into two groups along each index, top generalizing works extend 84.0% farther in reference space ($t$ = 7,400.9, $p < 2 \times 10^{-16}$, $df$ = 248,419,652), 15.5% farther in semantic space ($t$ = 4,050.1, $p < 2 \times 10^{-16}$, $df$ = 441,327,231), and are 114.0% more likely to receive cross-domain citations ($t$ = 2,549.5, $p < 2 \times 10^{-16}$, $df$ = 436,036,538) than bottom-group works (Fig. 2e).

Fourth, we validate against expert-recognized contributions and representative cases. The foundation index consistently outperforms both the generalization and disruption indices in identifying Nobel Prize-winning papers, major award-winning papers, concept-introducing papers, and the original research synthesized by review articles, categories widely regarded as paradigm-shifting. The generalization index, by contrast, preferentially identifies review articles, which synthesize and disseminate existing knowledge (Extended Data Figs. 4-5; Figs. S35–S39). Case studies qualitatively corroborate the quantitative results: high-foundation papers are landmark contributions that open research directions; high-generalization papers supply reusable, modular components, such as methods, instruments, datasets, tools, and software. These cases involve five highly disruptive papers, ranging from foundational to generalizational contributions (Fig. 2a), influential papers across seven domains spanning chemistry, computer science, economics, mathematics, medicine, physics, and sociology (Figs. S27–S33), as well as twelve exemplary papers covering reviews, devices, software, and theory with divergent F, E, and G values (Figs. S22–S25). Although the F, E, and G indices correlate with publication type, substantial variation remains within type, indicating that the indices capture differences in knowledge function beyond conventional publication categories.

**Historical shifts in the locus of innovation from field to system**

The temporal evolution of these contribution types reveals a broad structural shift in scientific knowledge production over the past 75 years, which separates into two eras of roughly equal duration (Fig. 3). The first, from the end of WWII until the early 1990s, is a period of disciplinary emergence beginning with field-founding papers like Watson and Crick's 1953 discovery of DNA's structure. During this period, science increasingly took the form of a nearly decomposable system: within-field foundations were laid, extensions built upon them, and disciplinary boundaries strengthened. The average foundation index declined from 0.059 in 1950 to 0.025 in 1990, a 57.6% decrease ($t$ = 20.6, $p < 2 \times 10^{-16}$, $df$ = 6,510), and the average extension index rose from 0.536 to 0.762, a 42.2% increase ($t$ = 49.5, $p < 2 \times 10^{-16}$, $df$ = 6,726). Generalization also decreased across this period, from 0.405 to 0.214, a 47.2% decrease ($t$ = 42.1, $p < 2 \times 10^{-16}$, $df$ = 6,707), with the emergence and fortification of disciplinary boundaries. The second period, beginning in approximately 1991, the year the World Wide Web was launched, marks a period of post-disciplinary recombination. The generalization index increases from 0.218 in 1991 to 0.382 in 2024 (a 75.2% increase, $t$ = 198.2, $p < 2 \times 10^{-16}$, $df$ = 159,955), while the foundation index decreases from 0.025 to 0.022 (a 12.0% decrease, $t$ = 14.8, $p < 2 \times 10^{-16}$, $df$ = 151,497) and the extension index drops from 0.757 to 0.596 (a 21.3% decrease, $t$ = 189.8, $p < 2 \times 10^{-16}$, $df$ = 159,025). These patterns are remarkably consistent across citation thresholds, citation windows, continuous variants of the indices, datasets (Extended Data Fig. 6), and scientific fields (Extended Data Fig. 7).

We interpret these trends as evidence of a shift in the locus of scientific innovation. Importantly, we do not equate highly generalizing works themselves with system innovation. Rather, the aggregate rise in the generalization index indicates a change in how scientific knowledge is consumed and reused: on average, scientific contributions increasingly function as modular knowledge components that can be adopted, combined, and repurposed across domains, rather than serving primarily as foundations for or extensions of research within their originating fields. It is this system-level shift in the organization and use of scientific knowledge, not generalization at the level of any individual work, that we characterize as the transition toward **system innovation**.

**The shift is not an artifact of community growth, databases, or publishing patterns**

Because the F, E, and G indices are computed from local citation neighborhoods, their longitudinal interpretation could in principle be contaminated by the dramatic growth of science: more papers, longer reference lists, citation inflation, denser citation neighborhoods for highly cited work, an increasingly fragmented landscape of venues and subfields, and time-varying database coverage. We address these concerns with explicit null models and coverage analyses.

We constructed randomized citation networks by swapping the cited endpoints of randomly selected pairs of citation links while approximately preserving the degree distribution of the citation graph within each publication year and either each publication venue or each publication domain (at the topic, subfield, and field levels of the OpenAlex taxonomy). Under this procedure, a citation from paper A to paper B can be reassigned only to a paper B′ published in the same year and in the same venue (or domain) as B. The reshuffling therefore preserves the number of papers published each year, the number of citations received and references made by every paper, and the annual venue and domain composition of the publishing landscape. Under this null model, growth in publication volume, reference-list length, citation counts, venue fragmentation, and field structure are all held fixed (see Methods).

The longitudinal trajectories of the F, E, and G indices in the randomized networks differ from the empirical network in two important respects (Fig. 3b; Fig. S14). First, the post-2010 increase in the generalization index is markedly steeper in the empirical network than in any randomized null model. Second, the randomized networks either exhibit no turning point in the generalization index transitioning from decline to increase when publication domains are preserved, or they display such a turning point only after 2000, at least a decade later than observed empirically. The venue-preserving null model does reproduce a qualitatively similar U-shape, indicating that changes in the venue landscape, including the growth of broad-scope and interdisciplinary venues, likely account for a non-trivial share of the post-2000s increase; but its transition arrives roughly 15 years late and its post-transition rise is markedly shallower than observed. Together, these comparisons indicate that the empirical transition toward generalization cannot be explained by growth in publication volume, citation and reference inflation, or the changing number and composition of venues and domains alone.

Database coverage receives similarly direct treatment. Among the 26,068,520 OpenAlex publications meeting our inclusion criterion ($\geq 5$ citations within five years), 89.9% have at least one recorded reference. Augmenting the citation network with Semantic Scholar links (an 8.7% increase in citation links) leaves the longitudinal trajectories essentially unchanged (Fig. S19), as does retaining papers with zero recorded references. Nevertheless, the transition point of the generalization index shifts approximately 15 years later and still appears earlier than its corresponding randomized network under this most conservative inclusion (Figs. S63–S64). Excluding all self-citations (11.2% of links) likewise leaves the qualitative patterns intact, while revealing that self-citations are disproportionately extensional (Fig. S20). Results replicate in the Web of Science, whose curated coverage has only 10.2% of papers lacking references (Fig. 3; Extended Data Fig. 6).

**A concurrent transformation in the composition of science**

If the rise of generalization reflects a genuine reorganization of scientific production, it should be visible not only in citation structure but in the content of what scientists produce. Using title keywords as proxies for categories of research output, we find a substantial shift in the composition of scientific publications (Fig. 4a). Publications that are inherently generalizing, such as review articles and papers introducing datasets, devices, software, and research tools, have increased markedly in prevalence, whereas papers associated with foundational or extensional contributions, including those centered on theories and hypotheses, have declined. Several categories exhibit pronounced changes in publication volume beginning in the early 1990s, coinciding with the reversal of the generalization index from steady decline to sustained increase. Because this compositional evidence derives from paper titles rather than citation links, it constitutes an independent signal of the transformation.

To disentangle compositional changes from shifts in citation practices within categories, we examine the evolution of the F, E, and G indices among papers belonging to the same keyword-defined category (Extended Data Fig. 8). Although the indices exhibit some temporal variation within individual categories, the magnitude of these within-category changes appears substantially smaller than the shifts associated with changes in the relative prevalence of papers across categories. The long-term evolution of the F, E, and G indices therefore appears to be driven primarily by changes in the composition of scientific output. In particular, the rise in generalizational research appears to reflect a reallocation of scientific effort toward reusable and modular research products, rather than a broad transformation of citation practices within particular categories of papers.

**The possible role of digital knowledge infrastructure**

We hypothesize that advances in information technology have facilitated the transition toward greater generalization in scientific knowledge production. Conceptually, technologies such as the Internet, search engines, social media, and, more recently, large language models substantially expand access to broad and heterogeneous bodies of knowledge [32–34]. By reducing the costs of discovering, accessing, and navigating knowledge beyond one's immediate domain, these technologies may facilitate the diffusion of ideas across disciplinary boundaries and make the recombination of knowledge from otherwise distant domains more feasible [35–37]. Empirically, the temporal patterns are consistent with this hypothesis. The transition from declining to increasing generalization occurs around the emergence of the World Wide Web in the early 1990s, after which the generalization index continues to rise alongside the subsequent diffusion of search engines, social media, and, most recently, large language models. To further examine this relationship, We present additional quasi-experimental evidence consistent with this hypothesis.

First, leveraging the Fulltext Sources Online dataset, which records the timing of online text availability for over 30,000 publication venues, we implement a stacked difference-in-differences (DiD) design. We compare the average generalization index of papers published in a given venue before and after it becomes available online, relative to a control group of randomly selected venues that remain offline during the observation period. The transition to online accessibility is followed by a modest but significant increase in the average generalization index (0.043 standard deviations; 95% CI: 0.004–0.081; Fig. 4b), a result consistent across observation windows and across alternative outcomes including cross-domain citation share and reference and semantic distance (Extended Data Fig. 9a). Because venues did not adopt online access at random (e.g., earlier adopters may differ in international reach, field composition, and editorial strategy) these estimates should be read as evidence of association consistent with facilitated cross-domain diffusion, not as definitive causal effects.

Second, using a dataset identifying authors' adoption of large language models (LLMs) in manuscript preparation[38], we apply a similar stacked DiD framework, comparing authors before and after first detected LLM use against authors with no detected adoption. Across three preprint databases, LLM adoption is followed by a consistent increase in generalizing references: 0.050 standard deviations for *arXiv* authors (95% CI: 0.030–0.071), 0.117 for *bioRxiv* (95% CI: 0.060–0.174), and 0.081 for *SSRN* (95% CI: 0.005–0.157) (Fig. 4c). These results are robust to alternative detection thresholds ($\tau = 0.05, 0.2$) and to alternative distance-based outcomes (Extended Data Fig. 9b–f). The LLM analysis warrants caution: adoption is inferred from an AI-detection procedure rather than directly observed, adopters may differ from non-adopters in many ways (e.g., research style, field, career stage, or language background), and outcomes are observed only at publication events (see Methods).

However, each piece of evidence has important limitations. First, although the turning point in the generalization index occurs around 1990 in our primary analysis (Fig. 3), this timing is sensitive to the treatment of papers with no references. When papers with zero references are included, the estimated turning point in OpenAlex shifts to approximately 2010 (Fig. S63), substantially later than the emergence of the Internet and the introduction of many modern information technologies. This sensitivity weakens any interpretation that directly attributes the temporal transition in generalization to the advent of the Internet. Second, our quasi-experimental analyses of online access and LLM adoption also face identification limitations. Treatment assignment is not random in either setting, and, in the case of LLM adoption, treatment status is inferred indirectly rather than directly observed. Consequently, unobserved differences between treated and untreated groups may contribute to the estimated effects.

Taken together, these analyses provide suggestive, rather than conclusive, evidence that advances in information technology have contributed to the increasing generalization of scientific knowledge. Neither the temporal correspondence nor the quasi-experimental evidence is sufficient to establish a causal relationship. We therefore interpret these findings as evidence consistent with our hypothesis and as motivation for further research to identify the causal role of information technologies in reshaping how scientific knowledge is accessed, recombined, and diffused across disciplinary boundaries.

**Discussion**

Our analysis documents a broad structural shift in how scientific knowledge is produced, with implications for how we understand, measure, and support scientific progress. Over the past 75 years, the balance of scientific contributions has shifted from within-field foundations and extensions toward cross-field generalizations: compressed, modular representations taken up and reused in contexts distant from their origin. This pattern is consistent with a transition in the architecture of science from a more decomposable system, organized around semi-autonomous disciplines, toward an integrated one where knowledge components travel more freely across fields. Among other implications, this structural account reframes the widely discussed decline in disruption scores[39]. Much of what scholars have registered as an historical decline in novelty instead reflects a shift in the locus of innovation, from within fields to across the system.

We are careful about what this shift represents and what it does not. Generalization is conceptually distinct from the creative foundational work that introduces fundamentally new ideas and drives paradigm shifts, which has been the focus of the scientific innovation literature. Our evidence indicates that generalizing work is not the primary source of breakthrough innovation: it is cited across many domains and rarely dominates (Fig. 2c), it is mentioned less frequently and less centrally within citing papers (Fig. 2g), and it is negatively associated with major scientific recognition (Extended Data Fig. 5 and Fig. S13). Rather than characterizing generalization as innovation itself, we interpret it as a generalized scientific catalyst: enabling infrastructure like methods, tools, datasets, software, and syntheses that facilitates innovation across diverse domains without constituting the paradigm-shifting anchor in any of them. Viewed this way, the patterns we observe, including the longitudinal changes in F, E, and G and the accompanying shift in the composition of published output, indicate that an increasing share of scientific effort is directed toward producing broadly enabling knowledge rather than field-specific conceptual breakthroughs. This does not represent a decline in creativity, however, but a reallocation of creative effort from forging within-field breakthroughs toward amplifying the efficiency, accessibility, and scalability of knowledge fusion across the system.

This structural account integrates diverse findings across the recombinant innovation tradition. Existing frameworks, from Weitzman's growth models[3] to Fleming's search theory[4] to Arthur's account of technological evolution[7], establish that breakthroughs arise from novel combinations of existing components. Empirical measures of atypical combination[9], combinatorial surprise[11,40], and related approaches[10] have provided evidence for this view. Our contribution identifies the supply-side mechanism by which ideas, once produced, are compressed into forms portable enough to be taken up and recombined across distant fields. In Simon's architectural vocabulary, foundations and extensions represent progress within the modules of a nearly decomposable system, while generalizations create and strengthen interfaces between modules. The growing prevalence of generalization thus reflects a system becoming more tightly coupled across its parts.

What drives this transition? Our evidence points to multiple, potentially interacting factors. Our quasi-experimental analyses suggest that digital "knowledge infrastructure", through which knowledge is produced, validated, and circulated, is a plausible contributor[41,42]. Pre-digital infrastructure, from disciplinary journals to conferences, reinforced the decomposable architecture of science by circulating communication within fields. Digital infrastructure, such as the web, search engines, preprint servers, social media and now large language models, reduces the cost of accessing and integrating knowledge, creating conditions for cross-domain exchange that Galison described as "trading zones"[43]. Prior work has largely focused on how such technologies augment individual scientists' productivity[38,44–47], but our findings also point to a system-level effect on the relative costs of within- versus between-field knowledge integration. Our analysis suggests that another driver may be the compositional shift in scientific output toward reviews, software, datasets, standardized methods, and tools. Digital infrastructure may act partly through this compositional channel, by preferentially facilitating the production and dissemination of reusable outputs. Institutional and policy forces likely contribute as well.

Science funders have increasingly prioritized mission-oriented and interdisciplinary research, pushing toward generalization independent of digital connectivity. Disentangling these factors is an important direction for future research.

The collective effort to compress and generalize scientific knowledge, together with the emergence of digital knowledge infrastructure, may offer a modern response to the "burden of knowledge" dilemma[48]. Although the scientific knowledge base continues to expand beyond what any individual can master, researchers are increasingly augmented by Web search, personalized recommendation systems, LLMs, and the production of generalization contributions. Rather than undergoing years of training to acquire the intellectual foundations of a distant discipline, they can increasingly access abstracted and generalized knowledge representations that distill transferable principles while filtering out much of the original disciplinary context. By lowering the cognitive and informational costs of acquiring and applying knowledge beyond one's area of expertise, these infrastructures facilitate cross-disciplinary exploration and knowledge recombination without requiring researchers to become specialists in every relevant field. As a result, they may restore some of the innovative capacity traditionally constrained by the growing burden of knowledge, enabling small teams without a comprehensive set of specialized expertise to explore, integrate, and recombine ideas across domains[49].

Despite the growing importance of generalizing contributions, our results reveal a significant misalignment between how science is produced and how it is evaluated. Major forms of scientific recognition, such as awards and honors, are positively associated with foundational and extensional contributions but negatively associated with those that generalize (Extended Data Fig. 5 and Fig. S13). We do not read this as evidence that the recognition system is "wrong". Foundational work merits the recognition it receives, but our evidence suggests that an increasingly prevalent and infrastructurally essential class of contributions falls outside what current evaluation systems are calibrated to detect. Activities such as synthesizing ideas, writing reviews, and developing broadly applicable tools play a critical role in enabling recombination but remain underrecognized within current incentive structures. Abbott[50] has argued that disciplinary boundaries undergo cyclical construction and dissolution, and resistance to cross-domain contributions is a recurring institutional pattern. Our evidence suggests that the current shift may be more sustained, supported by digital infrastructure designed to progressively lower the cost of knowledge compression and modular reuse. Generalizing works reconfigure the space of scientific attention gradually, without any single contribution claiming outsized credit. This makes their cumulative importance difficult to detect under evaluation systems calibrated to within-field impact. As the impact of scientific work becomes less confined within disciplines, existing evaluation systems including peer review and departmental promotion have been slow to adapt[51], and the scientific enterprise risks systematically undervaluing contributions essential to its continued progress.

The structural shift toward generalizing work resonates with research in cultural evolution on the mechanisms of cumulative knowledge. Theories of cumulative culture emphasize that sustained progress depends on the capacity to compress complex procedures into transmissible, modular representations, ratcheting cultural complexity through selection[52] and high-fidelity transmission[53]. Our findings suggest that the scientific community has been engaging in just this type of large-scale knowledge compression. This parallels how biological evolution produces increasingly sophisticated mechanisms for processing and integrating information across scales[54], and how artificial intelligence systems achieve their most impressive capabilities through learning compressed representations from diverse data[55]. The attention mechanism that revolutionized AI, which our analysis identifies as a paradigmatic generalization, exemplifies this compression process: a pattern identified in one domain that, once abstracted and modularized, proved transformative across computer vision[56], biology[57], and beyond[58]. This is the logic of recombinant innovation operating through knowledge compression, and our evidence suggests it has become an increasingly important mode of scientific advance at the system level[59,60].

Our study makes a methodological contribution by introducing a novel framework for decomposing scientific contribution into three distinct functional types. This framework not only enables a more nuanced characterization of innovation dynamics but also provides a quantitative basis for informing science policy and institutional design. For example, to characterize the global scientific supply chain, we compute the average foundation, extension, and generalization indices for papers across countries. The resulting landscape reveals distinct national specializations. The United States and the United Kingdom lead in the production of foundational works, whereas France and Germany excel in generating important refinements and extensions of existing foundations. In contrast, since the early 2000s, rapidly developing countries such as China and India have emerged as leading producers of generalizing works, which create value by compressing, synthesizing, and disseminating existing innovations, thereby facilitating broader downstream innovation and underscoring the importance of cross-domain synthesis in their scientific ascent[12,61] (Extended Data Fig. 10c). This pattern mirrors these countries' comparative strengths in industrial innovation, where value creation increasingly derives from the integration, adaptation, and scaling of existing technologies rather than from the invention of entirely new technological paradigms[62–64] (Extended Data Fig. 10b). These findings suggest how the proposed framework may uncover complementary national roles within the global innovation ecosystem, providing an empirical basis for designing differentiated research funding strategies, fostering international collaboration, and evaluating scientific contributions beyond conventional measures of novelty.

At the same time, our proposed indices provide an effective decomposition of the widely used disruption index[22,29]. We show that much of what the disruption index captures may reflect the extent to which a paper is used in a generalized, cross-domain manner, rather than its role in establishing a new scientific foundation. This reinterpretation invites a reframing of prior conclusions, including claims about the decline of disruptive science[39], the role of team size[22] in innovation, and the association between scientific aging and creativity[65]. For example, we find that tiny teams (e.g., of 1-2 members) are particularly effective at producing highly generalizable work, whereas small to medium-sized teams (approximately four members) are most likely to make foundational contributions (Extended Data Fig. 10a). We also observe a weak positive association between scientists' academic age and the foundation index of their publications, while a strong negative one for generalization[65]. These results underscore the importance of distinguishing between different modes of innovation when evaluating scientific progress and designing research systems.

Our study has natural limitations. First, our characterization of research as forming scientific foundations, extensions, and generalizations is necessarily coarse-grained, leaving heterogeneity within each category. For example, generalization papers cited in Methods sections differ from those cited in Discussion sections in both content and contribution. Second, our indices are computed from citation networks whose coverage varies across time, field, and database. Although our null models, database-augmentation, and zero-reference analyses indicate the central patterns are robust, citation-based indices should be compared across years and domains with care. Third, the evidence linking information technology to the transition toward system innovation remains suggestive rather than causal. Our quasi-experimental analyses of online access and LLM adoption rely on non-random treatment assignment and indirectly inferred treatment status, and the longitudinal correspondence between the rise in generalization and the emergence of the Internet is sensitive to sample inclusion criteria. Therefore, they only provide suggestive evidence consistent with this hypothesis and motivate further causal investigation. Finally, we study science alone, leaving extension to other domains of innovative production in technology and culture for future work.

Despite these limitations, our findings invite a reassessment of what it means for science to make progress. The generation of new knowledge has always been essential. Our evidence suggests that its compression and redistribution may be equally so. Science advances not only by seeing further within a field but by making what has already been seen legible, portable, and recombinant across the system. The shift has been tectonic, with fields drifting together beneath the surface over decades, too gradual for any single measure to register, yet fundamentally redrawing the geography of knowledge. The most important scientific contributions of the future may be recognized not by the fields they forge but by the boundaries they break.

## Methods

**Index definitions**. For each citation received by a focal paper, we count the number of the focal paper's references and the number of its other citing papers that are co-cited by the citing paper. A citation is classified as foundational if more citing papers than references are co-cited, extensional if more references than citing papers are co-cited, and generalizational if neither references nor citing papers are co-cited; ties are handled as described in the Supplementary Information (Supplementary Methods Section 2), which also defines the continuous variants used in robustness analyses (Extended Data Fig. 6e,l). A paper's F, E, and G indices are the proportions of its citations in each class, computed within a fixed post-publication citation window (five years in the main analyses; one year for the longitudinal analyses in Fig. 3).

**Data**. We analyze citation networks from OpenAlex (main analyses; 2025 and 2026 releases as noted), the Web of Science (replication), Semantic Scholar (full-text citation contexts and network augmentation), and Microsoft Academic Graph (replication of prior team-size and age results). Citation-purpose annotations come from ACT2[30] and Jurgens et al[31]. Online-availability timing comes from Fulltext Sources Online; LLM adoption from Kusumegi et al[38]; awards and honors from the Nobel laureate and major science award datasets described in the Supplementary Information. Unless noted, analyses include papers with at least one recorded reference and at least five citations within five years of publication; robustness to these filters, including retention of zero-reference papers, is reported in Extended Data Fig. 6 and Figs. S57–S64.

**Null models**. Randomized citation networks preserve degree distributions within publication year and venue (or domain at the topic, subfield, and field levels), as described in the Results and in the Supplementary Information.

**Statistics**. Two-sided Welch's *t*-tests unless otherwise noted; given the very large samples, we emphasize effect sizes and confidence intervals alongside p-values. Stacked DiD specifications, event-study estimates, clustering, and sensitivity analyses are described in the Supplementary Information.

**Preregistration**. This study was not preregistered. The foundation–extension–generalization typology was developed inductively from examination of citation-network structure, informed by established citation-based measures including the disruption index.

**Data and code availability**. Replication code and data are available at [repository URL], with documentation describing the data structure and scripts required to reproduce each figure. Raw data underlying analyses that rely on proprietary sources (*Web of Science*, *Fulltext Sources Online*) are available upon reasonable request, subject to applicable data-use restrictions.

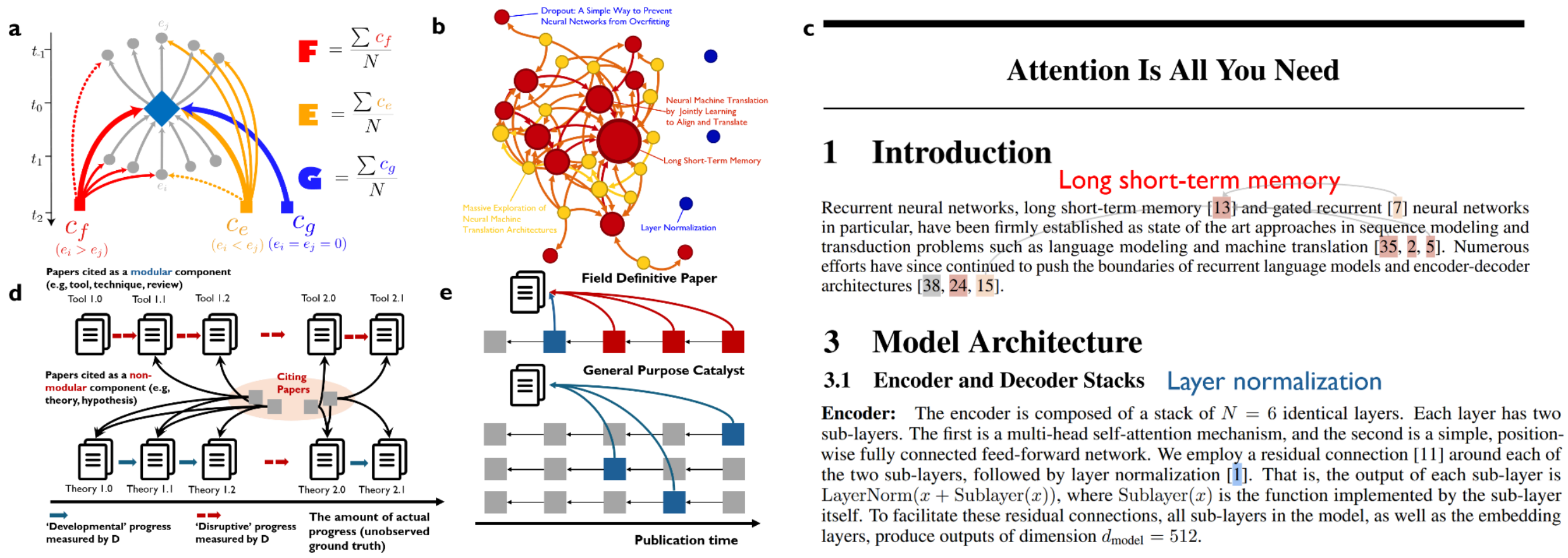

**Fig. 1 | Illustration of the foundation (F), extension (E), and generalization (G) index. a**, Conceptual illustration of the F, E, and G indices. Citations of a focal paper (blue diamond) can be (1) *Foundational* (red), in which the citing paper references more papers that cite the contribution (solid red edge) than the focal paper's references (dotted red edge), treating it as a foundational work; (2) *Extensional* (orange), in which the citing paper references more prior work the focal paper cites (solid orange edge) than those that reference it (dotted orange edge), treating it as an extension of other foundational work; and (3) *Generalizational* (blue), in which the citing paper does not reference any of the focal paper's references or citations, treating it like a generalized tool or background. The F, E, and G indices of a focal paper are defined as the proportions of its subsequent citations belonging to each category. **b**, An illustration of the citation network among all references of the *Attention Is All You Need* (2017) paper, where *Long Short-Term Memory* (1997, one of the previous best models in natural language processing that *Attention* largely replaced) and *Neural Machine Translation by Jointly Learning to Align and Translate* (2014, that first proposed the "Attention" mechanism) are cited by most articles referenced in *Attention* and serve as major foundations, while *Massive Exploration of Neural Machine Translation Architectures* (2017) cites the most other papers, as it systematically analyzes hyperparameter choices for neural machine translation models and provides guidance for subsequent model design. The *Layer Normalization* (2016) and *Dropout* (2014) papers largely did not cite or were cited in the reference list and contributed techniques to *Attention*. **c,** In-text references cited in the *Attention* paper. Node colors indicate the role of each cited work: foundational (red), extensional (orange), and generalizational (blue). References not captured in OpenAlex in gray. Directed gray edges denote references that also cite the LSTM paper (Reference 13). **d**, Diagram illustrating the *modular* nature of generalization contributions and their relationship to the disruption index. The squares denote works that cite the focal work and *paper emojis* represent the focal works. Horizontal distance between articles represents intellectual distance between adjacent publications. For works typically cited as objects of comparison, evaluation, or extension (e.g., theories or hypotheses), authors tend to discuss them together with their predecessors unless the newer work departs substantially from prior work. Consequently, whether a paper is cited independently of its predecessors serves as a proxy for the intellectual distance between the work and the research on which it builds. In contrast, modular knowledge components (e.g., software, methods, or algorithms) are often cited solely for their use, with their predecessors omitted regardless of the degree of intellectual advancement they represent. As a result, both the disruption index and the proposed generalization index favor modular works cited primarily for use rather than comparison, evaluation, or extension. **e,** Diagram illustrating foundational contributions as *field-defining* advances and generalization contributions as *catalysts* for scientific progress. Squares denote works that cite the focal work and the *paper emoji* represents the focal work. Publications within the same research topic tend to form densely connected citation networks. When a focal work exerts profound influence on a topic, it is cited by many downstream publications, making their citations predominantly foundational as these works cite one another and other related studies building upon the focal contribution (upper figure). In contrast, a focal work accumulates many

generalization citations only when it is cited by publications largely disconnected in the citation network, indicating influence across independent research trajectories (lower figure).

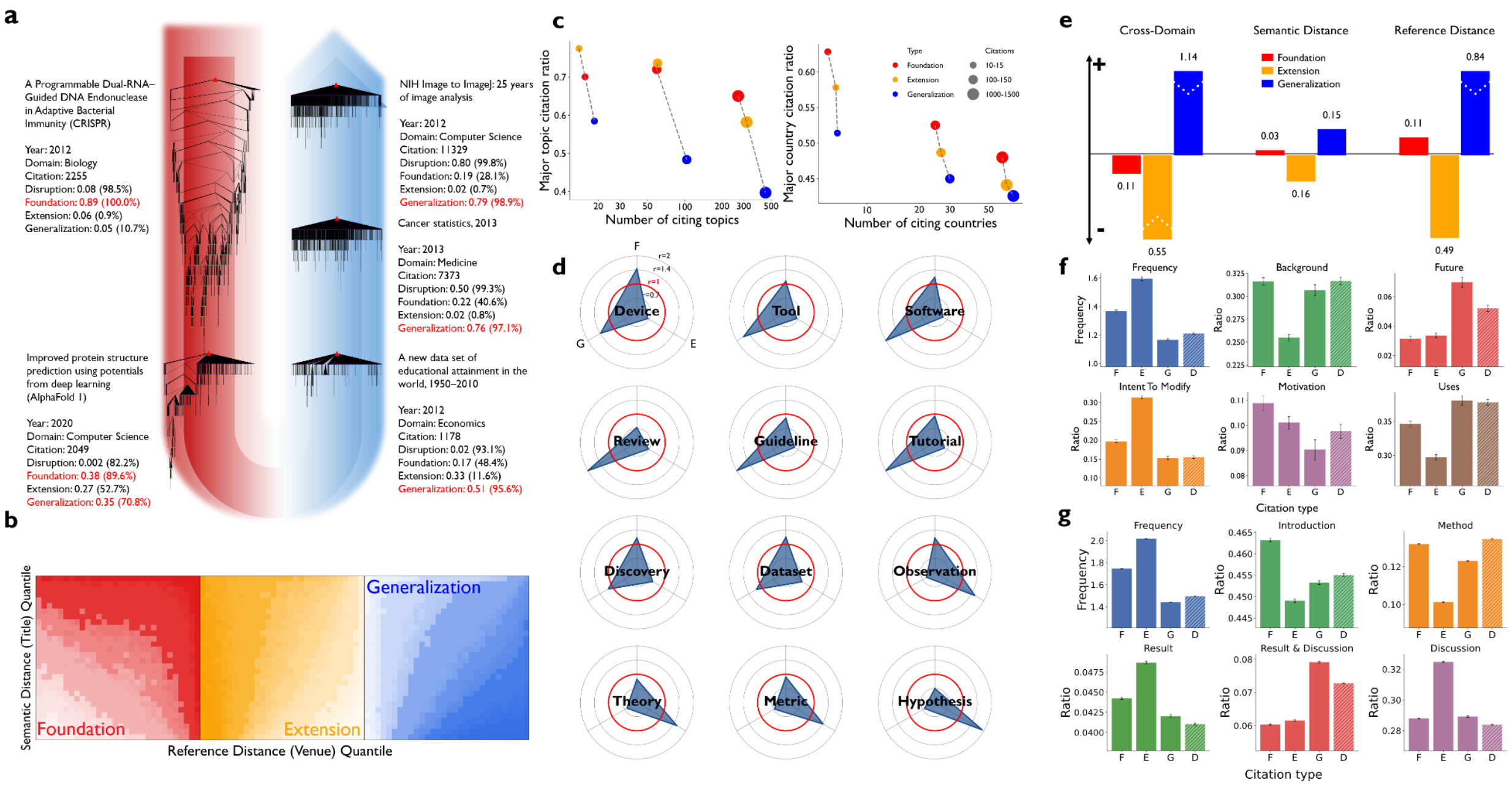

**Fig. 2 | Characterizing the foundation (F), extension (E), and generalization (G) indices. a,** examples of five "highly-disruptive" papers (measured by disruption index), ranging from the most foundational paper *A Programmable Dual-RNA–Guided DNA Endonuclease in Adaptive Bacterial Immunity*[66] (CRISPR, Nobel Prize in Chemistry 2020) towards the most generalizational, *NIH Image to ImageJ: 25 years of image analysis* (open source tools for analyzing scientific images). The value in parentheses denotes the quantile of the corresponding index for a given paper relative to all papers published in the same year and research domain that contain at least one reference and received at least 100 citations within five years of publication. Deep, narrow downstream citation patterns indicate that the focal work continues to be cited through generations of subsequent advances within the same research trajectory[67]. In contrast, shallow but broad downstream citation patterns indicate that the focal work is cited by many publications not connected within the citation network, and citations within any individual research trajectory diminish after only a few generations of subsequent progress. **b,** Relationship between the F, E, and G indices, within-paper reference, and semantic distance. It presents the average F (left), E (middle), and G (right) indices for papers stratified by reference distance and semantic distance, using 22,674,571 publications from *OpenAlex* spanning from 1951 to 2019 which have at least five citations within five years post-publication, and at least one reference with publication venue identified and two valid English tokens in their titles. The estimates control for publication year, reference count, and five-year citation count. Darker shading indicates higher values of the respective F, E, or G indices. The observed patterns are consistent with those reported in Panel a and are illustrated in Fig. S16. **c,** Citation diffusion of F, E, and G papers across domains and countries, with the top 10% in each category selected. In the left panel, we see that based on publications in *OpenAlex* with a primary topic, published 1945-2020, and containing at least one reference, F papers are cited most within their dominant field, but G papers the furthest in topic space. The $x$-axis represents the number of *OpenAlex* topics (from the lowest level of the hierarchy, comprising 4,516) from which at least one publication cites the selected papers, and the $y$-axis the proportion of citations originating from the single topic contributing the largest share of citations to the selected papers. In the right panel, the $x$-axis represents the number of countries from which at least one publication with an affiliated author cites the selected papers, whereas the $y$-axis represents the proportion of citations originating from

the country contributing the largest share of citations to the selected papers. Again, F papers are cited most within their dominant country, but G papers the furthest across countries. **d,** Word usage in paper titles, drawing from 23,448,431 publications in *OpenAlex* that have at least one reference and five citations within five years post-publication, published between 1945 and 2019. Papers are divided into two equal-sized groups based on their F, E and G indices. For a set of selected words, we compute the ratio of their occurrence in the upper half relative to the lower half. Values greater than 1 indicate higher prevalence in titles of papers with above-median index values. **e**, Relationship between F, E, and G citation types with alternative measures of interdisciplinarity, computed with citations from 159,809,661 publications in *OpenAlex*, published between 1951 and 2019. Each citation link (citing paper → cited, focal paper) is classified into one of the three citation types. For each group, we compute the difference between the average metric value within the group ($M$) and that of all remaining citations ($\overline{M}$): $\boldsymbol{\Delta} = \frac{M-\overline{M}}{\overline{M}}$. Positive values (bars above zero) indicate that citations of the given type occur at greater distances or have higher cross-domain ratios relative to the complement set, while negative values (bars below zero) indicate the opposite. **e**, Citation purposes across the four citation-context categories among natural language processing papers: F, E, G, and D (disruptive). Frequency denotes the average number of reference instances (i.e., mentions of a reference in the full text) per citation link (citing–cited paper pair), whereas ratio denotes the distribution of citation purposes across reference instances for citations in the four categories. Citation purposes were identified using a machine learning classifier that leverages the textual context surrounding each citation together with bibliographic features. The classifier was trained on the publicly available NLP citation-purpose dataset[31], comprising 83,494 citation links and 118,187 citation instances. The analysis was restricted to citation links for which both citing and cited papers could be matched to *OpenAlex* records and citation-purpose annotations were available. For simplicity, the *CompareOrContrast* and *Extends* categories were merged into a single *Intent to Modify* category; results for the original categories are presented in Fig. S17-S18. Results based on the ACT2 dataset[30], which adopts a similar taxonomy of citation purposes and provides machine-learning-based citation-purpose labels generated by a classifier trained on manually annotated citation instances spanning multiple disciplines. **f,** Location of references across the four citation-context categories. Frequency denotes the average number of reference instances (i.e., mentions of a reference in the full text) per citation link, and ratio denotes the distribution of reference locations within the full text. The analysis is based on 3,192,732 full-text papers from *Semantic Scholar* linked to *OpenAlex* via DOI and conforms to the IMRaD structure (Introduction, Methods, Results, and Discussion) based on section titles. After applying these selection criteria, the dataset comprises 15,260,936 citation links, corresponding to 26,910,728 citation instances. Citation locations were determined by identifying the IMRaD section in which each reference instance appears, and their distributions were compared across citations classified as foundational (F), extensional (E), generalizational (G), or disruptive (D).

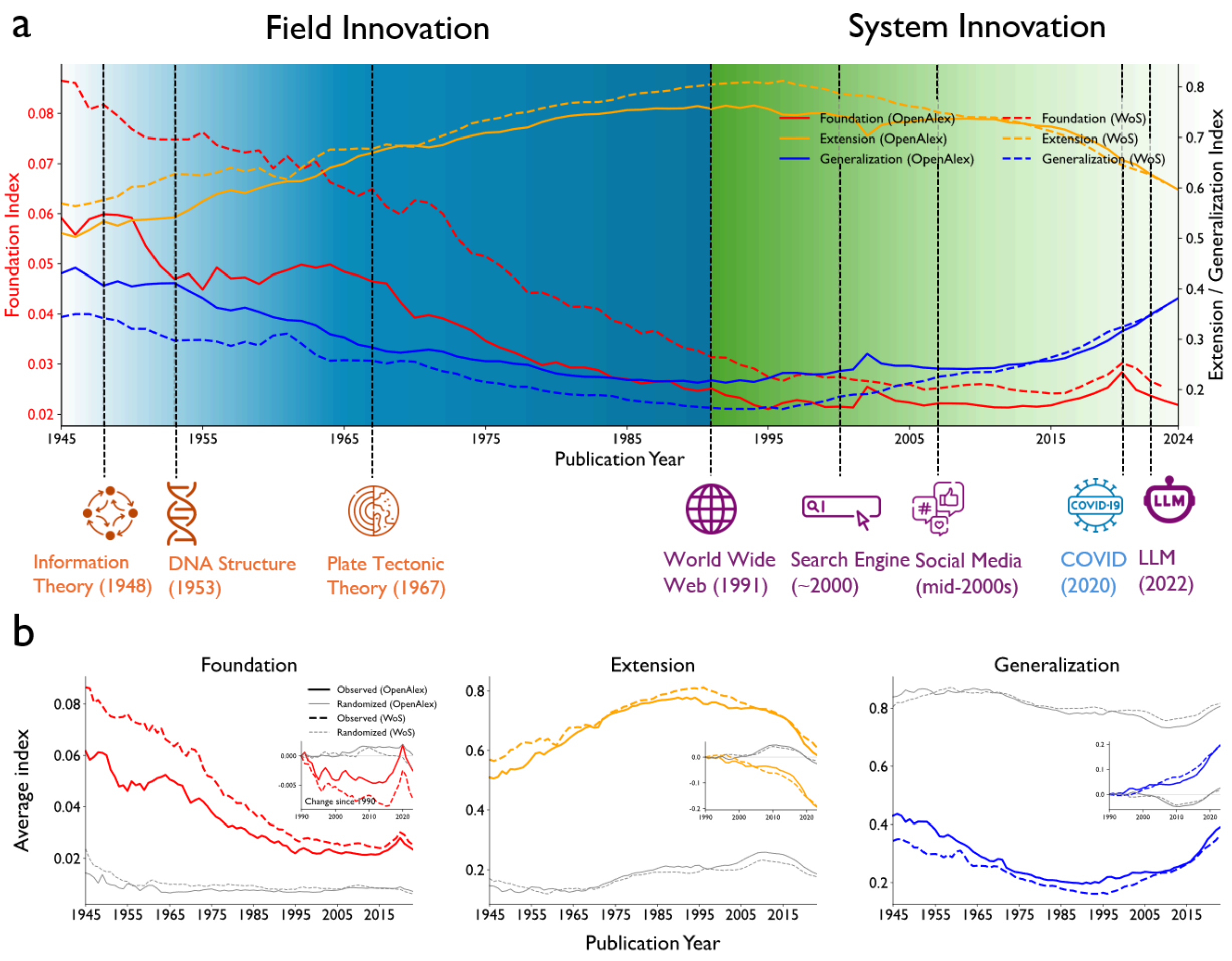

**Fig. 3 | Longitudinal patterns of field innovation (before 1990) and system innovation (after 1990). a**, Temporal evolution of the average foundation (F), extension (E), and generalization (G) indices for publications in the *Web of Science (WoS)* and *OpenAlex* datasets (2026 version). Indices calculated using citations received within one year of publication, restricting the analysis to papers with at least one reference and two citations during the first year post-publication. After filtering, the *WoS* dataset contains 16,977,067 valid articles between 1945 and 2023, whereas the *OpenAlex* dataset contains 28,128,459 valid articles between 1945 and 2024.

**b,** Longitudinal trajectories of the F, E, and G indices for papers in randomized citation networks. Curves show the corresponding trends in citation networks generated by randomly reshuffling citation links while approximately preserving the degree centrality of both citing and cited papers within each publication year and publication venue. Papers without a valid publication venue were excluded from the network analyses. The F, E, and G indices were computed using only citations accrued within the first year after publication and were restricted to papers published between 1945 and 2024 that received at least two citations and contained at least one reference within first year of publication, yielding 19,642,763 papers in *OpenAlex* and 16,959,643 papers in *WoS*. Results obtained using alternative randomization strategies, citation thresholds, and citation windows are presented in Fig. S14. The inset shows the relative changes in F, E, and G indices for the observed and randomized networks following the post-1990 transition toward generalization. The colored line represents the trend observed in the empirical network, and the gray lines the corresponding trends in the randomized networks.

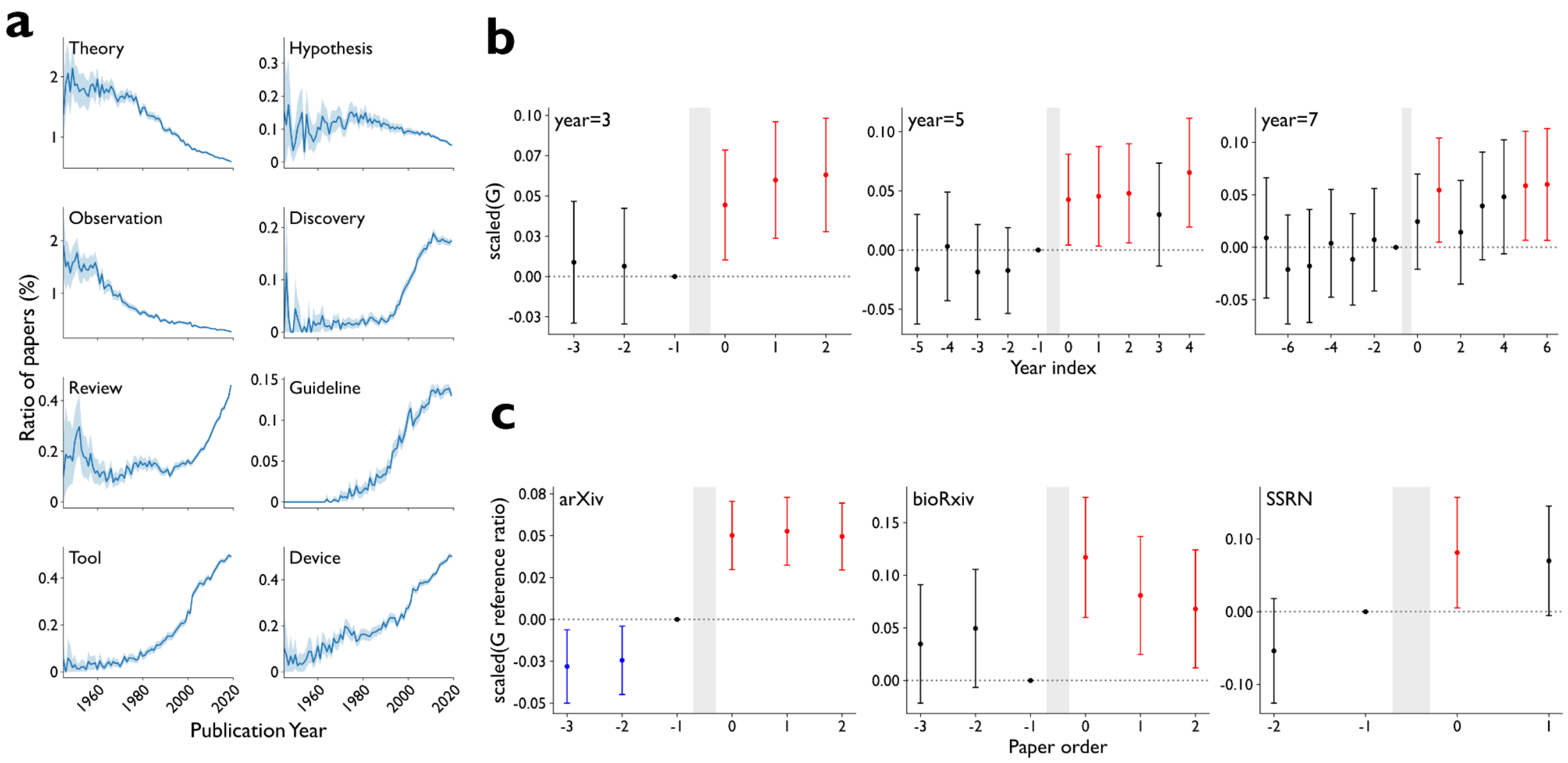

**Fig. 4 | Possible mechanisms underlying the decline in foundation (F) and extension (E) indices and the rise in the generalization (G) index. a,** Percentage of papers published in each year whose titles contain a given keyword, after converting both keywords and paper titles to lowercase. Percentages were calculated among 22,996,553 *OpenAlex* papers with English titles that contain at least one reference, received at least five citations within five years of publication, and were published before 2020. **b**, Publication venues' online coverage and its estimated influence on the G index. This investigates the effect of publication venues' adoption of online dissemination on the $G_5$ index, defined as the proportion of a paper's citations classified as generalizing within five years of publication. Specifically, we compare the average $G_5$ index of papers published in a given venue before and after the year the venue introduced online access to its publications, as identified in the *Fulltext Sources Online* dataset. The treatment group, comprising venues that adopt online dissemination, is contrasted with a control group of venues that remain offline throughout the observation period (there must be at least two papers published every year in the observation period for both treatment and control venues). To assess the robustness of our findings, we report results across multiple observation window lengths. There are 3,592 treatment venues with a valid control available when studying the intervention effect 3-year post-treatment (left figure), 3,020 treatment venues for year = 5 (middle figure), and 2,590 treatment venues for year = 7 (right figure). In the corresponding figures, red lines denote estimates statistically significantly greater than zero, whereas blue lines indicate estimates that are statistically significantly less than zero. Extended Data Fig. 9a presents the effect with alternative citation distance outcomes. **c**, Authors' adoption of LLMs in manuscript preparation and its influence on G citations. We compare the proportion of references classified as generalizational in papers written before and after an author adopts LLM-based editing tools. This is evaluated against a control group of authors who do not adopt LLMs by the end of the observation period. The analysis is conducted using papers from *arXiv*, *bioRxiv*, and *SSRN*, and all authors must have a valid number (at least three papers for *arXiv* and *bioRxiv* datasets and two papers for the *SSRN* dataset) of publications before and after estimated LLM adoption. There are 65,072 valid authors in the treatment group forthe *arXiv* dataset (left figure), 23,369 authors for the *bioRxiv* dataset, and 4,602 authors for the *SSRN* dataset. Similarly in all figures, red lines denote estimates statistically significantly greater than zero. Extended Data Fig. 9b-f presents the effect with alternative LLM adoption thresholds and citation distance outcomes.

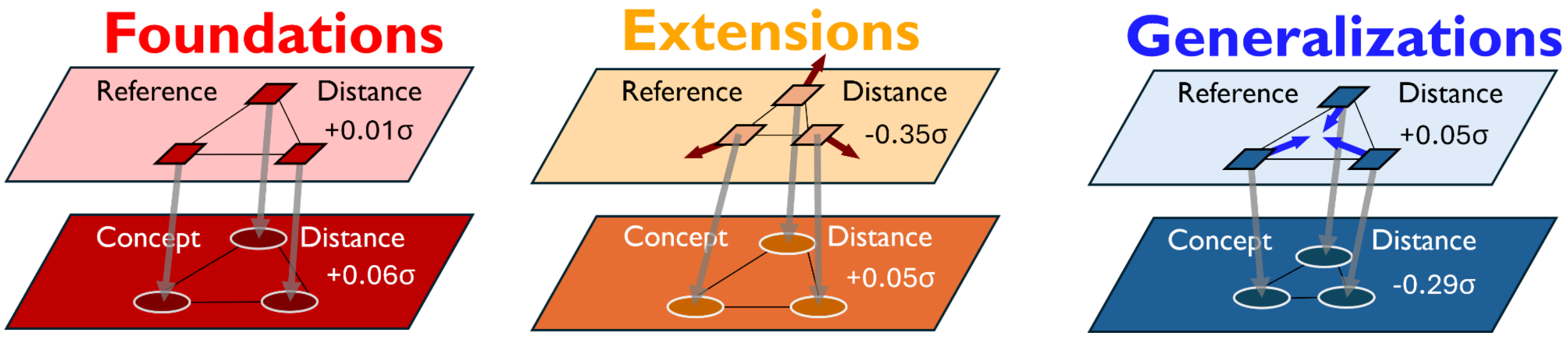

**Extended Data Fig. 1 | Relationship between the F, E, and G indices and within-paper reference and semantic distance.** This figure illustrates the relationship between reference distance and semantic distance in titles across foundation (F), extension (E), and generalization (G) papers. Using a sample of 1,634,576 papers indexed in OpenAlex and published in 2019—each with at least one reference and a minimum of five citations within five years post-publication—we construct an embedding space based on the co-occurrence of referencing venues and title tokens. This enables a visual representation of how reference distance and semantic distance vary relative to the overall average across all papers.

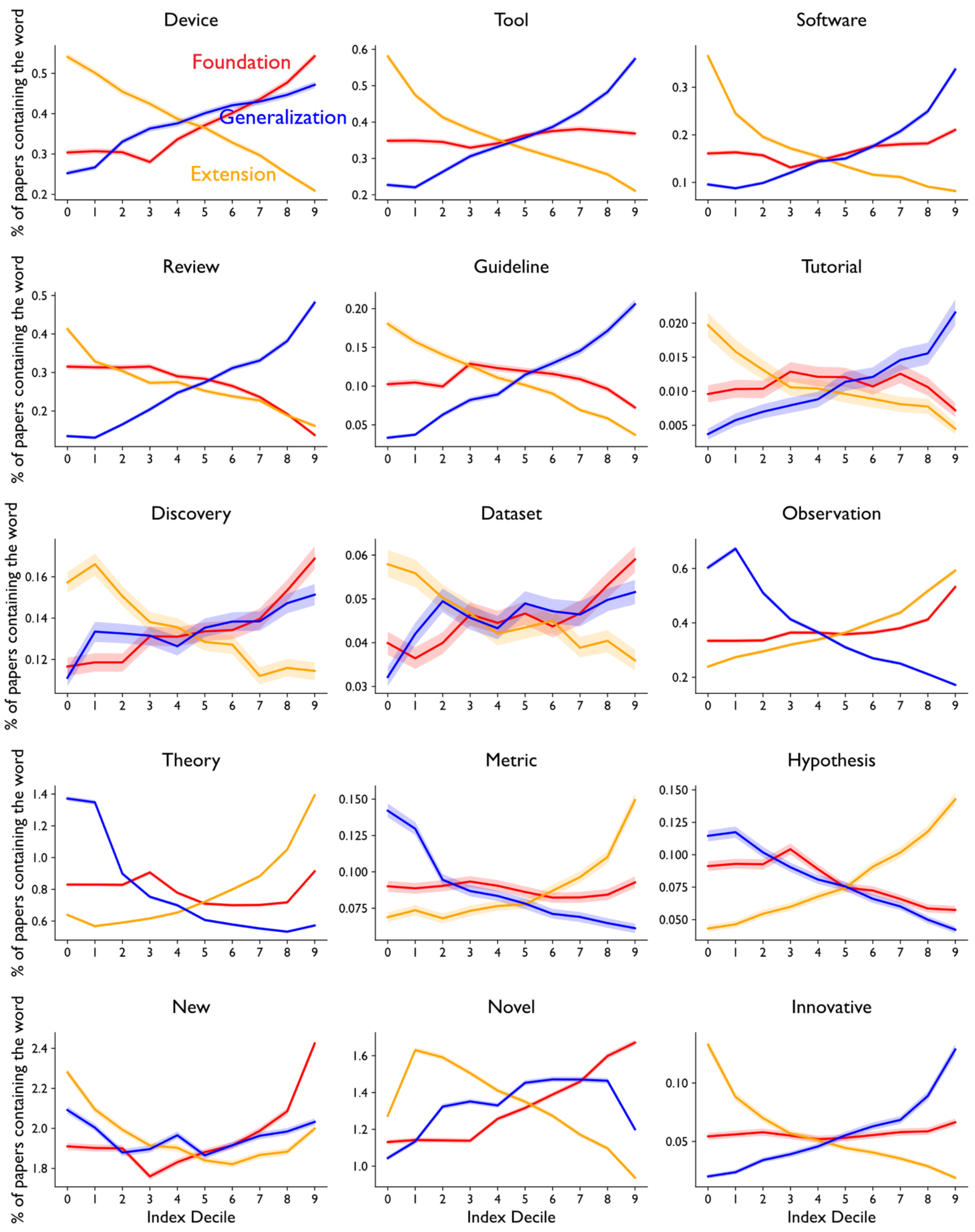

**Extended Data Fig. 2 | Percentage of appearance for given keywords in the titles of papers.** We divide all papers into ten equal-sized bins based on their F (red), E (yellow) and G (blue) indices, and present the percentage of papers in each decile that contain the corresponding keywords in their titles. The shaded area represents the 95% confidence interval obtained by bootstrapping; the plots are drawn based on 23,448,431 papers in OpenAlex that have at least one reference and received at least five citations within five years of publication, and published between 1951 and 2019.

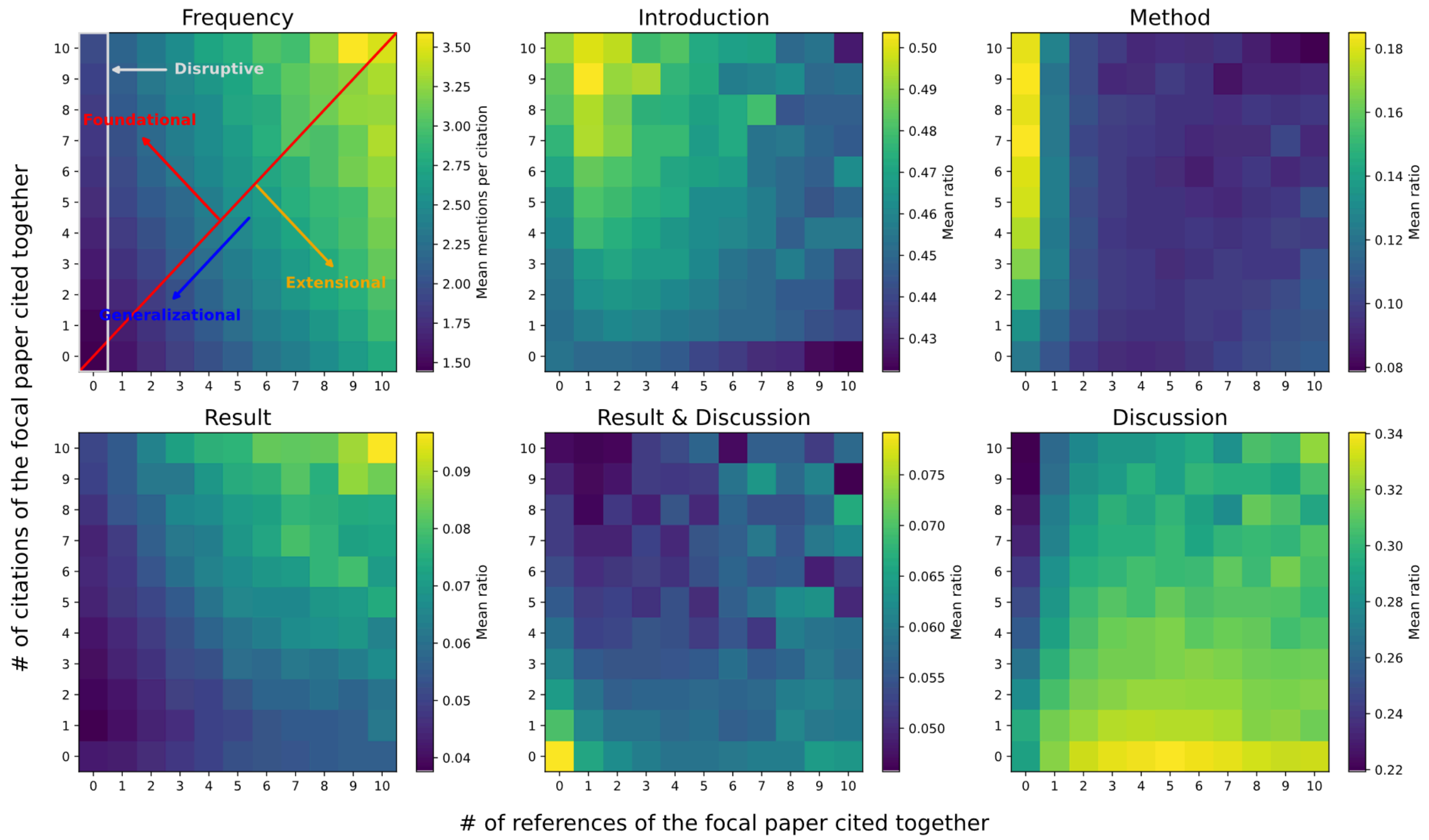

**Extended Data Fig. 3 | Distribution of reference locations according to the number of co-cited references and citations of the focal paper.** Citation links are grouped by the number of the focal paper's references and citations that are co-cited when the focal paper is cited. For each group, we report the distribution of citation instances (i.e., individual mentions of the focal paper within the text of a citing article) across different sections of the citing paper, together with the number of citation instances associated with each citation link. This figure is based on the same sample as Fig. 2g, comprising 14,702,790 citation links and their corresponding 25,181,703 citation instances. Citation links in which the focal paper is co-cited with more than ten of its references or more than ten of its citations were excluded.

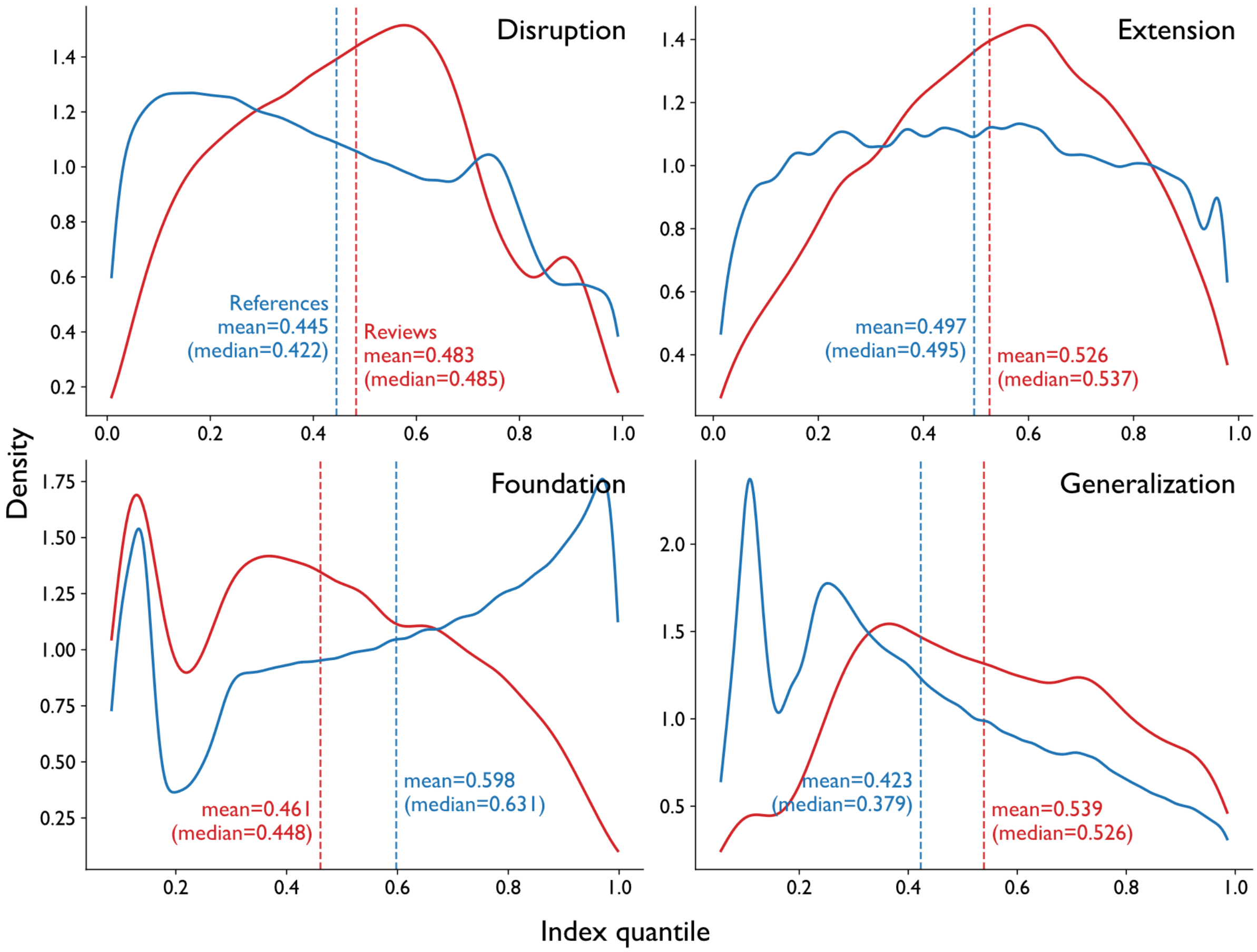

**Extended Data Fig. 4 | Foundation and generalization indices distinguish review articles from their references.** Review articles are identified in the Web of Science dataset as papers published in venues whose titles contain both the keywords "annual" and "review", following the approach of Wu et al. [22]. We compare the average disruption ($\frac{I-J}{I+J+K}$) and the proposed foundation ($\frac{F}{C}$), extension ($\frac{E}{C}$), and generalization ($\frac{G}{C}$) indices between review articles and the papers they cite (hereafter, references), which are presumed to represent influential prior contributions within their respective fields. The analysis includes 18,367 review articles and their 1,319,022 references that are classified as *review* or *journal* articles, were published before 2020, have an identified research domain, contain at least one reference, and received at least five citations within five years of publication. The D, F, E, and G indices are calculated using citations received during the same five-year post-publication window. To facilitate comparisons across publication years and research domains, all indices are converted to quantiles based on publication year and domain, such that each paper is ranked relative to contemporaneous publications within the same field. Our results differ from those reported by Wu et al. [22], and we explain the discrepancy in the SI Supplementary Results Section 21.

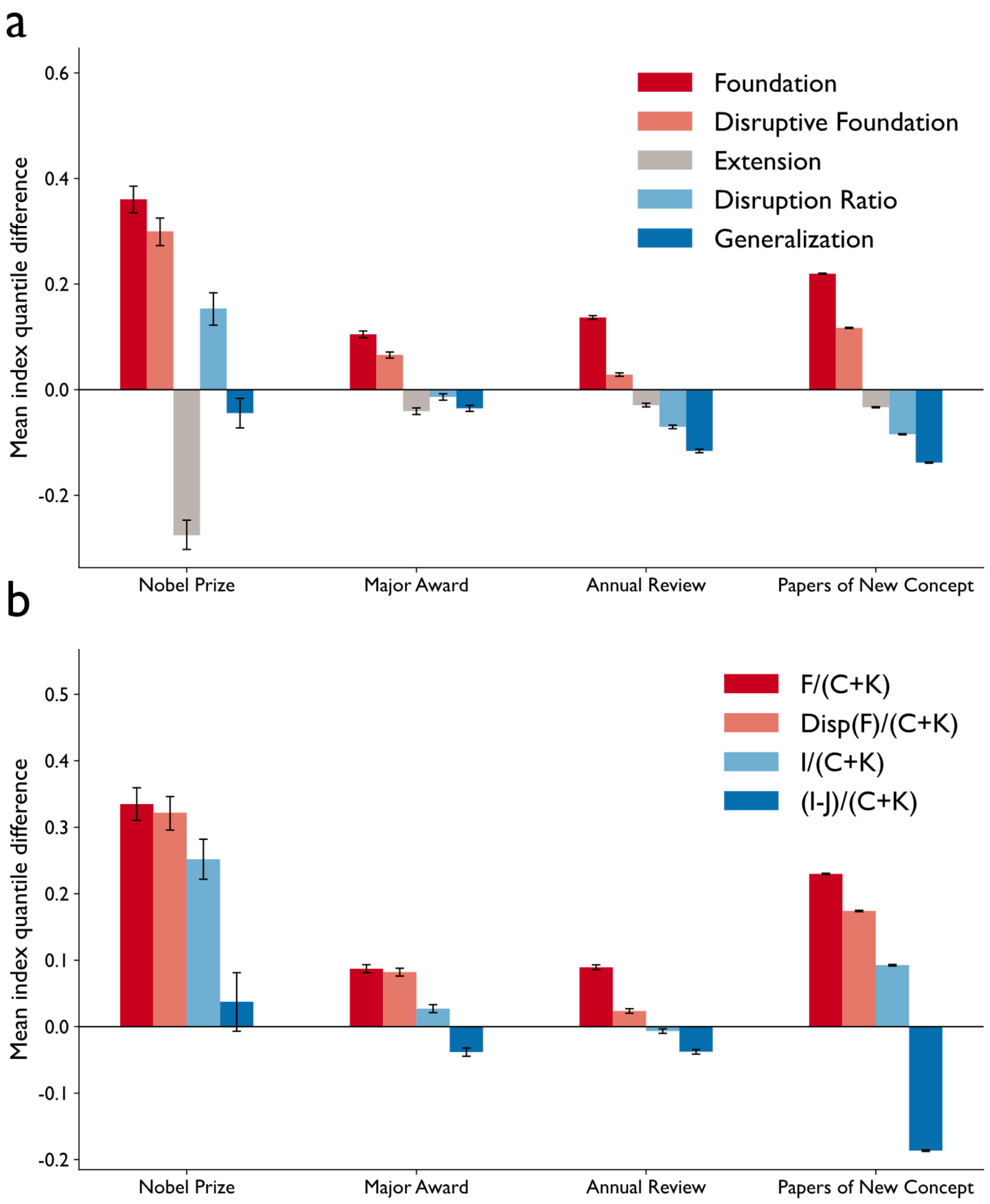

**Extended Data Fig. 5 | Comparison of the ability of indices based on foundational, disruptive foundation, extensional, disruptive, and generalizational citations to distinguish desired from undesired papers.** We construct four pairs of comparison groups: Nobel Prize–winning papers versus other papers, major award–winning papers versus other papers, references cited by review articles versus the review articles themselves, and papers introducing new concepts versus other papers. In each pair, the former group represents the desired papers, whereas the latter represents the undesired papers. For each comparison, we compute indices based on foundational, disruptive foundation, extensional (only in Panel a), disruptive, and generalizational citations and evaluate their ability to distinguish the two groups by measuring the difference in their quantile rankings, with quantile normalized by the year and domain of publication. A larger quantile difference indicates greater discriminative ability that favors the desired papers. This figure summarizes the results across all comparison groups and indices; the detailed results for each comparison group are presented in Figs. S35-38.

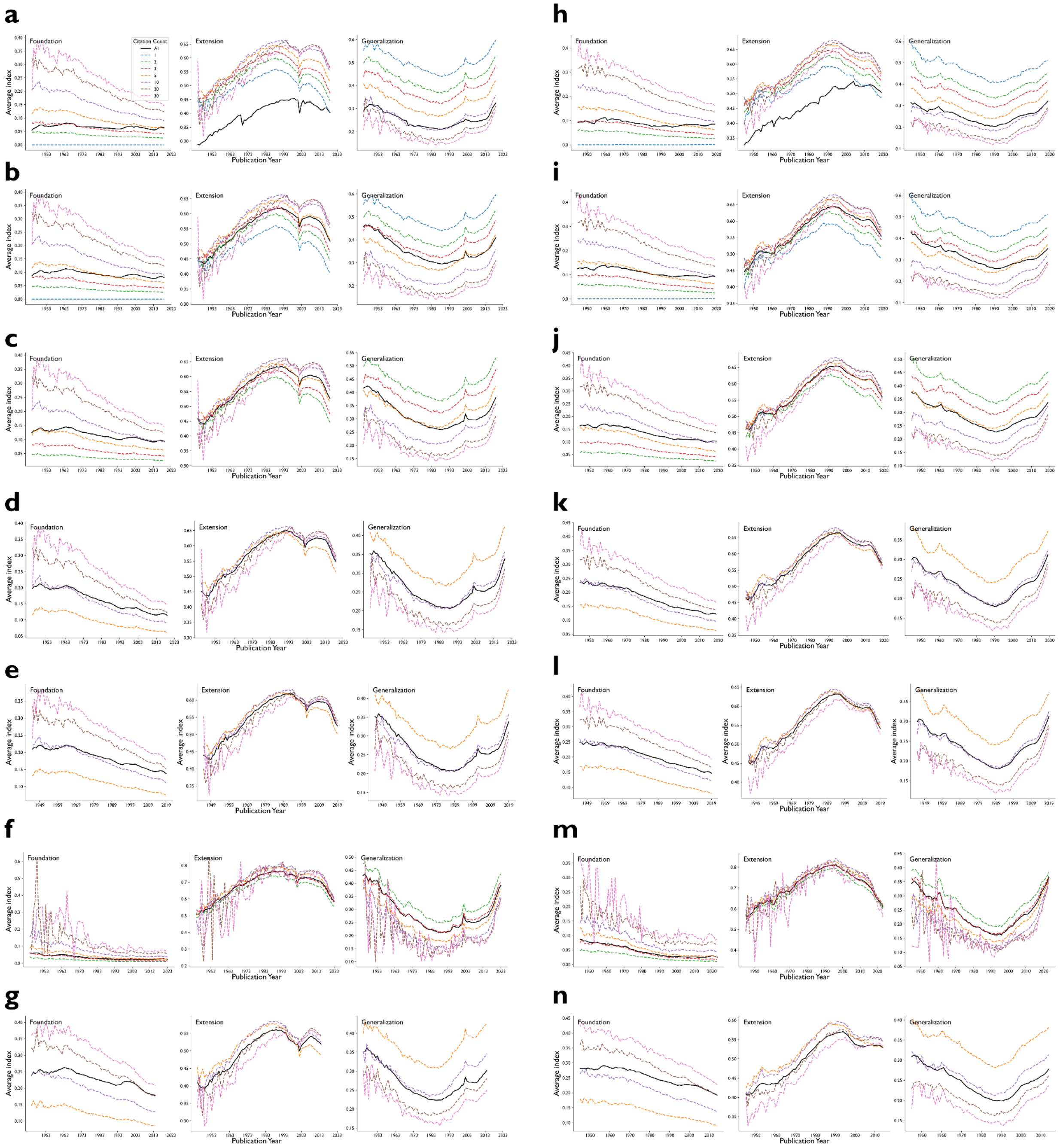

**Extended Data Fig. 6 | Longitudinal change of F, E, G indices across years, with different paper filters and computation specifications.** The plots are computed based on: **a**, Papers in OpenAlex with at least one reference, zero citations within one year of publication, and published before 2024 (the F, E, G indices are set to zero for papers with zero citations, 72,218,881 papers in total). **b**, Papers in OpenAlex with at least one reference, one citation within five years of publication, and published before 2020 (41,404,271 papers). **c**, Papers in OpenAlex with at least one reference, two citations within five years of publication, and published before 2020 (34,815,384 papers). **d**, Papers in OpenAlex with at least one reference, five citations within five years of publication, and published before 2020 (23,448,431 papers). **e**, Papers in OpenAlex with at least one reference, five citations within five years of publication, and published before 2020, computed with the continuous version of F, E, G indices, as specified in

the SI Supplementary Results Section 2 (23,448,431 papers). **f**, Papers in OpenAlex with at least one reference, two citations within one year of publication, and published before 2024 (22,200,492 papers). **g**, Papers in OpenAlex with at least one reference, five citations within ten years of publication, and published before 2015 (21,183,902 papers). **h**, Papers in WoS with at least one reference, zero citations within one year of publication, and published before 2024 (the F, E, G index are set to zero for papers with zero citation, 47,668,611 papers in total). **i**, Papers in WoS with at least one reference, one citation within five years of publication, and published before 2020 (32,318,645 papers). **j**, Papers in WoS with at least one reference, two citations within five years of publication, and published before 2020 (27,937,815 papers). **k**, Papers in WoS with at least one reference, five citations within five years of publication, and published before 2020 (19,026,615 papers). **l,** Papers in WoS with at least one reference, five citations within five years of publication, and published before 2020, computed with the continuous version of F, E, G indices, as specified in the SI Supplementary Results Section 2 (19,026,615 papers). **m**, Papers in WoS with at least one reference, two citations within one year of publication, and published before 2024 (17,032,355 papers). **n**, Papers in WoS with at least one reference, five citations within ten years of publication, and published before 2015 (17,805,346 papers).

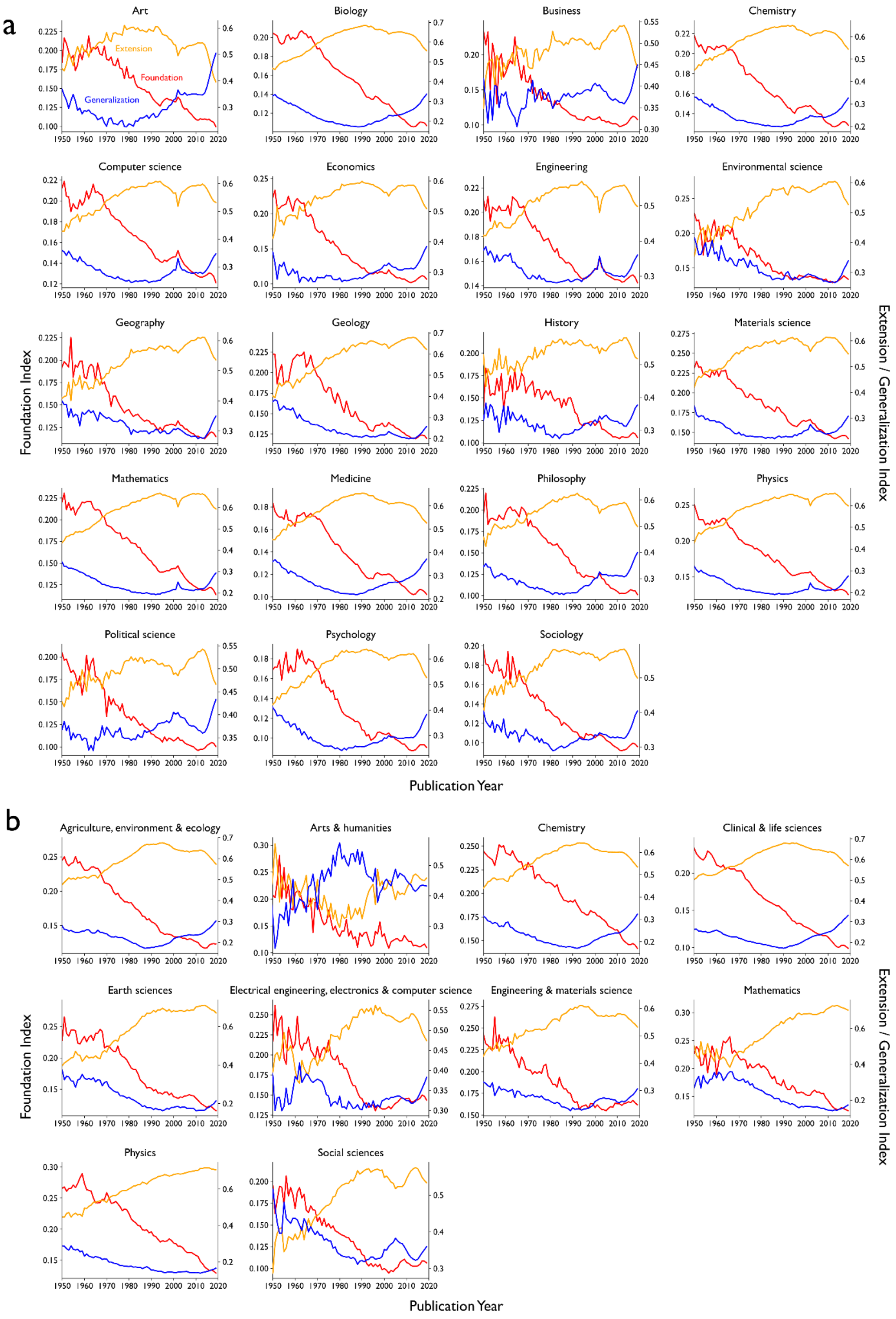

**Extended Data Fig. 7 | Longitudinal change of F, E, G indices for papers across domains in OpenAlex and WoS. a,** This figure presents the longitudinal change of F (red), E (yellow), G (blue) indices for papers in different domains. They are drawn based on 23,448,431 papers in *OpenAlex* with at least one reference and five citations within five years of publication, and that were published between 1945 and 2019. **b,** This figure presents the longitudinal change of F (red), E (yellow), G (blue) indices for papers in different domains. They are drawn based

on 19,026,615 papers in WoS with at least one reference and five citations within five years of publication, have a valid domain, and were published between 1945 and 2019.

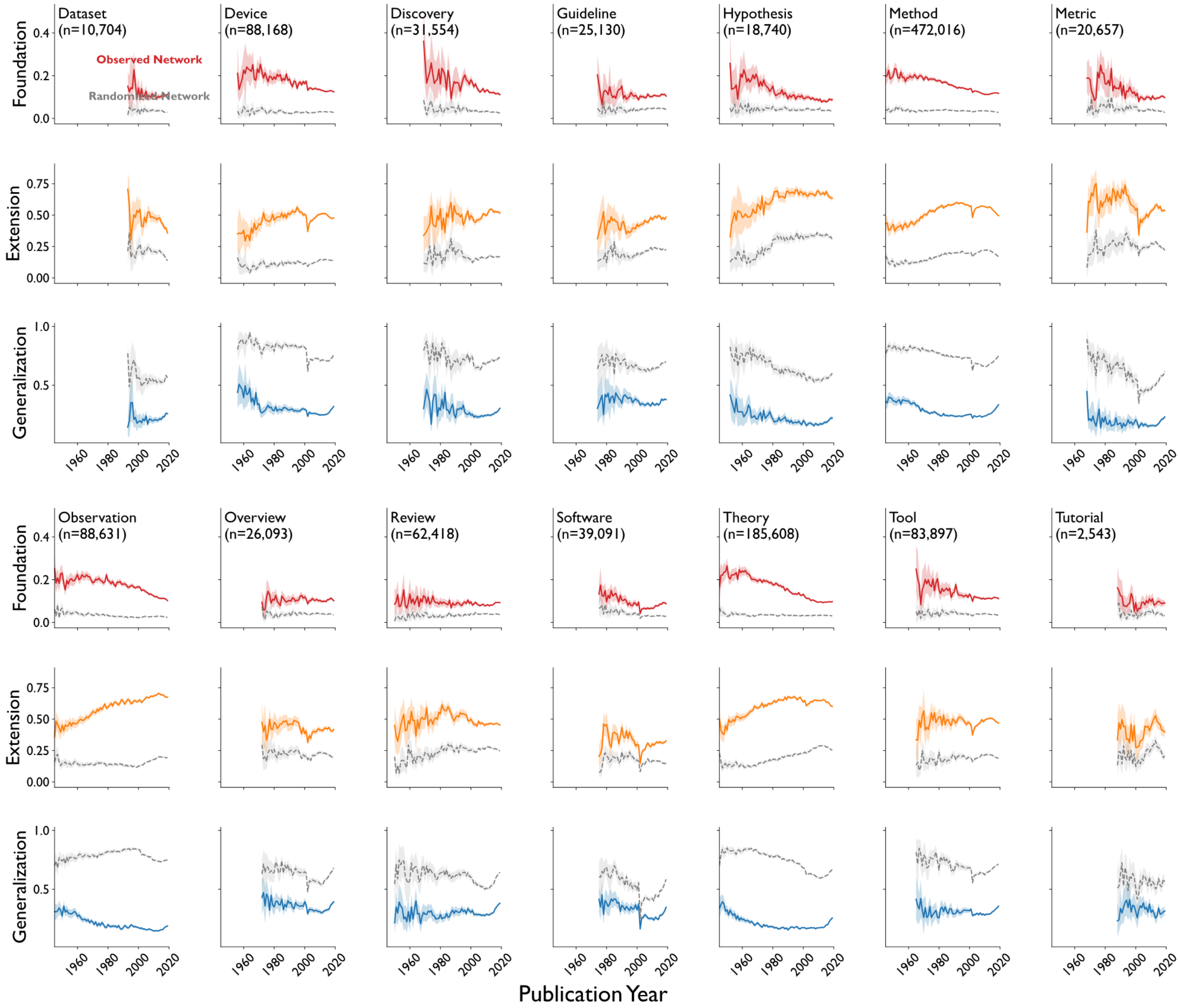

**Extended Data Fig. 8 | Longitudinal trends in the F, E, and G indices for papers containing different keywords in their titles.** The analysis was restricted to publications issued on or before 2019 that included at least one reference and accrued a minimum of five citations within five years of publication. The colored lines (red, orange, and blue) represent the F, E, and G indices, respectively, calculated from the observed citation network, whereas the gray lines denote the corresponding indices derived from a randomized network approximately preserving the degree distribution across publication year and venues. The sample size for each keyword-defined cohort is reported in the figure.

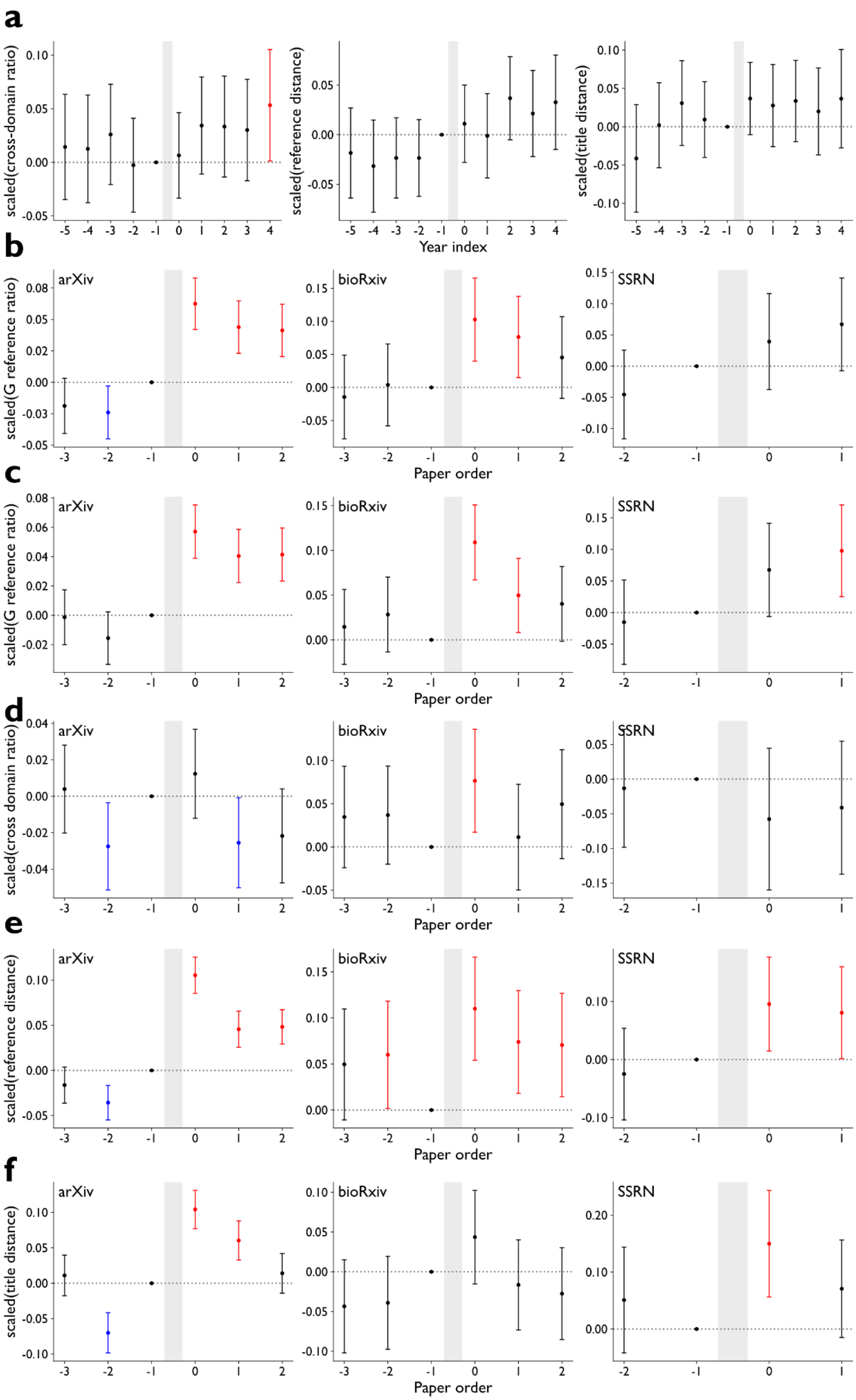

a
scaled(cross-domain ratio)
scaled(reference distance)
scaled(title distance)
Year index
b
arXiv
bioRxiv
SSRN
scaled(G reference ratio)
Paper order
c
arXiv
bioRxiv
SSRN
scaled(G reference ratio)
Paper order
d
arXiv
bioRxiv
SSRN
scaled(cross domain ratio)
Paper order
e
arXiv
bioRxiv
SSRN
scaled(reference distance)
Paper order
f
arXiv
bioRxiv
SSRN
scaled(title distance)
Paper order

**Extended Data Fig. 9 | Effects of paper online exposure (a) and authors' adoption of large language models (b–d) on the distant diffusion of papers and authors' propensity to cite distant references.** **a**, Changes in proportion of cross-domain citations from papers in focal venues, as well as the distance between papers in the venue and their references in both semantic and reference spaces, following the transition of a venue to online accessibility. The figures are based only on papers for which the corresponding distance metrics are computable (e.g, both the papers in the venue and their references must have domains available in the dataset to compute cross-domain citation ratio), and only venues with at least two papers published every year in the observation period are used in the analysis. Therefore, there are 2,655 unique venues in the treatment group with cross-domain citations as outcome variable (left figure), 2,872 venues with reference distance as outcome variable (middle figure), and 2,569 venues with semantic distance (in title) as outcome variable. **b-c**, Effects of large language model adoption on authors' propensity to employ generalized references, using alternative detection thresholds: $\tau = 0.05$ in Panel **b** (67,625 authors in the treatment group in *Arxiv*, 25,879 authors in *Biorxiv*, and 4,822 authors in *SSRN*), and $\tau = 0.2$ in Panel **c** (50,619 authors in the treatment group in *Arxiv*, 13,294 authors in *Biorxiv*, and 3,769 authors in *SSRN*), complementing the main result in Fig. 4c. **d-f**, Impact of large language model adoption on authors' tendency to cite distant papers ($\tau = 0.1$), measured by the proportion of cross-domain references (**d**, 65,070 authors in the treatment group in *Arxiv*, 23,369 authors in *Biorxiv*, and 4,602 authors in *SSRN*) and the distance between focal papers and their references in reference (**e**, 64,912 authors in the treatment group in *Arxiv*, 23,359 authors in *Biorxiv*, and 4,486 authors in *SSRN*) and semantic spaces (**f**, 64,940 authors in the treatment group in *Arxiv*, 23,349 authors in *Biorxiv*, and 4,548 authors in *SSRN*).

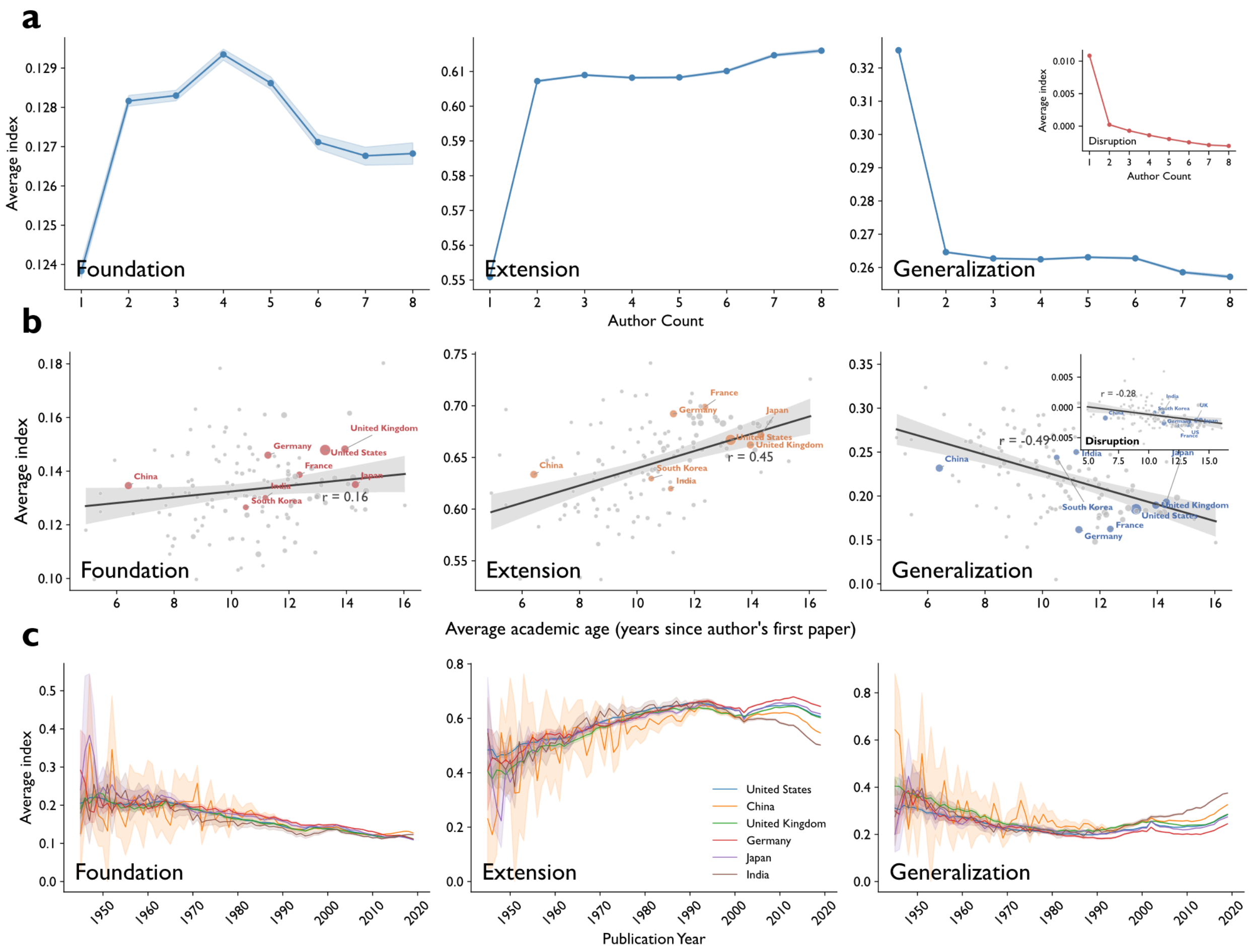

**Extended Data Fig. 10 | Relationships between team size, scientists' academic age, publication country, and the foundation, extension, and generalization indices.** The foundation, extension, generalization, and disruption indices in both panels are computed using citations received within five years of publication. The foundation, extension, and generalization indices are defined as the proportion of a paper's citations classified as foundational, extensional, or generalizational, respectively, whereas the disruption index follows the original definition of Wu *et al.*[22] These analyses reevaluate the findings presented in the seminal publications by Wu *et al*[22] and Cui *et al*[65]. **a,** Based on 21,308,068 OpenAlex publications published between 1945 and 2019 that have at least one reference, received at least five citations within five years of publication, and have between one and eight authors. **b,** Based on 12,899,274 Microsoft Academic Graph (MAG) publications published between 1945 and 2015 that have at least one reference and received at least five citations within five years of publication. Following Cui *et al.* [65], the analysis further includes only authors with at least three publications during the study period (6,418,966 authors) and countries with at least 500 publications (131 countries). **c**, Relationship between author team size and the F, E, and G indices of scientific papers. Analyses are restricted to 21,308,068 papers in OpenAlex that contain at least one reference, received at least five citations within five years of publication, were published between 1951 and 2019, and have identified authors. Shaded regions indicate 95% confidence intervals estimated by bootstrap resampling. We use OpenAlex rather than Microsoft Academic Graph (MAG; used in panel **b**) because its coverage extends to more recent publication years, allowing the relationship to be examined over a longer time period.

# Supplementary Information

**for**

**Generalization and the Rise of System-level Creativity in Science**

Hongbo Fang[†], James Evans[†‡]
[†]University of Chicago, [‡]Santa Fe Institute

# Supplementary Methods

**1. *Datasets***

**OpenAlex** constitutes one of the largest publicly accessible catalogs of scientific publications. At the time of this study, it offers comprehensive coverage of publication records through the end of 2024. To ensure comparability across publication types, we restrict our analysis to records categorized as articles (including both journal and conference articles) and preprints. This selection yields a corpus of 191,457,232 publications published between 1945 and 2024.

From this corpus, we apply three restrictions to construct the working dataset. First, we retain only publications that contain at least one reference. This criterion addresses a known data issue in OpenAlex [1], whereby a substantial fraction of publications are recorded as having zero references (115,377,910 papers, or 60.3% of the initial sample) despite in reality citing prior work. Including such records would bias trend estimates, as they would systematically exclude extensional citations. Second, we require that each publication receive at least five citations within five years of publication. Because the computation of the foundation, extension, and generalization (F, E, and G) indices depends on the citation behavior of the focal paper, a minimum level of citation activity is necessary to ensure the robustness of these measures. While we adopt a threshold of five citations in the main analysis, we also test alternative thresholds to validate the results, including a threshold as low as zero citations, all of which manifest a consistent pattern. Third, we restrict the publication period to 1945–2019. Although OpenAlex covers publications through 2024, our computation of the indices relies on a five-year citation window, necessitating truncation at 2019. We additionally explore alternative citation windows of varying lengths, with results reported in Extended Data Fig. 6. Applying these criteria yields a working dataset of 23,448,431 publications.

For selected analyses, we use a more recent OpenAlex release downloaded in 2026, which provides complete coverage of publication records through the end of 2025 and includes entities absent from earlier releases. As in the main analysis, we restrict the sample to journal articles, conference papers, and preprints. The resulting dataset comprises 240,742,877 publications published between 1945 and 2025.

**Web of Science (WoS)** is a commercially curated database of scientific publications, distinguished by its high-quality, consistent coverage of journal literature. Accordingly, we restrict our analysis to journal articles within WoS, and remove papers with only external dataset ids such as MEDLINE and BIOSIS, as they may not have their citation network covered. From this selection, we identify 55,434,109 publications published between 1945 and 2024. Applying the same criteria as for OpenAlex—at least one reference, at least five citations within five years of publication, and publication between 1945 and 2019—yields a final dataset of 19,026,615 publications.

**Semantic Scholar** is an open-access corpus of scientific publications that contains metadata for 227,693,772 papers and 2,538,028,308 citation links. Similar to OpenAlex, a substantial proportion of publications in Semantic Scholar are recorded as having no references (59.1%). To assess the robustness of our findings with respect to citation coverage, we integrate the citation network from Semantic Scholar with that of OpenAlex, thereby expanding the overall coverage of citation relationships.

Beyond bibliographic metadata, Semantic Scholar provides full-text data for a substantial subset of publications. Notably, it records the precise locations of citations within the full text of citing articles. The corpus includes 13,735,422 papers with available full-text information. These data enable the quantification of citation distributions across different sections of scientific articles and facilitate the examination of how citation placement varies across citation types.

**Microsoft Academic Graph (MAG)** is a large-scale scholarly knowledge graph developed by Microsoft Research that integrates publication records with rich metadata, including information on authors, institutions, publication venues, fields of study, citations, and affiliations. The final release of MAG was published in 2021. Owing to its high-quality author disambiguation, MAG serves as the primary data source for replicating the analyses of Cui et al. [2]. For this study, we restrict the dataset to journal articles and exclude authors with fewer than three publications. After applying these filtering criteria, the resulting dataset comprises 91,102,426 journal articles and 14,945,426 unique authors.

**Fulltext Sources Online** is a dataset that captures the temporal dynamics of online coverage across scientific publication venues [3]. It provides a comprehensive, time-resolved record of when different vendors initiate and terminate online access to publications from 37,139 venues. In this study, we leverage this dataset to examine how the release of online coverage influences the generalized diffusion of scholarly papers.

**LLM adoption data for arXiv, bioRxiv, and SSRN** were generously provided by the authors of [4]. In their study, the authors employ a text-based AI detection algorithm that compares the distribution of human-written text in papers published prior to the introduction of large language models (LLMs) with the distribution of tokens characteristic of LLM-modified text. New texts are then evaluated against these two reference distributions to estimate the probability that they were generated or assisted by an LLM.

The dataset spans publications from January 2018 to June 2024 and covers multiple disciplinary domains. The *arXiv* sample includes mathematics, physics, computer science, electrical engineering, quantitative biology, statistics, and economics; the *bioRxiv* sample covers biology and the life sciences; and the *SSRN* sample encompasses the social sciences, law, and the humanities. In total, the dataset provides LLM-assistance probability estimates for 513,921 papers from *arXiv*, 95,590 papers from *bioRxiv*, and 38,026 papers from *SSRN*. We utilize this dataset to investigate whether the adoption of LLMs influences the extent to which generalized citations are incorporated into the production of new scholarly work.

**The Publication Records for Nobel Laureates Dataset** [5] comprises a curated collection of papers most closely associated with the Nobel Prize–winning discoveries in Chemistry, Medicine, and Physics prior to 2016. In total, the dataset contains 661 unique publications with their DOIs available. We employ this dataset to validate our proposed metrics, with particular emphasis on assessing the robustness and interpretability of the foundation index.

**The Major Science Award Dataset** originates from [6], in which the authors compile a collection of 137 prizes and awards spanning the fields of biology (e.g, Acharius Award), chemistry (e.g, American Institute of Chemists Gold Medal), and medicine (e.g, Sedgwick Memorial Medal). This dataset is constructed by leveraging Wikipedia category pages for biology, medicine, and chemistry awards, followed by a manual curation process in which up to three papers most closely associated with each award are identified. We again employ this dataset to validate our proposed metrics, with particular emphasis on the F Index.

**The Citation Purpose Dataset in Natural Language Processing (*NLP Citation Purpose Dataset* for short)** [7] contains 234,448 citation links, each representing a cited–citing paper pair, and all of their 318,351 citation instances corresponding to individual mentions of cited papers within citing papers. For each citation instance, a citation purpose is assigned using a machine learning–based classifier that leverages contextual text surrounding the citation along with other features. The classifier categorizes each citation instance into one of six purpose types: Background (the cited work provides relevant contextual information for the citing work), Motivation (the cited paper illustrates need for data, goals, or methods for the citing work), Uses (the citing work uses the method, data, etc from the cited paper), Extension (the citing work extends the methods, data, etc from the cited work), Compare or Contrast (the citing work express similarity/differences to the cited work), and Future (the cited work is a potential avenue for future work). The classification model was trained and validated using expert-annotated labels.

**Academic Citation Typing 2 (ACT2) Dataset**[8] comprises 4,000 publicly available citation instances (3,919 citation links) collected from scholarly articles across multiple disciplines, together with their corresponding citation-purpose labels. The dataset adopts the same taxonomy as *NLP Citation Dataset*, classifying citations into six categories: Background, Motivation, Compare_Contrast, Use, Extension, and Future. Citation purposes were assigned using a machine-learning classifier that leverages the textual context surrounding each reference, along with additional citation-related features. Unlike the NLP Dataset, ACT2 does not include every occurrence of each citation link; instead, it consists of a randomly sampled subset of citation instances. Consequently, the ratio of citation links to citation instances differs between the two datasets.

**2.** ***Operationalization of Foundation (F), Extension (E), and Generalization (G) Index***

We begin by operationalizing the proposed framework at the level of individual citations. For each citation to a focal paper, we assign three indicator variables representing whether the citation is (i) **foundational** ($c_f$), (ii) **extensional** ($c_e$), or (iii) **generalizational** ($c_g$), such that: $c_f + c_e + c_g = 1$.

As illustrated in Fig. 1a, consider a focal paper (depicted as the blue diamond in the middle) and one of its citing papers. Let $e_i$ denote the number of *other citations of the focal paper* that this citing paper also cites, and let $e_j$ denote the number of *references of the focal paper* that the citing paper also cites. Based on the relative magnitudes of $e_i$ and $e_j$, we classify the citation as follows:

1. **Foundational citation**: if $e_i > e_j$, the citing paper builds primarily on other works that cite the focal paper, suggesting the focal paper is treated as a foundation. In this case, we assign $c_f = 1, c_e = 0, c_g = 0$.
2. **Extensional citation**: if $e_j > e_i$, the citing paper builds primarily on the focal paper's references, suggesting the focal paper is treated as an extension of prior work. In this case, we assign

$c_f = 0, c_e = 1, c_g = 0.$

3. **Generalizational citation**: if $e_i = e_j = 0$, the citing paper neither cites the focal paper's references nor its other citations. This suggests the focal paper is used as a tool, component, or background without engaging its intellectual lineage and related contexts. In this case, we assign $c_f = 0, c_e = 0, c_g = 1$
4. **Borderline case**: if $e_i = e_j > 0$, the citation draws equally from the focal paper's references and citations. In this case, we assign $c_f = 0.5, c_e = 0.5, c_g = 0$

Next, at the paper level, we aggregate these citation-level indicators to construct the three indices. Specifically, for a focal paper with N total citations, we have:

$$F = \frac{\Sigma c_f}{N}, E = \frac{\Sigma c_e}{N}, \text{ and } G = \frac{\Sigma c_g}{N}$$

Thus, F, E and G represent the proportions of a focal paper's citations that are classified as foundational, extensional, and generalizational, respectively.

In this study, we use a discrete classification scheme in which each citation is categorized as either *foundational*, *extensional*, or *generalizational*. Nevertheless, alternative continuous formulations can also be considered. For example, one may define the foundational and extensional weights as $c_f = \frac{e_i}{e_i + e_j}$, and $c_e = \frac{e_j}{e_i + e_j}$, respectively, for citations where either $e_i$ or $e_j$ is positive. Additionally, our baseline definition classifies a citation as generalizational only when the citing paper does not cite any references or citing papers of the focal cited paper. This criterion may be overly restrictive, as some shared references or citations may arise purely by chance. A normalized definition may therefore provide a more robust measure. For instance, let paper Y denote the cited paper and paper X the citing paper. A citation from X to Y may be classified as generalizational when both $e_i$ and $e_j$ are below their respective average values across all references cited by (X). Formally, this condition can be expressed as $((e_i)_{XY} < \frac{\Sigma (e_i)_{XZ}}{N}$, $(e_j)_{XY} < \frac{\Sigma (e_j)_{XZ}}{N}$, where *N* is the number of X's references and Z iterates over all references of X). As a robustness check, we also compute the longitudinal changes in the F, E and G indices using the continuous definitions of $c_f$ and $c_e$, with the result presented in Extended Data Fig. 6e and 6l.

### 3. *Visualizing Papers' Downstream Citations*

The citation networks surrounding papers with high foundation index values differ markedly from those surrounding papers with high generalization index values. Fig. S1 illustrates these contrasting patterns. In Panel a, paper A represents a highly foundational paper, whereas in Panel b, paper A represents a highly generalizational paper.

In Panel a, paper A is cited by numerous subsequent papers (e.g., Paper C), which not only cite paper A but also cite other papers that build upon it (e.g., paper B). This pattern generates a densely connected downstream citation structure. By contrast, in Panel b, paper A is cited by many papers (e.g., papers B and C) that cite paper A independently but rarely cite one another, resulting in a set of largely parallel citation trajectories.

We construct multilayer citation networks based on the citation relationships among all papers that cite a focal paper. Specifically, treating the focal paper as the root node (depth = 0), we define depth = 1 papers as those that cite the focal paper but do not cite any of its citing papers. More generally, for any depth *t*, papers at depth *t+1* are defined as those that cite at least one paper at depth *t*, while citing no papers deeper than depth *t*. This procedure transforms the citation network shown in the first column of Fig. S1 into the layered structures illustrated in the third column (with the second column as the intermediate step), where the focal paper (e.g, paper A) is positioned at the center of the network.

A multilayer downstream citation tree indicates that the downstream citations of a focal paper form interconnected clusters of research through successive citation relationships. In such cases, the focal paper continues to be cited and remains embedded within scholarly discourse across multiple generations of scientific development. Consequently, foundational papers tend to generate downstream citation networks with greater depth and a larger number of interconnected layers.

As illustrated in Fig. S2, in sociology, *The Strength of Weak Ties* [9] (hereafter *Weak Ties*) initiated a sustained line of inquiry into the role of weak ties in social networks. Building on this foundation, *The Diversity–Bandwidth Trade-off* [10] (hereafter *Diversity–Bandwidth*) examined the trade-off between information diversity afforded by weak ties and structural holes, and the high information bandwidth enabled by strong, intensive interactions. Extending this line of research, *The Strength of Long-Range Ties in Population-Scale Social Networks* [11] (hereafter Long-Range Ties) demonstrated that social networks can simultaneously exhibit weak structural connectivity and strong interaction intensity. Because both *Diversity–Bandwidth* and *Long-Range Ties* contribute to the same theoretical framework established by *Weak Ties*, they both retain citations to the focal paper. Under our construction of downstream citation networks (Fig. S1), these papers therefore form a two-layer downstream citation structure for *Weak Ties*. In contrast, *Linear Models with Spatially Distributed Data: Spatial Disturbances or Spatial Effects?* [12] (hereafter *Spatial Effects*) is cited in *Diversity–Bandwidth* primarily to justify a statistical methodology rather than to advance the theoretical discussion on weak ties. Because this methodological citation is not carried forward in subsequent studies within the same research trajectory, *Spatial Effects* would be expected to exhibit only a one-layer downstream citation network in this line of research. A similar pattern can be observed in computer science. *ImageNet Classification with Deep Convolutional Neural Networks* [13] (*AlexNet*), *Very Deep Convolutional Networks for Large-Scale Image Recognition* [14] (*VGG*), and *Deep Residual Learning for Image Recognition* [15] (ResNet) represent successive milestone contributions to deep neural network architectures for computer vision. When *AlexNet* is treated as the focal paper, *VGG* and *ResNet* form a two-layer downstream citation network that reflects the continued development of the same research direction. By contrast, *Deep Convolutional Filter Banks for Texture Recognition and Segmentation* [16] (*Texture Recognition*) is cited in *VGG* only to demonstrate the model's effectiveness for texture recognition. Consequently, this citation is not propagated along the research trajectory of deep neural network model development and does not contribute to a multilayer downstream citation network.

This structure suggests that the foundational paper serves as the basis for one or more sustained and interconnected research trajectories, remaining influential throughout successive stages of knowledge accumulation. Similar qualitative observations have been discussed in recent studies, although they have not been directly quantified[17] . In contrast, generalizational papers are typically cited by multiple parallel streams of research that exhibit relatively limited interaction with one another. While such papers may diffuse ideas broadly across diverse domains, their influence within any particular research trajectory is often weaker and more transient. As a result, their impact tends to diminish after one or two generations of subsequent research, producing citation networks that are comparatively shallow and less interconnected.

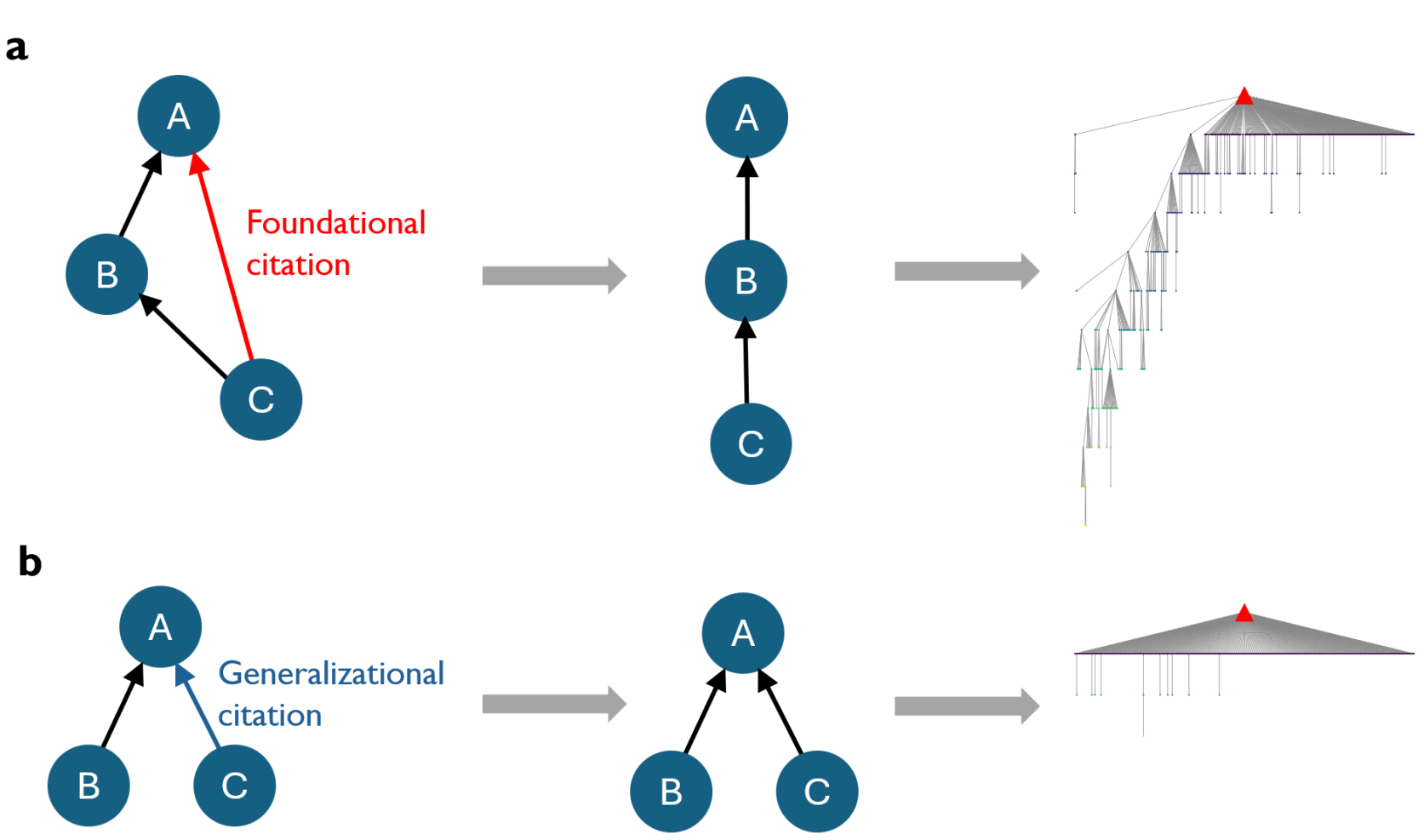

**Fig. S1 | Illustration of the construction of downstream citation networks.** Panel **a** depicts the downstream citation network of a foundational paper, characterized by a deep, interconnected, and multilayer structure. Panel **b** depicts the downstream citation network of a generalizational paper, characterized by multiple parallel citation streams with limited cross-citation and relatively shallow depth.

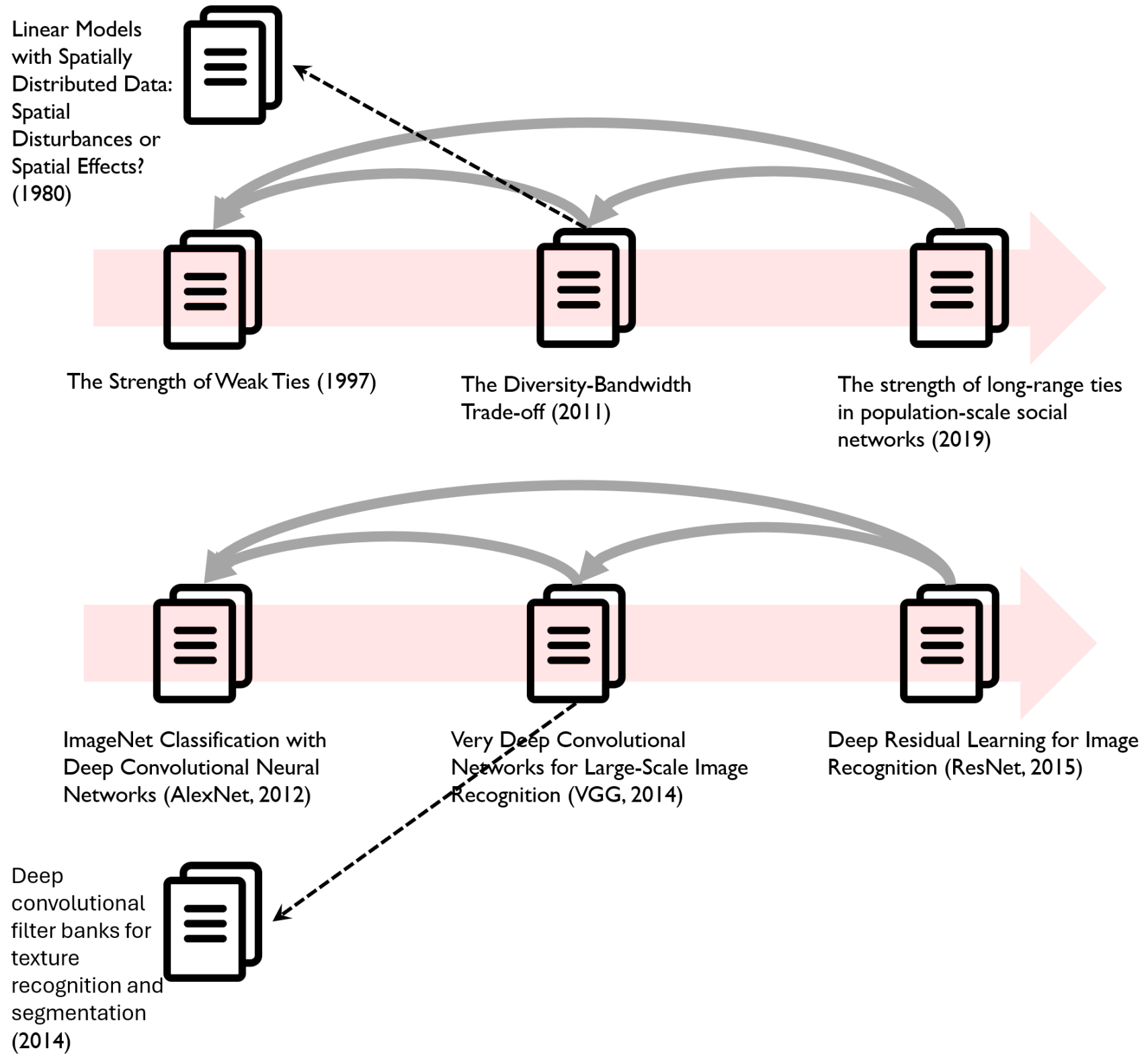

**Fig. S2 | Multilayer downstream citation networks illustrating the continuation of research directions.** Two representative research trajectories are shown: sociology (upper panel) and computer science (lower panel). Red arrows indicate the progression of time and the development of a research direction. Gray arrows denote citation relationships between papers that contribute to the same research trajectory, whereas black dotted arrows indicate citations to references that are not essential to the trajectory and are not retained in subsequent developments. In the sociology example, *The Strength of Weak Ties* [9]

represents a foundational work whose downstream citations form a multilayer citation network. By contrast, *Linear Models with Spatially Distributed Data: Spatial Disturbances or Spatial Effects?* [12] is a non-central work whose downstream citations are confined to a single layer. Similarly, in the computer science example, *ImageNet Classification with Deep Convolutional Neural Networks* [13] serves as a foundational work with a multilayer downstream citation network, whereas *Deep Convolutional Filter Banks for Texture Recognition and Segmentation* [16] is a non-central work whose downstream citations are restricted to a single layer.

### 4. *Identification of Paper Domains*

The domain of a paper is used for two purposes: (i) to evaluate whether a citation occurs between papers from different domains, and (ii) to examine the longitudinal dynamics of the foundation (F), extension (E), and generalization (G) indices across scientific domains.

In *OpenAlex,* we use *concepts* as proxies for domains. Concepts are assigned to papers based on their titles, abstracts, and the titles of their publication venues [18]. OpenAlex contains more than 65,000 concepts organized in a hierarchical tree structure. For our analysis, we focus on the 19 top-level concepts (level = 0). On average, each paper in our sample is associated with 2.69 concepts (with a median of two).

As an alternative domain classification, we also identify a paper's *top domain(s)* based on the *scores* attached to each assigned concept. Each concept is associated with a score that quantifies the strength of its connection to the paper. Beginning at the top level of the hierarchy (level = 0), we iterate over all levels to evaluate the scores of assigned concepts and exclude those with scores lower than any others. For levels below the top (level > 0), we compute the score of a top-level concept by summing the scores of assigned concepts of its children in that level. This algorithm identifies the domain(s) with the highest overall score while prioritizing higher-level classifications. The procedure is illustrated with pseudocode in Table S1. Using this approach, most papers are assigned to exactly one top domain. Although ties across scores at all levels may occasionally result in more than one top domain assigned to one paper, this occurs only rarely in our sample (only 52 papers).

In *Web of Science (WoS)*, we use *macro_citation_topic* as the proxy for domain. This represents the highest level of a three-layer hierarchical classification of research areas, derived from citation network structures[19]. Each paper is assigned to exactly one macro_citation_topic. In our sample, 106,319 papers (0.56%) lack an assigned macro_citation_topic.

**Input:** Paper p

```
candidate set C = AllTopConcepts(p)

C = RemoveLowScore(TopScore(C)) # remove concepts at top level where score is lower than others

for i in 1 to 5: # starting from top to bottom, iterate over concepts in each level (5 is the highest)

    c2score = {}

    for c in C:
```

```
        sum_score = 0
        for c_ip in Child(c): # use the score of children of c as a proxy for the score of c at level i
            sum_score += Score(c_ip)
        c2score[c] = sum_score
    C = RemoveLowScore(c2score) # remove concepts if their scores at this level is lower than others
```

**Output:** Set of Top Concepts C

**Table S1 | Pseudo code to illustrate the identification of top domains.**

**5. *Computation of Disruption at the Citation Level***

We adapt the disruption index, originally defined at the paper level, to measure disruption at the level of individual citations. As illustrated in Fig. S3, consider a citation from paper B to paper A. We first identify all citations received by paper B within five years of its publication. Among these, we count:
j: the number of citations to paper B that also cite paper A;
i: the number of citations to paper B that do not cite paper A;
k: the number of citations that cite paper A without citing paper B.
Using these quantities, the disruption score for the citation B→A can be expressed as $D = \frac{i-j}{i+j+k}$, with alternative formulations provided in Fig. S12a. This citation-level measure captures the extent to which paper B is used independently of paper A.

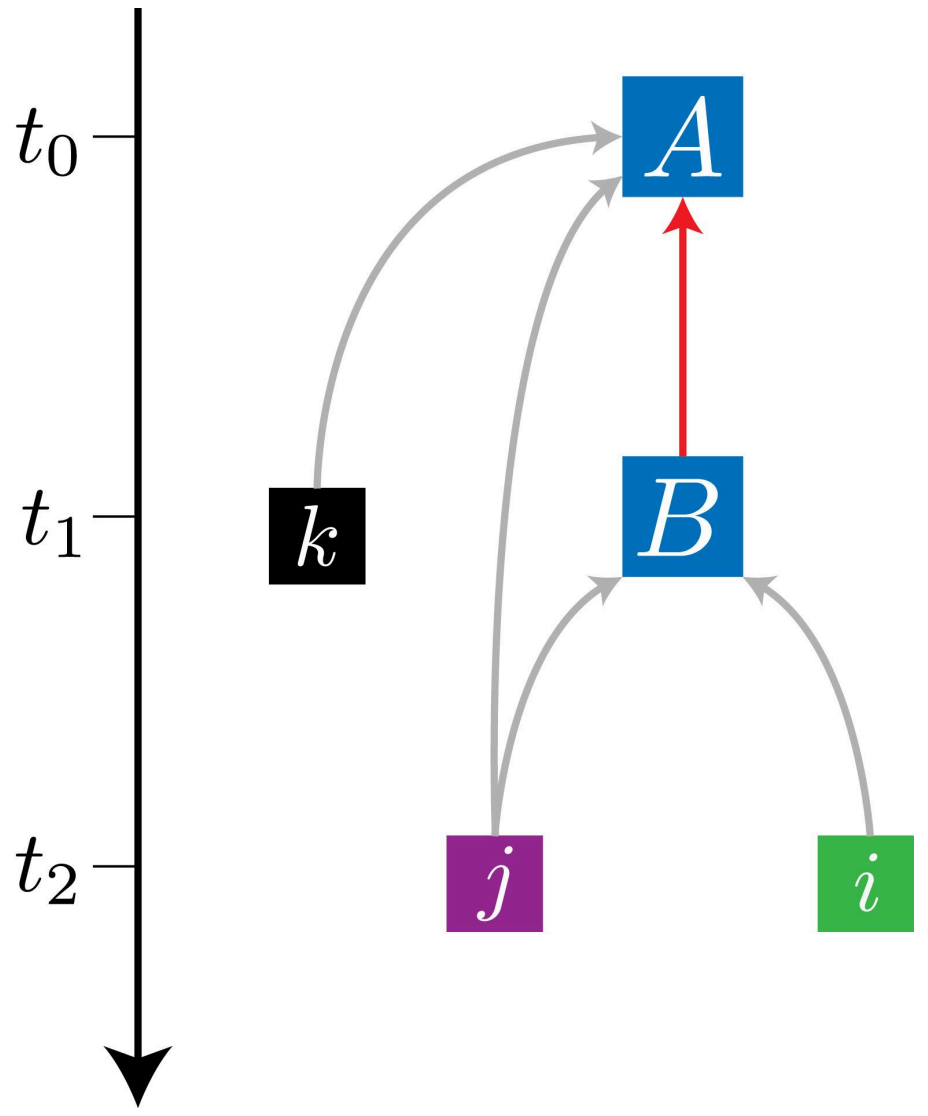

**Fig. S3 | Illustration of disruption computation at the citation level.** Paper B, published at time $t_1$, cites paper A, which was published at time $t_0$. A subsequent paper *i* constitutes a disruptive citation of B if it cites B without citing any of B's references (e.g., A). In contrast, a subsequent paper *j* represents a developmental citation if it cites both B and A concurrently.

**6. *Computation of the Longitudinal Change of F, E, and G***

We compute the annual averages of the foundation (F), extension (E), and generalization (G) index across all papers published in a given year. Following established practice [20,21], we restrict the calculation to citations received within X years of publication. This restriction allows the indices to be comparable across cohorts in each year. Given that our period of observation ends at the end of 2024, we can only compute the longitudinal change until year 2024 - X. For example, when X=1, the series ends in 2023, since citation records are complete only through 2024.

Our main analyses employ X=5, a widely used threshold [21] that balances the trade-off between data availability and allowing sufficient time for citation accumulation. To capture more recent dynamics, particularly in the large language model era, we also report results with X=1. In addition, we present results with X=10 as a robustness check.

Because the computation of our indices rely on a paper's citations to evaluate its role in the knowledge network, presumably the accuracy of the estimates increases with the number of citations a paper receives. Accordingly, our primary analyses include only papers with at least five citations within the X-year window, and we also analyze all papers with more than one citation in the same period as a robustness check.

Finally, we note that OpenAlex fails to identify references for a non-trivial proportion of papers, which may bias the estimation of citation-based metrics [1]. To address this limitation, we exclude from our analyses all papers with no identified references, and compare OpenAlex results with those from the Web of Science.

**7. *Correlating the Novelty-related Metrics with Award Winning Papers***

The *Publication Records for Nobel Laureates* dataset [5] and the *Major Science Award* dataset [6] provide curated lists of papers that have received major awards in medicine, chemistry, physics, and biology. We link Nobel Prize–winning papers to records in OpenAlex using DOI identifiers, while other award-winning papers are matched to OpenAlex using PMID identifiers.

We conduct logistic regression analyses separately for Nobel Prize–winning papers and for papers associated with other major awards, as well as separately across domains. The dependent variable is a binary indicator equal to 1 if a paper is associated with an award and 0 otherwise. The primary independent variables are the three indices proposed in this study (F, E, and G), along with other measures of novelty, including the disruption index [20] and atypicality [22]. As expected, citation counts are positively associated with the likelihood of receiving a major award; accordingly, the number of citations is included as a control variable. We also control for the number of references—given its mechanical relationship with the construction of F, E, G, and the disruption index—as well as a fixed effect variable for publication period. All models are estimated separately by domain.

The analysis sample is restricted to papers within each discipline that include at least one reference and have accumulated a minimum of five citations within five years of publication. Because the number of award-winning papers is extremely small relative to the number of non-award-winning papers, the estimation is susceptible to the rare-events problem [23], which can bias coefficient estimates. This issue is particularly pronounced when publication-year fixed effects are included, as they substantially increase the number of parameters to be estimated. To mitigate this problem, publication-year fixed effects are replaced with five-year publication-period fixed effects. In addition, publication-period fixed effects corresponding to periods in which no papers received the award under study are omitted, as all observations within such periods share the same negative outcome.

Overall, the regressions estimating the probability of receiving a Nobel Prize are based on 779,426 papers in Medicine (including 35 Nobel Prize-winning papers), 2,006,730 papers in Chemistry (73 Nobel Prize-winning papers), and 1,769,926 papers in Physics (100 Nobel Prize-winning papers). An additional 62 Nobel Prize-winning articles are classified under OpenAlex concepts as one of Medicine, Chemistry, or Physics, but differ from the disciplinary domains recorded in the Nobel Laureates Dataset [5]. These articles are therefore excluded from the domain-specific analyses but retained in the pooled analysis across all domains. The regressions estimating the probability of receiving other major scientific awards are based on 4,704,977 papers in Biology (3,990 award-winning papers), 5,611,286 papers in Chemistry (1,367 award-winning papers), and 3,916,616 papers in Medicine (1,223 award-winning papers). All publication data are drawn from the OpenAlex dataset.

**8. *Computation of Semantic Distance***

As shown in Fig. S4, we compute two primary types of semantic distance. Fig. S4a illustrates the *within-paper distance*. For each paper, we first identify the centroid of all word tokens by averaging their embeddings. The within-paper distance is then defined as the average cosine distance between each token embedding and the centroid. This measure captures the extent to which the combination of words or scientific concepts in the focal paper resembles conventional combinations used in prior work versus representing a novel or surprising combination.

Fig. S4b illustrates the *cross-paper distance*. Here, the centroid of all tokens in a paper serves as a proxy for the paper-level embedding. The cross-paper distance is calculated as the cosine distance between the embeddings of two papers that are connected by a citation link. This measure quantifies the textual dissimilarity between a citing paper and the paper it references.

To preprocess the text, we employ the FastText model [24] to identify papers with English-language titles, and the *en_core_web_sm* model in *spaCy* to tokenize these titles. Because the vocabulary of scientific writing evolves over time, we adopt a dynamic embedding approach using a sliding window. Specifically, embeddings are trained on a rolling five-year corpus (stride = one year). For instance, the semantic distance of a paper published in 2010 is computed using embeddings trained on texts from 2004–2009. We use the Skip-Gram model implemented in the *gensim* package, with a context window size of 2 and an embedding dimension of 128.

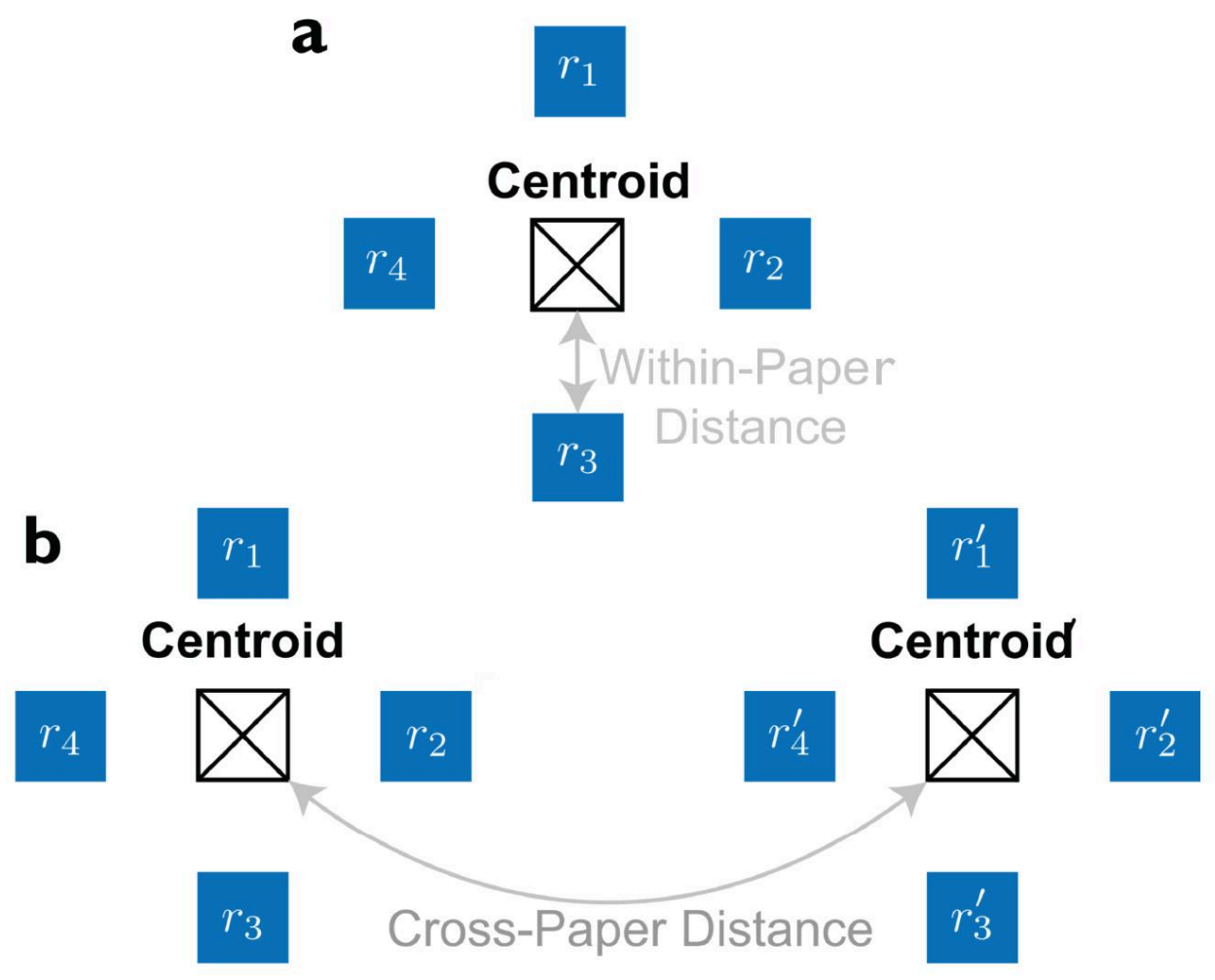

**Fig. S4 | Illustration of within and cross paper distances.**

We construct two groups of embeddings: one based on paper titles and the other on abstracts. For title-based embeddings, we train models annually from 1951 to 2019 (papers published prior to 1945 are excluded, and five years of prior text are required to construct embeddings). For abstract-based embeddings, models are trained annually from 1986 to 2019, as abstracts from earlier years are frequently missing or incomplete.

To assess robustness, we validate results across different hyperparameter settings. Specifically, we repeat training with three random seeds (6, 42, 100) and two embedding dimensions (128 and 256). Across all specifications, the resulting patterns remain qualitatively consistent.

### 9. *Computation of Reference Distance*

In parallel with semantic distance, we compute *within-paper* and *cross-paper* reference distances using dynamic embeddings of both papers and their publication venues. Papers (or venues) that are frequently cited together are positioned closer in the embedding space. Each paper's publication venue is identified through its *primary_location* field (e.g., conference or journal).

The *within-paper reference distance* is defined as the average cosine distance between the embedding of each referencing paper and the centroid of these embeddings. The *cross-paper reference distance* is defined as the cosine distance between the centroids of two sets of references. While it is possible to compute the cross-paper reference distance by comparing the embedding of the focal paper's own publication venue instead of the centroid embedding of its references, we adopt the centroid-to-centroid approach for consistency with the computation of semantic distance.

The dynamic embedding procedure employed here mirrors the parameterization used for the semantic distance analysis. Specifically, we adopt a sliding window of five years (with a stride of one year) and train embeddings using the Skip-Gram model with an embedding dimension of 128. The key distinction lies in the choice of context window size. Because references within a paper lack an inherent ordering, we set the context window size to a sufficiently large value (100) to ensure equal treatment of all references. To avoid bias from papers with extremely long reference lists, we restrict the training set to papers with no more than 200 references (allowing 100 references on either side of the focal item). Under this criterion, 30,707 papers—accounting for 0.13% of the dataset—are excluded from the embedding training process. These papers are reintroduced in downstream analyses when computing within- and cross-paper distances.

**10. *Computation of the Breadth and Concentration of Influence***

To evaluate the breadth and concentration of influence of foundational, extensional, and generalizational papers, we first computed the quantile rankings of the foundation, extension, and generalization indices for all papers in the OpenAlex 2025 and 2026 releases, relative to other papers published in the same year and within the same domain. In the OpenAlex 2025 release, domains were identified using the *top_domain* entity, whereas in the OpenAlex 2026 release, they were identified using the *primary field* entity. Papers ranking in the top 10% of the foundation, extension, or generalization index within their publication year and domain were classified as foundational, extensional, or generalizational papers, respectively.

Using the OpenAlex 2026 release, among papers receiving 10–15 citations, 699,221 (550,977 in the 2025 release), 705,765 (554,837 in the 2025 release), and 699,814 (550,337 in the 2025 release) papers were identified as belonging to the top 10% in the foundation, extension, and generalization indices, respectively. Among papers receiving 100–150 citations, the corresponding numbers were 42,767 (22,251 in the 2025 release), 42,735 (22,247 in the 2025 release), and 42,767 (22,243 in the 2025 release), respectively. Among papers receiving 1,000–1,500 citations, 783 (373 in the 2025 release), 782 (373 in the 2025 release), and 783 (373 in the 2025 release) papers were selected, respectively.

For each selected paper, we quantified the breadth of topical influence by computing the number of distinct topics represented by its citing papers and then averaging this quantity across all selected papers of the same type (i.e., foundational, extensional, or generalizational). To assess the concentration of topical influence, we identified the topic contributing the largest number of citations to each selected paper and calculated the proportion of citations originating from that topic relative to the paper's total citations. These proportions were then averaged across all selected papers within each paper type.

We similarly evaluated the breadth and concentration of influence at the country level. Specifically, we computed the average number of countries represented by the affiliations of authors citing each selected paper as a measure of geographic breadth. We then identified the country contributing the largest number of citations to each paper and calculated the proportion of citations originating from that country, averaging these proportions across papers of the same type to quantify geographic concentration.

Because a citing paper may be assigned to multiple topics and may include authors affiliated with multiple countries, broader influence does not necessarily imply lower concentration of influence within a particular topic or country. Consequently, breadth and concentration capture complementary dimensions of scientific influence rather than representing opposite ends of the same continuum.

### 11. *Computation of Reference Atypicality and Its Longitudinal Change*

Proposed in [22], atypicality is a metric that quantifies the extent to which a paper cites other works from distant venues. Conceptually, it is similar to within-paper reference distance. A comparison of its longitudinal change is not a simple task. Fig. S5a illustrates how we approach this. First, we select all publications with a valid venue label (we use the *primary_location* entity and remove venues such as *arXiv* and *Zenodo* as they do not correspond to real publication venues) in OpenAlex before a given year X, and compute the z-score of co-appearance between every pair of venues based on their references (indicating to what extent that two venues are cited together compared to random chance before year X). For all papers published after year X, we use a sliding window to divide them into samples in each period of ten years (e.g, period 1, period 2, and period 3 in Fig. S5a), and we compute the average atypicality of papers in each period after year X.

This procedure ensures that the atypicality of all papers is computed using a common z-score distribution derived from venue pairs, enabling consistent comparisons across papers. The resulting atypicality measure can be interpreted as the extent to which future papers cite combinations of venues that are historically distant from one another.

A limitation of this approach is that some venues cited by future papers may not exist within the historical period used to construct the venue-level z-scores. Consequently, citations involving such venues must be excluded from the atypicality calculation, reducing data coverage. For example, among the 84,276,569 citation links made by the 3,429,091 papers published in 2019, 5,859,812 citations point to papers without valid venue information; these citations are excluded from atypicality calculations regardless of the computational approach. Of the remaining 78,416,757 citations, when venue-level z-scores are constructed using co-citation data prior to 1960, 61,003,357 citations (77.8%) point to papers published in venues that did not exist before 1960 and therefore lack corresponding z-scores. As a result, these citations must be excluded from the computation. In contrast, when venue-level z-scores are constructed using co-citation data prior to 2000, only 17,198,406 citations (21.9%) point to valid venues that were not established before 2000 and are therefore omitted.

In our analysis, we construct venue-level z-scores using co-citation data accumulated before six cutoff years—1960, 1970, 1980, 1990, 2000, and 2010—and report the corresponding results separately to facilitate longitudinal comparisons across historical periods.

Our approach is different from the approach used by Park et al[25] (See Fig. S5b), and this difference also leads to an opposite pattern of results. We describe the difference between our approaches and theirs, and explain they generate different results in Supplementary Results Section 21.

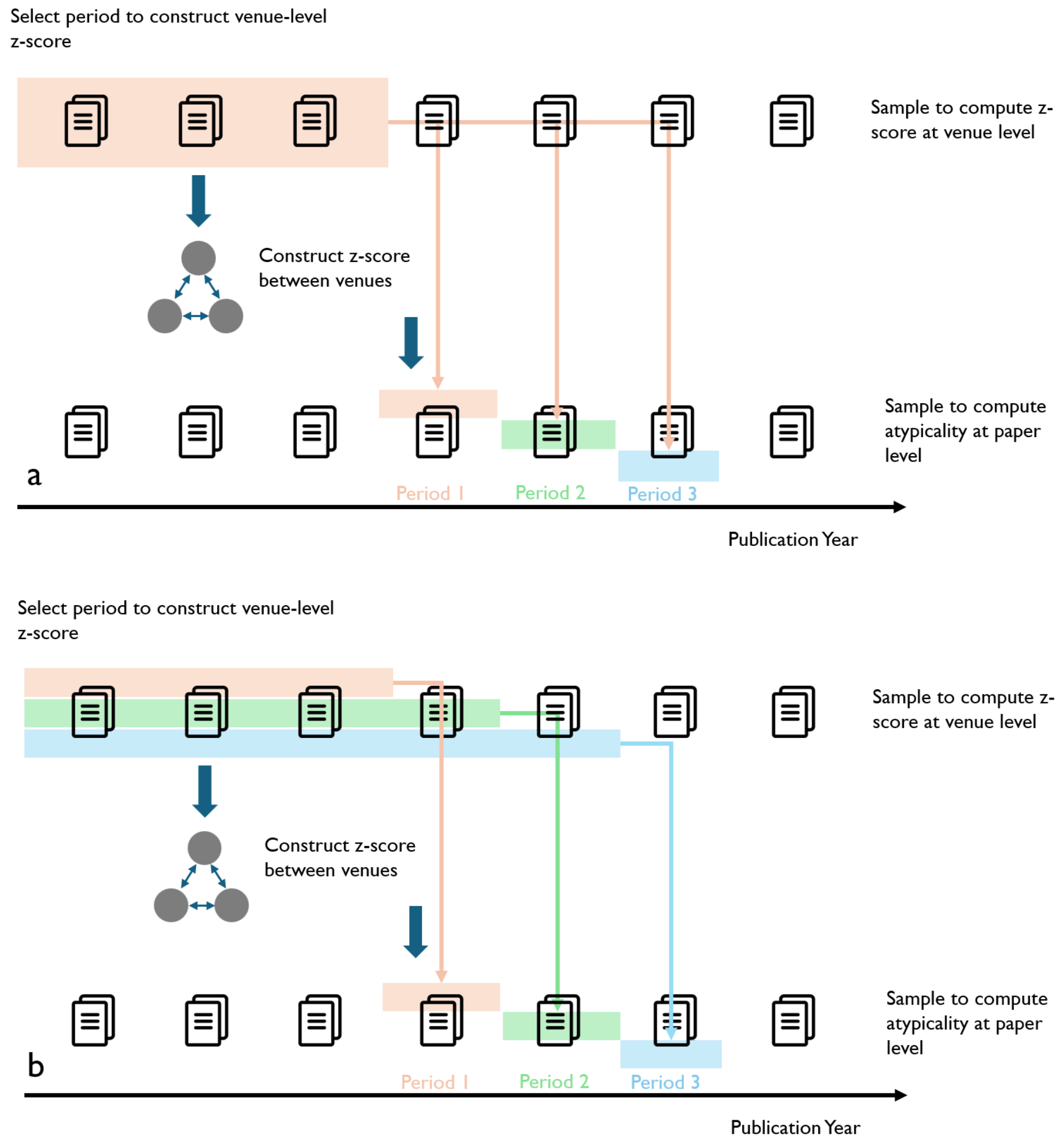

**Fig. S5 | Illustration of approaches to compute longitudinal change in atypicality. a,** The approach used by our paper. **b,** The approach used by Park et al [25].

## 12. *Identification of Reference Location in the Full Text*

As illustrated in Fig. S6, the Semantic Scholar dataset parses the position of every occurrence of a reference annotation (e.g., “[3]” in Fig. S6). Each reference annotation is linked to its corresponding bibliographic entry. We use regular expressions to extract DOI strings (e.g., *10.1371/journal.pone.0090501* in Fig. S6) from the bibliography and thereby associate each reference annotation with a specific cited paper.

Because full-text data are not available for all publications and only references containing DOIs can be identified using this approach, the resulting reference-location data exhibit imbalanced coverage across scientific domains. Fig. S7 shows the domain-wise coverage of reference positions. By comparing the positions of reference annotations with those of section headers, we identify the specific sections in which each reference is cited.

### 13. *Identification of Reference Location in Different Sections*

We infer the purpose of each citation based on the section of the paper in which it appears. Consistent with prior work, we employ section-specific keywords to identify sections, as citations in different sections tend to fulfill distinct rhetorical functions[7]. We adopt the IMRaD (Introduction, Method, Results, and Discussion)[26] framework as a set of predefined section categories, a structure that is also prevalent in our dataset. The complete set of section types, along with the corresponding keywords used for their identification, is presented in Table S2. Furthermore, as illustrated in Fig. S5, the positional information of each section header is systematically recorded in the dataset.

Given the positions of section headers and reference occurrences, we determine the section to which each specific reference occurrence belongs. A key challenge in this process is that the dataset does not preserve the hierarchical structure of section headers; consequently, sections and subsections are not distinguishable. As illustrated in Fig. S6, the red-boxed reference "*Lehnart et al., 2010*" appears within the *Background* section (which is detected as *Introduction*). However, because the subsection *Social Sexual Norms* is treated as an independent section, the reference is incorrectly attributed to *Social Sexual Norms*, which is classified as an unrecognized section type.

One possible solution is to ignore all section headers unless they can be detected using the keywords listed in Table S2. However, this approach introduces a substantial number of false positives in assigning reference positions, as papers vary widely in structure and format. For example, in a review paper where only the *Introduction* section is detectable under our keyword-based method, all references appearing after the *Introduction* header would be assigned to that section. In practice, however, these references may occur in distinct parts of the paper and serve different rhetorical functions.

Therefore, we exclude all section headers that cannot be detected using our keyword set and retain only papers in which all five section types (see Table S2) are successfully identified. These five section types are required to appear in the canonical order: *Introduction*, *Method*, *Results*, *Discussion*, and *Conclusion* (with the exception that *Results* and *Discussion* may be merged into a single section). After this step, 3,192,732 papers remain for analysis. For each reference occurrence, we compare its position in the text with the positions of the detected section headers and assign it to a section if it appears between two consecutive detected headers, as illustrated in Fig. S7.

We note that the Semantic Scholar dataset provides citation-purpose labels derived from a machine-learning–based analysis of reference contexts[27]. We choose not to use these labels for two main reasons. First, citation purpose is aggregated at the level of paper pairs (i.e., between a citing paper and a cited paper), which biases the classification toward references that appear more frequently within a paper, as multiple occurrences increase the likelihood of detecting multiple purposes. Second, the dataset categorizes citation purposes into only three classes—*background information*, *method*, and *result comparison*. While these categories largely correspond to citations appearing in the *Introduction*, *Method*, and *Results* sections under our approach, they do not capture other purposes such as *discussion of implications*, which represents a distinct function of more generalized citations.

| Section types | Keywords used for detection (include at least one of the following keywords) |
|---|---|
| *Introduction* | Introduction; Background |
| *Method* | Method; Data; Procedure; Experiment |
| *Result* | Result; Outcome |
| *Discussion* | Discussion; Implication |
| *Conclusion* | Conclusion |

**Table S2 |** The types of sections identified with keywords used for detection. Note that some papers merge the *Result* and *Discussion* section together, therefore we also have a '*Result and Discussion*' section for those papers.

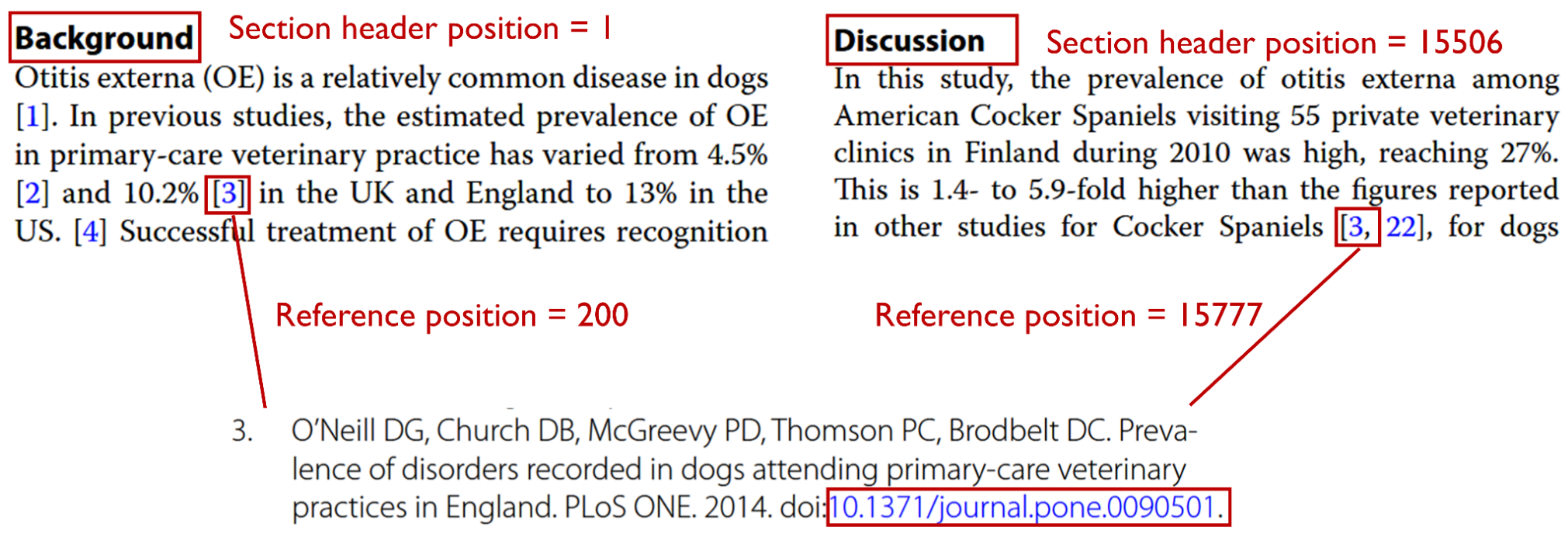

**Fig. S6 | Example of section header and reference locations.**

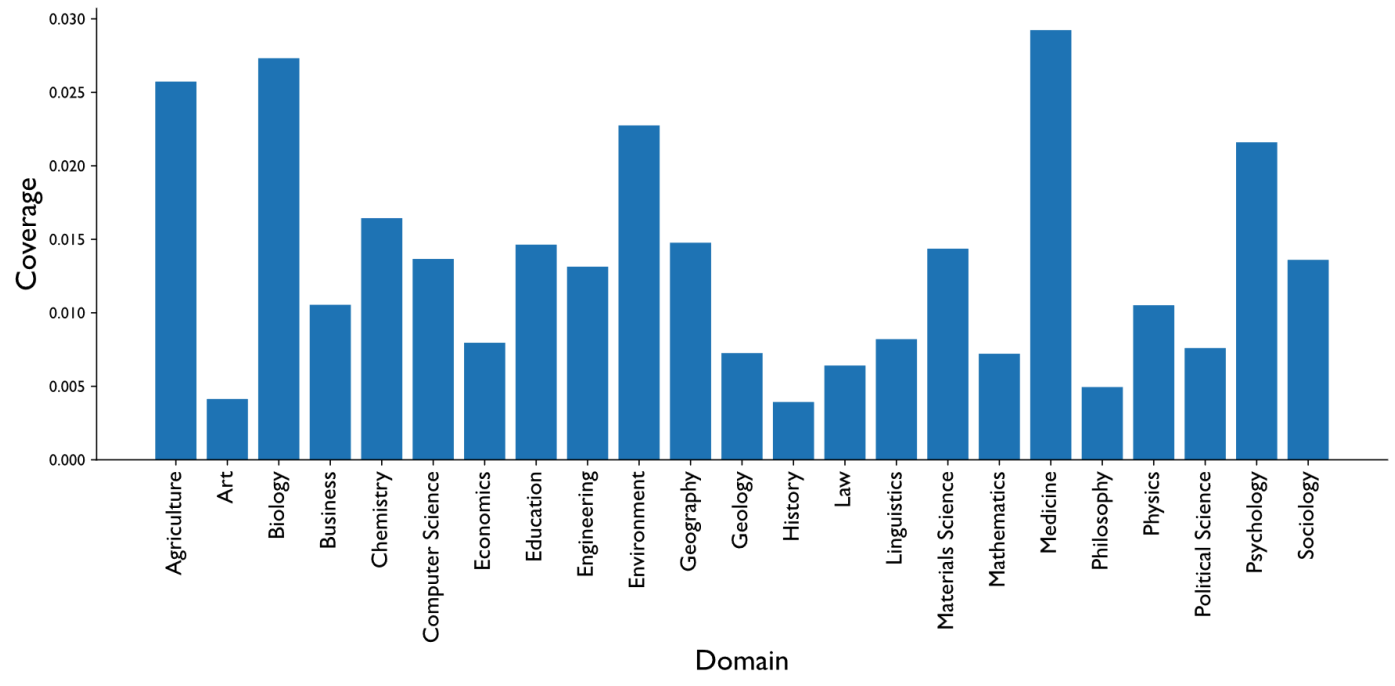

**Fig. S7 | Citation coverage across scientific domains.** The figure reports the proportion of citations captured by the full-text dataset across scientific domains. Coverage is calculated using the 2,538,028,308 citation links available in *Semantic Scholar* and is defined as the percentage of all citations to papers within a given domain for which the full text of the citing article is available and the precise location of the citation can be identified. "Agriculture" refers to Agricultural and Food Sciences, and "Environment" refers to Environmental Science.

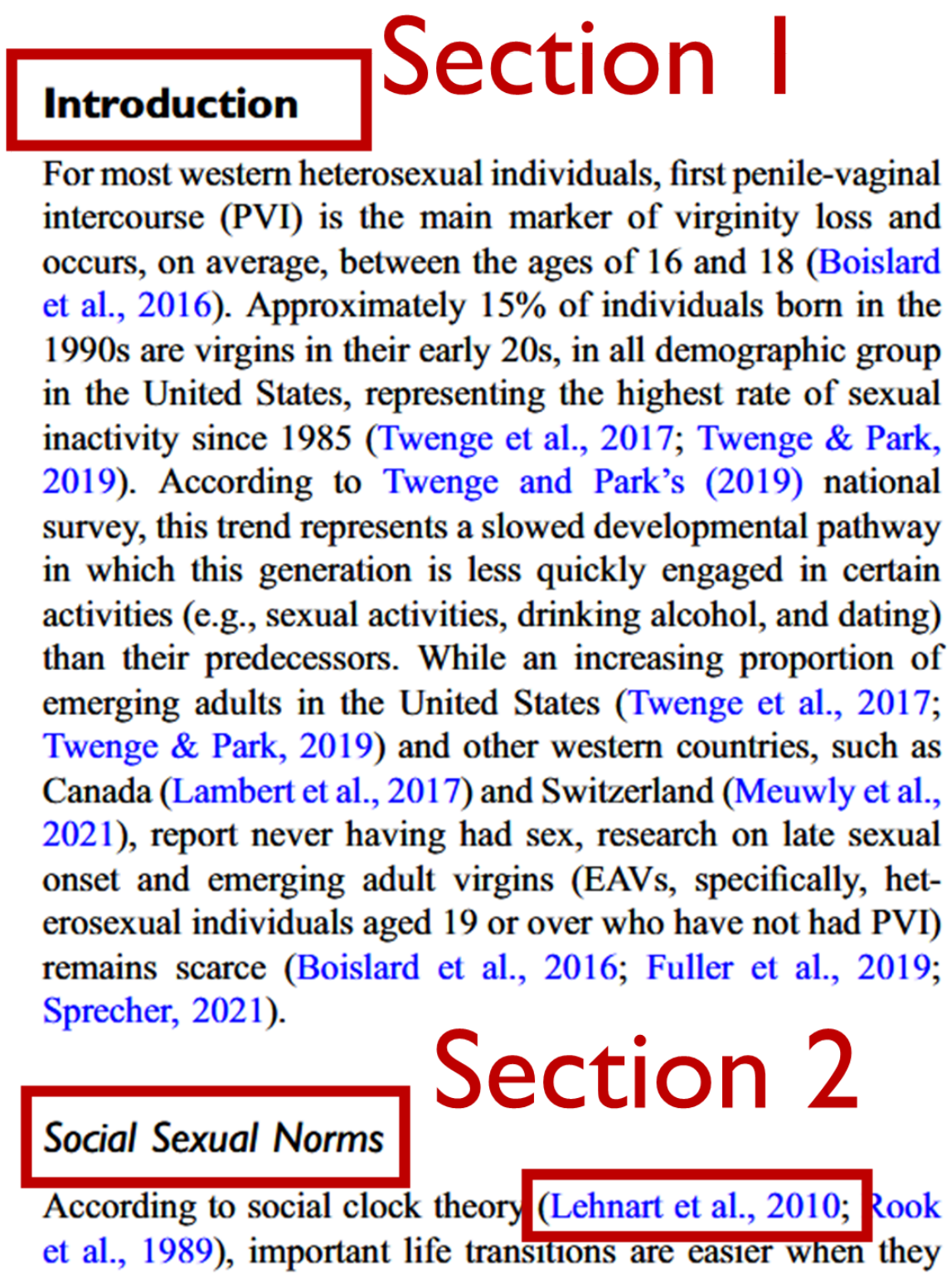

**Introduction**

For most western heterosexual individuals, first penile-vaginal intercourse (PVI) is the main marker of virginity loss and occurs, on average, between the ages of 16 and 18 (Boislard et al., 2016). Approximately 15% of individuals born in the 1990s are virgins in their early 20s, in all demographic group in the United States, representing the highest rate of sexual inactivity since 1985 (Twenge et al., 2017; Twenge & Park, 2019). According to Twenge and Park's (2019) national survey, this trend represents a slowed developmental pathway in which this generation is less quickly engaged in certain activities (e.g., sexual activities, drinking alcohol, and dating) than their predecessors. While an increasing proportion of emerging adults in the United States (Twenge et al., 2017; Twenge & Park, 2019) and other western countries, such as Canada (Lambert et al., 2017) and Switzerland (Meuwly et al., 2021), report never having had sex, research on late sexual onset and emerging adult virgins (EAVs, specifically, heterosexual individuals aged 19 or over who have not had PVI) remains scarce (Boislard et al., 2016; Fuller et al., 2019; Sprecher, 2021).

*Social Sexual Norms*

According to social clock theory (Lehnart et al., 2010; Rook et al., 1989), important life transitions are easier when they

**Fig. S8 | Example of the ambiguous section and subsection header.**

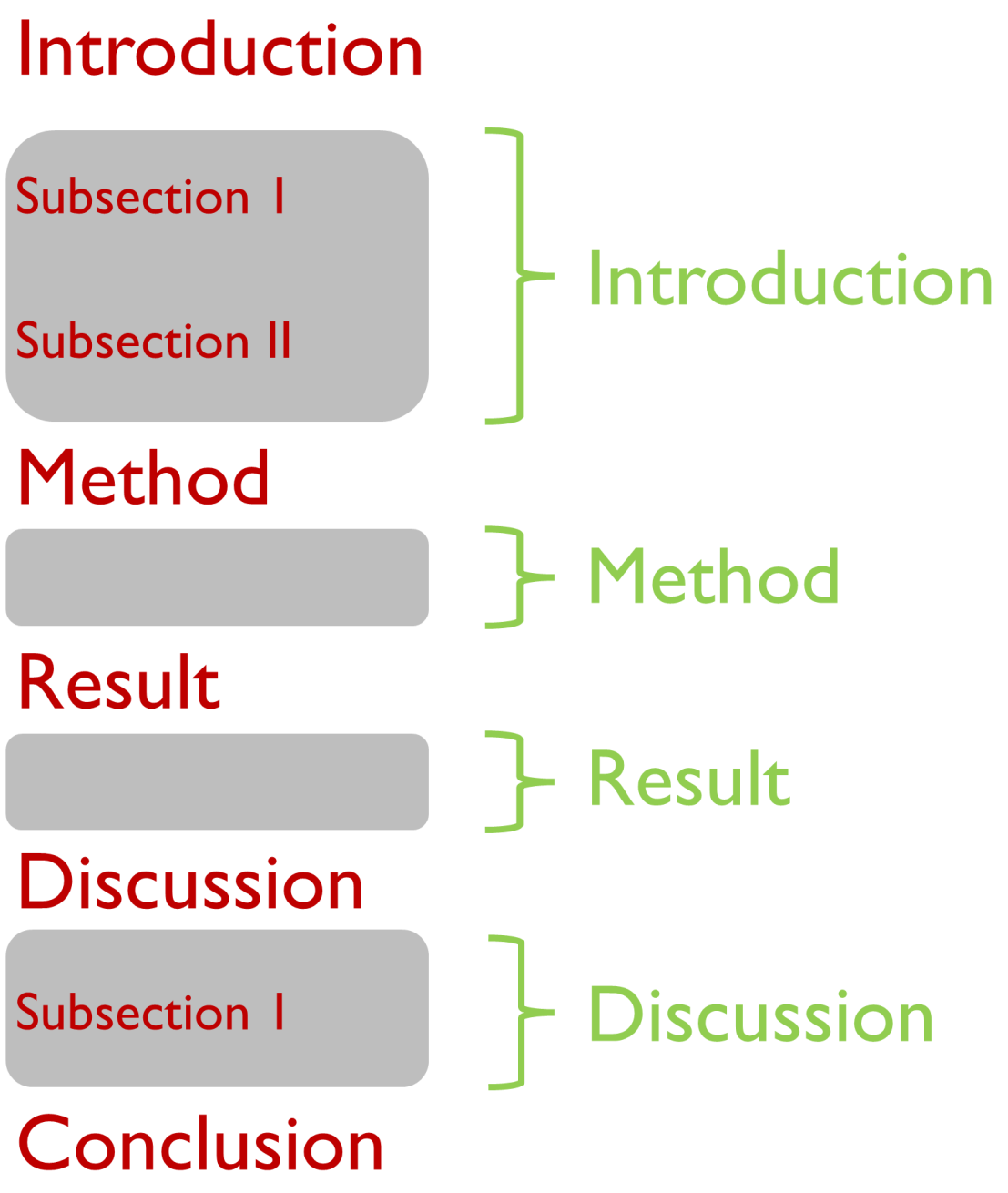

**Fig. S9 | Illustration of the identification of different sections.**

## 14. ***Mapping F, E and G Indices with Citation Purposes***

We aligned paper identifiers from the OpenAlex dataset with papers in both the NLP Citation Purpose Dataset and the *ACT2 Dataset* by converting paper titles to lowercase and performing exact title matching across datasets. A citation was considered valid only if (i) both the citing and cited papers could be uniquely matched to a single OpenAlex identifier, enabling precise identification of the citation within OpenAlex, and (ii) the corresponding citation relationship was also recorded in OpenAlex. Applying these criteria resulted in the retention of 83,494 citation links (35.6% of the original dataset) and 118,187 citation instances (37.1% of the original dataset) from the *NLP Citation Purpose Dataset*. For the *ACT2 Dataset*, we retained 1,996 citation links (50.9% of the original dataset) and 2,039 citation instances (51.0% of the original dataset). The citations in the *NLP Citation Purpose Dataset* are are made by papers in natural language processing, and we plot the distribution of domains of the citing paper for citation instances in ACT2 dataset in Fig. S10.

For each retained citation, we classified its citation type as foundational, extensional, or generalizational based on the full citation graph in OpenAlex, and we analyzed the distribution of citation-purpose categories across the three citation types (foundational, extensional, and generalizational).

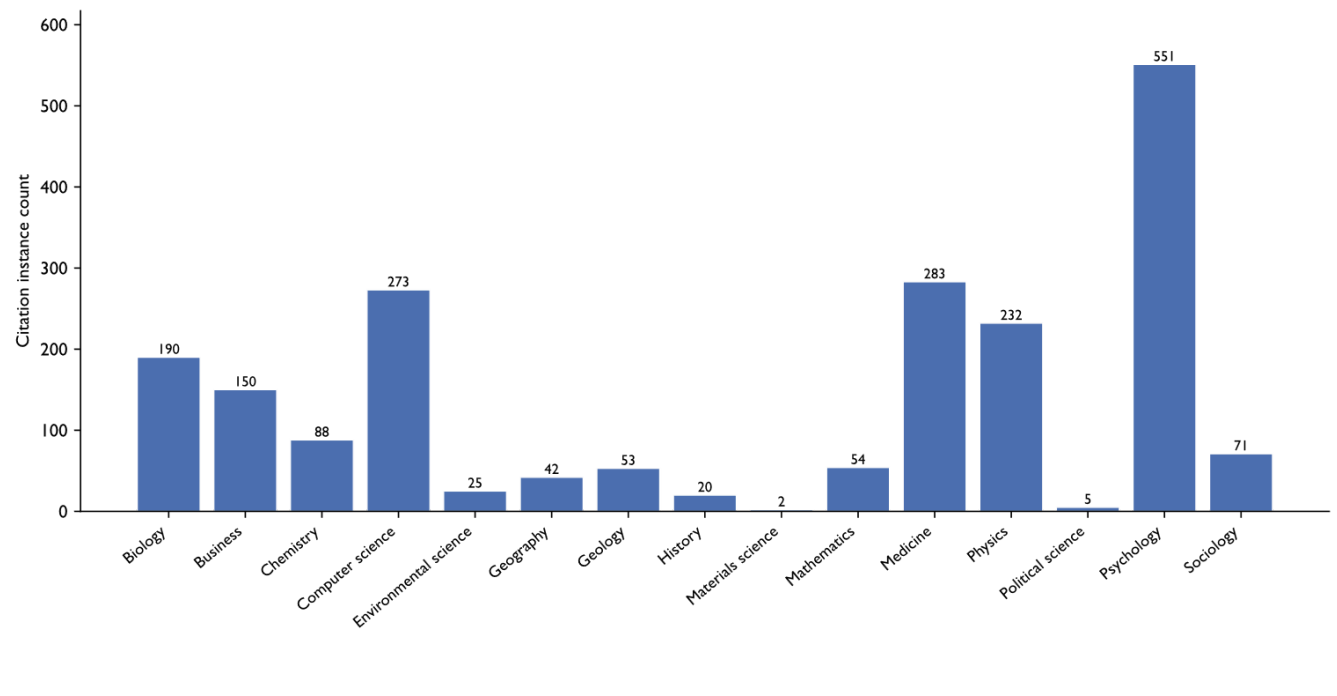

**Fig. S10 | Distribution of the domains of citing papers for the 2,039 citation instances in the ACT2 dataset that were successfully mapped to OpenAlex.** Domains were assigned based on the *top_domain* classification as described in the Supplementary Methods Section 4.

## 15. ***Identification of Papers that Introduce New Concepts***

King et al. [28] proposed an approach for identifying academic concepts based on citation patterns. Their method is founded on the assumption that an academic concept is typically introduced by a single seminal paper, after which subsequent papers adopting the concept tend to cite the original work in a highly concentrated manner. Based on this intuition, they define the concept score, **ForeCite**, as:

$$ForeCite(G_t) = max_{p \in G_t} log(f_t^p + 1) * \frac{f_t^p}{f_t}$$

where $G_t$ denotes the citation network of all papers containing term $t$, $f_t^p$ is the number of papers that both cite paper $p$ and contain term $t$, and $f_t$ is the total number of papers containing term $t$. A higher ForeCite score indicates that citations associated with term $t$ are more strongly concentrated on a single paper, suggesting that the term is more likely to represent a well-defined academic concept.

In our analysis, we extract the titles and abstracts of all English-language papers in the Semantic Scholar dataset that have received at least three citations, following the practice of King et al. [28]. From each abstract, we extract both unigram and bigram terms. We retain only bigrams with a positive pointwise mutual information (PMI) score, where PMI is defined as

$$PMI = log(\frac{P(t_1, t_2)}{P(t_1) * P(t_2)})$$

with $P(t_1)$ and $P(t_2)$ denoting the probabilities of observing terms $t_1$ and $t_2$ in the corpus, respectively, and $P(t_1, t_2)$ denoting the probability of observing the two terms together. A positive PMI indicates that the two terms co-occur more frequently than expected under independence, suggesting that they form a meaningful compound term (e.g., *machine learning*). We further exclude terms that appear in fewer than 10 papers.

Next, we compute the ForeCite concept score for each unigram and bigram using the formulation proposed by King et al. [28]. This procedure yields concept scores for 1,665,729 terms. We classify terms with a concept score greater than 0.5 as academic concepts, resulting in 362,282 identified concepts. As a robustness check, we also adopt a more stringent threshold of 2, which identifies 19,193 academic concepts.

For each paper (p) containing term (t), we additionally compute a paper-specific concept score,

$$log(f_t^p + 1) * \frac{f_t^p}{f_t}$$

which measures the extent to which paper $p$ contributed to the introduction of term $t$. A higher score indicates a stronger contribution. Based on this measure, we consider three alternative definitions of a paper that introduces an academic concept: (i) the paper with the highest paper-specific concept score for term $t$; (ii) the five papers with the highest paper-specific concept scores for term $t$; and (iii) the earliest published paper containing term $t$ that subsequently received at least three citations from papers also containing term $t$. To illustrate these identification strategies, we randomly select ten academic concepts identified using the 0.5 concept-score threshold and report the titles of the corresponding concept-introducing papers under each definition in Table S3 (only the top three papers with the highest concept scores are shown owing to space limitations).

| **Academic concept** | **Paper with the highest concept score** | **Paper with the second highest concept score** | **Paper with the third highest concept score** | **Earliest paper that introduces the concept** |
|---|---|---|---|---|
| federate learning | Communication-Efficient Learning of Deep Networks from Decentralized Data | Federated Optimization in Heterogeneous Networks | Advances and Open Problems in Federated Learning | Communication-Efficient Learning of Deep Networks from Decentralized Data |
| socioemotional wealth | Socioemotional Wealth and Business Risks in Family-controlled Firms: Evidence from Spanish Olive Oil Mills | Socioemotional Wealth in Family Firms | The Bind that Ties: Socioemotional Wealth Preservation in Family Firms | Socioemotional Wealth and Business Risks in Family-controlled Firms: Evidence from Spanish Olive Oil Mills |
| multifractal detrende | Multifractal Detrended Fluctuation Analysis of Nonstationary Time Series | Multifractal detrended cross-correlation analysis for two nonstationary signals. | Introduction to Multifractal Detrended Fluctuation Analysis in Matlab | Multifractal Detrended Fluctuation Analysis of Nonstationary Time Series |
| electric spring | Electric Springs—A New Smart Grid Technology | General Steady-State Analysis and Control Principle of Electric Springs With Active and Reactive Power Compensations | Reduction of Energy Storage Requirements in Future Smart Grid Using Electric Springs | Electric Springs—A New Smart Grid Technology |
| mixmatch | MixMatch: A Holistic Approach to Semi-Supervised Learning | ReMixMatch: Semi-Supervised Learning with Distribution Alignment and Augmentation Anchoring | DivideMix: Learning with Noisy Labels as Semi-supervised Learning | MixMatch: A Holistic Approach to Semi-Supervised Learning |
| angiomatoid nodular | Sclerosing Angiomatoid Nodular Transformation (SANT): Report of 25 Cases of a | Sclerosing Angiomatoid Nodular Transformation (SANT) of the spleen: Case report | Sclerosing angiomatoid nodular transformation of the spleen. | Sclerosing Angiomatoid Nodular Transformation (SANT): Report of 25 Cases of a |
| | Distinctive Benign Splenic Lesion | and review of the literature. | | Distinctive Benign Splenic Lesion |
| quasimap | Stable quasimaps to GIT quotients | Moduli stacks of stable toric quasimaps | Wall-crossing in genus zero quasimap theory and mirror maps | The Laumon's resolution of Drinfeld's compactification is small |
| fragrance ingredient | Criteria for the Research Institute for Fragrance Materials, Inc. (RIFM) safety evaluation process for fragrance ingredients. | Application of the expanded Creme RIFM consumer exposure model to fragrance ingredients in cosmetic, personal care and air care products | Use of an aggregate exposure model to estimate consumer exposure to fragrance ingredients in personal care and cosmetic products. | Contact allergy: predictive testing of fragrance ingredients in humans by Draize and Maximization methods. |
| move morphable | A new topology optimization approach based on Moving Morphable Components (MMC) and the ersatz material model | Doing Topology Optimization Explicitly and Geometrically—A New Moving Morphable Components Based Framework | Explicit structural topology optimization based on moving morphable components (MMC) with curved skeletons | Doing Topology Optimization Explicitly and Geometrically—A New Moving Morphable Components Based Framework |
| wiedemann spectrum | Expert consensus document: Clinical and molecular diagnosis, screening and management of Beckwith-Wiedemann syndrome: an international consensus statement | Characteristics Associated with Tumor Development in Individuals Diagnosed with Beckwith-Wiedemann Spectrum: Novel Tumor-(epi)Genotype-Phenotype Associations in the BWSp Population | The effectiveness of Wilms tumor screening in Beckwith-Wiedemann spectrum | Expert consensus document: Clinical and molecular diagnosis, screening and management of Beckwith-Wiedemann syndrome: an international consensus statement |

**Table S3 |** Randomly selected academic concepts and the papers introducing them, identified using different metrics.

## 16. *Difference-in-Differences Analysis of Papers' Online Coverage and Generalized Citation*

*The Fulltext Sources Online* dataset provides detailed records of the timing at which academic venues obtained online coverage, encompassing over 30,000 venues. We link these venues to corresponding records in OpenAlex using ISSN identifiers. To ensure the validity of the analysis, we restrict the sample to venues that obtained online coverage after 1990, marking the advent of the World Wide Web. Earlier records are excluded, as they may reflect false detections or alternative forms of access (e.g., dial-up services) that fall outside the scope of this study.

We further refine the sample by selecting venues that published at least two papers annually during the interval spanning $k$ years before and $k$ years after the introduction of online coverage. This criterion ensures sufficient publication activity to support robust pre- and post-treatment comparisons. For each treated venue, the year in which it first became accessible online is defined as the treatment year ($t$). A control venue is then randomly selected from the set of venues with recorded online coverage timing, subject to the condition that it does not receive online coverage by year $t + k$. Control venues must also meet the same publication activity requirement over the interval $[t-k, t+k]$.

For both treated and control groups, we compute the $G_5$ index for individual papers and aggregate these measures by their relative year with respect to the treatment event. The $G_5$ index is defined as the G-index calculated using citations accrued within five years post-publication. Notably, for papers published prior to the introduction of online coverage, their $G_5$ values may still be partially influenced by the treatment due to the multi-year citation accumulation window. However, papers published after the treatment are expected to exhibit the most pronounced effects of online accessibility. To estimate the causal impact of online coverage, we implement a difference-in-differences (DiD) framework. The formal specification of the model is as follows:

$$O_{tv} \sim T_v * I(t) + X_{vt} + I(t) + C(v) + C(t)$$

where $O_{tv}$ denotes the average $G_5$ index of papers published in venue $v$ in year $t$. The treatment indicator $T_v$ equals 1 if venue $v$ belongs to the treated group and 0 otherwise. The vector $X_{vt}$ includes control variables such as the average number of citations received within five years of publication, the average number of references, and the total number of papers published in venue $v$ in year $t$. The term $I(t)$ represents a set of indicator variables for time relative to the treatment event. We further include venue fixed effects $C(v)$ and year fixed effects $C(t)$ to account for unobserved heterogeneity across venues and over time. Standard errors are clustered at the venue level.

To further assess robustness, we replicate our analysis using three alternative distance-based measures: (i) the proportion of citations that are cross-domain, and the average distance between the focal paper and its citations in the (ii) reference space, and (iii) the semantic space, and the results are presented in Extended Data Fig. 9a.

**17. *Difference-in-Differences Analysis of LLM Adoption and Generalized References***

We use the *LLM adoption in arXiv, bioRxiv and SSRN dataset* provided by the authors of [4], and link these to papers in *OpenAlex* via *DOI* and *arXiv Id*. Following the methodology outlined in [4], we adopt a threshold of $\tau = 0.1$ to classify a paper as involving LLM-assisted editing, with $\tau = 0.05$ and $\tau = 0.2$ used for robustness checks. For each author in the sample, we assume that once LLM use is adopted, the author continues to employ LLMs in subsequent papers. Accordingly, authors with at least one paper identified as using LLM assistance are assigned to the treatment group, and the publication date of their first such paper is defined as the intervention time. Authors with no observed LLM adoption are assigned to the control group, for whom we randomly assign a pseudo-treatment time between January 1, 2023 (approximately corresponding to the release of ChatGPT), and June 1, 2024, the endpoint of our dataset.

We then conduct a Difference-in-Differences (DiD) analysis to estimate the effect of LLM adoption on the proportion of generalized references, defined as references in which a paper is cited without any of its own references or citing papers being co-cited. Our underlying hypothesis is that LLM use facilitates cross-domain knowledge diffusion, thereby increasing the prevalence of generalized referencing practices.

Unlike productivity measures, which can be calculated at regular time intervals (e.g., monthly) as in [4], our outcome variable is only observable at the time of publication. Consequently, the conventional time dimension used in standard DiD frameworks is not directly applicable. Instead, for each author, we order their publications chronologically and analyze the proportion of generalized references in papers published before and after the treatment event. Specifically, we index papers relative to the treatment: paper order i-1 denotes the $i^{th}$ paper published after treatment, while order -i denotes the $i^{th}$ paper published prior to treatment.

The formal specification of the DiD model is as follows:

$$O_{ao} \sim T_{a} * I(O) \ + X_{ao} + I(O) \ + C(a) \ + C(t)$$

Here, $O_{ao}$ denotes the average proportion of generalized references for author *a* at publication order *o*. The indicator $T_{a}$ equals 1 for treated authors and 0 otherwise. $X_{ao}$ controls for the average number of references, which is mechanically related to the outcome. $I(O)$ is a dummy variable indicator for publication order relative to the author's first treated paper. We include author fixed effects $C(a)$ and monthly publication-time fixed effects $C(t)$ to control for unobserved heterogeneity. Standard errors are clustered at the author level.

To ensure comparability between treatment and control groups, we restrict the sample to authors who have published at least k papers both before and after the treatment (or pseudo-treatment) period. Owing to limited sample sizes—particularly in the *bioRxiv* and Social Science Research Network (SSRN) datasets—we adopt source-specific thresholds: k=2 for *SSRN* and k=3 for both *bioRxiv* and *arXiv*. Increasing the value of k leads to a substantial reduction in sample size and a corresponding loss of statistical significance, but the direction of the estimated effects remains qualitatively consistent. Similarly, we replicate our analysis using alternative distance-based metrics, including the proportion of references that are cross-domain, and the average distance between the focal paper and its references in the reference and semantic spaces. The results are presented in Extended Data Fig. 9b-f.

**18. *Constructing a Null Model of Longitudinal Evolution of F, E, and G Indices***

The observed longitudinal trends in the foundational (F), extensional (E), and generalization (G) indices may be influenced by temporal changes in publication volume and citation behavior. To control for these potential confounding factors, we construct a randomized citation-network null model for the *OpenAlex* and *WoS* datasets, and compare its properties with those of the empirical citation network.

We first exclude all papers that lack a valid publication venue label (we use *primary_location* entity in OpenAlex and *source_title* in WoS), including papers associated with unidentified venues as well as preprint repositories or websites (e.g., *arXiv* and *Zenodo*). Preprint repositories and websites were identified through manual inspection of the 50 venues with the largest numbers of associated papers. Next, we partition all citation links in the empirical network into groups defined by the four-tuple *(cited year, cited venue, citing year, citing venue)*, which captures the publication year and venue of both the cited and citing papers. Within each group, we perform (*2*log(E)*E*) random edge swaps, where (E) denotes the number of citation links in the group. During each swap, two citation links are selected uniformly at random, and their cited papers are exchanged while preserving the citation-group membership of both links.

This rewiring procedure preserves the number of citations that each paper directs to papers published in a given year and venue, while randomizing the identities of the cited papers. As a result, the reshuffled network retains the temporal and venue-specific citation structure of the original network but removes paper-level citation preferences. Consequently, comparisons between the empirical and randomized networks enable us to assess whether the observed patterns in the F, E, and G indices arise beyond what would be expected from temporal and venue-related citation constraints alone.

We performed a random reshuffling of citation links in the citation networks derived from the OpenAlex and WoS datasets. After removing papers without valid venue information, along with all citation links involving those papers, the OpenAlex network contained 1,403,402,602 citation links, representing a reduction of 15.0% from the original network. In contrast, the WoS network contained 1,063,382,720 citation links after preprocessing, with only four citation links removed from the original network. For each dataset, we generated a single randomly reshuffled network. This choice was motivated by our experimental observations that the longitudinal trajectories of the F, E, and G indices remain highly consistent across different random reshuffling realizations, indicating that the results are robust to the specific randomization instance.

As an alternative randomization strategy, we used a more recent OpenAlex release (downloaded in 2026). We excluded papers lacking annotations for their primary topic (4,516 topics), subfield (244 subfields), or field (26 fields), as well as all citation links involving these papers. After filtering, the dataset contained 220,622,105 papers and 2,192,381,683 citation links, reduced from 248,771,620 papers and 2,227,632,465 citation links in the original dataset. We then performed constrained random reshuffling, allowing two citation links to be swapped only when the cited papers were published in the same year and belonged to the same research domain. The domain constraint was applied separately at three levels of the OpenAlex taxonomy: topic, subfield, and field.

Our approach builds upon the citation-network randomization procedure introduced in [22], with an important modification. Specifically, we constrain citation reshuffling not only by the publication years of the cited and citing papers, but also by their publication venues. This additional constraint is necessary because randomizing citations solely within publication-year groups substantially disrupts the network's underlying community structure. In exploratory analyses, such unconstrained reshuffling caused nearly all citations to be classified as generalizational, yielding an average G index greater than 0.995 for papers published in every year. By preserving venue-level citation patterns, our null model maintains a more realistic degree of community organization and disciplinary clustering, thereby providing a more appropriate baseline against which to evaluate the observed evolution of the F, E, and G indices.

Furthermore, the proposed reshuffling procedure should be regarded as a conservative baseline for assessing the increase in the G index observed after the 1990s. The null model implicitly controls for changes in citation distance over time, an increasing trend in the observed network as shown in Fig. S21. This consideration is particularly important because generalizational citations are disproportionately likely to occur between papers from distant research domains, as demonstrated in Fig. 2e. Consequently, part of the observed increase in the G index may be attributable to the growing tendency of papers to cite work from more distant areas of science. By preserving the temporal and venue-specific citation structure associated with this shift, our null model accounts for an important source of variation that could otherwise inflate estimates of the growth in generalizational citations. Therefore, any residual increase in the G index relative to the randomized baseline represents a particularly stringent test of changes in citation behavior over time.

**19. *Combining the Citation Network between OpenAlex and Semantic Scholar***

A non-trivial proportion of publications in the OpenAlex dataset contain no recorded references [1]. Although we exclude these publications from our main analysis, we construct an augmented citation network as a robustness check. Specifically, we match publications across the OpenAlex and Semantic Scholar datasets using shared DOIs, and then supplement the Semantic Scholar citation network with citation links that are present in OpenAlex but absent from Semantic Scholar. This augmentation adds 221,733,165 citations to the original 2,538,028,308 links—an increase of 8.7 percent—and reduces the share of papers with zero references from 59.1 percent to 56.1 percent. Although these differences are modest, the augmented network arguably constitutes the most comprehensive available record of citation relationships among papers. We therefore recompute the longitudinal trajectories of the F, E, and G indices using the augmented dataset, and the resulting estimates serve as a robustness check for our main findings.

The choice to augment the Semantic Scholar citation network using citation links from OpenAlex, rather than augmenting OpenAlex with links from Semantic Scholar, is largely arbitrary. Because the objective is to improve citation coverage by integrating information across the two databases, we do not expect the direction of augmentation to materially affect the results.

**20. *Constructing a Citation Network with No Self-Citation***

Self-citation practices have evolved over time within the scientific community [25], potentially influencing the temporal patterns observed in our F, E, and G indices. As a robustness validation, we construct a self-citation-free citation network using OpenAlex data, where we remove all citation links for which the citing and cited papers share at least one author.

Following this procedure, the resulting network contains 1,467,092,225 citation links, representing an 11.2% reduction from the 1,651,960,324 citation links in the original network. We then recompute the longitudinal trajectories of the F, E, and G indices using this self-citation-free network, which enables us to evaluate their temporal trends independent from the change in self-citation pattern.

**21. *Computing the Longitudinal Evolution of F, E, and G Indices by Country***

To examine the scientific progress of different countries, we extract author information for each paper from OpenAlex, including the institutional affiliations and their corresponding countries. We then compute the longitudinal evolution of the average F, E, and G indices for papers associated with each country. The analysis is restricted to papers that contain at least one reference and have accumulated a minimum of five citations within five years of publication.

### 22. *Computing the Relationship between Team Size and F, E, and G Indices*

One of the most influential empirical findings based on the disruption index is that small teams are more likely to produce disruptive work than large teams [29]. Given the close conceptual and empirical relationship between our proposed metrics and the disruption index, we revisit this result by computing the average F, E, and G indices for papers across different team sizes, as measured by the number of authors.

# Supplementary Results

### 1. *Distribution of Foundation, Extension, and Generalization Indices*

We examine the distribution of the foundation (F), extension (E), and generalization (G) indices for all papers that meet the following criteria: at least five citations within five years of publication, at least one reference, and publication between 1945 and 2019. Results are shown for OpenAlex (Fig. S11a) and Web of Science (Fig. S11b).

In *OpenAlex*, 7,359,074 papers (31.4%) have a foundation index of zero, 707,591 papers (3.0%) have an extension index of zero, 1,531,735 papers (6.5%) have an extension index of one, 3,650,045 papers (15.6%) have a generalization index of zero, and 317,012 papers (1.3%) have a generalization index of one. Across all papers, the average foundation index is 0.13 (median = 0.09), the average extension index is 0.60 (median = 0.63), and the average generalization index is 0.27 (median = 0.20).

In *Web of Science*, 5,414,253 papers (28.5%) have a foundation index of zero, 378,892 papers (2.0%) have an extension index of zero, 1,244,297 papers (6.6%) have an extension index of one, 3,335,358 papers (17.6%) have a generalization index of zero, and 141,282 papers (0.07%) have a generalization index of one. The average foundation index is 0.14 (median = 0.11), the average extension index is 0.61 (median = 0.64), and the average generalization index is 0.24 (median = 0.20).

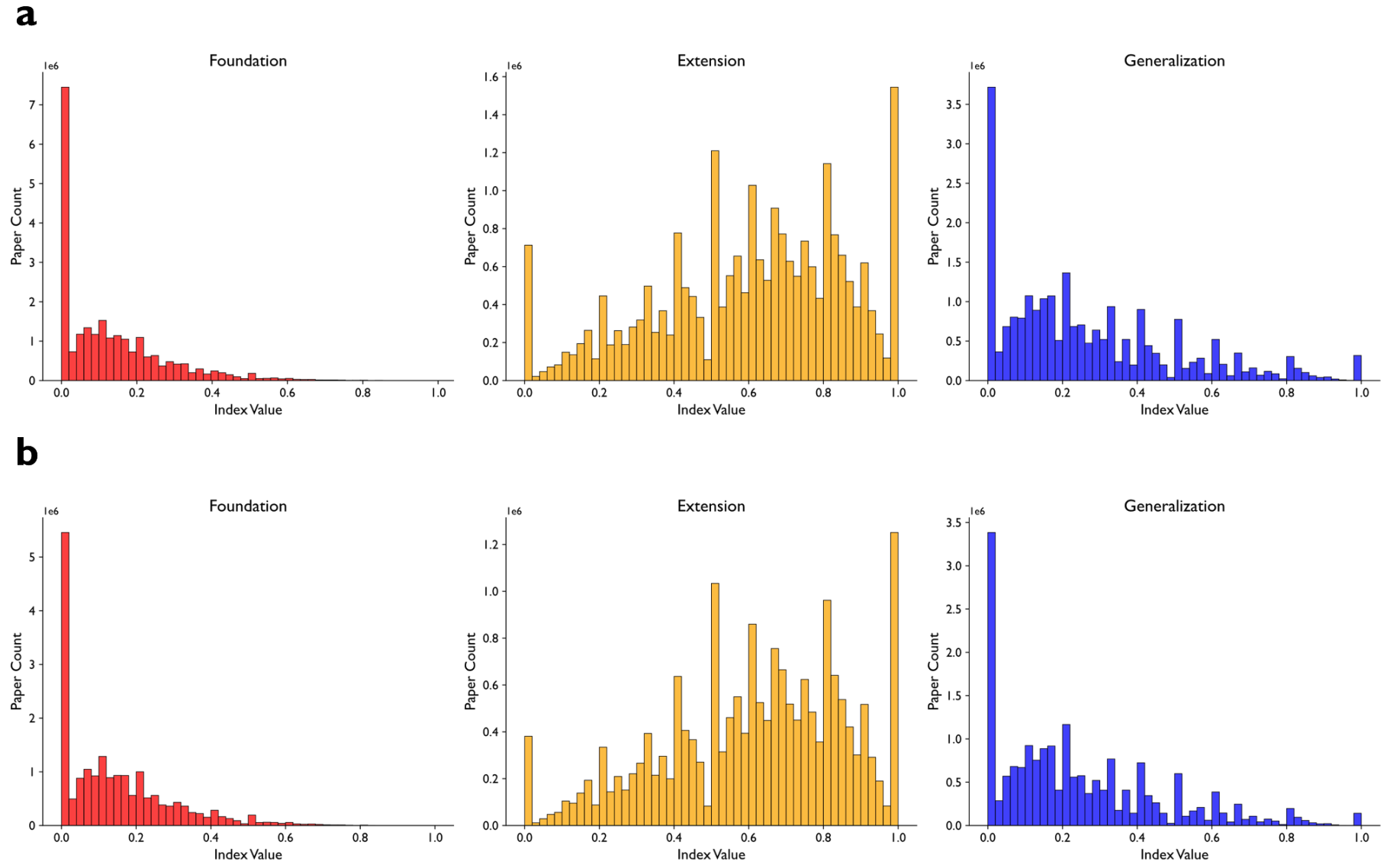

**Fig. S11 | Distribution of F, E, G index for all papers published between 1945 and 2019.** They are computed with papers that have at least one reference, and received at least five citations within five years of publication in *OpenAlex* (Panel a, 23,448,431 papers) and *Web of Science* (Panel b, 19,026,615 papers).

**2.** ***Validation of Distance-Related Properties of Foundational, Extensional, and Generalizational Citations***

Following the analysis in Fig. 2e, we use alternative metrics to evaluate the distance-related properties of foundational, extensional, and generalizational citations. The results are presented in Fig. S12.

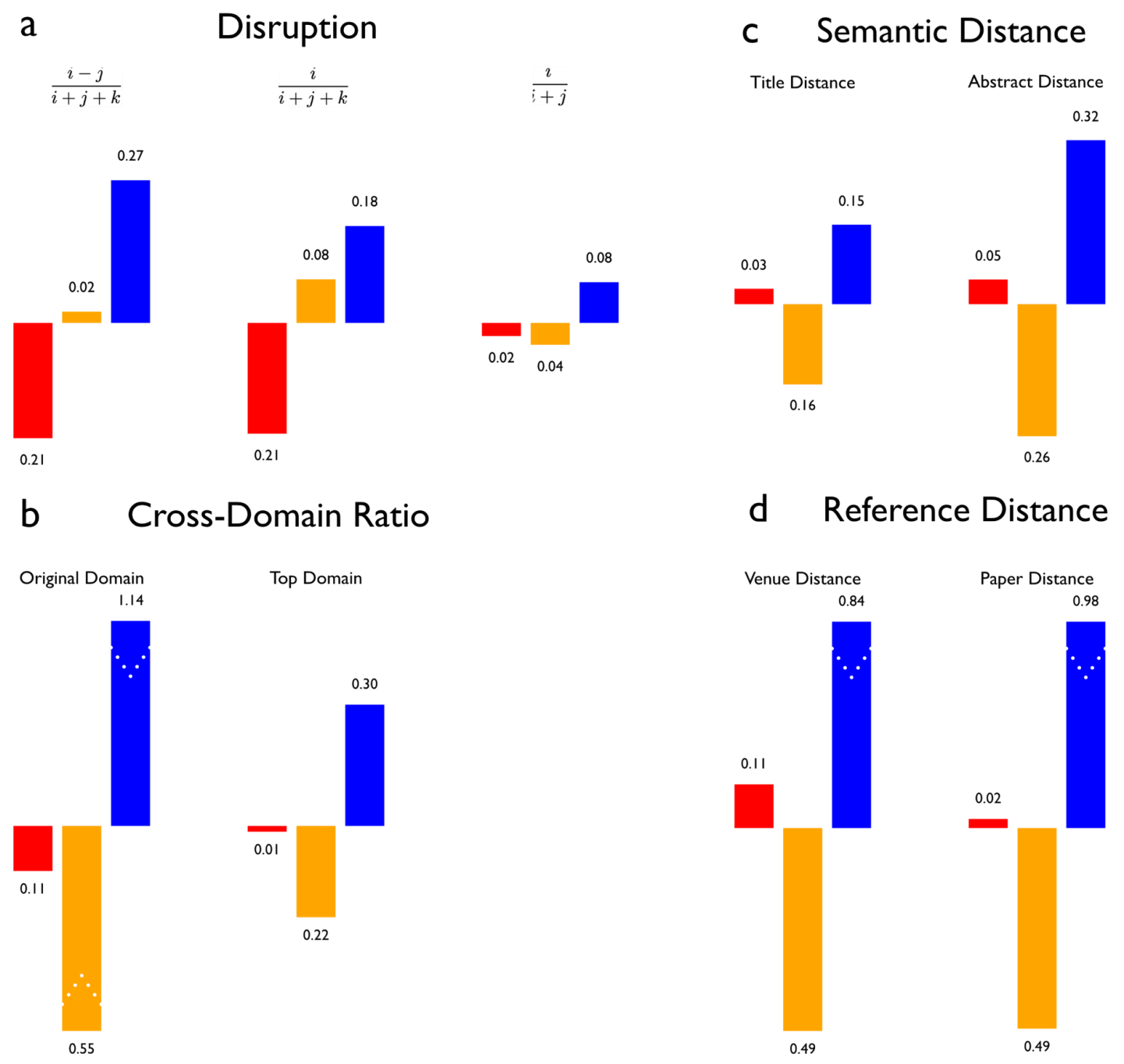

**Fig. S12 | Relationship between F, E, G citations and alternative measurements of disruption (a), the percentage of cross-domain citations (b), the semantic (c), and the reference distances (d).** Quantifies the relative difference between the average value of the F (red), E (yellow), and G (blue) indices for a given type of citation and that for the other citations (e.g, average disruption for generalizing citations and others), restricted to the citations among 159,809,661 publications in OpenAlex that are published between 1951 and 2019. Bars pointing upward indicate that the average metric for the given citation type is higher than for the other citations, and vice versa.

**3.** ***Relationship between F, E, G Indices and Likelihood of Winning a Major Award***

Panel **a** of Fig. S13 presents the estimated effects of our proposed metrics, alongside established measures of innovation, on the likelihood that a paper is associated with a Nobel Prize, controlling for the number of citations and references, publication year, and scientific domain. Across all domains examined, the F index is positively and statistically significantly associated with the probability of winning a Nobel Prize. Substantively, a one standard deviation increase in the F index corresponds to an increase in the odds of winning a Nobel Prize ranging from approximately 58% ($e^{0.459}$ -1≈0.58) to 104% ($e^{0.713}$ -1≈1.04). In contrast, the E index is consistently negatively associated with the odds of receiving a Nobel Prize across all domains. The G index and other novelty measures exhibit comparatively weak and inconsistent predictive power. These findings are consistent with our theoretical expectation that Nobel Prizes are more likely to recognize work that establishes new scientific foundations.

Panel **b** of Fig. S13 presents the effects of various metrics on the likelihood of receiving major scientific awards across disciplines, excluding the Nobel Prize, controlling for the number of citations and references, publication year, and scientific domain. In contrast to the Nobel Prize which tends to recognize more foundational research, major awards in Biology are more likely to honor highly extensional work. The G-index exhibits either no statistically significant association with award receipt or a negative association, as observed in Biology. Similarly, the disruption index and reference atypicality generally exhibit weak or statistically insignificant associations, with Medicine representing a notable exception, where both metrics are positively associated with award receipt. On the other hand, semantic atypicality is positively associated with award receipt and is statistically significant in all tested domains, suggesting that it is a robust predictor of scientific recognition across disciplines.

Taken together, the two panels reveal several noteworthy patterns. First, the Nobel Prize—arguably the most prestigious scientific distinction—tends to recognize foundational contributions that establish the intellectual basis for subsequent research, rather than work that ranks highest on measures of surprise and atypicality [22,30]. A prominent example is Albert Einstein, who was awarded the Nobel Prize for his explanation of the photoelectric effect rather than for his theory of relativity. In contrast, other major disciplinary awards recognize a broader spectrum of scientific contributions, encompassing both foundational discoveries and significant extensions of existing knowledge. This distinction suggests that scientific awards may serve complementary roles in recognizing different forms of scholarly contribution. Collectively, these awards therefore incentivize a diverse range of scientific advances, from foundational discoveries that redirect subsequent research trajectories to innovative developments that meaningfully extend established bodies of knowledge.

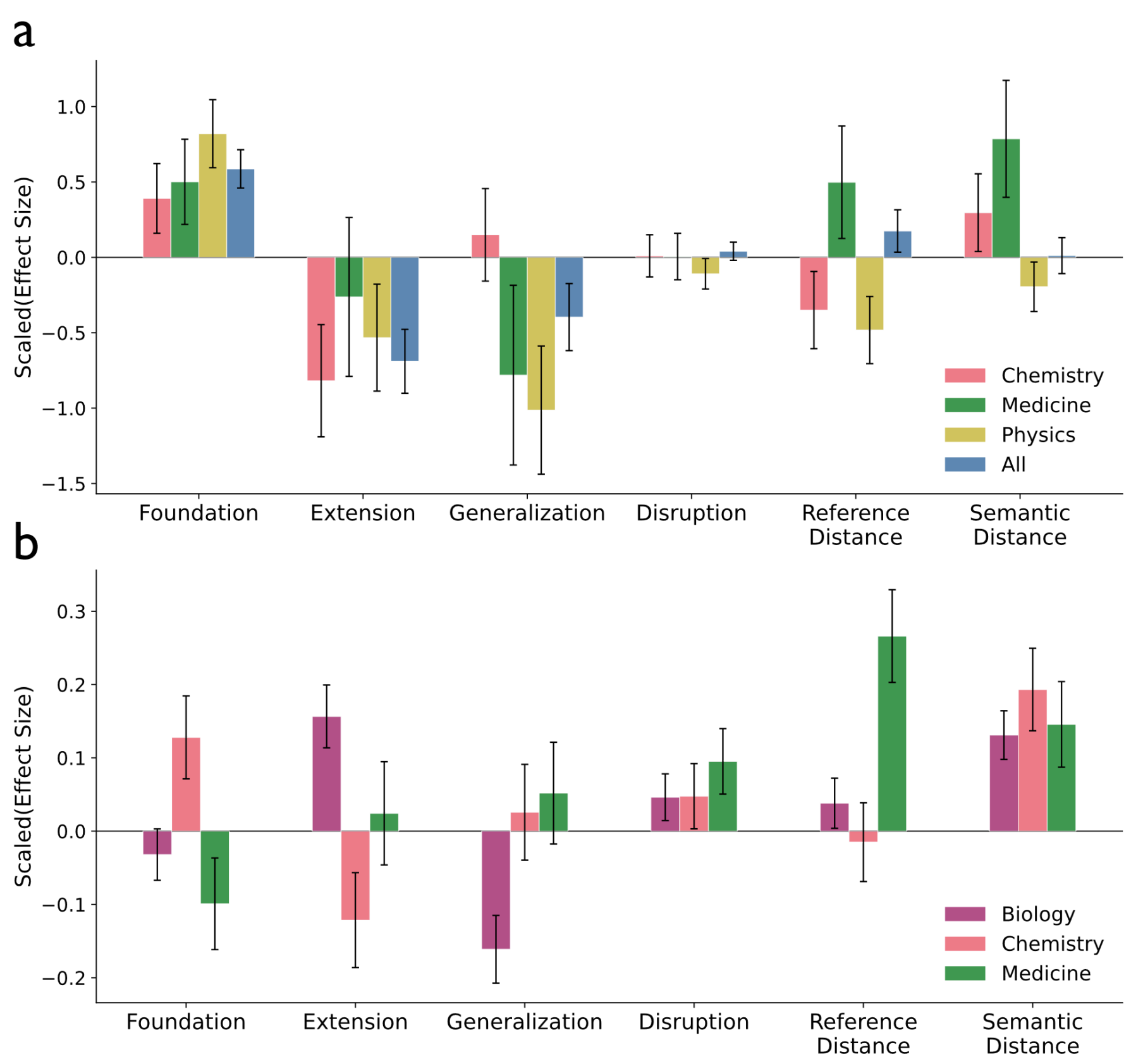

**Fig. S13 | Effects of our proposed and other novelty measurements on the likelihood of winning major awards.** Controlling for the paper's number of citations (within five years of publication), references, and the year of publication, **a,** Effect of a paper's E, F, G index, disruption index, and atypicality in reference and semantic spaces on the likelihood of the paper receiving a Nobel Prize across various domains, and **b,** Corresponding effects of other major awards, excluding the Nobel Prize, across different domains.

**4.** ***Longitudinal Change of Foundation, Extension, and Generalization Indices***

We analyze the longitudinal dynamics of the foundation (F), extension (E), and generalization (G) indices across multiple subsamples of papers (Extended Data Fig. 6), with specific sample selection criteria described in the caption. In all plots, we additionally stratify the trends by citation count to assess heterogeneity across papers of varying impact.

Across both datasets, and consistent with the main results presented in Fig. 3a, the indices exhibit robust temporal dynamics that can be broadly characterized by two distinct phases. The first phase, spanning approximately from 1950 to the early 1990s, reflects a period of field innovation. During this phase, the foundation and generalization indices decline, while the extension index rises. This pattern suggests the consolidation of disciplinary boundaries and a sustained emphasis on building upon field-defining works introduced at the outset of this period. The second phase, beginning in the 1990s and continuing to the present, reflects a shift toward system-level innovation. Here, both the foundation and extension indices decline, whereas the generalization index increases. This shift indicates that scientific contributions within individual fields increasingly exert influence beyond their home domains, serving as intellectual resources for the construction of new knowledge across diverse areas.

We further analyze yearly trends of the F, E, and G indices by domain (Extended Data Fig. 7), and we find that there is overall a consistent two-phase pattern of evolution across all fields, although notable differences are also present.

In most natural sciences (e.g., Chemistry, Biology, Medicine) and Computer Science, we observe the canonical trajectory: an increase in the extension index from 1950 to the early 1990s followed by decline, an inverted trend in the generalization index (decline until the 1990s followed by steady growth), and a persistent decrease in the foundation index. These patterns are consistent in both OpenAlex and Web of Science.

In the social sciences (e.g., Business, Sociology), the extension index increases from 1950 through the 1990s, remains relatively stable between the 1990s and 2000s, and then experiences a sharp rise until around 2010 followed by a sharp decline.

The earth sciences (e.g., Geology, Geography) display dynamics broadly similar to those of the social sciences. Extension rises rapidly from 1950 to the 1990s, stabilizes during the 1990s to 2000s, and subsequently increases until 2010 before undergoing a marked decline.

Taken together, these results highlight that while the directional shifts of F, E, and G indices are broadly consistent across fields, the timing and magnitude of these changes vary substantially across disciplinary domains.

### 5. *Longitudinal Change of Foundation, Extension, and Generalization Indices against the Null Model*

Fig. 3b presents the longitudinal evolution of the F, E, and G indices for both the observed citation network (after excluding papers and citations lacking valid venue information) and a randomly shuffled citation network in OpenAlex (upper graph) and WoS (lower graph), preserving the degree distribution across publication years and venues. Across both datasets, the G index in the randomized network declines continuously until 2010, after which it begins to increase. Notably, the turning point of the G index occurs roughly 20 years later in the randomized network than in the observed network, suggesting that factors beyond changes in publication venue and reference counts contributed to the earlier rise of the G index in the empirical citation network. Furthermore, although the G index increases after 2010 in both observed and randomized networks, the rate of increase appears steeper in the observed network. Direct comparisons of slopes should be interpreted with caution, however, because the G indices of the observed and randomized networks are of different scales. Overall, these results indicate that the increase in the G index since the 1990s (particularly between 1990 and 2010) cannot be attributed solely to changes in the number of publications or references per paper.

The foundation index (F) remains largely stable after 1965 in the randomized networks of both datasets, whereas it continues to decline in the observed networks. This divergence suggests a persistent reduction in the foundation index that cannot be explained by shifts in publication volume or reference counts alone. Similarly, the extension index (E) exhibits an earlier decline in the observed networks than in the randomized counterparts, indicating that the observed decrease in extension-oriented citation behavior is likewise independent of changes in publication and referencing practices.

Fig. S14 presents the results obtained using alternative constructions of the reshuffled network and different subsample selection strategies. These results consistently confirm the qualitative differences in the changes of the F, E, and G indices between the empirical and randomized networks. This finding suggests that the observed shift from field-level innovation to system-level innovation cannot be fully explained by changes in publication volume or other evolving citation practices.

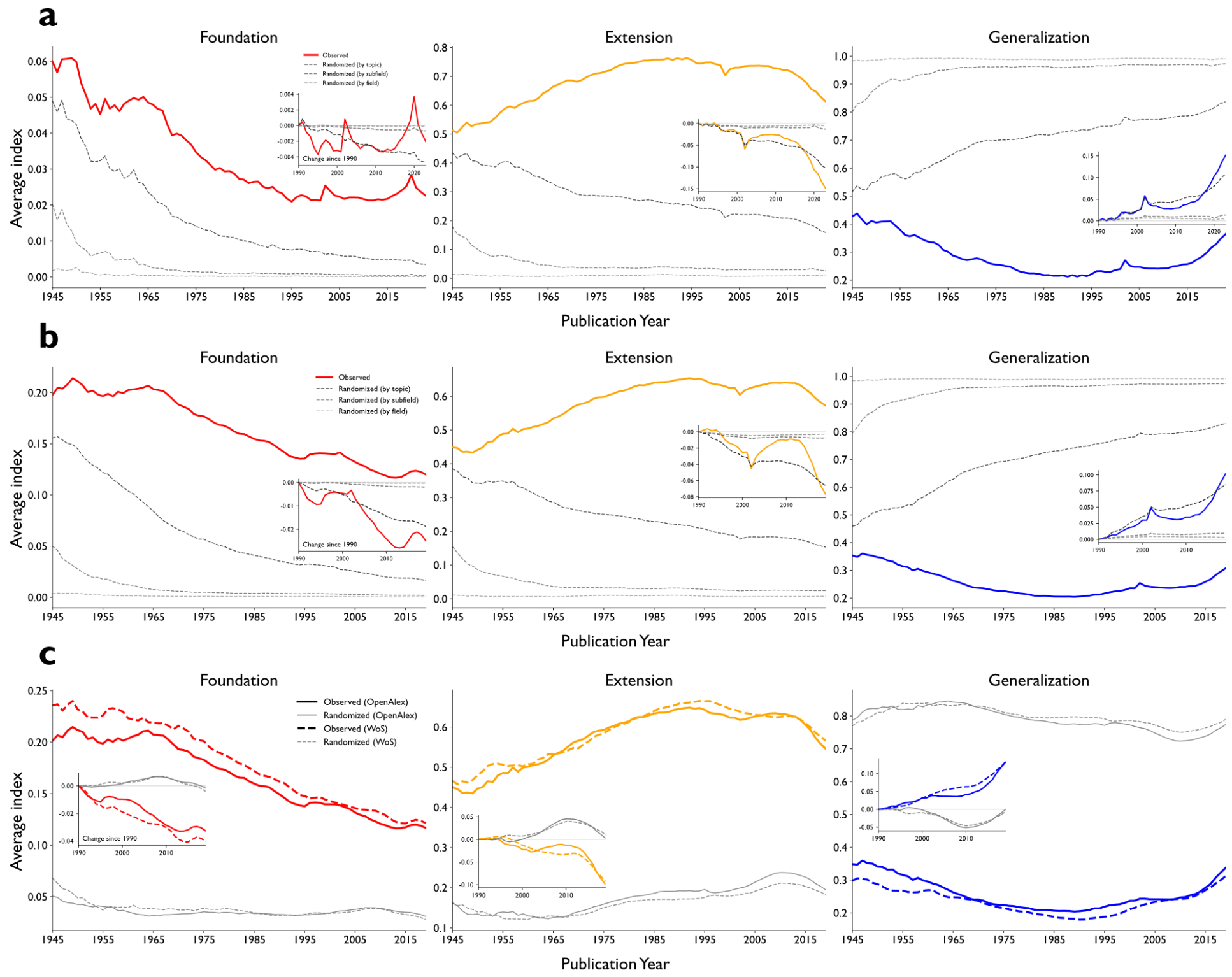

**Fig. S14 | Longitudinal changes in the F, E, and G indices in the observed and randomized citation networks. a,** Results based on 26,004,796 publications in OpenAlex (2026 version) with a valid primary topic assignment, published between 1945 and 2024, containing at

least one reference, and receiving at least two citations within one year of publication. The F, E, G indices are computed with citations received in the same one-year window after publication. The randomized networks preserve the citation distribution by publication year and across three levels of disciplinary classification (topic, subfield, and field). **b,** Results based on 26,314,324 publications in OpenAlex (2026 version) with a valid primary topic assignment, published between 1945 and 2020, containing at least one reference, and receiving at least five citations within five years of publication. The F, E, G indices are computed with citations received in the same five-year window after publication. The randomized networks preserve the citation distribution by publication year and across the same three levels of disciplinary classification. **c,** Results based on 21,060,389 publications in OpenAlex, and 18,957,843 publications in WoS that have a valid publication venue assignment, published between 1945 and 2020, containing at least one reference, and receiving at least five citations within five years of publication. The F, E, G indices are computed with citations received in the same five-year window after publication. The randomized networks preserve the citation distribution by publication year and publication venue. In all panels, the colored lines represent the longitudinal trajectories of the observed networks, whereas the gray lines represent those of the randomized networks. Insets show the relative changes in the F, E, and G indices for the observed and randomized networks after 1990, the inflection point at which the generalization index begins to increase.

### 6. *Semantic Validation of the Foundation, Extension, and Generalization Indices*

We validate the interpretation of the foundation, extension, and generalization indices by examining the frequency of word appearances in paper titles. As shown in Extended Data Fig. 2, we partition all papers into ten equal-sized bins based on their scores in each index, and then calculate the proportion of papers that contain a given word in their titles across bins.

**Reusable components.** Words associated with reusable components (e.g., *tool*) tend to appear more frequently in titles of papers with higher generalization scores. For example, the word *software* appears in only 0.096% of papers in the bottom decile of the generalization distribution (95% CI 0.092%-0.010%), but rises by 251% to 0.337% in the top decile (95% CI 0.330%-0.345%, $t$=56.2, $p < 2\times10^{-16}$, $df$ = 3579305.2 with Welch's $t$-test). The foundation index shows a weaker and more heterogeneous effect. For instance, the word *device* increases in prevalence from 0.304% in the bottom decile (95% CI 0.296%-0.311%) to 0.543% in the top decile (95% CI 0.534%-0.552%, $t$=39.8, $p < 2\times10^{-16}$, $df$ = 4346342.7 with Welch's $t$-test), whereas the word *tool* shows only a negligible rise, from 0.349% (95% CI 0.341%-0.356%) to 0.369% (95% CI 0.361%-0.376%, $t$=3.6, $p = 2.7\times10^{-4}$, $df$ = 4686036.3 with Welch's $t$-test). By contrast, highly extensional papers are substantially less likely to include such terms: the appearance rate of *device*, *tool*, and *software* each decreases by at least 60% from the bottom to the top decile of the extension index.

**Review-related words.** Terms characteristic of review-type papers (*review*, *guideline*, *tutorial*) are strongly associated with generalization. Each exhibits at least a 269% increase in appearance likelihood from the bottom to the top generalization decile. Conversely, their prevalence declines as papers move toward higher foundation or extension scores.

**Words related to the creation of science foundation.** The presence of the terms *observation* and *dataset* in paper titles appears to correlate with the emergence of new scientific foundations. On average, *observation* appears in 0.334% (95% CI: 0.327%–0.342%) of papers in the bottom foundation decile, compared to 0.532% (95% CI: 0.523%–0.541%) in the top decile—representing a statistically significant increase of 59.3% ($t$ = 32.6, $p < 2 \times 10^{-16}$, df = 4458904.5). Similarly, the term *dataset* appears in 0.040% (95% CI: 0.037%–0.043%) of papers in the bottom decile, rising to 0.059% (95% CI: 0.056%–0.062%) in the top decile, corresponding to a 47.5% increase (t = 9.3, $p < 2 \times 10^{-16}$, df = 4521251.8). These patterns suggest that the introduction of novel observations (e.g., the observation of gravitational waves) or the creation of new datasets (e.g., ImageNet) may serve as foundational elements that open the door to new scientific exploration.

**Innovation-related words.** Words reflecting novelty (*new*, *novel*, *innovative*) are most often found in foundational or generalized papers. Foundational papers show higher rates of *new* (1.910%, 95% CI 1.892%-1.927% → 2.424%, 95% CI 2.404%-2.444%, $t$=38.3, $p < 2\times10^{-16}$, $df$ = 4627328.6 with Welch's $t$-test) and *novel* (1.131%, 95% CI 1.117%-1.145% → 1.671%, 95% CI 1.655%-1.687%, $t$=49.8, $p < 2\times 10^{-16}$, $df$ = 4526100.7 with Welch's $t$-test), while generalized papers are more likely to include *innovative* (0.020%, 95% CI 0.018%-0.022% → 0.129%, 95% CI 0.124%-0.133%, $t$=43.1, $p < 2\times10^{-16}$, $df$ = 3060068.6 with Welch's $t$-test). All three terms are least common among highly extensional papers, though nonlinear patterns emerge. For example, the prevalence of *new* decreases from 2.280% (95% CI 2.262%-2.300%) in the bottom decile of extension to 1.821% (95% CI 1.803%-1.838%) in the 50–60% decile , before rebounding slightly to 2.000% (95% CI 1.981%-2.017%) in the top decile.

**Analytical refinement.** Words denoting analytical refinements (*theory*, *metric*, *hypothesis*) appear more frequently in extensional papers but less frequently in generalized ones. For instance, the proportion of papers containing *theory* increases from 0.639% (95% CI 0.630%-0.650%) in the bottom decile of extension to 1.393% (95% CI 1.377%-1.407%, $t$=81.4, $p < 2\times10^{-16}$, $df$ = 4131578.8 with Welch's $t$-test) in the top decile. In contrast, *theory* appears in 1.372% (95% CI 1.356%-1.387%) of papers in the bottom decile of generalization but only 0.572% (95% CI 0.563%-0.582%, $t$=88.3, $p < 2\times10^{-16}$, $df$ = 4020814.6 with Welch's $t$-test) in the top decile.

### 7. *Alternative Metrics for Validating Foundation, Extension, and Generalization at the Citation Level*

The computation of the foundation, extension, and generalization indices at the paper level relies on the identification of corresponding citation links. To validate these classifications, we compare them with other established metrics that quantify the "interdisciplinarity" of citations. Fig. S12 illustrates these comparisons. For each citation type (i.e., foundational, extensional, or generalizational), we compute the average value of all citations of a given type, and compare it against the average value for all remaining citations. The relative difference is expressed as $dif = \frac{M - \overline{M}}{\overline{M}}$, where $M$ denotes the mean value for the focal citation group and $\overline{M}$ the mean for the rest (all values are positive across the metrics computed in our sample).

**Disruption.** In Fig. S12a, we examine three variants of disruption. We find that generalizational citations are consistently more "disruptive" than others. For example, using the original disruption index $\frac{i-j}{i+j+k}$, the average disruption of generalizational citations is 0.33744 (95% CI: 0.33740–0.33747), compared with 0.26670 (95% CI: 0.26668–0.26671) for the remaining ones, representing a significant 27% increase. The results for foundational citations depend on the specific formulation of disruption. When k (the total citations to the reference of the focal paper) is included in the denominator, foundational citations are significantly less disruptive than others by a large margin (0.24433 vs. 0.31096, 95% CIs:

0.24430–0.24436 and 0.31093–0.31098, respectively). When k is excluded (using $D = \frac{i}{i+j}$, the difference remains but is far smaller (0.88788 vs. 0.91009, a 2% difference). Extensional citations exhibit only small differences relative to the baseline across all disruption variants.

**Cross-domain citation.** In Fig. S12b, we assess interdisciplinarity using two domain-identification schemes. The "Original Domain" metric defines a paper's domain as the set of all assigned level-0 concepts, and a citation is classified as cross-domain if the citing and cited papers share no overlap. The "Top Domain" metric uses only the highest-scoring domains (see Methods), with overlap again determining whether a citation is cross-domain. Both approaches yield qualitatively similar results: generalizational citations are substantially more likely to cross domain boundaries, while extensional and foundational citations tend to remain within-domain, and such effect is strongest for extensional links. For example, under the Top Domain metric, 42.773% of generalizational citations are cross-domain (95% CI: 42.768%–42.779%), compared to 33.006% of other citations (95% CI: 33.002%–33.009%). In contrast, only 30.172% of extensional citations are cross-domain (95% CI: 30.167%–30.177%), compared with 38.819% for non-extensional citations (95% CI: 38.816%–38.823%).

**Semantic distance.** In Fig. S12c, we measure the semantic distance between citing and cited papers at the time of citation using dynamic text embeddings (see *Supplementary Methods*). Considering both title- and abstract-based semantic distances, we find that, on average, generalizational citations connect semantically distant papers (average title distance 0.15363, with 95% CI 0.15362-0.15364 for generalizational citation, versus 0.13303, 95% CI 0.13302-0.13303 for non-generalizational citations; average abstract distance 0.04662, with 95% CI 0.04661-0.04664 for generalizational citation, versus 0.03534, 95% CI 0.03533-0.03534 for non-generalizational citations), followed by foundational citations (average title distance 0.14168, with 95% CI 0.14167-0.14169 for foundational citation, versus 0.13753, 95% CI 0.13753-0.13754 for non-foundational citations; average abstract distance 0.03968, with 95% CI 0.03967-0.03969 for foundational citation, versus 0.03786, 95% CI 0.03785-0.03787 for non-foundational citations), whereas extensional citations tend to connect semantically proximate papers (average title distance 0.12395, with 95% CI 0.12395-0.12396 for extensional citation, versus 0.14690, 95% CI 0.14689-0.14690 for non-extensional citations; average abstract distance 0.03174, with 95% CI 0.03173-0.03175 for extensional citation, versus 0.04274, 95% CI 0.04273-0.04275 for non-extensional citations).

**Reference distance.** Similarly, in Fig. S12d, we compute the reference distance between citing and cited papers at the time of citation using dynamic embeddings of papers and publication venues (see *Methods*). Across all three distance measures, a consistent pattern emerges: generalizational citations link papers that are distant (average venue distance 0.14720, with 95% CI 0.14718-0.14721 for generalizational citation, versus 0.08000, 95% CI 0.07999-0.08000 for non-generalizational citations; average paper distance 0.11112, with 95% CI 0.11111-0.11114 for generalizational citation, versus 0.05615, 95% CI 0.05614-0.05615 for non-generalizational citations), extensional citations link proximate papers (average venue distance 0.06098, with 95% CI 0.06097-0.06099 for extensional citation, versus 0.11988, 95% CI 0.11987-0.11989 for non-extensional citations; average paper distance 0.04433, with 95% CI 0.04432-0.04433 for extensional citation, versus 0.08618, 95% CI 0.08617-0.08619 for non-extensional citations), and foundational citations occupy an intermediate position (average venue distance 0.10220, with 95% CI 0.10219-0.10221 for foundational citation, versus 0.09240, 95% CI 0.09239-0.09241 for

non-foundational citations; average paper distance 0.07011, with 95% CI 0.07010-0.07012 for foundational citation, versus 0.06859, 95% CI 0.06859-0.06860 for non-foundational citations).

Taken together, these results demonstrate that the foundation, extension, and generalization classifications align with established structural properties of citations: generalizational links are more likely to cross disciplinary boundaries, and connect more distant ideas; extensional links remain within established domains and closer neighborhoods; and foundational links occupy an intermediate position between the generalizational and extensional links.

### 8. *Regression Validation of the Longitudinal Change in Foundation, Extension, and Generalization*

Because the foundation, extension, and generalization indices of a paper may be confounded by its number of references and received citations, we conduct regression analyses to adjust for these factors. Specifically, we regress each index on *Year since X* (the difference between a paper's publication year and a fixed baseline year), while controlling for reference count and citation count. To capture temporal heterogeneity, we split the sample at the identified phase transition point (approximately 1991, see Fig. 3a). Results for papers published between 1945 and 1990 are reported in Table S4, and those for papers published between 1991 and 2019 are reported in Table S5.

Overall, the findings corroborate the longitudinal patterns reported in the main text. During the earlier phase (1945–1990), the foundation and generalization indices exhibit significant declines, with the foundation index decreasing by an average of 0.0009 per year and the generalization index by 0.0008 per year, while the extension index increases by 0.002 per year. In contrast, in the later phase (1991–2019), the generalization index rises markedly (0.006 per year), accompanied by declines in both the foundation (0.0005 per year) and extension (0.005 per year) indices. These results demonstrate that the temporal dynamics of the indices remain robust even after accounting for citation- and reference-based confounders.

| Independent Variables | Dependent Variables | | |
|---|---|---|---|
| | Foundation | Extension | Generalization |
| Reference (log) | -0.075 *** | 0.178 *** | -0.103 *** |
| | (0.005) | (0.003) | (0.003) |
| Citation (log) | 0.124 *** | -0.089 *** | -0.035 *** |
| | (0.005) | (0.006) | (0.004) |
| Year since 1945 | -0.0009 *** | 0.002 *** | -0.0008 *** |
| | (0.00009) | (0.0002) | (0.0002) |
| Top Domain | X | X | X |
| Observations | 2,677,596 | 2,677,596 | 2,677,596 |
| Adjusted R2 | 0.283 | 0.293 | 0.195 |
| Note: | | | *p<0.05; **p<0.01; ***p<0.001 |

**Table S4 | Regression analysis on the longitudinal change of foundation, extension, and generalization Index, Phase I (1945-1990).** The regression analysis is run on all papers having at least one reference, five citations (within five years of publication), and published between 1945 and 1990.

| Independent Variables | Dependent Variables | | |
|---|---|---|---|
| | Foundation | Extension | Generalization |
| Reference (log) | -0.059 *** | 0.177 *** | -0.118 *** |
| | (0.004) | (0.007) | (0.007) |
| Citation (log) | 0.083 *** | -0.048 *** | -0.035 *** |
| | (0.004) | (0.005) | (0.003) |
| Year since 1991 | -0.0005 ** | -0.005 *** | 0.006 *** |
| | (0.0002) | (0.0004) | (0.0003) |
| Top Domain | X | X | X |
| Observations | 20,654,793 | 20,654,793 | 20,654,793 |
| Adjusted R2 | 0.239 | 0.263 | 0.221 |
| Note: | | | *p<0.05; **p<0.01; ***p<0.001 |

**Table S5 | Regression analysis on the longitudinal change of foundation, extension, and generalization Index, Phase II (1991-2019).** The regression analysis is run on all papers having at least one reference, five citations (within five years of publication), and published between 1991 and 2019.

### 9. *Regression Validation of the Relationship Between F, E, and G Indices and Cross-Paper Distances*

As shown in Fig. 2e, generalizational papers tend to be cited by papers that are more distant in both semantic and reference space, followed by foundational papers, and, lastly, extensional papers. We further validate this finding through regression analysis, reported in Table S6.

After controlling for the number of references and citations, we find that the effect is most pronounced for reference distance. On average, each one-standard-deviation increase in cited reference distance corresponds to a 0.408 standard-deviation increase in the generalization index and a 0.019 standard-deviation increase in the foundation index, while decreasing the extension index by 0.387 standard deviations. In contrast, the effect of cited semantic distance is weaker. The only statistically significant relationship is with the foundation index, where each standard-deviation increase in cited semantic distance is associated with a 0.007 standard-deviation increase in the foundation index—a much smaller effect compared to that of reference distance.

We also replicate the regression analysis across domains, with the estimated coefficients presented in Fig. S15, Panels c and d. The results reveal a consistent pattern for reference distance: being cited by papers with more distant references is strongly associated with a higher generalization index, followed by the foundation index, and a lower extension index. The relationship with semantic distance is more nuanced. In most domains, being cited by semantically distant papers is positively associated with the generalization index, although computer science, economics, mathematics, and psychology represent notable exceptions. Moreover, in 10 of the 19 domains, semantically distant citations exert a larger effect on the foundation index than on the extension index, whereas in 8 domains the opposite holds. In one domain, the coefficients are statistically indistinguishable.

Overall, our findings confirm that papers with a high generalization index are most likely to be cited by papers that are distant in reference space, followed by papers with a high foundation index, whereas papers with a high extension index are more often cited by papers that are local in reference space. The relationship between the F, E, and G indices and the semantic distance of citing papers is more complex. Although the majority of evidence indicates that generalizational papers are cited by the most semantically distant papers, followed by foundational papers and, lastly, extensional papers, the effect size is substantially smaller than that observed for reference distance. Moreover, considerable variation exists across domains, suggesting that the semantic-distance effect is less robust and context-dependent.

| Independent Variables | Dependent Variables | | |
|---|---|---|---|
| | Foundation (Scaled) | Extension (Scaled) | Generalization (Scaled) |
| Reference (log) | -0.446 *** | 0.475 *** | -0.249 *** |
| | (0.008) | (0.008) | (0.005) |
| Citation (log) | 0.644 *** | -0.217 *** | -0.148 *** |
| | (0.016) | (0.005) | (0.006) |
| Cross-paper Semantic (Title) Distance (Scaled) | 0.007 * | -0.011 | 0.008 |
| | (0.003) | (0.008) | (0.008) |
| Cross-paper Reference (Venue) Distance | 0.019 *** | -0.387 *** | 0.408 *** |
| (Scaled) | | | |
| | (0.003) | (0.008) | (0.010) |
| Publication Year | X | X | X |
| Observations | 22,502,446 | 22,502,446 | 22,502,446 |
| Adjusted R2 | 0.240 | 0.339 | 0.297 |
| Note: | | | *p<0.05; **p<0.01; ***p<0.001 |

**Table S6 | Relationship between foundation, extension, and generalization index and the cross-paper distances.** The regression analysis is run on all papers having at least one reference, five citations (within five years of publication), and published between 1951 and 2019.

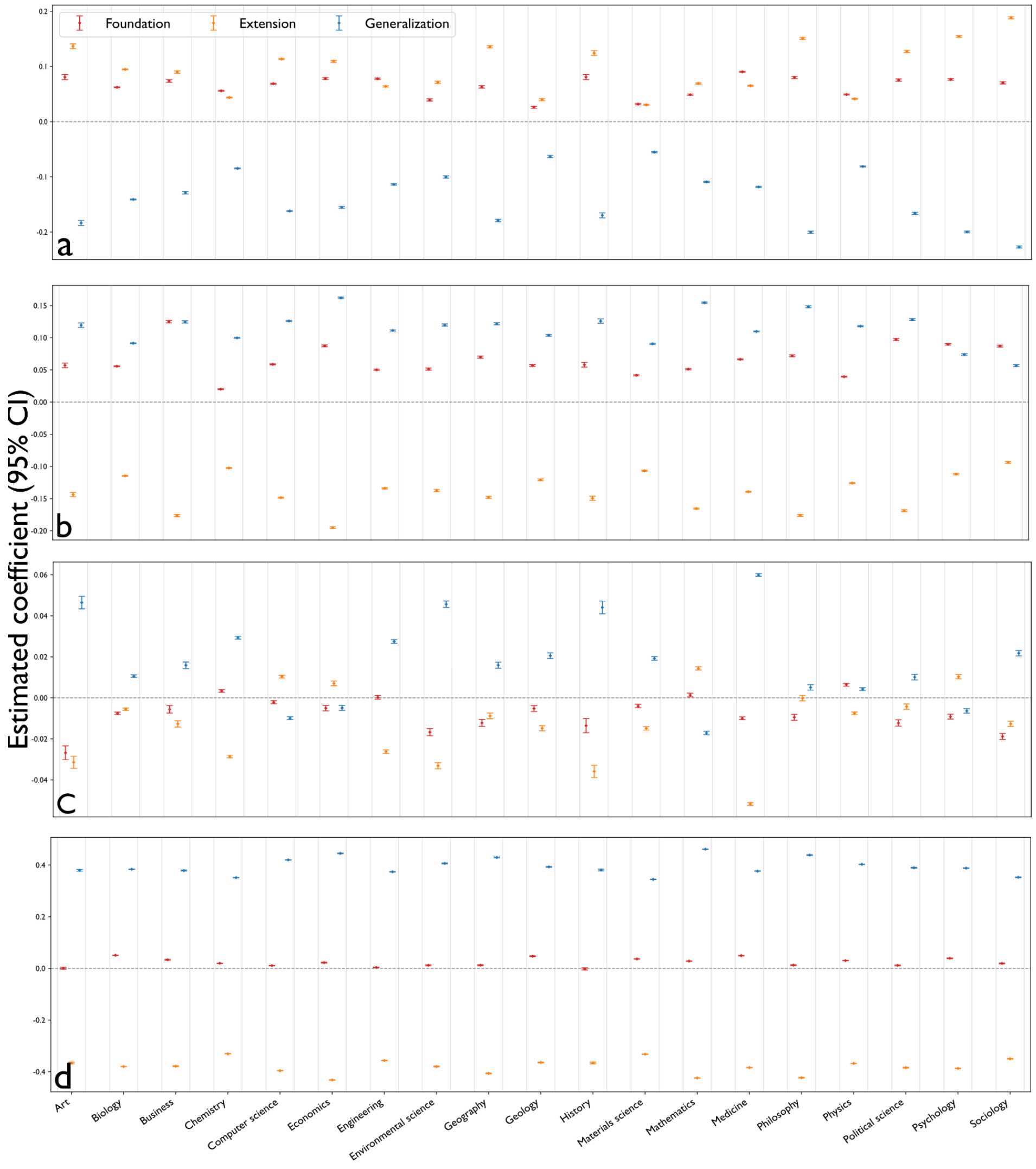

**Fig. S15 | Estimated coefficients of within- and cross-paper semantic and reference distances on the F, E, and G indices across domains. a,** Within-paper semantic distance. **b,** Within-paper reference distance. **c,** Cross-paper semantic distance. **d,** Cross-paper reference distance. Model specifications are identical to those reported in Table S6 (cross-paper distance) and Table S7 (within-paper distance), with both outcome variables (F, E, and G indices) and distance measures standardized to enable comparison of effect sizes.

## 10. *Validation of the Relationship Between Indices and Within-Paper Distances*

We report in panel b of Fig. 2 a central finding of this study: generalization papers tend to employ words that are semantically proximate to one another while citing works that are distant in the reference space. By contrast, extensional papers draw upon semantically distant words while referencing a proximate pool of works. To validate this pattern, we employ alternative measures of semantic and reference distance to examine the relationship between a paper's foundation (F), extension (E), and generalization (G) indices and its within-paper semantic and reference distances. The results, presented in Fig. S16, are consistent with the main finding reported in Fig. 2b, with all alternative metrics exhibiting similar patterns. A complementary regression analysis in Table S6 further substantiates these findings: on average, a one–standard deviation increase in within-paper semantic distance is associated with significant increases in the F index (0.085 standard deviations) and the E index (0.068 standard deviations), but with a 0.124 standard deviation decrease in the G index. Conversely, a one–standard deviation increase in reference distance corresponds to a 0.04 standard deviation increase in the F index, a 0.123 standard deviation increase in the G index, and a 0.135 standard deviation decrease in the E index. Together, these results reaffirm the core pattern identified above.

Finally, we extend the regression analysis presented in Table S7 to individual domains, with results reported in Fig. S15 (Panels a and b). Across most domains, extensional papers employ the most semantically distant words, followed by foundational papers and then generalizational papers. By contrast, in the reference space, generalizational papers tend to cite the most distant works, followed by foundational papers and finally extensional papers. The effects of both semantic and reference distance on the F index are consistently positive, indicating that foundational papers are characterized by simultaneously high within-paper semantic distance and high within-paper reference distance.

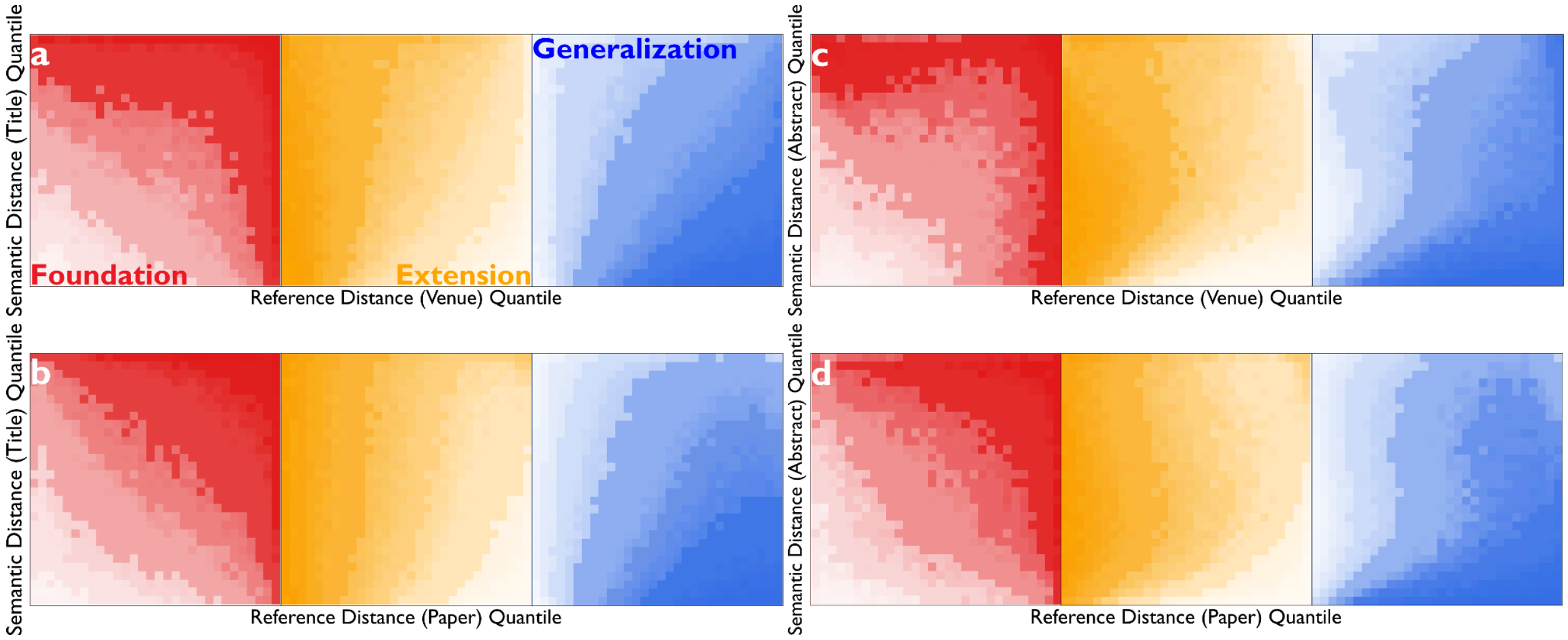

**Fig. S16 | Relationship between F,E,G indices and within-paper semantic and reference distances.** Displays association between papers' F (red), E (yellow), and G (blue) indices and their average within-paper distances. **a**, Association computed using title semantic distance and venue-based reference distance, computed based on 22,674,571 publications in OpenAlex that are published between 1951 and 2019, have at least five citations within five years post publication, and at least one reference with identified venues and two valid English tokens in their titles. **b**, Association computed using title semantic distance and paper-based reference distance, computed based on

22,674,571 publications in OpenAlex that are published between 1951 and 2019, have at least five citations within five years post publication, and at least one reference with identified venues and two valid English tokens in their titles. **c**, Association computed using abstract semantic distance and venue-based reference distance, computed based on 13,722,721 publications in OpenAlex that are published between 1986 and 2019, that have at least five citations within five years post publication and at least one reference with identified venues, and an abstract with at least two valid English tokens within. **d**, Association computed using title semantic distance and paper-based reference distance, computed based on 13,722,721 publications in OpenAlex that are published between 1986 and 2019, that have at least five citations within five years post publication and at least one reference with identified venues, and an abstract with at least two valid English tokens within. Opacity reflects the median value of the corresponding index across papers with different distances. All plots adjusted for publication year, number of references, and citation count. Details of the data and methods underlying these plots can be found in the SI *Supplementary Results: Validation of the Relationship Between Indices and Within-Paper Distances*.

| Independent Variables | Dependent Variables | | |
|---|---|---|---|
| | Foundation (Scaled) | Extension (Scaled) | Generalization (Scaled) |
| Reference (log) | -0.519 *** | 0.457 *** | -0.188 *** |
| | (0.022) | (0.031) | (0.033) |
| Citation (log) | 0.646 *** | -0.209 *** | -0.156 *** |
| | (0.015) | (0.006) | (0.006) |
| Valid Reference (log) | 0.037 * | 0.286 *** | -0.332 *** |
| | (0.019) | (0.028) | (0.025) |
| Valid Token (log) | -0.125 *** | -0.155 *** | 0.242 *** |
| | (0.005) | (0.033) | (0.036) |
| Within-paper Semantic (Title) Distance (Scaled) | 0.085 *** | 0.068 *** | -0.124 *** |
| | (0.002) | (0.008) | (0.009) |
| Within-paper Reference (Venue) Distance (Scaled) | 0.040 *** | -0.135 *** | 0.123 *** |
| | (0.0006) | (0.009) | (0.010) |
| Publication Year | X | X | X |
| Observations | 22,493,287 | 22,493,287 | 22,493,287 |
| Adjusted R2 | 0.245 | 0.234 | 0.182 |

| Note: | *p<0.05; **p<0.01; ***p<0.001 |
|---|---|

**Table S7 | Relationship between foundation, extension, and generalization index and the within-paper distances.** The regression analysis is run on all papers having at least one reference, five citations (within five years of publication), and published between 1951 and 2019.

### 11. *Evaluating the Breadth and Concentration of Influence*

Fig. 2c presents the breadth and concentration of influence of foundational, extensional, and generalizational papers. Overall, foundational and extensional papers exhibit relatively narrow and concentrated patterns of influence, whereas generalizational papers exhibit broader and more diffuse influence across both research topics and countries.

At the topic level, among papers receiving 10–15 citations, foundational papers are cited by an average of 15.64 topics (95% CI, 15.62–15.65), with the most-citing topic accounting for 70.06% (95% CI, 70.00%–70.11%) of all citations. Extensional papers are cited by an average of 13.97 topics (95% CI, 13.96–13.98), with the top-citing topic contributing 77.50% (95% CI, 77.45%–77.54%) of citations. In contrast, generalizational papers are cited by an average of 18.61 topics (95% CI, 18.60–18.63), while the top-citing topic accounts for only 58.46% (95% CI, 58.41%–58.51%) of citations. The same pattern persists among highly cited papers. For example, among papers receiving 1,000–1,500 citations, foundational papers are cited by an average of 270.41 topics (95% CI, 260.02–280.89), with the top-citing topic accounting for 65.04% (95% CI, 63.33%–66.69%) of citations. Extensional papers are cited by an average of 317.03 topics (95% CI, 305.95–329.33), with the leading topic accounting for 58.19% (95% CI, 56.33%–59.93%) of citations. By comparison, generalizational papers are cited by an average of 450.29 topics (95% CI, 433.89–467.13), while the top-citing topic contributes only 39.69% (95% CI, 37.87%–41.56%) of citations.

A similar pattern is observed at the country level, indicating that generalizational papers also have broader and more geographically diffuse influence. Among papers receiving 10–15 citations, foundational papers are cited by authors affiliated with an average of 5.98 countries (95% CI, 5.97–5.99), with the top-citing country accounting for 62.83% (95% CI, 62.77%–62.89%) of citations. Extensional papers are cited by an average of 6.65 countries (95% CI, 6.64–6.66), with the leading country contributing 57.82% (95% CI, 57.76%–57.87%) of citations. In contrast, generalizational papers are cited by authors from an average of 6.79 countries (95% CI, 6.78–6.80), while the top-citing country accounts for only 51.42% (95% CI, 51.36%–51.48%) of citations. This pattern likewise holds for highly cited papers. Among papers receiving 1,000–1,500 citations, foundational papers are cited by authors from an average of 59.83 countries (95% CI, 58.18–61.58), with the top-citing country accounting for 48.00% (95% CI, 46.55%–49.52%) of citations. Extensional papers are cited by authors from an average of 63.43 countries (95% CI, 61.66–65.48), with the leading country accounting for 44.12% (95% CI, 42.58%–45.78%) of citations. In comparison, generalizational papers are cited by authors from an average of 69.26 countries (95% CI, 66.99–71.39), while the top-citing country contributes only 42.59% (95% CI, 40.79%–44.51%) of citations.

Taken together, these findings indicate that foundational and extensional papers tend to exert influence within a relatively concentrated set of topics and countries, whereas generalizational papers diffuse their influence across a substantially broader range of scientific domains and geographic regions.

**12.** ***Alternative Metrics for Assessing Effect of Online Venue Coverage on the Distant Diffusion of Papers***

Extended Data Fig. 9a presents the difference-in-differences (DiD) analysis, where the outcome variables include the ratio of cross-domain citations (left panel), the average distance of citations from the focal papers in reference space (middle panel), and in semantic space (right panel). Although many of the estimated effects are statistically insignificant or only marginally significant (i.e., at the 0.1 significance level), all distance-based citation metrics exhibit a consistent upward trend. This pattern suggests that online coverage may facilitate the cross-domain diffusion of scientific publications.

**13.** ***Alternative Metrics for Evaluating the Impact of LLM Adoption on the Use of Distant References***

We adjust the threshold for LLM adoption detection to $\tau = 0.05$ and $\tau = 0.2$, and present the corresponding results from the same difference-in-differences (DiD) analysis in Panels b and c of Extended Data Fig. 9. Additionally, we modify the outcome variables to include: (i) the ratio of cross-domain references in papers, and (ii) the average distance between the focal paper and its references in both reference and semantic space. The estimated coefficients for these specifications are reported in Panel d-f of Extended Data Fig. 9.

Overall, we identify a pattern consistent with our main findings in Fig. 4c for the outcomes based on reference and semantic space distances: following the adoption of LLMs, authors tend to cite more distant references in their subsequent papers. In contrast, the effects on the percentage of cross-domain references are mixed and largely statistically insignificant. One possible explanation is that our coarse-grained classification of domains (e.g., *Computer Science* versus *Biology*) may not adequately capture more nuanced shifts toward distant referencing.

**14.** ***Different Types of Citations and Their Purposes to Cite***

Fig. 2f presents the distribution of citation purposes for foundational, extensional, generalizational, and disruptive citations. To assess the robustness of these patterns, Figs. S17-S18 present the corresponding distributions using an alternative categorization of citation purposes and citations derived from different datasets.

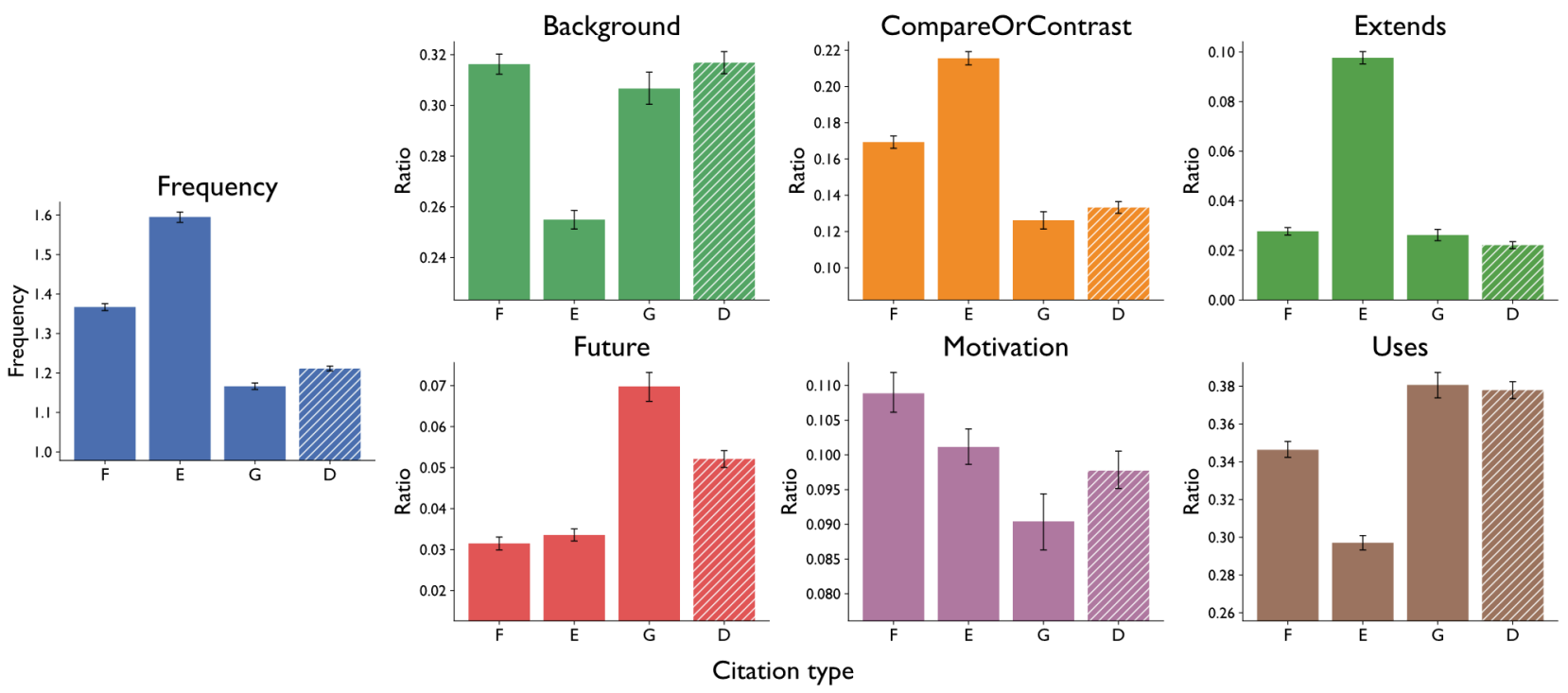

**Fig. S17 | Citation purposes across the four citation-context categories among natural language processing papers: foundational, extensional, generalizational, and disruptive**. This figure presents the same analysis as Fig. 2f, except that the original citation-purpose categories are shown. Specifically, the *CompareOrContrast* and *Extends* categories are presented separately

rather than being merged into the *Intent to Modify* category.

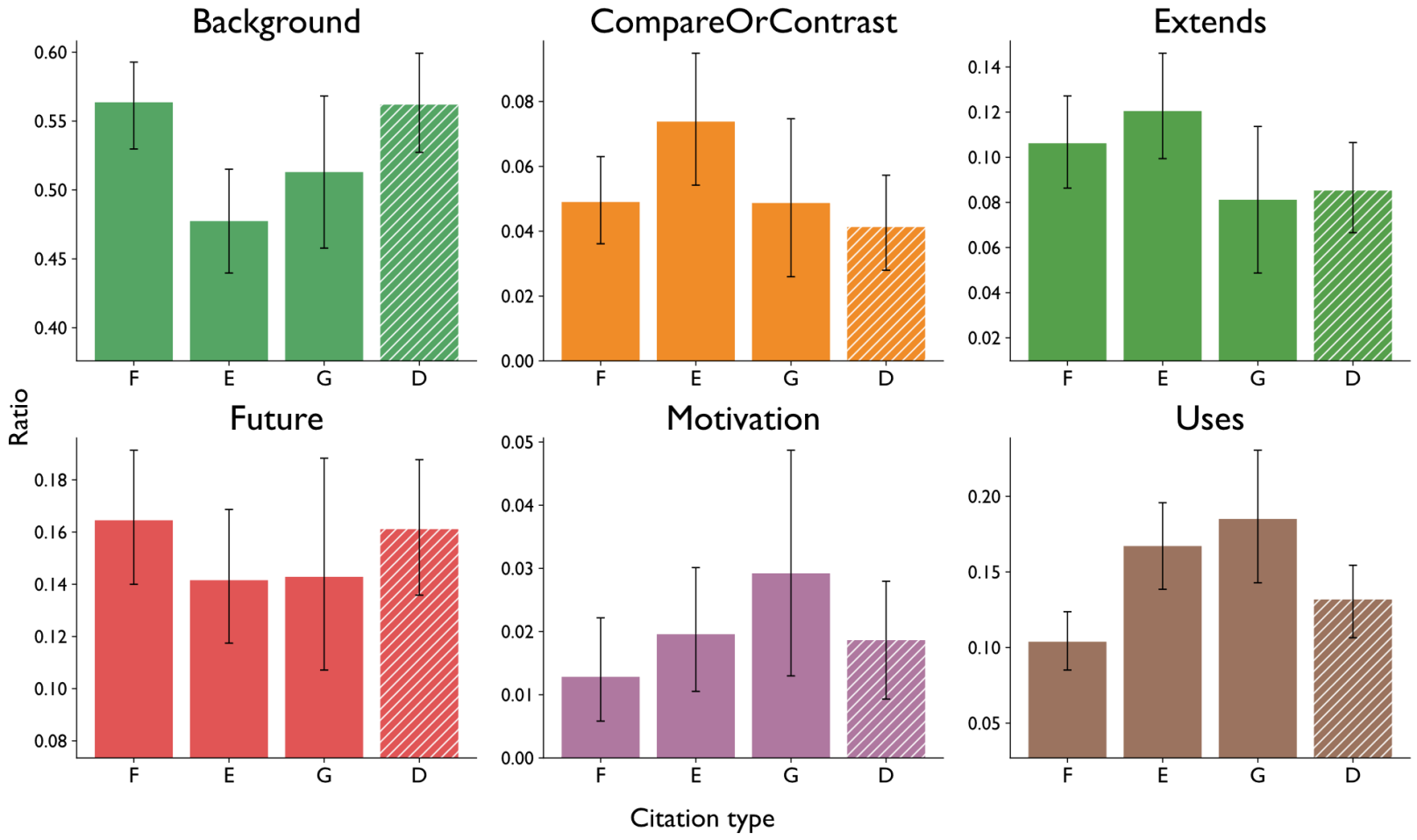

**Fig. S18 | Citation purposes across the four citation-context categories in papers from diverse domains: foundational, extensional, generalizational, and disruptive.** Ratio denotes the distribution of citation purposes across reference instances for citations in the four citation-context categories. Citation purposes were identified using a machine learning classifier that leverages the textual context surrounding each citation together with bibliographic features. The classifier was trained on the publicly available ACT2 citation-purpose dataset [8], comprising 1,792 citation links and 1,829 citation instances. The analysis was restricted to citation links for which both the citing and cited papers could be matched to OpenAlex records and for which citation-purpose annotations were available.

### 15. *Longitudinal Trends in F, E, and G Indices for Different Types of Papers*

First, we use keywords appearing in paper titles as proxies for paper types (e.g., papers containing the term *review* are predominantly review articles) and examine changes in their prevalence over time. The selected keywords are drawn from Fig. 2d, where our prior analyses indicate whether papers containing a given keyword are more likely to be foundational, extensional, or generalizational in nature. This analysis assesses the extent to which the observed temporal shifts in the F, E, and G indices can be explained by changes in the composition of published research, given that different types of papers tend to exhibit systematically different F, E, and G profiles.

As shown in Fig. 4a, keywords associated with more generalizational forms of research, such as *review*, *tool*, and *software*, have become increasingly prevalent over time. For example, the proportion of papers containing the keyword *review* increased from 0.103% (95% CI: 0-2.57%) in 1945 to 0.459% (95% CI: 0.449%-0.469% ) in 2019, representing a 345.6% increase. In contrast, keywords more closely associated with foundational and extensional research, such as *theory* and *observation*, exhibit declining trends. For example, the proportion of papers containing keyword *theory* decreased from 1.386% (95% CI: 0.873%-1.899%) in 1945 to 0.596% (95% CI: 0.585%-0.608%) in 2019, representing a 60.0% decrease. Taken together, these patterns suggest that shifts in the relative prevalence of different paper types may contribute substantially to the observed changes in the average F, E, and G indices.

The evidence presented in Extended Data Fig. 8 further indicates that temporal changes within papers of the same type may also contribute to the observed shifts in the average F, E, and G indices, although the magnitude of these within-type changes appears considerably smaller than that associated with shifts in the relative prevalence of different paper types. Specifically, we calculate the average F, E, and G indices within subsamples of papers containing the same title keyword, thereby approximately holding paper type constant. We find that these indices continue to exhibit temporal changes within keyword-defined paper types. For example, among papers containing the word *review* in the title, the average generalization index increased from 0.318 (95% CI: 0.281-0.358) in 1945 to 0.380 (95% CI: 0.374-0.386) in 2019, representing a 19.5% increase.

Part of these within-type changes may, in turn, reflect broader shifts in the scientific publication and citation environment. To assess this possibility, we compare the empirical trends with those observed in randomized citation networks that preserve temporal changes in publication volume, citation inflation, and the number and composition of publication venues. Under this randomized benchmark, the temporal increase in the generalization index becomes substantially less pronounced. Specifically, the average generalization index among review papers increased from 0.618 (95% CI: 0.576-0.660) in 1945 to 0.644 (95% CI: 0.638-0.651) in 2019, corresponding to a 4.2% increase. These results suggest that the long-run shifts in the F, E, and G indices reflect both changes in the composition of published research and changes occurring within comparable paper types, with some portion of the latter potentially attributable to broader changes in publication volume, citation practices, and the structure of publication venues.

Using keywords in paper titles as proxies for paper types provides an imperfect approximation, as the presence of particular keywords may not correspond precisely to the underlying nature of the research. Moreover, changes in keyword prevalence over time may partly reflect evolving conventions in scientific writing and terminology rather than substantive shifts in the composition of research. These limitations highlight the need for further analyses using more direct and comprehensive measures of paper types to evaluate the robustness of the reported patterns. Nevertheless, the current evidence suggests that the observed temporal shifts in the F, E, and G indices may reflect both changes in the composition of different types of papers and changes in these indices within the same keyword-defined categories, with compositional shifts appearing to account for the larger effect.

## 16. *Longitudinal Trends in F, E, and G Indices in the OpenAlex-Semantic Scholar Combined Network*

We construct an *OpenAlex-Semantic Scholar* combined citation network to increase the dataset coverage, and compute the longitudinal trends in F, E and G indices in this network with results shown in Fig. S19. The result presents a similar pattern to Fig. 3a.

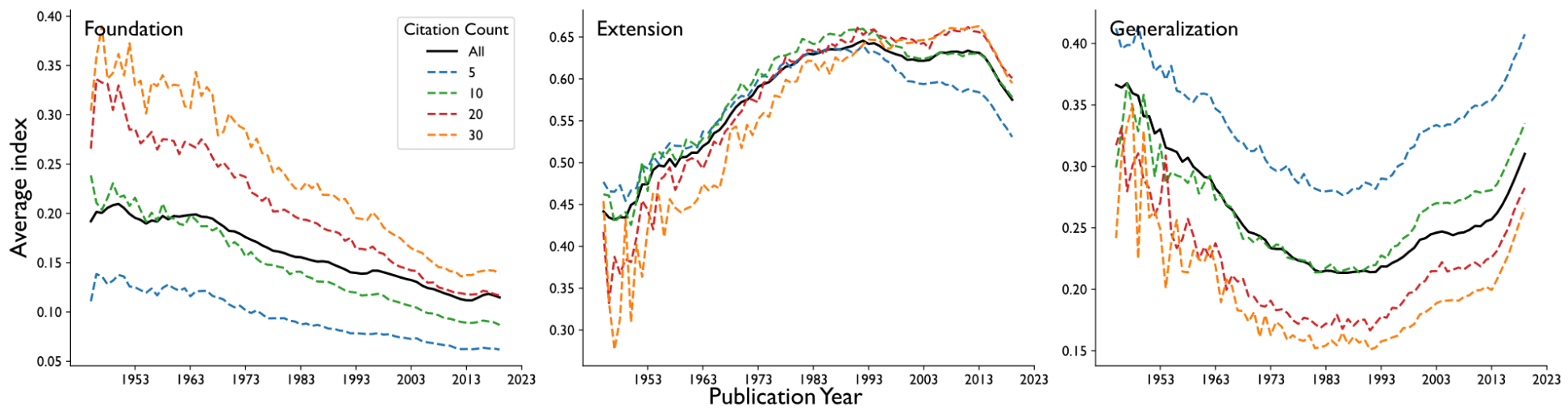

**Fig. S19 | Longitudinal change in the F, E, and G indices computed on the combined OpenAlex–Semantic Scholar citation**

**network.** The analysis is restricted to the 31,001,497 papers that have at least one reference, received at least five citations within five years of publication, and were published no later than 2019.

## 17. *Longitudinal Trends in F, E, and G Indices in the Self-Citation Free Network*

Fig. S20 presents the longitudinal trends in F, E and G indices in a self-citation free network, where we found that the trends in the F, E and G indices align with the main result reported in Fig. 3a.

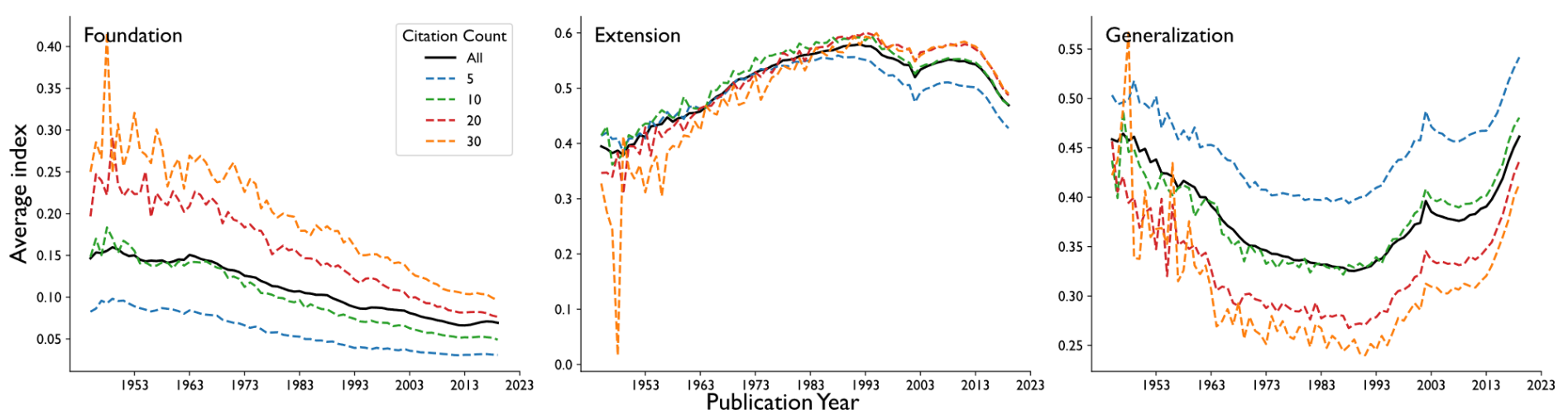

**Fig. S20 | Longitudinal change in the F, E, and G indices computed on the citation network in OpenAlex after removing self-citation links.** The analysis is restricted to the 19,368,030 papers that have at least one reference, received at least five citations within five years of publication, and were published no later than 2019.

## 18. *Longitudinal Trends in Reference Atypicality*

We plot the average atypicality of papers published across different time periods using a fixed venue-level z-score for each subplot, as described in the Supplementary Methods Section 11. The results are presented in Fig. S21.

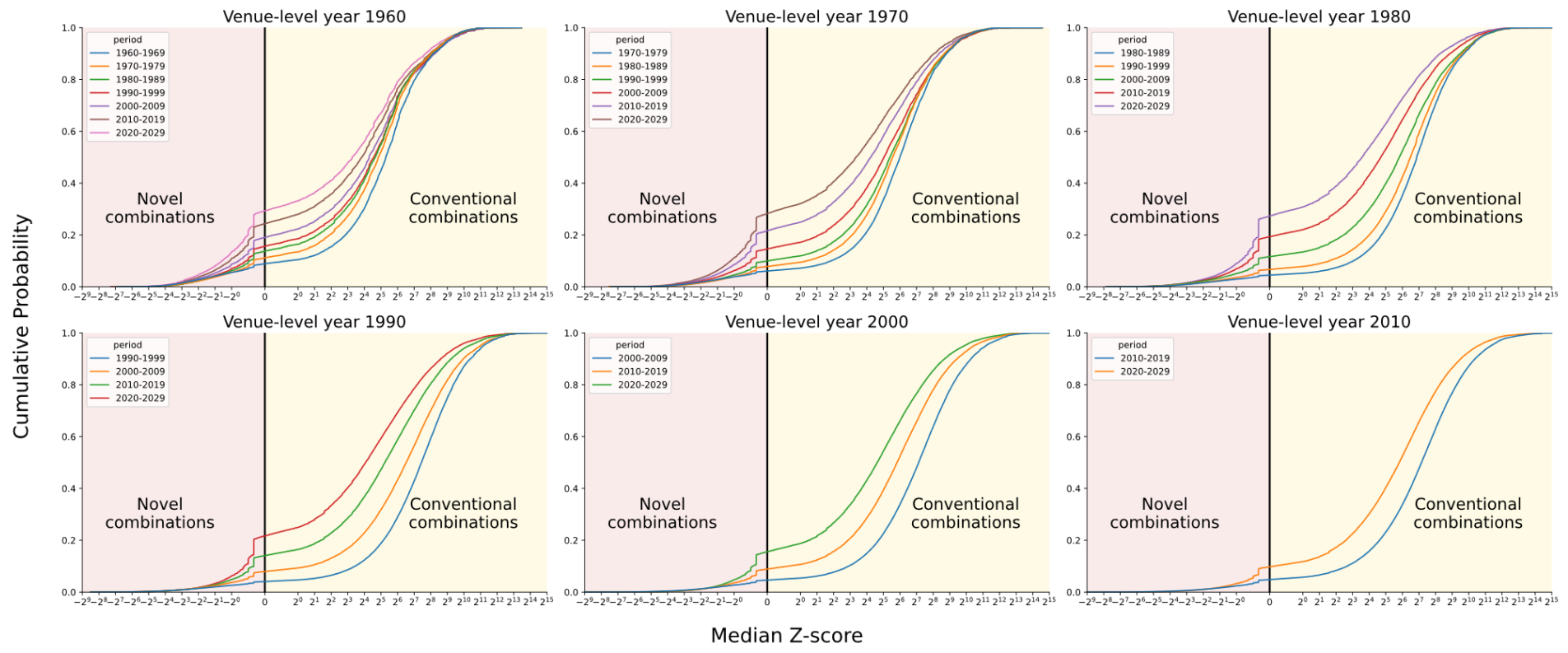

**Fig. S21 | Changes in atypicality under fixed venue-level z-score computation.** For each subplot, venue-level z-scores are computed using papers published before year *t* (shown in the title) and held fixed thereafter. Atypicality distributions are then calculated for papers published in successive 10-year periods following *t*, as indicated in the legend. The figure illustrates the temporal evolution of atypicality relative to a constant venue-specific reference distribution.

## 19. *Exemplary Papers of the Same Type with Different Foundation, Extension, and Generalization Indices*

While our F, E, and G indices correlate with the type of articles, they also capture variations between papers of the same type. Here we present example papers of the same type, and published in the same domain and in similar years, yet with very different F, E and G indices. Across Figs. S22-S25, the left to right graphs represent papers that have a high foundation, extension, and generalization index respectively, with the values reported in parentheses indicating the quantile rank of each index among papers receiving at least 100 citations within five years of publication published in the same year and research domain.

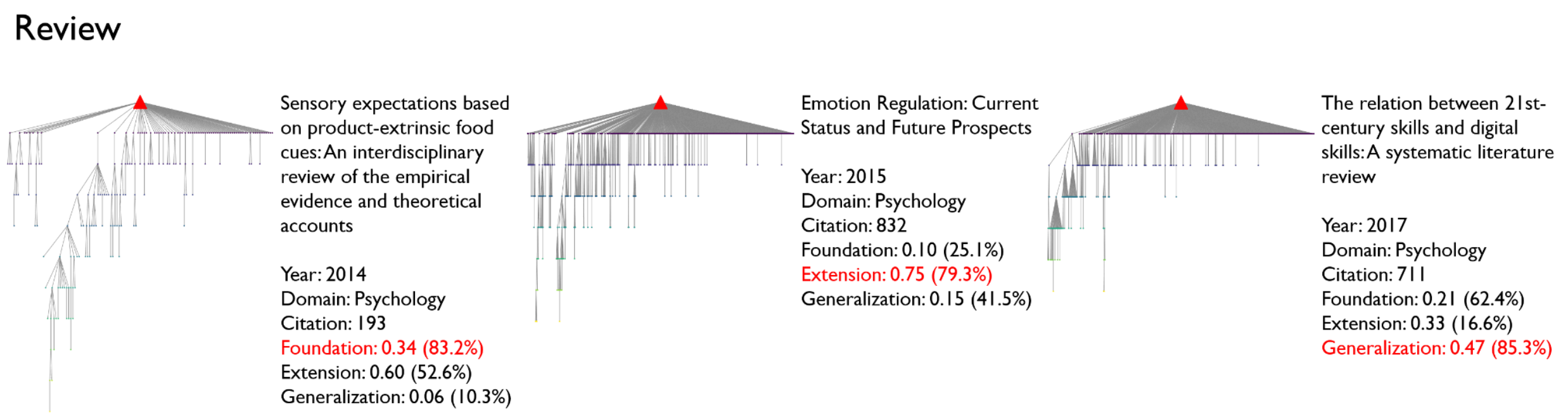

**Fig. S22** | Example papers that are review articles, but have high foundation (left), extension (middle), and generalization (right) indices.

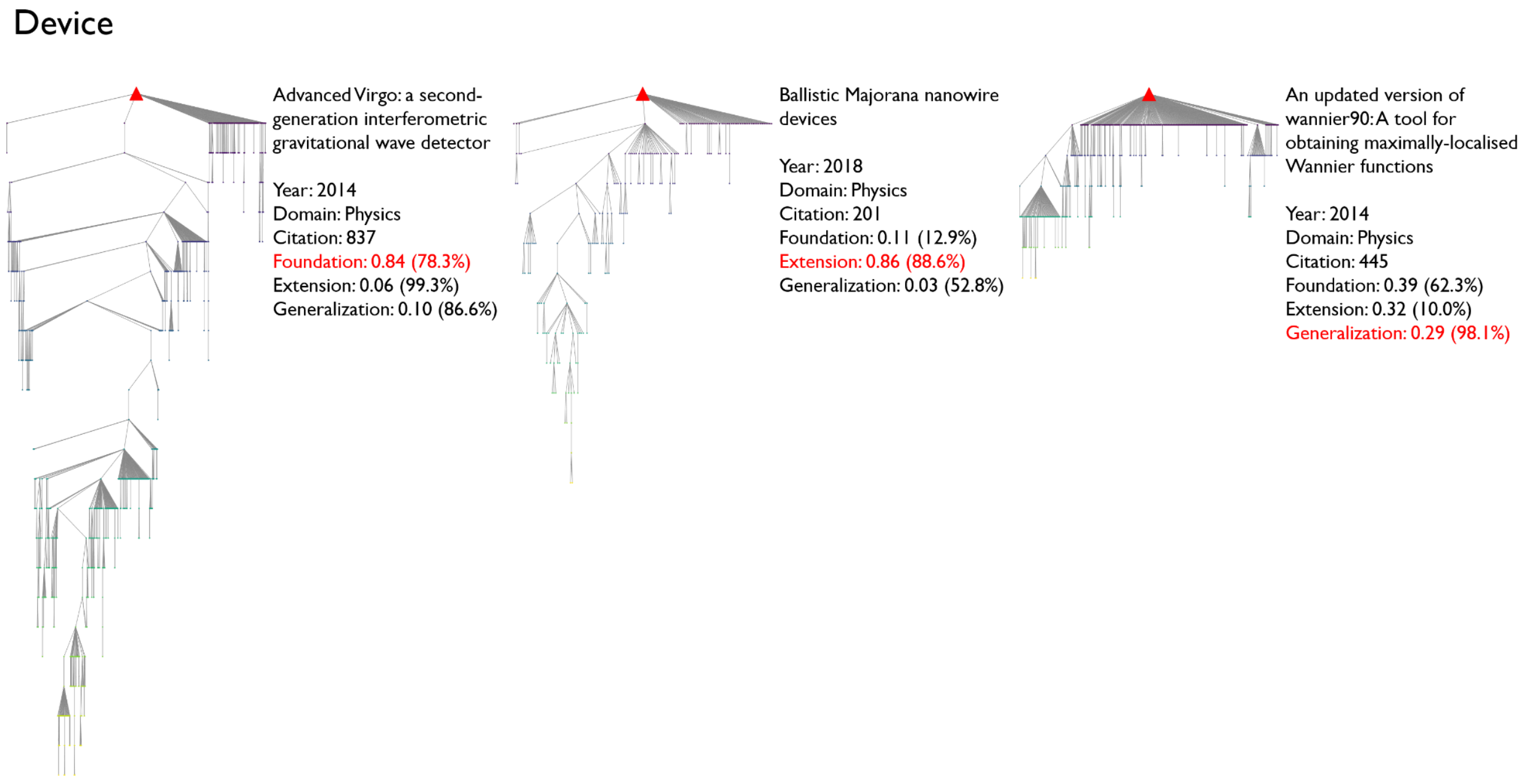

**Fig. S23** | Example papers that introduce a device, but have high foundation (left), extension (right), and generalization index (right).

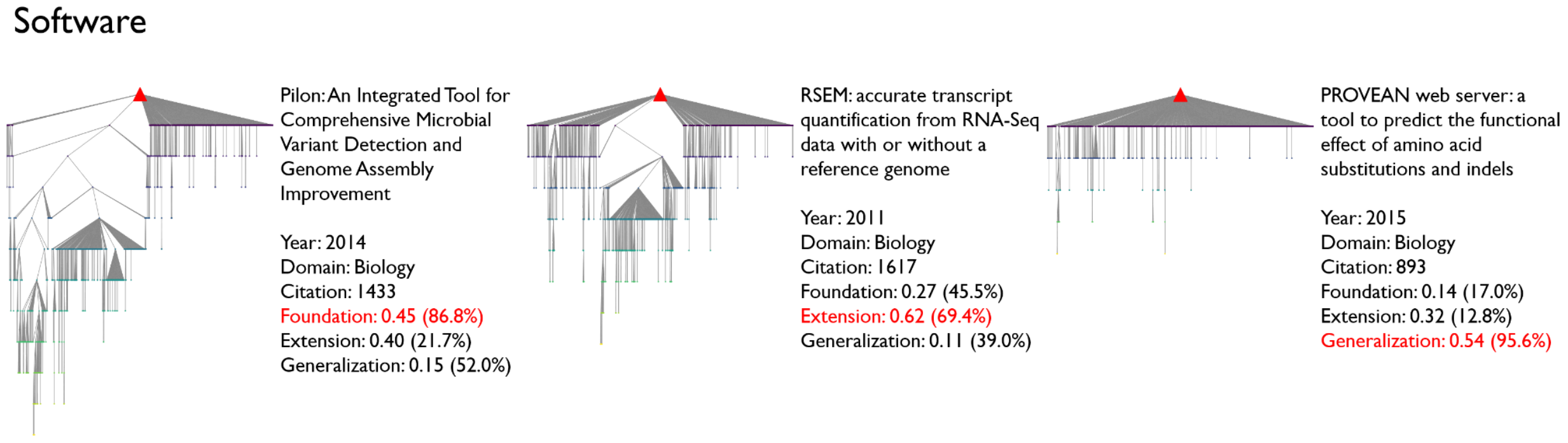

**Fig. S24** | Example papers that introduce a software tool, but have high foundation (left), extension (right), and generalization index (right).

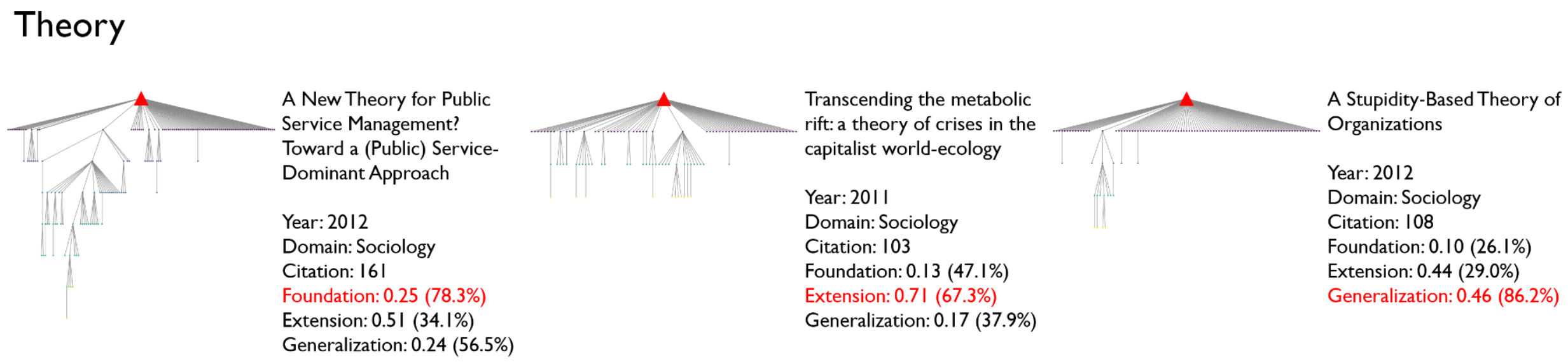

**Fig. S25** | Example papers that are proposing a new theory, but have high foundation (left), extension (right), and generalization index (right).

## 20. *Comparing the Disruption Index with the Proposed Foundation and Generalization Indices*

**Conceptual Debate on the Difference between the Disruption Index, and Our Proposed Foundation, Generalization Index.** *The disruption index does not uniquely measure "disruption", and our refactorization of the citation structure into the foundation and generalization index does.* When the disruption index was first introduced to bibliometric research, it was described as a metric that distinguishes works that *launched new streams of research* from those that primarily *elaborated upon possibilities that had previously been posed* [29]. Subsequent conceptual and empirical developments have focused predominantly on its *displacement* property [31,32]—that is, the extent to which a focal paper can be cited independently of its predecessors, rather than on its purported ability to identify works that genuinely *disrupt* existing research trajectories and give rise to new ones.

Seminal discoveries such as the elucidation of the DNA double-helix structure [33] and the introduction of AlexNet [13] illustrate what scientific disruption entails. These contributions were not merely improvements over previous work in terms of explanatory power or predictive performance; they fundamentally redirected the trajectory of their respective fields. The discovery of DNA structure shifted genetics research from observing inheritance patterns toward understanding molecular mechanisms, while AlexNet drove the transition from manually engineered image features to end-to-end deep learning. In both cases, earlier research paradigms rapidly declined as entirely new research agendas emerged.

In contrast, many highly influential works play a fundamentally different role. Annual cancer statistics reports (e.g., [34]), introductory overviews of 5G technology [35], or the introduction of the *k*-means clustering algorithm [36], all have had substantial scientific impact without establishing distinct research fields of their own. Instead, these works facilitate research across numerous existing domains by synthesizing knowledge, providing widely adopted methodological tools, or serving as authoritative references. Researchers cite these publications not because they redefine the trajectory of their disciplines, but because they provide essential resources that accelerate progress along established research directions, acting as generalized *catalysts* for scientific discovery rather than originators of fundamentally new trajectories.

As we demonstrate empirically below, the disruption index conflates these two distinct types of scientific contributions, with most of the papers it identifies belonging to the latter category. This ambiguity poses a challenge to the conceptual interpretation and empirical application of the index. Although the disruption index is widely characterized as a measure of innovation that *disrupts* [2,25,29,31,32,37–41], our results suggest that it predominantly captures broadly applicable, general-purpose contributions that facilitate scientific progress without necessarily redefining existing research trajectories.

Our proposed foundation and generalization index explicitly distinguishes between these two classes of scientific contributions. Specifically, the foundation index captures field-defining innovations that establish new research trajectories, whereas the generalization index captures general-purpose contributions that diffuse knowledge and accelerate progress along existing ones. This conceptual distinction, illustrated in Fig. 1e, is further substantiated by empirical analyses presented in the following sections.

*How a cited paper is used explains the disruption index.* Per the operationalization of the disruption index, a focal paper often cited independent of its predecessors is an indication of displacement over its predecessors[32]. Although this interpretation is plausible in many cases, we argue that citation purpose provides an alternative explanation for why a paper is cited independently of its references.

When a paper is cited because it is used as a generalized component, method, or tool in subsequent work, its intellectual lineage is often omitted either because it is less relevant to how the cited work is applied or it is sufficiently distant that the authors are not aware of it. In contrast, when a paper is cited as the object of comparison, evaluation, or extension, authors are more likely to discuss its intellectual lineage because its predecessors establish the conceptual foundation or competing approaches against which the new work is positioned. In such cases, references to predecessors help both authors and readers understand the development of the focal idea. Consequently, the absence of a cited paper's predecessors is more likely to reflect that those predecessors are *irrelevant* or *invisible* to the new work rather than reflecting *displacement* of the cited work's references. This behavior aligns with the view that knowledge is often operationally modularized. When researchers employ an established method, they typically cite an existing article or protocol rather than reconstructing its historical and conceptual development, because those details are ordinarily unnecessary for the immediate research task[42,43]. A similar organizational principle has been recognized in other complex intellectual systems. In software engineering, for example, Parnas[44] proposed the principle of information hiding, which advocates concealing implementation details from users who simply use a software module while exposing them only to those who modify or extend it.

Fig. 1c illustrates this distinction using the “Transformer paper” (*Attention Is All You Need*[45]). The “Transformer paper” cites the “LayerNorm paper” (*Layer Normalization*[46]) as a functional component of the proposed architecture. Although one could discuss the intellectual lineage of the “LayerNorm paper” (e.g, “BatchNorm paper” [47]), doing so would contribute little to explaining the Transformer’s central innovation. The “LayerNorm paper” is cited because it is used, rather than because its intellectual development is central for the new innovation. Consequently, its predecessors are omitted, not because they have been *displaced*, but because they are either *irrelevant* to the citing paper’s purpose or they are unknown.

A contrast that better aligns with the conceptual claim of *displacement* is provided by the citations to the “NMT paper” (*Neural Machine Translation by Jointly Learning to Align and Translate*[48]) and the “LSTM paper” (*Long Short-Term Memory*[49]). The “NMT paper” introduced the attention mechanism that forms the conceptual foundation of the Transformer. Nevertheless, under the disruption index, it is classified as a non-disruptive citation because the “Transformer paper” cites it together with its predecessor, the “LSTM paper”. This occurs because the attention mechanism proposed in the “NMT paper” was originally designed as a complementary component to the LSTM architecture rather than as a standalone model. Consequently, discussing the intellectual contribution of the “NMT paper” naturally entails acknowledging the LSTM framework within which the attention mechanism was originally developed. By contrast, the “LSTM paper” is cited because it provides the immediate intellectual context for the “NMT paper” and, in turn, the Transformer. At the same time, predecessors of the “LSTM paper,” such as the “Recurrent Neural Network paper” [50], have only a remote intellectual connection to the attention mechanism and therefore contribute little to understanding the Transformer’s central innovation. Consequently, these earlier predecessors are not included as references, causing the “LSTM paper” to be classified as a disruptive citation under the disruption index.  This citation pattern reflects the type of intellectual *displacement* that the disruption index is intended to capture.

Differences in citation purposes are associated with the properties and types of the focal paper (i.e, the work to be cited). For example, papers introducing software, methods, algorithms, reviews, or tutorials are more likely to be cited for their direct use. We refer to these generalizations as *modular works* because they can be readily incorporated as components in the development of new ideas. In contrast, papers proposing models or theories are more frequently cited for comparison, evaluation, or extension. This distinction helps explain previously reported findings that methods tend to exhibit higher disruption index values than theories[29,51].

Our proposed foundation and generalization index differs from the disruption index in its characterization of foundational influence. Specifically, the foundation index does not require a foundational citation to satisfy the disruption criterion; that is, the focal paper need not be cited independently of its referenced predecessors. As a result, citations to both the original "LSTM paper" and the "NMT paper" are classified as foundational citations. Although this approach provides less discrimination between milestone works that establish a new research domain and those that fundamentally transform an existing one, it has the advantage of avoiding an undue preference for modular works. In contrast, the generalization index adopts a criterion analogous to that of the disruption index by requiring the focal paper to be cited without its references. This requirement excludes citations that primarily reflect incremental extensions and instead captures instances in which the focal work is adopted independently or generalized to new contexts. Consequently, the generalization index places greater emphasis on modular contributions, similar to the disruption index.

**Correlation Between Foundation, Extension, Generalization Indices and Disruption Index**. At the paper level, we examine the relationships between the proposed F, E and G indices and disruption (D) index. In Fig. S26, papers are grouped into deciles according to their D-index values (x-axis), and the y-axis reports the mean F, E, G indices within each decile.

Among the three indices, the generalization (G) index exhibits the strongest association with disruption, displaying a positive and monotonic relationship ($r = 0.37$, $p < 2\times10^{-16}$, $df = 23448429$). The foundation (F) index also shows a statistically significant, albeit much weaker, association with disruption ($r = 0.05$, $p < 2\times10^{-16}$, $df = 23448429$), characterized by a U-shaped pattern across disruption deciles. In contrast, the extension (E) index is negatively associated with disruption ($r = -0.37$, $p < 2\times10^{-16}$, $df = 23448429$), indicating that papers with higher disruption scores tend to exhibit lower levels of extension.

At the citation level, among citations accumulated within five years of publication for papers published before 2020 in OpenAlex that contain at least one reference, 121,949,681 (73.96%) of the 164,894,258 *disruptive* citations (i.e., citations in which the focal paper is cited without any of its references being co-cited) are also *generalizational* citations (i.e., citations in which the focal paper is cited without any of its references or citing papers being co-cited), demonstrating a strong empirical correspondence between disruptiveness and generalization.

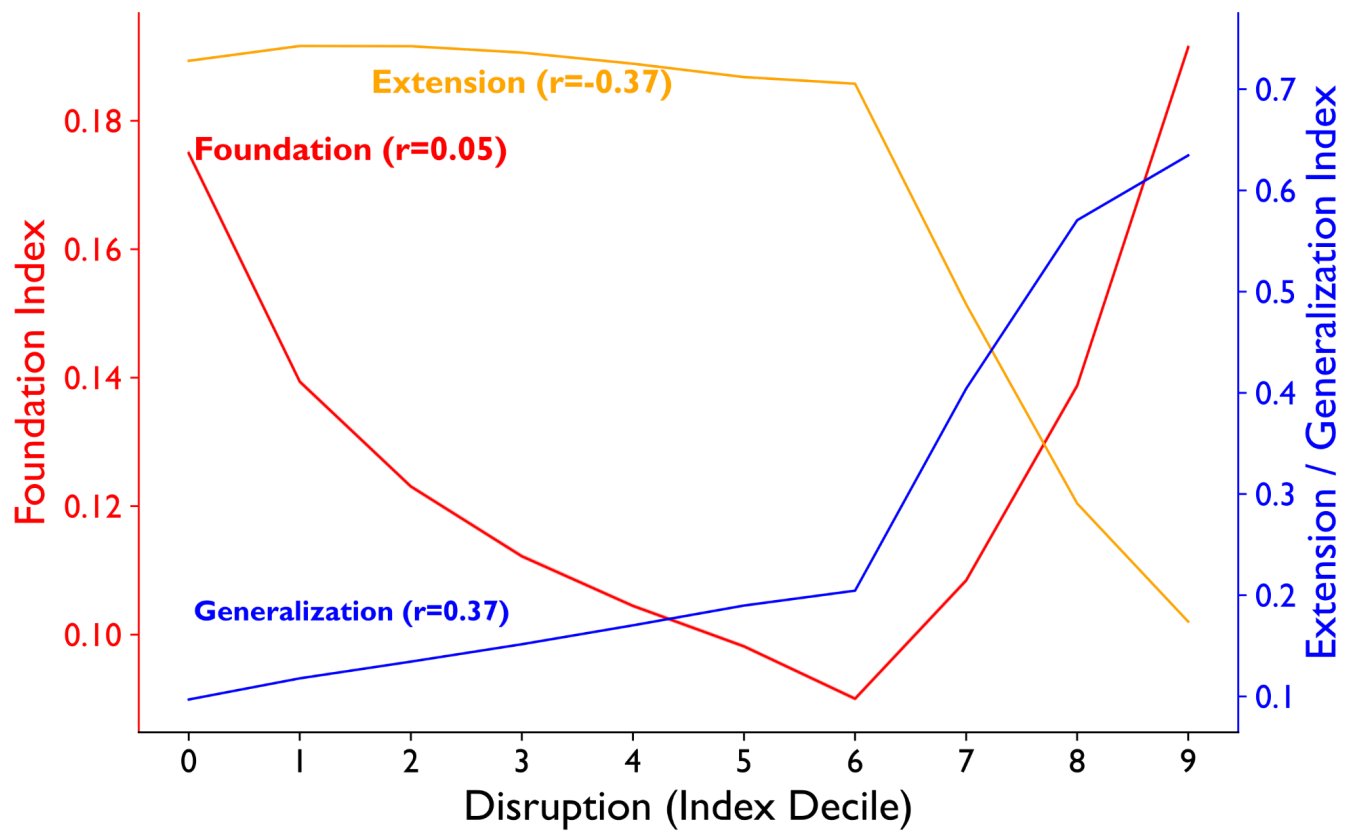

**Fig. S26 | Relationship between the F, E, and G indices and the disruption (D) index.** The figure is based on 23,448,431 papers in the OpenAlex database that contain at least one reference, received at least five citations within five years of publication, and were published between 1945 and 2020.

**Location of Reference for Disruptive, Foundational, Extensional, and Generalizational Citations.** Extended Data Fig. 3 presents the distribution of citation locations according to the co-citation configuration of a cited paper, defined by the numbers of its references and citations that are co-cited in the citing paper. These two dimensions are also used to classify citations as disruptive, foundational, extensional, or generalizational.

As shown in the frequency panel, the number of in-text mentions of a cited paper increases as more of both its references and its citations are co-cited. The average number of reference instances rises from 1.442 (95% CI: 1.440–1.444) when the paper is cited with neither its references nor its citations to 3.463 (95% CI: 3.107–3.861) when it is cited with ten references and ten citations. This pattern suggests that generalizational citations are mentioned less frequently throughout the citing paper and are therefore more likely to serve a peripheral, rather than central, role in the development of the new study.

Regarding the location of reference instances, a notable discontinuity is observed in the proportion of citations appearing in the Method section for disruptive citations. When a paper is cited as a disruptive citation, 13.42% (95% CI: 13.39%–13.44%) of its reference instances occur in the Method section, compared with only 10.35% (95% CI: 10.33%–10.37%) for non-disruptive citations. This finding suggests that citation disruptiveness may primarily capture methodological contributions rather than identifying work that is conceptually distinct from its predecessors.

Reference instances in the Introduction section exhibit a strong association with foundational citations. Specifically, the probability that a citation reference appears in the Introduction increases as the number of co-cited references decreases and the number of co-cited citations increases. The proportion of reference instances in the Introduction rises from 42.19% (95% CI: 41.77%–42.64%) when a paper is cited with ten references and no citations to 50.36% (95% CI: 49.68%–50.99%) when it is cited with one reference and nine citations, the highest value observed in Extended Data Fig. 3. Because citations in the Introduction are commonly used to describe the historical development and current state of a research field [52,53], this pattern supports the interpretation that foundational citations are more likely to represent prior work on which new research explicitly builds. In contrast, disruptiveness has little effect on whether reference instances appear in the Introduction. Disruptive citations are mentioned in the Introduction in 45.45% (95% CI: 45.41%–45.48%) of reference instances, compared with 45.65% (95% CI: 45.62%–45.68%) for non-disruptive citations. The negligible difference suggests that disruptiveness does not distinguish citations serving as the conceptual foundation of subsequent research, but instead may reflect other forms of scholarly influence, such as methodological adoption.

We also found that the proportion of reference instances appearing in the Discussion section increases as fewer of a cited paper's subsequent citations are co-cited. For example, 28.92% (95% CI: 28.88%–28.96%) of reference instances appear in the Discussion section when a paper is cited without any of its references or citations, compared with only 21.93% (95% CI: 21.44%–22.43%) when it is cited without any of its references but with ten of its subsequent citations. Because the Discussion section is typically used to interpret findings, discuss implications, and outline future research directions [53], this pattern suggests that citing a paper without citing its subsequent citations reflects limited relevance of the later developments built upon that work to the citing study.

Overall, these findings support the interpretation that foundational citations represent prior work that serves as the intellectual foundation for new research. In contrast, the disruptive nature of a citation appears to be more indicative of methodological contributions that can be adopted independently of their intellectual antecedents, rather than of a genuine departure from the existing citation lineage. Finally, generalizational citations appear to capture both the modular nature of the cited work and the limited relevance of its subsequent development to the citing study. This pattern suggests that citing papers employing generalizational citations may pursue research directions that diverge from those taken by the majority of subsequent studies citing the same work.

**Comparative Case Studies of Foundation, Generalization and Disruption Indices through Selected Examples of Influential Papers Across Domains .** To further elucidate the conceptual distinctions between the proposed foundation and generalization indices, as well as their relationship to the established disruption index, we examine a set of highly cited papers spanning multiple research domains (Figs. S27-S33). In all figures, citation counts refer to the number of citations received within five years of publication, and all indices are calculated using citation data from the same five-year post-publication window. Values reported in parentheses indicate the quantile rank of each index among highly cited papers (i.e., those receiving at least 100 citations within five years of publication) published in the same year and research domain. The accompanying diagrams illustrate the citation networks formed by papers citing the focal paper during the five-year post-publication period, constructed following the procedure described in the Supplementary Methods Section 3.

The figures are organized into three columns. The first column presents papers that introduced foundational shifts within their respective fields but did not initiate entirely new research domains. Consequently, these papers are not identified as disruptive by the disruption index but are classified as foundational by our foundation index. The second column highlights papers whose contributions are primarily modular and generalizational in nature. These papers are identified as generalizational by our generalization index and are likewise classified as disruptive by the D index. The third column contains papers that both introduced disruptive advances and subsequently served as foundational building blocks for future research. Such contributions are recognized by both the disruption and foundation indices.

We first discuss the distinct characteristics captured by our foundation and generalization indices and compare them with the conventional Disruption index. The foundation index identifies milestone contributions that exert substantial influence on subsequent research, even when the focal work is not the first to initiate a new research direction. For example, *Canagliflozin and Cardiovascular and Renal Events in Type 2 Diabetes* [54] in medicine (Fig. S31) is not the earliest study examining the effects of SGLT2 inhibitors on patients with elevated cardiovascular risk. Nevertheless, as one of the earliest large-scale clinical trials, it provided compelling evidence that fundamentally shaped subsequent research. Likewise, *Rich Feature Hierarchies for Accurate Object Detection and Semantic Segmentation* [55] in computer science (Fig. S28) did not introduce deep convolutional neural networks, but successfully adapted them to object detection, producing a dramatic improvement over previous non-neural-network approaches and establishing a new foundation for later work. Such influential milestone papers are not identified by the disruption index but are effectively captured by our foundation index. Notably, many of these landmark articles are not merely absent from the top of the disruption ranking—they are often placed near the very bottom. For example, *Observation of Gravitational Waves from a Binary Neutron Star Inspiral* [56], which inaugurated the era of multi-messenger astronomy, falls within the bottom 1% of papers ranked by the disruption index. These seminal papers receive even lower disruption scores than papers with virtually no citations. This counterintuitive result arises because highly cited but predominantly *developmental* papers typically receive negative disruption scores of large magnitude, whereas uncited or rarely cited papers have disruption scores close to zero. Consequently, when disruption scores are aggregated (e.g., by taking the mean), highly influential developmental work can reduce the average more than papers with negligible scientific impact. This problematic statistical property has been noted previously [57,58], but has received relatively little attention in the mainstream literature.

In contrast, the generalization index captures papers that are cited as modular building blocks within downstream research, providing auxiliary support rather than serving as the primary conceptual foundation. Representative examples include review articles (e.g., *NIH Image to ImageJ: 25 Years of Image Analysis* [59]; Fig. S28), dataset or data-compilation papers (e.g., *A New Data Set of Educational Attainment in the World, 1950–2010* [60]; Fig. S29), and methodological papers (e.g., *Fast Stable Restricted Maximum Likelihood and Marginal Likelihood Estimation of Semiparametric Generalized Linear Models* [61]; Fig. S30). These works are indispensable research resources and are frequently reused across diverse contexts, yet they typically neither introduce paradigm-shifting ideas nor embody the primary scientific innovation. Nevertheless, because downstream papers often cite them without citing many of their references, the disruption index frequently assigns them high disruption scores. Our generalization index separates such broadly reusable contributions from papers that genuinely disrupt.

These examples also illustrate the complementary roles of the disruption and foundation indices. A paper that is simultaneously highly disruptive—being cited independently of its predecessors—and highly foundational—serving as the central intellectual basis upon which subsequent work is built—is considerably more likely to represent a genuine paradigm shift. The exemplar papers shown in the third columns of Figs. S27–S33 illustrate this intersection, combining disruptive characteristics with enduring foundational influence. Despite this complementarity, a straightforward combination of the two metrics does not necessarily improve the identification of paradigm-shifting work relative to the foundation index alone. As demonstrated by the empirical analysis below, the foundation index by itself provides a more effective indicator of paradigm shifts than a simple combination of the two measures.

## Chemistry

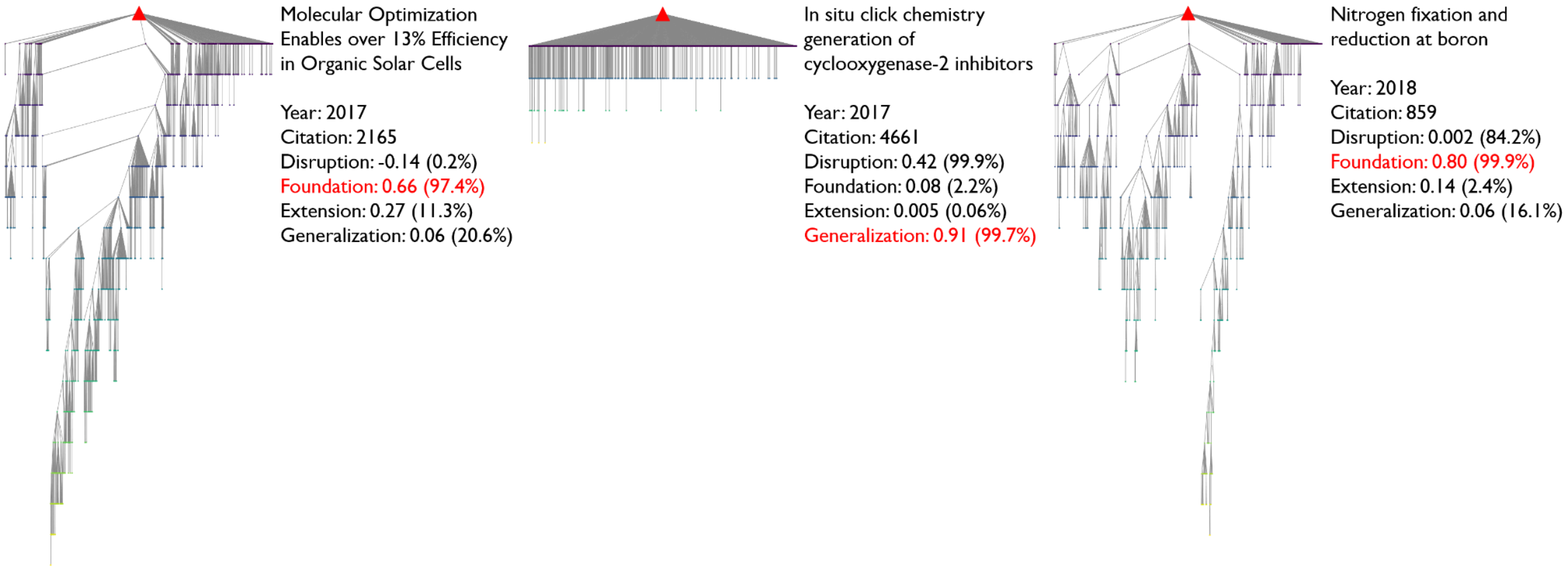

**Fig. S27 |** *Molecular Optimization Enables over 13% Efficiency in Organic Solar Cells*[62] is widely regarded as a landmark study in the development of organic solar cells (OSCs). Prior to this work, the highest-performing OSCs were predominantly based on fullerene acceptors. This study provided some of the earliest compelling evidence that non-fullerene acceptors (NFAs) could achieve power conversion efficiencies surpassing those of state-of-the-art fullerene-based devices, thereby validating the potential of NFAs and accelerating the field's transition toward NFA-based OSC architectures. This paper is classified as non-disruptive according to the D index, as 42.3% of subsequent studies citing it also cite its key antecedent, *An Electron Acceptor Challenging Fullerenes for Efficient Polymer Solar Cells* [63], which introduced ITIC—one of the first high-performance NFAs that established the competitiveness of non-fullerene acceptors relative to fullerene-based systems. The middle column highlights *In situ click chemistry generation of cyclooxygenase-2 inhibitors* [64], which introduced a novel target-guided drug discovery strategy for the identification of cyclooxygenase-2 (COX-2) inhibitors through in situ click chemistry. Finally, *Nitrogen Fixation and Reduction at Boron* [65] challenged the long-standing paradigm that molecular nitrogen activation is exclusive to transition metals by demonstrating that a boron-based system can bind and reduce $N_2$, thereby expanding the fundamental understanding of main-group element reactivity.

## Computer Science

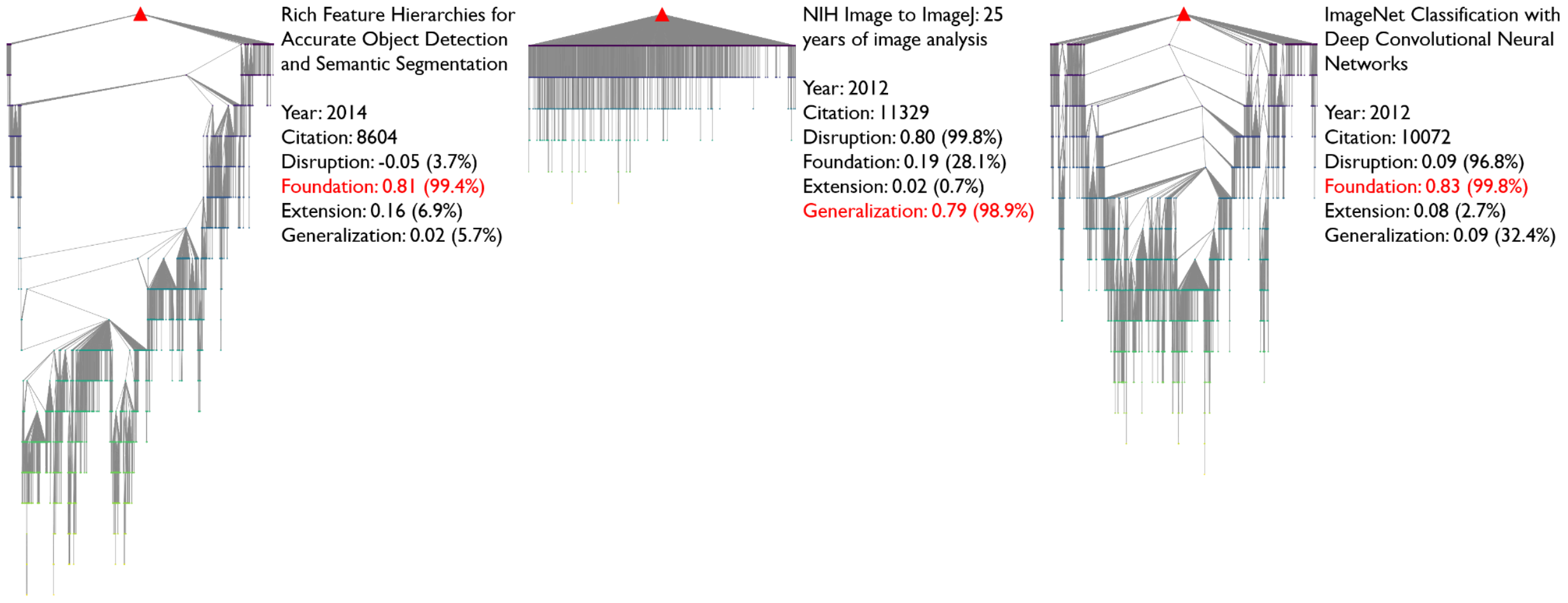

**Fig. S28 |** In computer science, *Rich Feature Hierarchies for Accurate Object Detection and Semantic Segmentation* [55], commonly known as R-CNN, is widely recognized as one of the first studies to successfully apply deep convolutional neural networks to object detection. By substantially outperforming traditional, non-neural-network-based approaches, it established a new paradigm for object detection. According to the disruption index, however, the paper is classified as non-disruptive because 41.3% of publications citing R-CNN also cite *ImageNet Classification with Deep Convolutional Neural Networks*[13], which introduced the deep convolutional neural network architecture that served as the backbone of the R-CNN framework. The middle column highlights *NIH Image to ImageJ: 25 Years of Image Analysis* [59], a review article that chronicles the development and impact of the *Image* and *ImageJ* software platforms. Finally, *ImageNet Classification with Deep Convolutional Neural Networks* [13], commonly known as AlexNet, is widely regarded as the milestone work that initiated the modern era of deep learning in computer vision.

## Economics

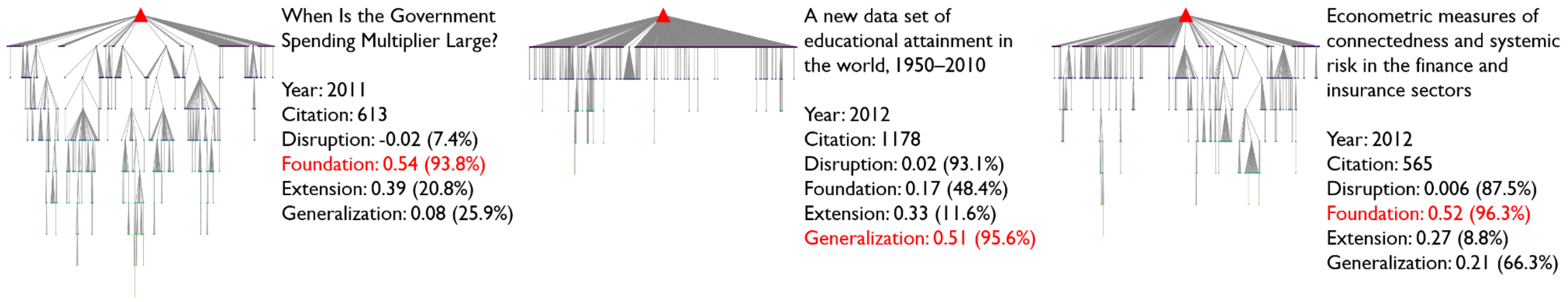

**Fig. S29 |** In economics, *When Is the Government Spending Multiplier Large?*[66] is a landmark paper in the fiscal policy literature. It demonstrates that the effectiveness of fiscal stimulus depends critically on monetary policy conditions, particularly when interest rates are constrained by the zero lower bound, challenging the earlier prevailing view that government spending multipliers are generally small or close to one due to crowding-out effects and monetary policy responses. Despite its substantial influence, the paper is classified as non-disruptive according to the disruption index because 23.2% of publications citing it also cite *Understanding the Effects of Government Spending on Consumption*[67], one of its key references and a seminal study on the macroeconomic effects of government spending. The middle column highlights *A New Data Set of Educational Attainment in the World, 1950–2010* [60], which introduced a comprehensive, internationally comparable panel dataset of educational

attainment that has become a standard resource in empirical economics. Finally, *Econometric Measures of Connectedness and Systemic Risk in the Finance and Insurance Sectors* [68] made a seminal contribution to the systemic risk literature by establishing a network-based framework for measuring financial interconnectedness, shifting the focus from the risk of individual institutions to the propagation of risk across the financial system.

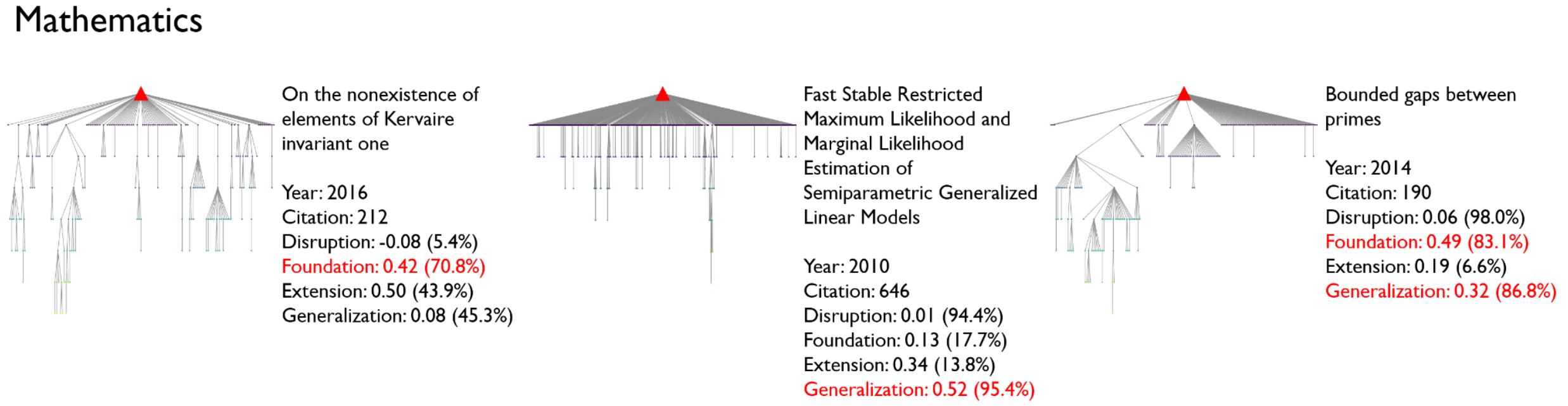

**Fig. S30 |** In mathematics, *On the Nonexistence of Elements of Kervaire Invariant One* [69] is a landmark breakthrough in algebraic topology that resolved the decades-old Kervaire invariant one problem. The paper is classified as non-disruptive according to the disruption index because 22.2% of publications citing it also cite *Real-Oriented Homotopy Theory and an Analogue of the Adams–Novikov Spectral Sequence* [70], one of its key references that laid the theoretical groundwork for the eventual solution of the Kervaire invariant one problem. The middle column highlights *Fast Stable Restricted Maximum Likelihood and Marginal Likelihood Estimation of Semiparametric Generalized Linear Models* [61], an influential statistics paper that introduced efficient methods for parameter estimation in generalized additive models. The methods developed in this work were subsequently incorporated into the widely used *mgcv* package in R. Finally, *Bounded Gaps Between Primes* [71] achieved a landmark breakthrough in analytic number theory by providing the first major progress toward the Twin Prime Conjecture in decades.

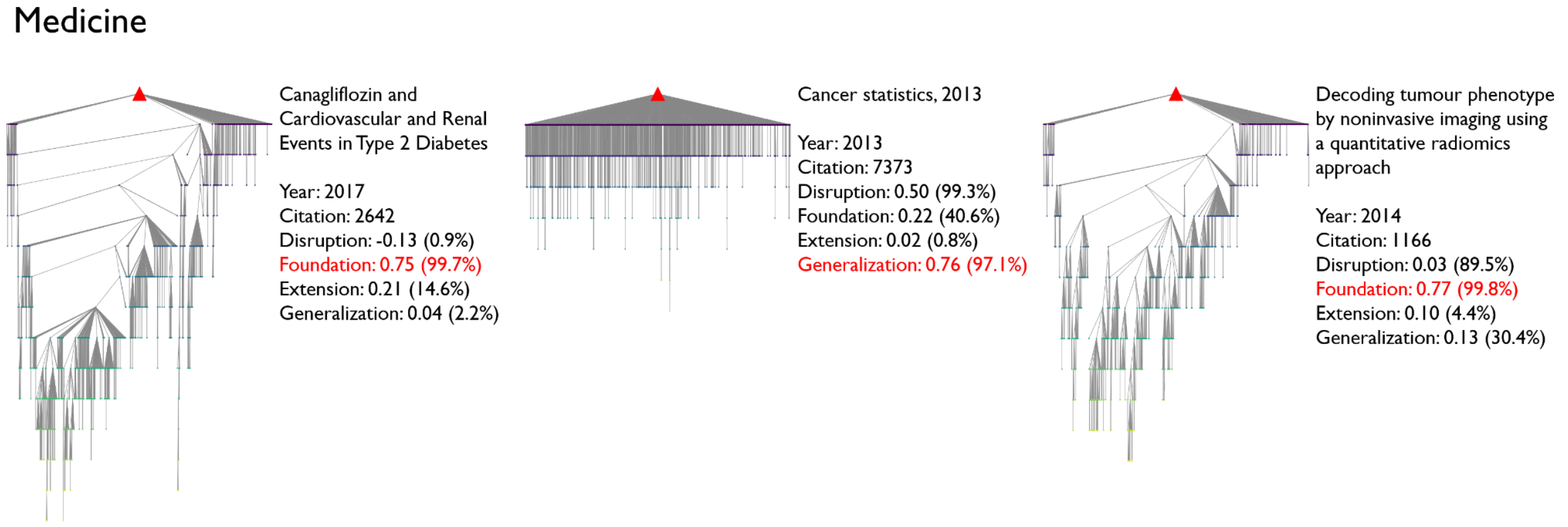

**Fig. S31 |** In medicine, *Canagliflozin and Cardiovascular and Renal Events in Type 2 Diabetes* [54] reports the results of a large-scale clinical trial evaluating the effects of canagliflozin, an SGLT2 inhibitor used to treat type 2 diabetes, in patients at high cardiovascular risk. Unlike earlier studies, which primarily focused on glycemic control, this work demonstrated the drug's cardiovascular and renal benefits. The paper is classified as non-disruptive according to the disruption index because 77.0% of publications citing it also cite *Empagliflozin, Cardiovascular Outcomes, and Mortality in Type 2 Diabetes* [72], a key reference that first established the cardiovascular and mortality benefits of another SGLT2 inhibitor, empagliflozin. The middle column highlights *Cancer Statistics, 2013* [34], part of the annual

*Cancer Statistics* series published by the American Cancer Society. These reports provide comprehensive statistics on cancer incidence, mortality, and survival in the United States and are widely cited throughout the biomedical literature. Finally, *Decoding Tumour Phenotype by Noninvasive Imaging Using a Quantitative Radiomics Approach* [73] is a field-defining paper that established radiomics as a quantitative imaging paradigm for characterizing tumor phenotypes.

Physics

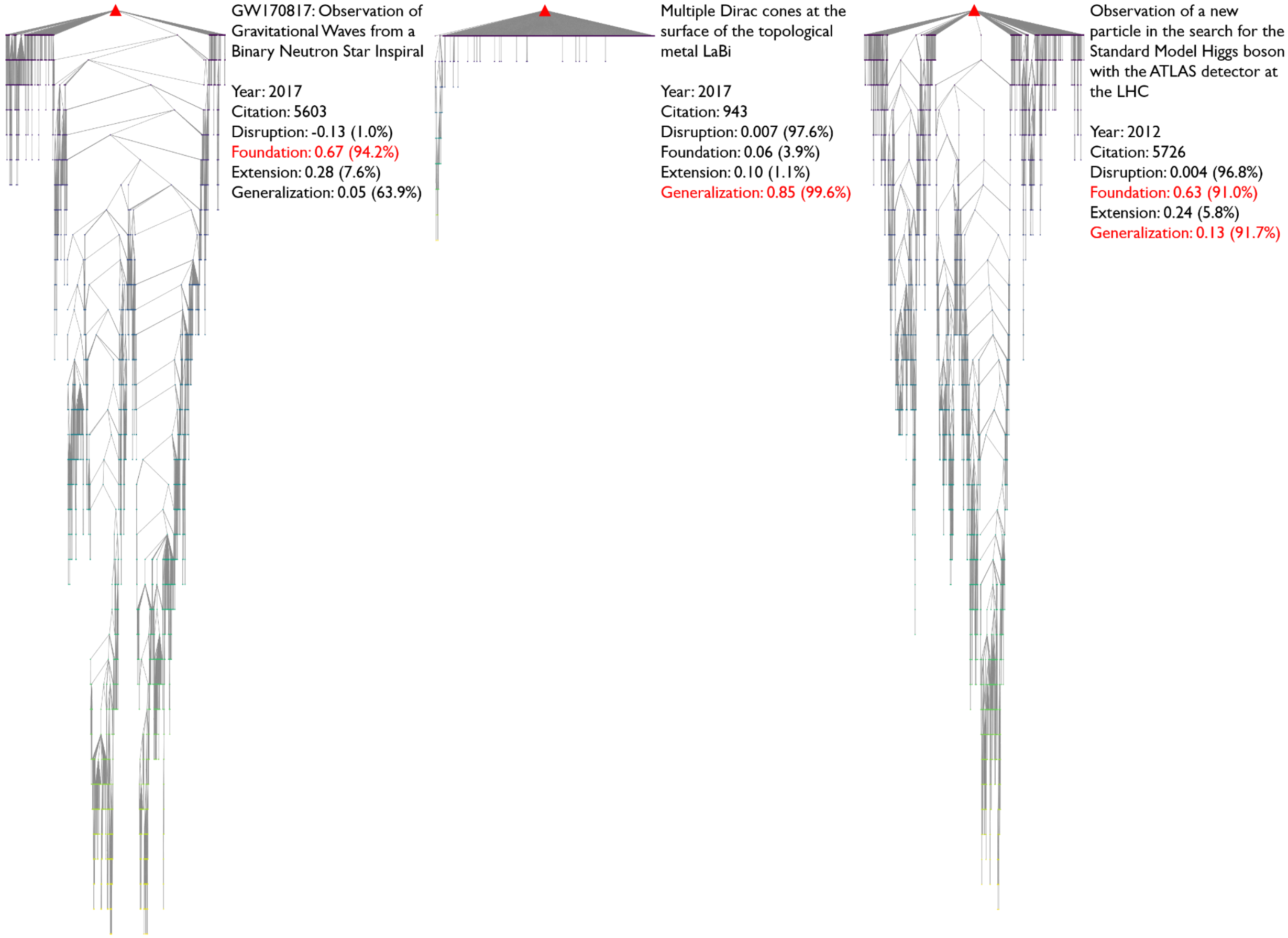

**Fig. S32 |** In physics, *Observation of Gravitational Waves from a Binary Neutron Star Inspiral* [56] reports the first observation of gravitational waves from a binary neutron star merger, marking the beginning of the era of multi-messenger astronomy through the combined detection of gravitational-wave and electromagnetic signals. The paper is classified as non-disruptive according to the disruption index because 39.0% of publications citing it also cite *Observation of Gravitational Waves from a Binary Black Hole Merger* [74], which reported the first direct detection of gravitational waves and established the observational foundation for subsequent discoveries. The middle column highlights *Multiple Dirac Cones at the Surface of the Topological Metal LaBi* [75], which experimentally confirmed the existence of multiple Dirac surface states in LaBi, providing important evidence for its topological electronic structure. Finally, *Observation of a New Particle in the Search for the Standard Model Higgs Boson with the ATLAS Detector at the LHC* [76] reports the discovery of a new particle consistent with the Standard Model Higgs boson, representing one of the defining achievements of modern particle physics and a central contribution to the 2013 Nobel Prize in Physics.

# Sociology

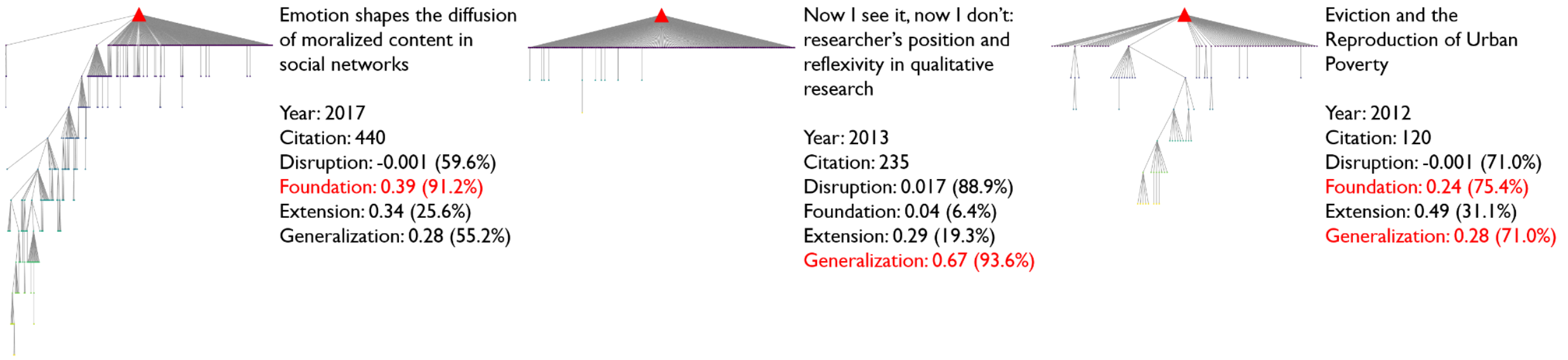

**Fig. S33 |** In sociology, *Emotion Shapes the Diffusion of Moralized Content in Social Networks* [77] made an important contribution to computational social science and moral psychology by demonstrating that moral-emotional language substantially increases the diffusion of political content in online social networks. The study integrated the roles of emotion, morality, and information diffusion, which had often been examined separately in earlier work. The paper is classified as moderately disruptive and highly foundational. Its downstream influence remains closely connected to prior research, as many publications citing it also cite *Experimental Evidence of Massive-Scale Emotional Contagion through Social Networks* [78], one of its key references that established the large-scale effects of emotion on behavior in online social networks. The middle column highlights *Now I See It, Now I Don't: Researcher's Position and Reflexivity in Qualitative Research* [79], an influential methodological article that examines how researchers' social positions, personal experiences, and insider–outsider status shape reflexive qualitative inquiry. Finally, *Eviction and the Reproduction of Urban Poverty* [80] is a landmark study demonstrating that eviction is widespread in low-income urban neighborhoods, particularly among Black women, and that it is both a consequence of and a driver of persistent poverty.

**Empirical Evaluation of Foundation, Generalization and Disruption through a Controlled Comparison with 'Disruptive Foundation'**. We introduce the *disruptive foundation citation* to clarify the distinctions among *foundation*, *generalization*, and the *disruption index*. A disruptive foundation citation is defined as a citation in which the focal paper is cited without any of its references being co-cited (thereby satisfying the criterion for disruptiveness), while at least one of the focal paper's own citations is co-cited (thereby indicating that the focal paper serves as a foundation for subsequent work). Comparing the index computed from *foundation citations* with that computed from *disruptive foundation citations* isolates the contribution of disruptiveness—namely, citations in which none of the focal paper's references are co-cited—while holding the foundational nature of the citation constant. Conversely, comparing the index based on *disruptive foundation citations* with that based on *disruptive citations* isolates the role of *generalization*—that is, whether none of the focal paper's subsequent citations are co-cited—while controlling for disruptiveness (Fig. S34).

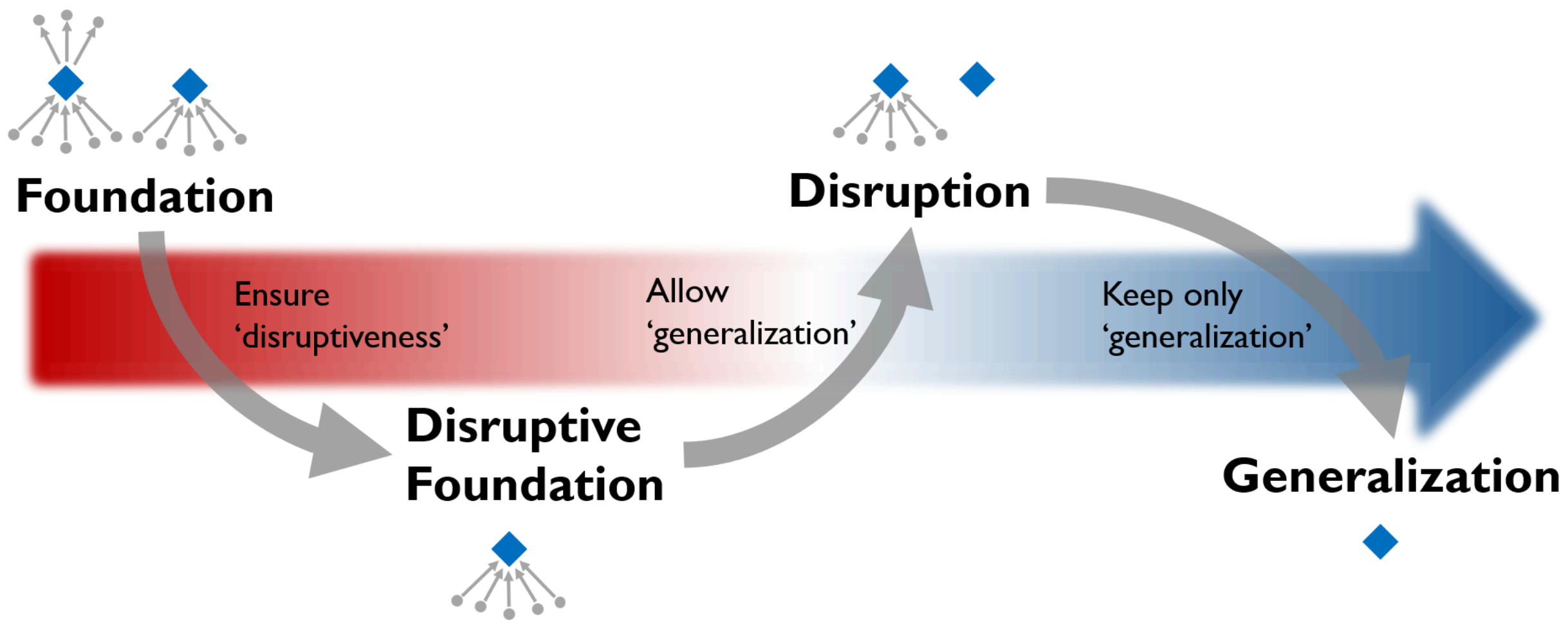

**Fig. S34 | Illustration of foundational, disruptive foundation, disruptive, and generalizational citations.** We construct four categories of citations for comparison. As defined previously, *foundational citations* refer to cases in which the cited paper is co-cited with more of its references (i.e., papers cited by the focal paper) than with its citations (i.e., papers citing the focal paper). In contrast, *generalizational citations* refer to cases in which the cited paper is cited without co-citing either its references or its citations. Restricting foundational citations to those that are also *disruptive*—that is, citations made without co-citing the cited paper's references—yields *disruptive foundation citations*. Combining disruptive foundation citations with generalizational citations produces *disruptive citations*, corresponding to the original definition underlying the disruption index. Accordingly, the four citation categories—foundational, disruptive foundation, disruptive, and generalizational—represent progressively increasing degrees of citation generalization.

Next, we conduct two sets of comparisons. The first compares the proportion of a focal paper's citations that are classified as foundation citations ($\frac{F}{C}$, where $F$ denotes the number of foundation citations and $C$ the total number of citations received by the focal paper), disruptive foundation citations ($\frac{Disp(F)}{C}$, where $Disp(F)$ denotes the number of disruptive foundation citations), disruptive citations ($\frac{I}{C}$, where $I$ denotes the number of disruptive citations), and generalizational citations ($\frac{G}{C}$, where $G$ denotes the number of generalizational citations). The second compares normalized variants of these measures, defined as $\frac{F}{C+K}$, $\frac{Disp(F)}{C+K}$, $\frac{I}{C+K}$, $\frac{I-J}{C+K}$, the latter (i.e, $\frac{I-J}{C+K}$) corresponding to the original disruption index [29]. Here, $K$ denotes the total number of citations received by all references cited by the focal paper, while $J$ denotes the number of non-disruptive citations received by the focal paper. The denominator, $C + K$, is identical to that used in the original disruption index [29], enabling direct comparison across metrics.

We compute each index for all papers in the dataset and evaluate its ability to distinguish (i) award-winning papers from non-award-winning papers, (ii) papers that introduce new concepts from those that do not, and (iii) articles cited by review articles from the review articles themselves. These categories are expected to capture highly influential and paradigm-shifting research and have been previously used as external validation benchmarks for the disruption index [29,31].

Extended Data Fig. 5a presents the aggregated performance of $\frac{F}{C}$, $\frac{Disp(F)}{C}$, $\frac{I}{C}$ and $\frac{G}{C}$. Detailed validation results using alternative definitions and threshold values are provided in Figs. S35 and S36. Extended Data Fig. 2b presents the aggregated performance of the normalized measures, including $\frac{F}{C+K}$, $\frac{Disp(F)}{C+K}$, $\frac{I}{C+K}$ and $\frac{I-J}{C+K}$. The corresponding validation results under alternative definitions and threshold values are reported in Figs. S37 and S38.

As an additional robustness check, we repeat the analysis using only papers published before 2015 and compute each index based on citations accumulated through the end of 2025, thereby ensuring a citation window of at least ten years for every paper. The aggregated quantile differences from this analysis are presented in Fig. S39. This analysis addresses recent concerns that the significance of some disruptive research may emerge only over longer citation windows [81]. The results show that variants of the disruption index ($\frac{I}{C}$ or $\frac{I-J}{C+K}$) assign, on average, lower index quantiles to the desired groups than to their undesired counterparts across all validation categories except Nobel Prize-winning papers. Moreover, across all categories, the differences in index quantiles between the desired and undesired groups are substantially smaller for the disruption-index variants than for the corresponding foundation-index variants ($\frac{F}{C}$ or $\frac{F}{C+K}$), indicating that the latter more consistently distinguish the desired groups from their undesired counterparts.

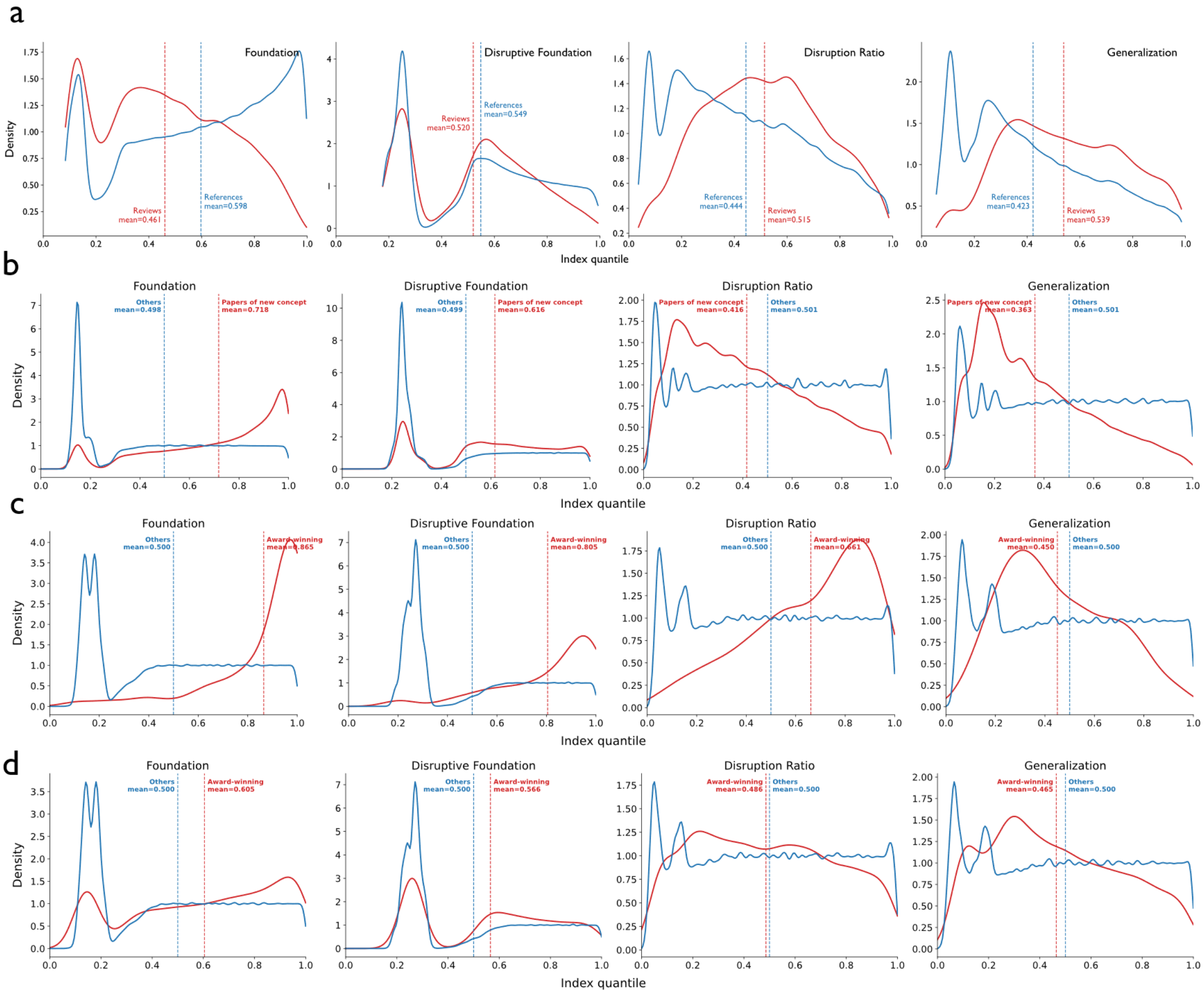

**Fig. S35 | Comparison of the ratios of foundation citations, disruptive foundation citations, disruptive citations, and generalizational citations on their ability to distinguish paradigm-shift papers from others.** The ratios of foundation citations, disruptive foundation citations, disruptive citations, and generalizational citations are calculated as the proportion of each citation type among all citations received by a paper. These metrics are evaluated for their ability to distinguish papers likely to be innovative and drive paradigm shifts from other papers. In each panel, the distributions show the quantile distribution of each metric within each cohort. Quantiles are normalized by publication year and research domain, such that each paper is ranked relative to all papers published in the same year and domain. The average quantile for each cohort is labelled. Unless otherwise specified, all panels are based on papers published between 1945 and 2020, with an identified research domain, at least one reference, and at least five citations received within five years of publication. **a,** Results are computed using the same review and reference cohorts as in Extended Data Fig. 4. **b,** Results are computed using publication records from Semantic Scholar. Papers introducing new concepts are identified as described in the Supplementary Methods Section 15, and we select academic concepts for all concepts with their concept score > 0.5. The dataset comprises 229,526 papers introducing new concepts and 27,335,337 papers in the *Others* category. **c,d,** Results are computed using OpenAlex. Nobel Prize papers are identified using the dataset of Li *et al.* [5], and papers associated with other major awards are identified using Foster *et al.* [6]; both datasets are matched to OpenAlex via DOI or PMID. **c,** The dataset includes 382 Nobel Prize papers meeting the inclusion criteria and 23,447,981 papers in the *Others* category. **d,** The dataset includes 7,853 papers associated with other major awards meeting the inclusion criteria and 23,440,510 papers in the *Others* categor

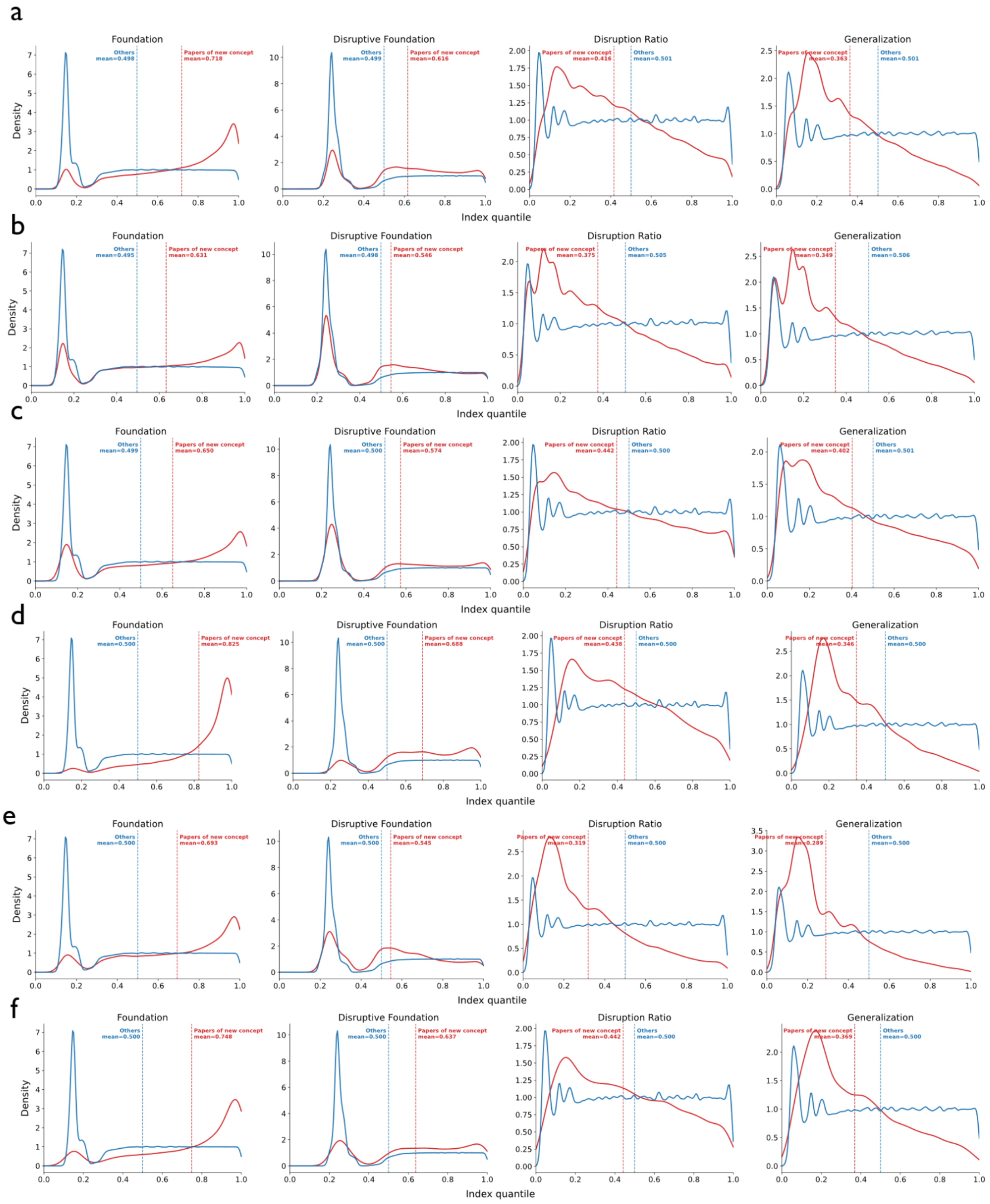

**Fig. S36 | Distribution of index quantiles for papers introducing new concepts versus other papers. Index quantiles are computed using $\frac{F}{C}$, $\frac{Disp(F)}{C}$, $\frac{I}{C}$ and $\frac{G}{C}$.** All analyses are based on publications in the Semantic Scholar dataset published between 1945 and 2020 that have an identified research domain, at least one reference, and at least five citations within five years of publication. All indices are calculated using citations received within

the same five-year window. Papers introducing new concepts are identified as described in Supplementary Methods Section 15. **a,** Papers are classified as introducing new concepts if the concept is considered an academic concept (concept score > 0.5) and the paper has the highest concept score among all papers introducing that concept (229,526 concept-introducing papers; 27,335,337 other papers). **b,** Papers are classified as introducing new concepts if the concept is an academic concept (concept score > 0.5) and the paper ranks among the top five highest concept scores for that concept (998,335 concept-introducing papers; 26,566,528 other papers). **c,** Papers are classified as introducing new concepts if they are the earliest publication (by publication date) introducing an academic concept (concept score > 0.5) (183,436 concept-introducing papers; 27,381,427 other papers). **d,** Same as a, but using a concept score threshold of > 2.0 (13,481 concept-introducing papers; 27,551,382 other papers). **e,** Same as b, but using a concept score threshold of > 2.0 (60,108 concept-introducing papers; 27,504,755 other papers). **f,** Same as c, but using a concept score threshold of > 2.0 (12,436 concept-introducing papers; 27,552,427 other papers).

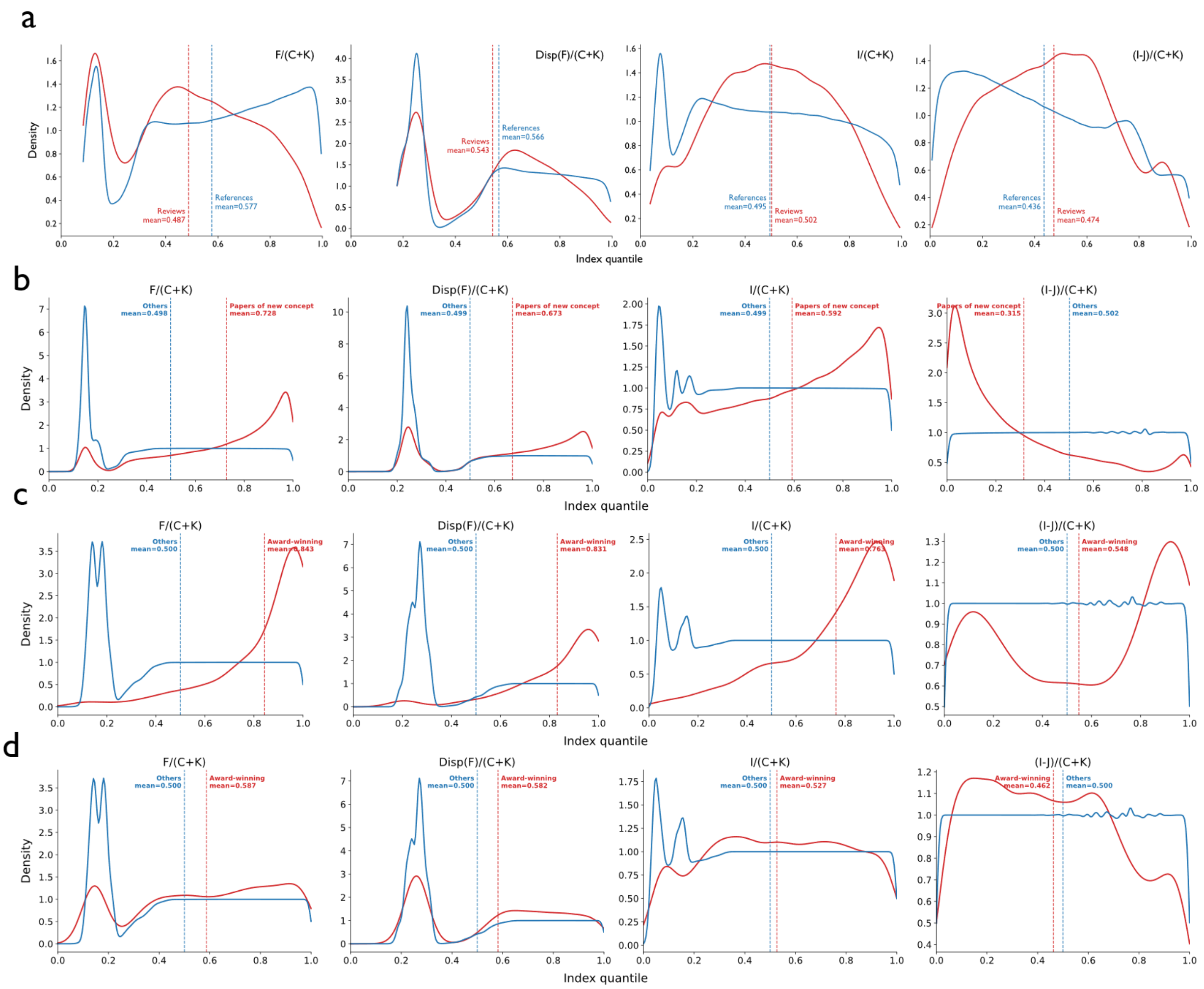

**Fig. S37 | Comparison of the $\frac{F}{C+K}$, $\frac{Disp(F)}{C+K}$, $\frac{I}{C+K}$, $\frac{I-J}{C+K}$ indices on their effect to distinguish paradigm-shift papers.** The samples analyzed in each panel are identical to those used in Fig. S35.

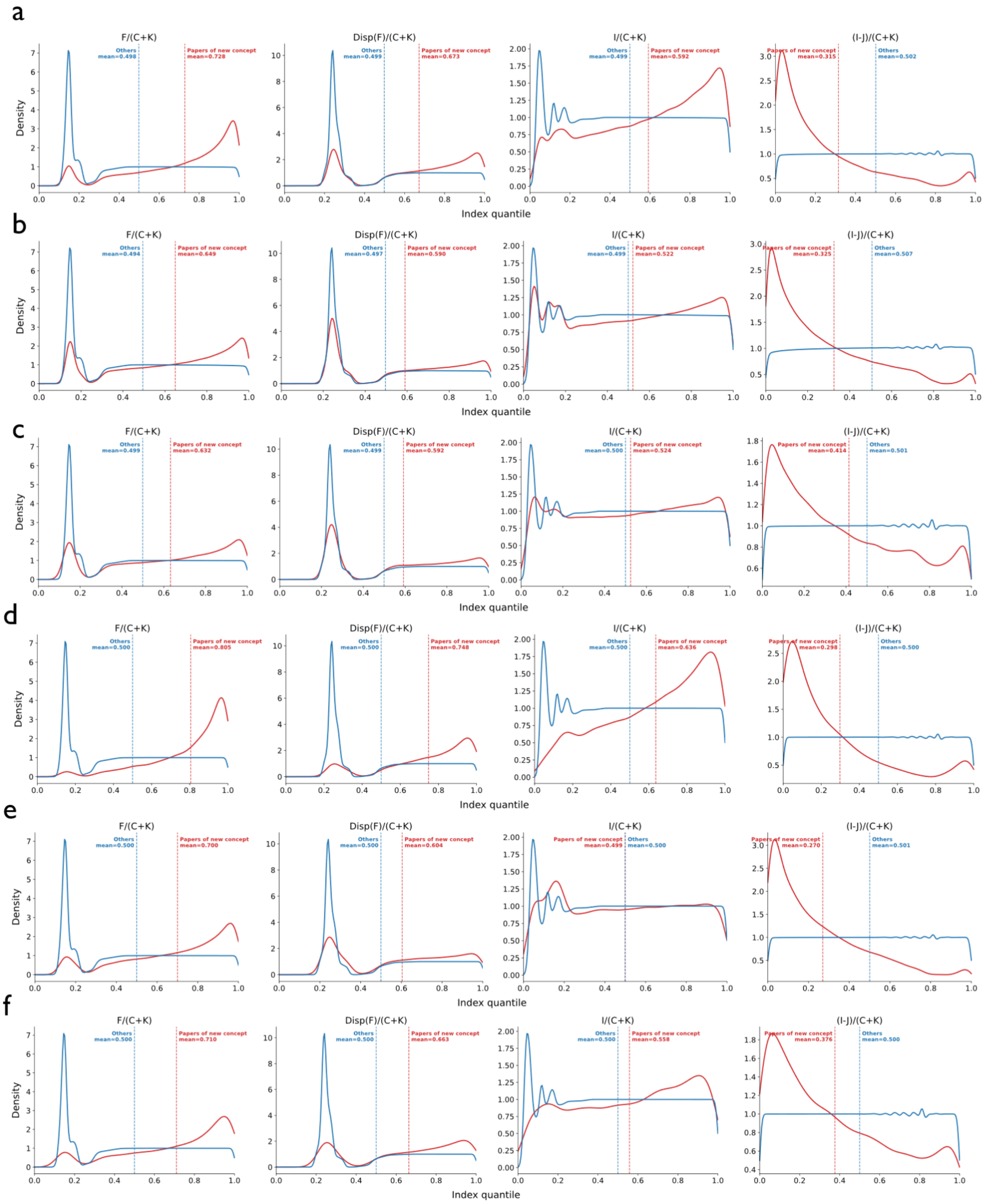

**Fig. S38 | Distribution of index quantiles for papers introducing new concepts versus other papers. Index quantiles are computed using $\frac{F}{C+K}$, $\frac{Disp(F)}{C+K}$, $\frac{I}{C+K}$ and $\frac{I-J}{C+K}$.** The sample of papers identified as introducing new concepts is identical to that used in the corresponding panels of Fig. S36.

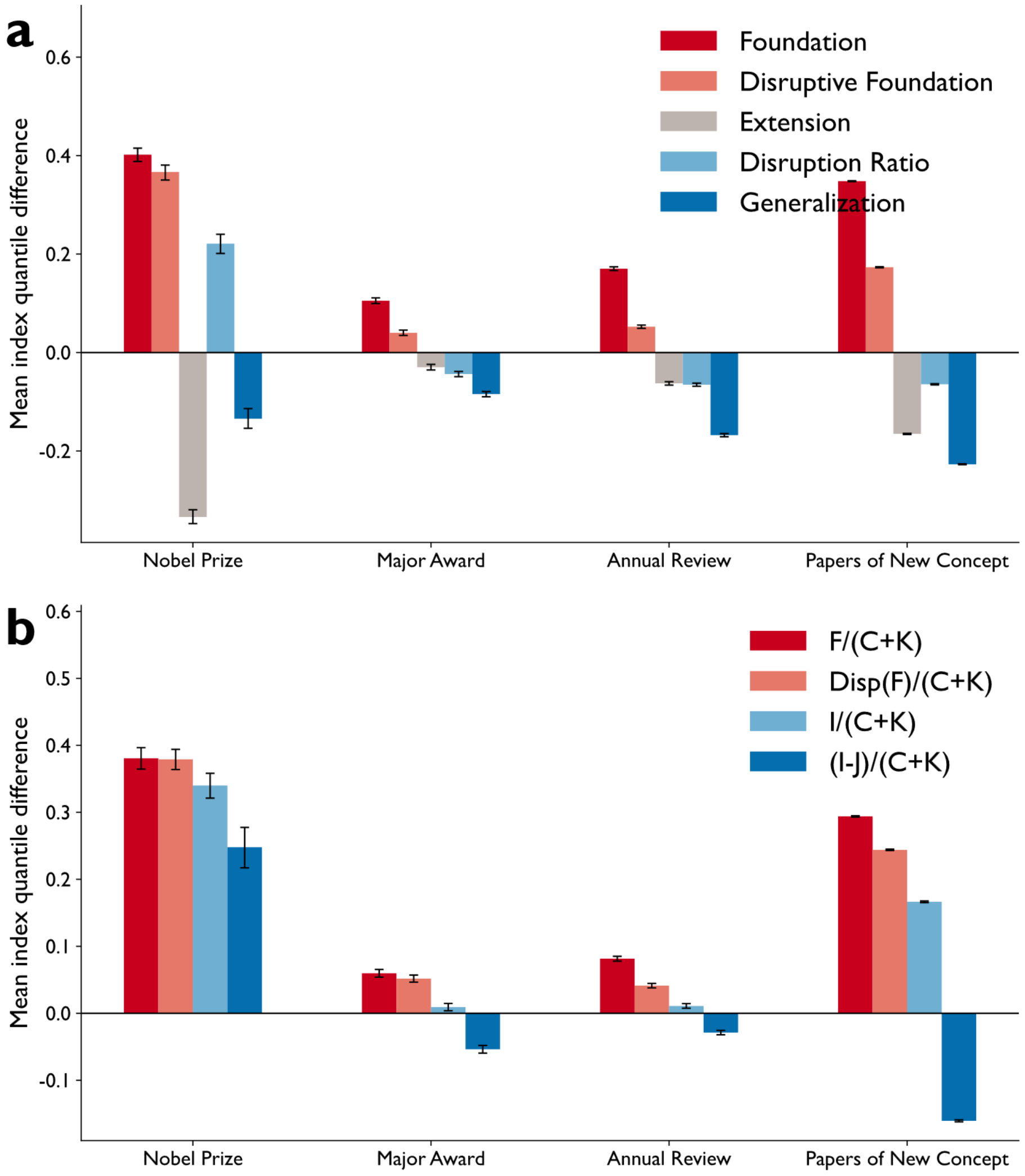

**Fig. S39 | Comparison of the ability of citation-based indices to distinguish desired from undesired papers, with a citation window of at least ten years.** We compare indices based on foundational, disruptive foundation, extensional, disruptive, and generalizational citations using papers published before 2015 and citations accumulated through the end of 2025. We repeat the analysis presented in Extended Data Fig. 5, restricting the sample to papers published before 2015 and computing each index using citations accumulated through the end of 2025. This restriction ensures that every paper has at least ten years to accumulate citations. As in Extended Data Fig. 5, we further restrict the analysis to papers with at least one reference and at least five citations by the end of 2025. The y-axis shows the mean difference in normalized quantile rank between the desired and undesired groups, with quantile ranks normalized by publication year and domain. The analysis includes 385 Nobel Prize-winning articles and 24,516,395 articles in the corresponding undesired group, and 8,930 articles associated with major scientific awards and 24,507,850 articles in the corresponding undesired group, based on OpenAlex data; 1,359,604 articles cited as references by 20,590 review articles, with the review articles constituting the corresponding undesired group, based on Web of Science data; and 175,316 papers introducing new concepts and 27,213,272 papers in the corresponding undesired group, based on Semantic Scholar data.

Across all validation tests, we found that the metrics based on foundational citation ($\frac{F}{C}$ and $\frac{F}{C+K}$) are better at distinguishing paradigm-shifting papers from all other papers, measured by the difference in average index quantiles between the paradigm-shift papers and the others, followed by the metrics based on the disruptive foundation citation ($\frac{Disp(F)}{C}$ and $\frac{Disp(F)}{C+K}$), and then by the disruptive-citation ($\frac{I}{C}$, $\frac{I}{C+K}$ and $\frac{I-J}{C+K}$), with the generalization index $\frac{G}{C}$ having the least effect. This suggests that the disruptive and generalizing nature of citations each degrade performance in isolating papers that drive paradigm-shifts. As discussed conceptually, we believe this pattern arises because the generalization dimension primarily captures papers that facilitate innovation across diverse domains rather than embodying the innovation itself. Similarly, the disruptive dimension tends to overweight modular contributions, such as methods and software, which are widely adopted independently of their predecessors. Both types of contributions are less likely to constitute paradigm-shifting work.

We also note a discrepancy between the average index quantile of Nobel Prize–winning papers obtained in our analysis and that reported by Wu et al. [29]. Wu et al. reported an average disruption index quantile of approximately 98% for Nobel Prize–winning papers, whereas our analysis yields a substantially lower average quantile of 55% (Fig. S37). The reasons underlying this discrepancy are discussed in Section 21 of the Supplementary Results.

**Re-evaluating the Results Derived from the Disruption Index with Foundation, Extension, and Generalization Indices**. We re-evaluate the results derived from the disruption index in two milestone papers, the effect of small teams on scientific innovation by Wu et.al[29], and the effect of scientists' aging on their creativity by Cui et al[2]. Extended Data Fig. 10 presents the pattern observed with the foundation, extension, and generalization index (i.e, the percentage of a paper's citation being foundational, extensional, or generalizing citations), with the replication result using the disruption index presented in the inset. In addition, Fig. S40 presents the patterns based on the quantiles of the foundation, extension, generalization, and disruption indices, while Fig. S41 reports the corresponding patterns based on the quantiles of the normalized indices (i.e, $\frac{F}{C+K}$, $\frac{Disp(F)}{C+K}$, $\frac{I}{C+K}$, $\frac{I-J}{C+K}$). In Fig. S42-S43, we further examine patterns based on the normalized-index quantiles with all indices computed using citations accumulated at least ten years of publication, through the end of the observation period (2025 for OpenAlex and 2020 for Microsoft Academic Graph). This addresses recent concerns that the value of research produced by small teams may not be fully recognized over shorter citation windows [81].

The result suggests that the reported decline in disruption as team size grows [29], and the countries with senior scholars tend to be less disruptive on average [2], are mostly effects of generalization, and may be very different from their ability to generate creative and paradigm-shift papers. We report that a medium-sized team of approximately four authors (measured by raw value as in Extended Data Fig. 10) produces the papers with the highest foundation index, or that papers' foundation index increases as team size grows (measured by index quantile as in Figs. S40-S41). Additionally, there is no significant correlation between the average academic age for authors in a given country, and the average foundation index of the works they produce (See Figs. S40-S41), and countries such as India and China have a high generalization score, while the United Kingdom and United States have a high foundation index.

These patterns are generally robust when the analysis is restricted to papers with citation windows of at least ten years. One notable exception concerns academic age: the mean quantiles of $\frac{F}{C+K}$, $\frac{Disp(F)}{C+K}$, $\frac{I}{C+K}$, $\frac{I-J}{C+K}$ all decline with increasing academic age, with the disruption-related measures ($\frac{I}{C+K}$ and $\frac{I-J}{C+K}$) exhibiting the strongest negative associations. Taken together, these findings suggest that relationships previously identified using the disruption index may mostly capture variation in generalization rather than the capacity to produce foundational or paradigm-shifting research. They therefore motivate a reconsideration of how the disruption index is interpreted as a measure of paradigm-shifting creativity and, more broadly, a re-evaluation of prior evidence on the determinants of scientific innovation.

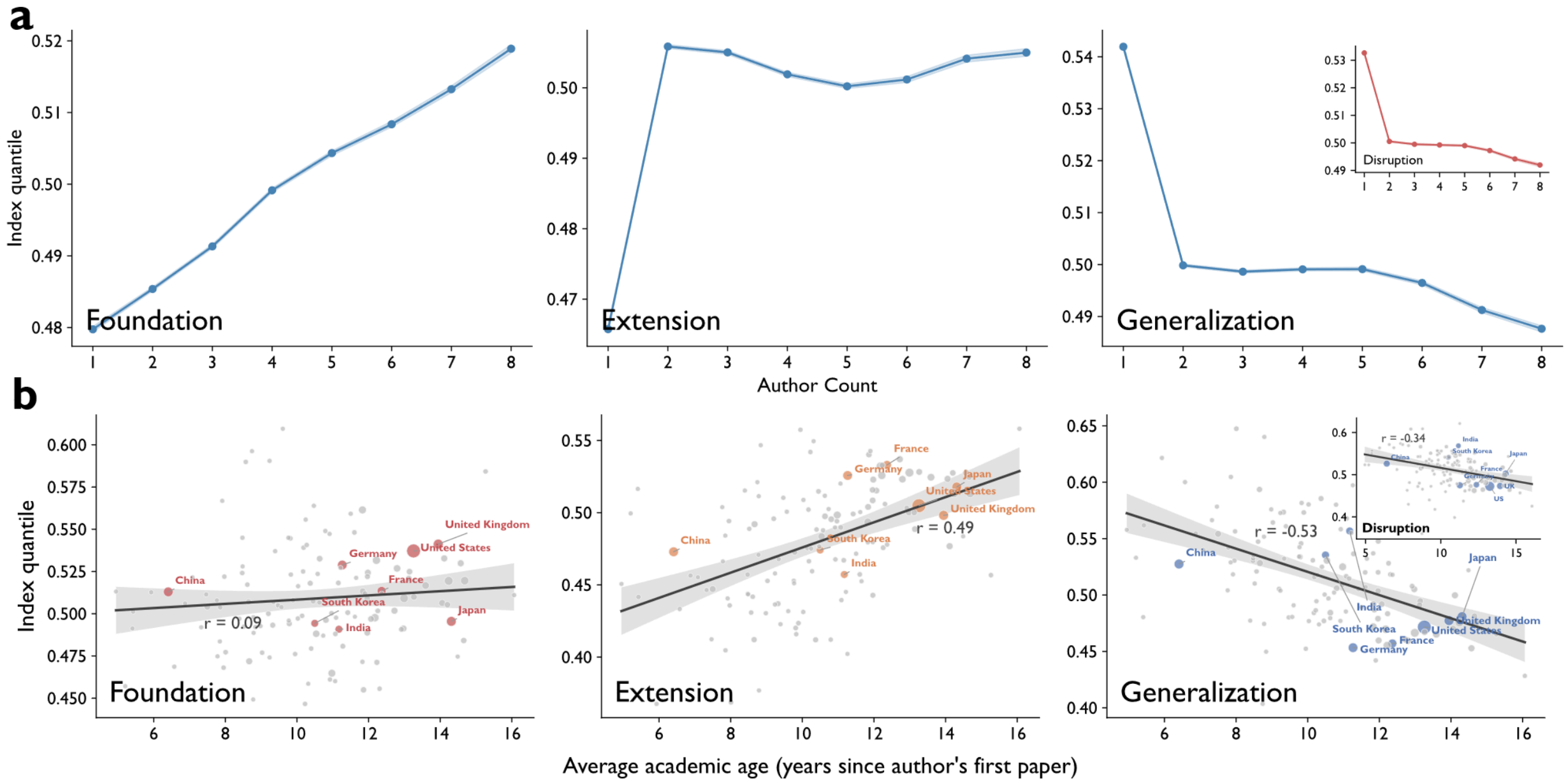

**Fig. S40 | Re-evaluating the "small team" [29] and "aging scientist" [2] findings using the quantiles of the foundation, extension, generalization, and disruption indices.** The foundation, extension, generalization, and disruption indices in both panels are calculated using citations accrued within the five years of publication. Index quantiles are normalized by publication year and domain, representing the relative standing of each paper among all publications from the same year and domain. **a,** Analysis based on 21,308,002 OpenAlex publications published between 1945 and 2019 that contain at least one reference, received at least five citations within five years of publication, have an assigned publication domain, and include between one and eight authors. **b,** Analysis based on 12,899,274 Microsoft Academic Graph (MAG) publications published between 1945 and 2015 that contain at least one reference, received at least five citations within five years of publication, and have an assigned publication domain. Following Cui et al. [2], the analysis is further restricted to authors with at least three publications during the study period (6,418,966 authors) and countries with at least 500 publications (131 countries).

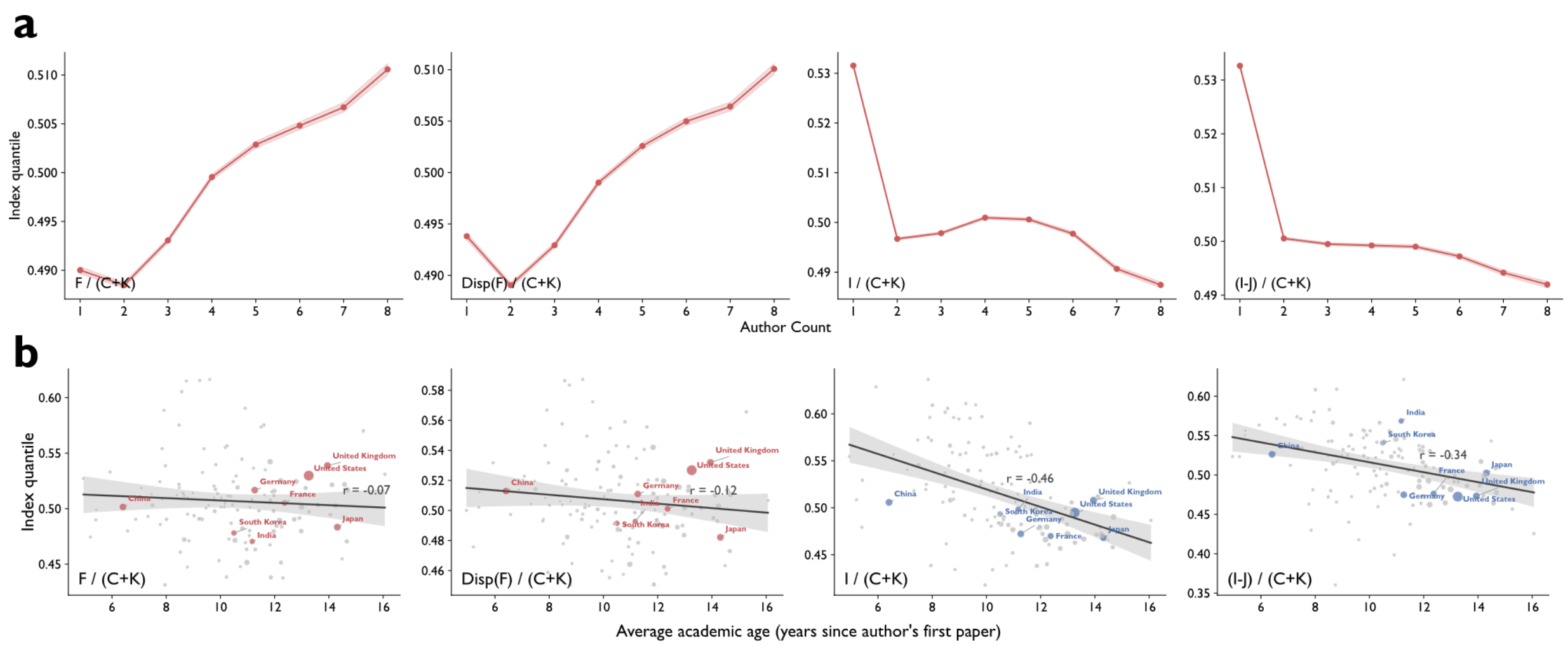

**Fig. S41 | Re-evaluating the "small team" [29] and "aging scientist" [2] findings using the quantiles of the $\frac{F}{C+K}$, $\frac{Disp(F)}{C+K}$, $\frac{I}{C+K}$, $\frac{I-J}{C+K}$ indices. |** The analysis is based on the same sample of publications used in Fig. S39. We do not report results based on the raw values of the $\frac{F}{C+K}$, $\frac{Disp(F)}{C+K}$, $\frac{I}{C+K}$, and $\frac{I-J}{C+K}$ indices because these measures are strongly correlated with publication year, driven by the disproportionate temporal growth of $K$ relative to the other components of the indices, as documented by Park et al[25] and explained by Petersen et al[82]. Moreover, variables such as average team size are also strongly associated with publication year[83], making publication year an important confounding factor in interpreting the observed patterns. Accordingly, all results are presented using publication year- and domain-normalized index quantiles.

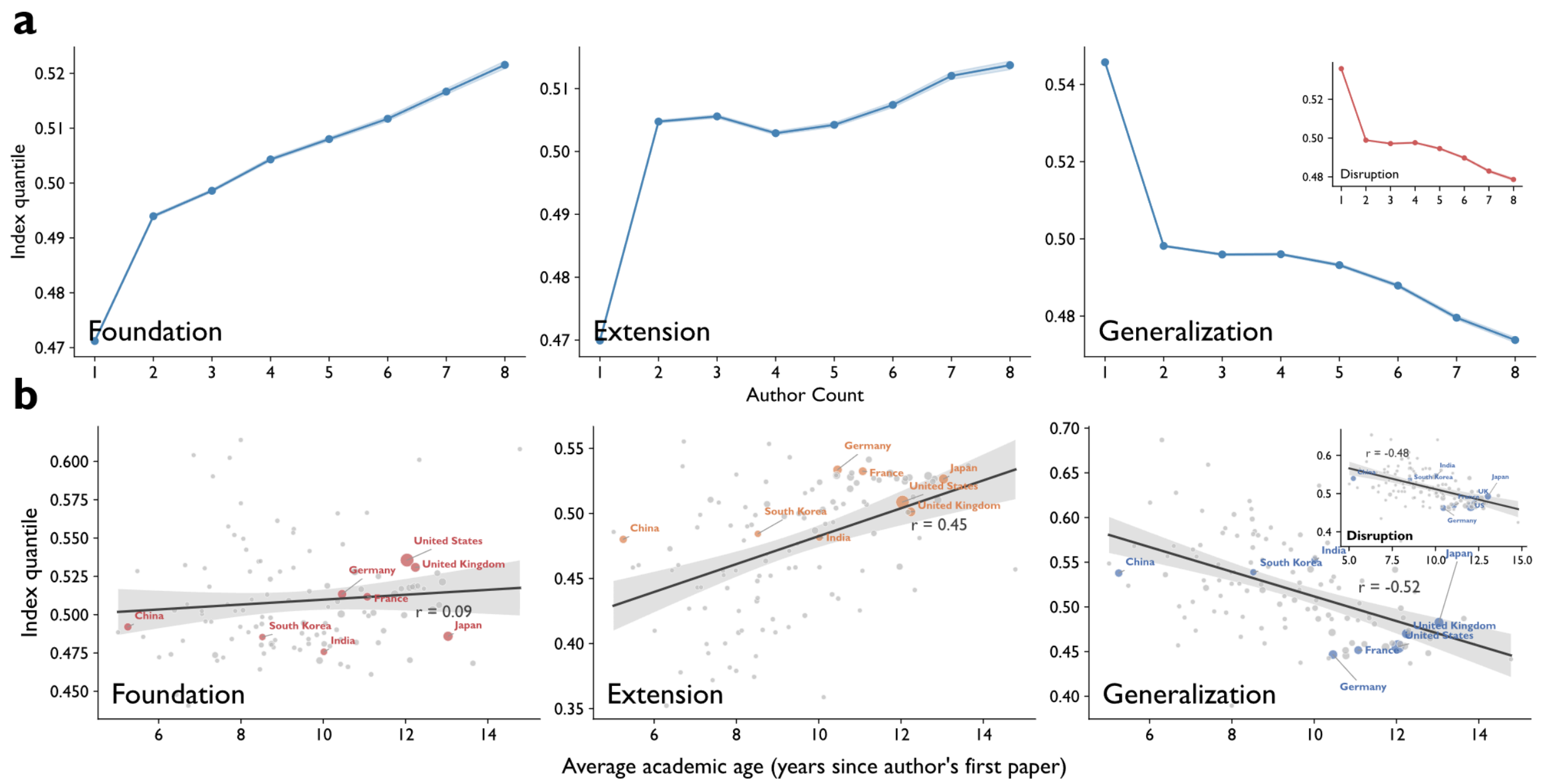

**Fig. S42 | Re-evaluating the "small team" [29] and "aging scientist" [2] findings using the quantiles of the foundation, extension, generalization, and disruption indices, with a citation window of at least ten years. a,** The analysis includes 23,143,225 papers in OpenAlex published between 1945 and 2014, each with at least one reference and at least five citations accumulated through the end of 2025. All other specifications are the same as

those in Fig. S40a. **b,** The analysis includes 15,222,928 papers in Microsoft Academic Graph published between 1945 and 2009, each with at least one reference and at least five citations accumulated through the end of 2021. All other specifications are the same as those in Fig. S40b.

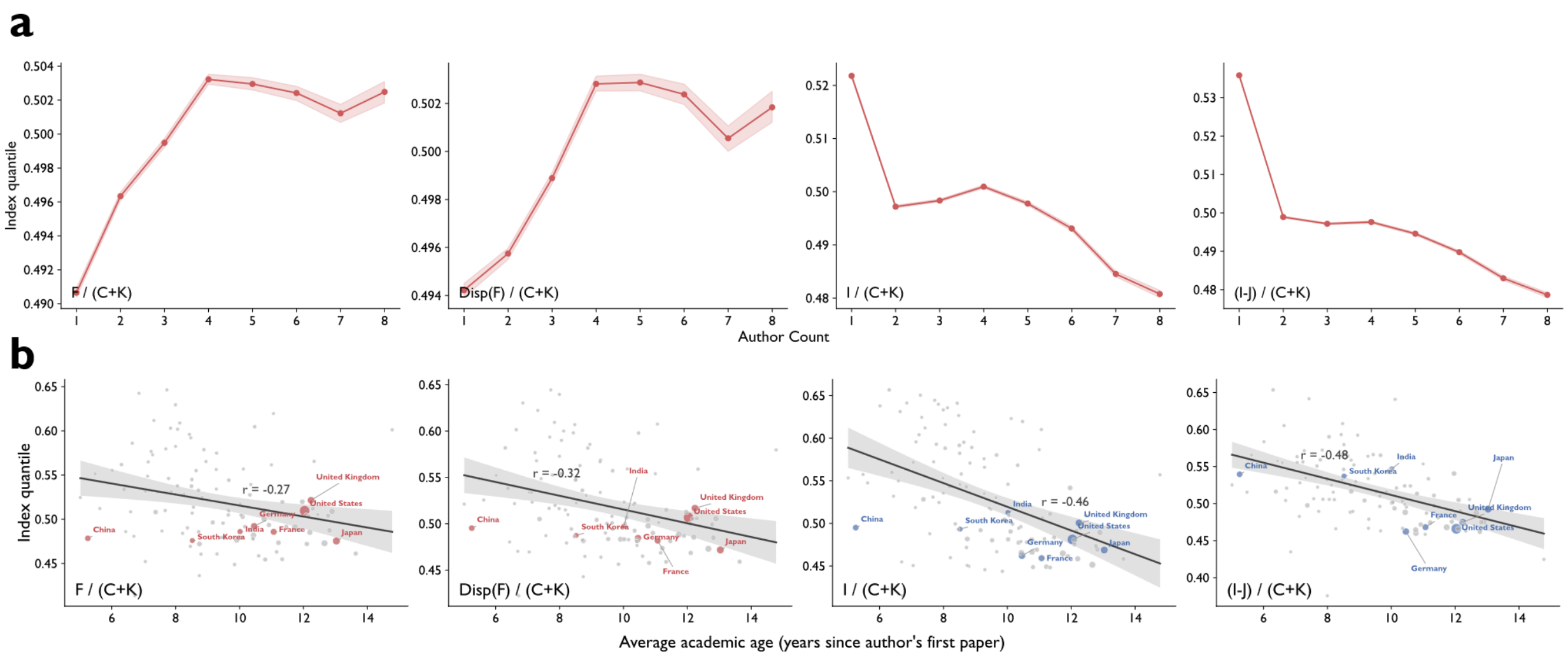

**Fig. S43 | Re-evaluating the "small team"** [29] **and "aging scientist"** [2] **findings using the quantiles of the** $\frac{F}{C+K}$, $\frac{Disp(F)}{C+K}$, $\frac{I}{C+K}$, $\frac{I-J}{C+K}$ **indices, with a citation window of at least ten years. a,** The analysis includes 23,143,225 papers in OpenAlex published between 1945 and 2014, each with at least one reference and at least five citations accumulated through the end of 2025. All other specifications are the same as those in Fig. S41a. **b,** The analysis includes 15,222,928 papers in Microsoft Academic Graph published between 1945 and 2009, each with at least one reference and at least five citations accumulated through the end of 2021. All other specifications are the same as those in Fig. S41b.

## 21. *Reconciling Our Results with Conflicting Evidence*

**The Disruptive Nature of the Nobel Prize–Winning Papers.** In the seminal study that introduced the disruption index to the study of science, Wu et al. [29] reported that Nobel Prize–winning papers have an average disruption index that places them within the top 2% of all papers. However, the original study does not provide sufficient methodological detail to fully reconstruct this calculation. We therefore present two replication approaches that help explain the discrepancy between their reported findings and those obtained in our analysis.

First, we compute the disruption index using citations accumulated during the first five years following the publication of each paper. Although this citation window is not explicitly specified in the original study, a five-year window is commonly employed in studies using the disruption index [25,82]. We match Nobel Prize–winning papers to their corresponding Web of Science (WoS) records using either DOI or exact title matching after converting titles to lowercase and removing special characters. We then calculate the mean disruption index across Nobel Prize–winning papers and determine its percentile relative to all other papers published in the same venues and over the same period as the Nobel

Prize–winning cohort. Our analysis is restricted to Nobel Prize–winning papers published between 1945 and 2009, whereas Wu et al. [29] consider papers published between 1902 and 2009, because we do not calculate disruption scores for papers published before 1945.

The results presented in Fig. S44a indicate that, under a five-year citation window, Nobel Prize–winning papers have a mean disruption index of 0.02, placing the mean within approximately the top 7% of all other papers. In contrast, the median disruption index is −0.006, a value typically characterized as developmental [2], and corresponds to the 51.8th percentile among other papers. These results suggest that the relatively high mean disruption score of Nobel Prize–winning papers is driven by a subset of exceptionally disruptive works and therefore does not characterize the broader distribution of Nobel Prize–winning papers. Indeed, the observation that a majority of Nobel Prize–winning papers have negative disruption scores complicates the interpretation of Nobel Prize–winning research as broadly disruptive and contrasts with the validation evidence reported by Wu et al. [29].

As an alternative replication strategy, we compute the disruption index using all citations accumulated through the end of our observation period (2025). We consider this approach more likely to approximate the procedure employed by Wu et al. [29]. Wu et al. restrict their analysis to papers published no later than 2009, while their dataset extends through 2014, thereby ensuring a minimum citation window of five years for all papers. Consequently, the use of any fixed citation window longer than five years would introduce right-censoring for papers published toward the end of the study period. A plausible alternative, and one that may better approximate their procedure, is therefore to adopt a common endpoint for the observation period and compute the disruption index using all citations accumulated by each paper up to that endpoint.

Fig. S44b presents the results obtained using this approach. The mean disruption index for Nobel Prize–winning papers is 0.167, placing these papers, on average, within the top 2.3% of all other papers; the median falls within the top 6.4%. These estimates are substantially closer to those reported by Wu et al. [29].

Nevertheless, aggregating disruption scores using either the mean or median raises an important methodological concern because these statistics combine disruption indices calculated over citation windows of different lengths. The corresponding percentile rankings are subject to the same limitation, as Nobel Prize–winning papers are compared with other papers whose disruption indices may likewise have been calculated using heterogeneous citation windows.

A more appropriate approach, in our view, is therefore to calculate the percentile rank of the disruption index for each Nobel Prize–winning paper relative to all papers published in the same year and venue, and subsequently calculate the mean and median of these paper-level percentile ranks. This procedure ensures that each Nobel Prize–winning paper is compared with papers published in the same year and, consequently, observed over the same citation window. This is also the procedure employed in our analysis presented in Extended Data Fig. 5.

As shown in Fig. S45, when the disruption index is computed using a five-year citation window, Nobel Prize–winning papers have a mean percentile rank of 52.4% and a median percentile rank of 56.6% relative to other papers published in the same year and venue. Even when the disruption index is computed using all citations accumulated through 2025, providing each paper with at least 15 years of citation history—a substantially longer citation window than that typically used in milestone-based analyses employing the disruption index [25,82]—the mean and median percentile ranks increase to only 73.7% and 88.8%, respectively. These values remain below the corresponding statistics reported by Wu et al. [29]. Moreover, they are lower than those obtained using different versions of our proposed foundation index, under which Nobel Prize–winning papers attain a mean percentile rank of 84.9% (or 84.0%) and a median percentile rank of 93.1% (or 94.4%).

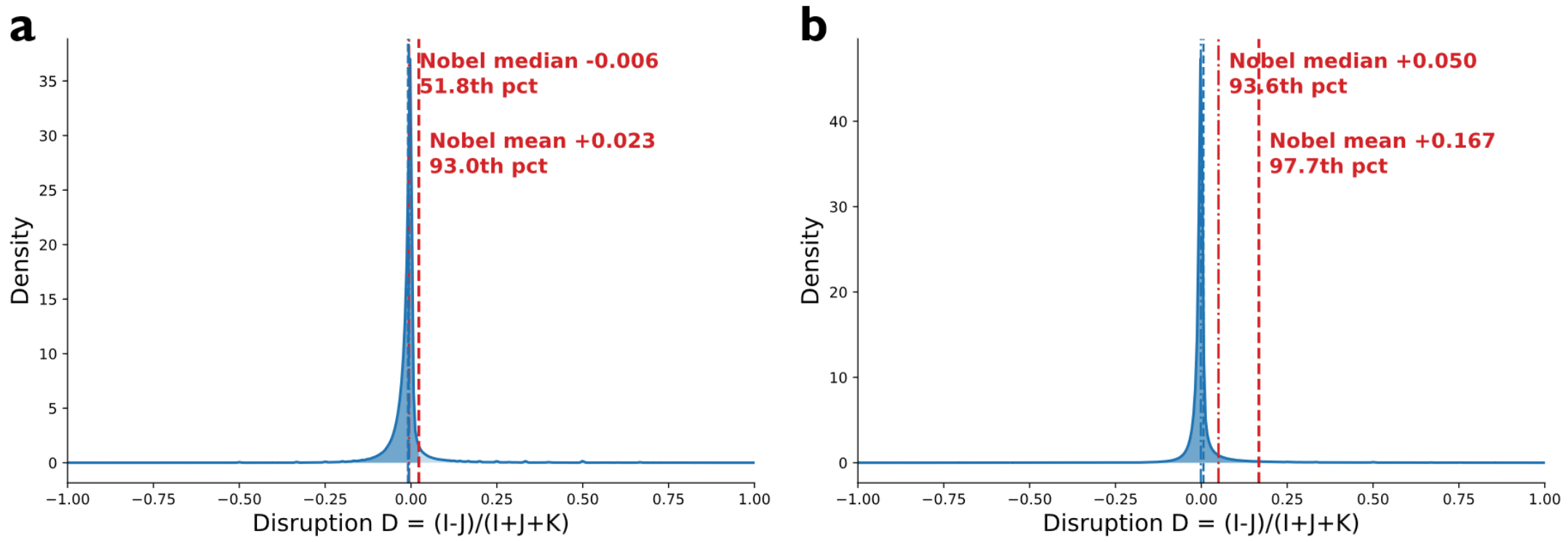

**Fig. S44 | Disruption index distributions for Nobel Prize–winning articles. a,** Distribution of the disruption index among 309 matched Nobel Prize–winning articles published between 1945 and 2009, each with at least one reference and one citation within five years of publication. Disruption indices are computed using citations accumulated within five years of publication. The mean and median disruption indices of the Nobel Prize–winning articles are shown together with their percentile ranks relative to 2,217,674 other articles published between 1945 and 2009 in the same set of 95 venues and satisfying the same requirement of at least one reference and one citation within five years of publication. **b,** Distribution of the disruption index among 311 matched Nobel Prize–winning articles published between 1945 and 2009, each with at least one reference and one citation accumulated through the end of 2025. Disruption indices are computed using all citations accumulated through 2025. The mean and median disruption indices of the Nobel Prize–winning articles are shown together with their percentile ranks relative to 2,301,518 other articles published between 1945 and 2009 in the same set of 96 venues and satisfying the same requirement of at least one reference and one citation through 2025. All data are from the Web of Science (WoS) dataset.

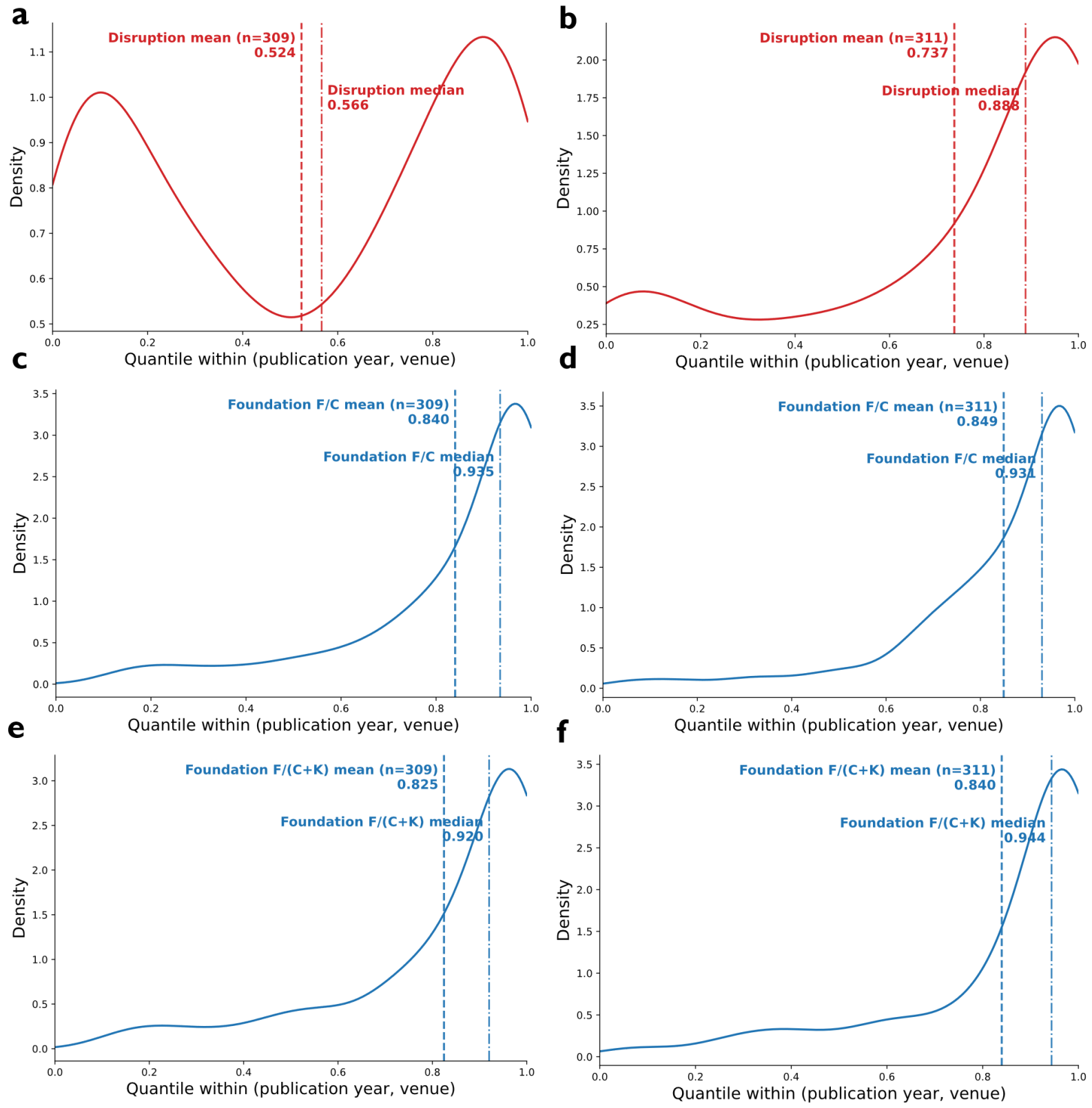

**Fig. S45 | Distribution of index quantiles for Nobel Prize–winning articles relative to other papers. All indices are computed using data from the Web of Science (WoS) dataset. For each Nobel Prize–winning article, the index quantile is calculated relative to comparison articles published in the same year and venue and satisfying the same citation and reference criteria; the figure reports the mean and median of these article-level quantiles. a,** Mean and median disruption-index ($\frac{I-J}{I+J+K}$) quantiles for 309 Nobel Prize–winning articles published between 1945 and 2009, each with at least one reference and

one citation within five years of publication. Each article's quantile is calculated relative to eligible articles published in the same year and venue, drawn from a comparison set of 169,421 articles satisfying the same citation and reference criteria. The disruption index is computed using citations accumulated within five years of publication. **b,** Mean and median disruption-index quantiles for 311 Nobel Prize–winning articles published between 1945 and 2009, each with at least one reference and one citation accumulated through the end of 2025. Each article's quantile is calculated relative to eligible articles published in the same year and venue, drawn from a comparison set of 174,571 articles satisfying the same citation and reference criteria. The disruption index is computed using citations accumulated through the end of 2025, providing at least 15 years of citation accumulation for all articles. **c,** Mean and median foundation-index ($\frac{F}{C}$) quantiles for the 309 Nobel Prize–winning articles and corresponding comparison set of 169,421 articles described in Panel a. Each article's quantile is calculated relative to eligible articles published in the same year and venue, and the foundation index is computed using citations accumulated within five years of publication. **d,** Mean and median foundation-index quantiles for the 311 Nobel Prize–winning articles and corresponding comparison set of 174,571 articles described in Panel b. Each article's quantile is calculated relative to eligible articles published in the same year and venue, and the foundation index is computed using citations accumulated through the end of 2025, providing at least 15 years of citation accumulation for all articles. **e,** The dataset specification is the same as Panel c, with the foundation-index computed as $\frac{F}{C+K}$ **f,** The dataset specification is the same as Panel d, with the foundation-index computed as $\frac{F}{C+K}$.

**The Developmental Nature of Review Papers.** In the same article, Wu and others[29] also reported that review articles exhibit lower disruption scores than the papers they review (cite) on average. This finding was presented as a validation of the disruption index, as it is consistent with the assumption that review articles are generally more developmental in nature than the research they synthesize. Together with the narrative that review articles tend to be cited alongside the works they review, this conclusion has been widely accepted in the disruption literature with limited further scrutiny[31,84]. However, our results suggest the reverse. Among the three indices examined in this study, the disruption (D) index exhibits the strongest association with the generalization (G) index – an index where papers with the word '*review*' in their titles (manual inspection suggests that these are mostly review articles) tend to rank high. If review articles are indeed less disruptive than the works they synthesize as computed by the D index, these two observations would appear to be inconsistent and should not simultaneously hold.

To investigate this discrepancy, we replicated the validation analysis reported by Wu et al. [29]. Our sample consists of all publications with doctype either *article* or *review*. To identify review articles, we first selected journals whose titles contain both the words "*annual*" and "*review*", as such venues exclusively or predominantly publish review articles. This procedure yielded 73 candidate journals. After manually examining each venue, we excluded three journals that were not primarily review-oriented (*Annual Review of Applied Linguistics*, *Annual Review of Cybertherapy and Telemedicine*, and *Lloyds Bank Annual Review*). We then treated all articles published in the remaining venues as review articles and considered all references cited by these articles as the papers being reviewed.

For both review articles and the papers they review, disruption scores were calculated using citations received within five years of publication. Publications that received no citations during this period were excluded. The final sample comprises 22,406 review articles and 1,580,751 reviewed papers. This sample differs slightly from that reported by Wu et al. [29], who analyzed 48 review venues, 22,672 review articles, and 1,338,808 reviewed papers. These discrepancies may stem from differences in cutoff dates (they restricted their analysis to publications from 1954 to 2014, whereas we included all publications prior to 2020), and filter criteria (we retained only records classified as *article* or *review*, whereas they may have included other document types, such as *editorials*, which we believe should be excluded). Despite these differences, we consider our replication to be sufficiently comparable to the original study, as the resulting sample sizes are broadly similar. More importantly, we are able to reproduce the key empirical findings reported by Wu et al. [29], as shown in Fig. S46a.

The review papers exhibit an average disruption index of −0.001 (reported as −0.0009 in [29]), which is lower than the average disruption index of 0.018 for their referenced papers (reported as 0.0008 in [29]). However, after excluding all articles with zero references—removing 125 review papers and 39,256 referenced papers, corresponding to only a 0.56% reduction in the review-paper sample and a 2.48% reduction in the referenced-paper sample—the relationship **reverses** (Fig. S46b). Under this specification, review papers have an average disruption index of −0.006, slightly higher than that of the referenced papers (−0.007).

The disproportionate effect of removing less than 2.5% of the observations arises from the definition of the disruption index, ($D = \frac{I-J}{I+J+K}$). By construction, papers with zero references receive a disruption index of 1, because these papers have no predecessors ($K$=0) and all citations are automatically classified as disruptive ($J$=0). Such observations therefore constitute extreme **high-leverage points**. This issue is particularly consequential because the overall magnitude of disruption scores is small: among all review and referenced papers in our sample, the mean absolute disruption index is only 0.032. Consequently, a disruption score of 1 is more than thirty times larger than the typical magnitude observed in the data and can substantially distort the sample mean.

Most zero-reference papers likely result from limitations in dataset coverage and should reasonably be excluded from the analysis. Even if those papers were not excluded, we should report metrics that are less sensitive to extreme values such as the median of the distribution (Fig. S46a), or converting the raw value into quantiles (Fig. S46c). Both results suggest that the review papers are more disruptive than their references, a result that contradicts the conceptual definition of disruption and their own validation in [29].

As a robustness check, we replicated the analysis using two datasets: OpenAlex and WoS, and employed two alternative approaches to identify review articles. First, we utilized the document-type classifications provided by each database, both of which include a dedicated *review* category. Second, we classified a publication as a review article if the term "*review*" appeared in its title.

To assess the validity of the title-based approach, we randomly sampled 50 papers containing the word "*review*" in their titles and manually evaluated them. The first author independently examined each paper and determined whether it constituted a genuine review article. Of the sampled papers, 39 (78%) were identified as true review articles, indicating a reasonably high level of precision for this method.

Using the title-based approach, we identified 428,645 review articles, approximately twenty times the number identified using the method proposed by Wu et al[29]. This substantial increase suggests that the title-based approach is complementary to that of Wu et al[29], primarily by improving recall, albeit at the cost of reduced precision (We expect the precision of Wu et al.'s approach to be close to perfect, as it relies on venues that predominantly publish review articles. Conducting a comparable manual validation of the papers identified by Wu et al.'s [29] method would be more challenging, as many of these publications have short or ambiguous titles, making it difficult and time-consuming to verify the exact articles and their review status).

Across all identification approaches and experimental settings in both datasets, review articles generally exhibit higher disruption scores than the works they cite. This finding is consistent with our re-evaluation and contrasts with the results reported by Wu et al [29]. The sole exception occurs in the OpenAlex dataset when review articles are identified using the database-provided document type classification. In this case, the referenced articles show a marginally higher disruption score than the review articles themselves (−0.001 versus −0.002, respectively). However, this relationship reverses when the raw disruption scores are transformed into quantiles.

To further account for temporal and disciplinary variation in disruption scores, we computed disruption quantiles after grouping publications by publication year, as well as by both publication year and research domain. Across all such specifications, review articles consistently rank higher in disruption than the articles they reference. These results suggest that the observed pattern is robust to alternative datasets, review-identification methods, and normalization procedures, providing strong evidence that review articles are, on average, more disruptive than their cited antecedents. Furthermore, the proposed F and G indices provide a clearer distinction between review articles and the works they cite. Across datasets and identification approaches, review articles consistently exhibit higher G-index values, whereas their referenced articles tend to exhibit higher F-index values. This pattern validates that the proposed index framework effectively decomposes the disruption index into interpretable components that capture distinct functional roles within the scientific literature.

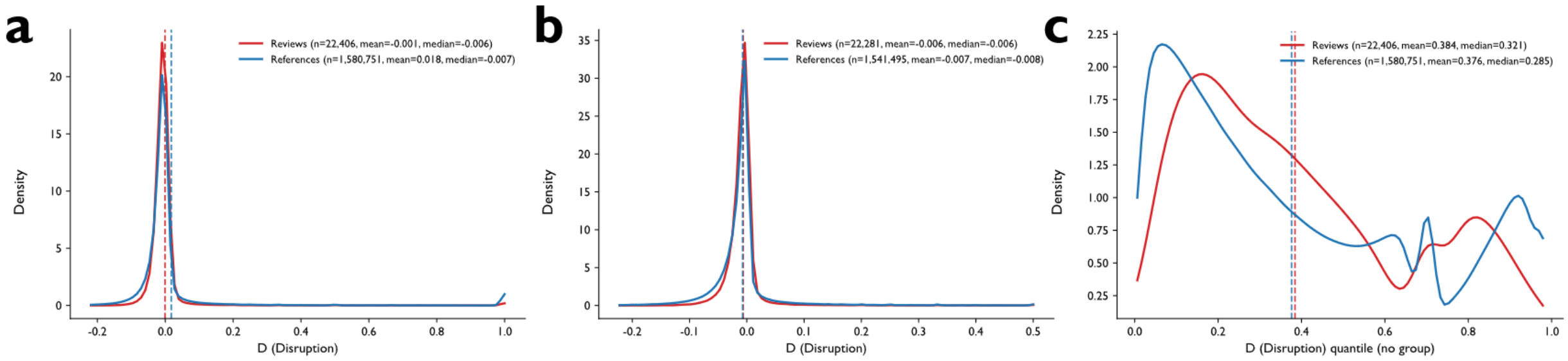

**Fig. S46 | Disruption index distributions for review articles and their referenced works.** Review articles are identified based on publication venues, following the method of Wu et al. [29]. Panels **a** and **c** include publications receiving at least one citation within five years of publication, and published no later than 2019, with disruption scores calculated from citations accrued during the same five-year window. Panel **a** shows raw disruption index values, while Panel **c** shows disruption quantiles relative to all WoS publications receiving at least one citation within five years of publication. Panel **b** is restricted to WoS publications with at least one reference and at least one citation within five years of publication, and published no later than 2019. Dotted horizontal lines indicate cohort means, which are also reported in the legend along with the sample size in each cohort.

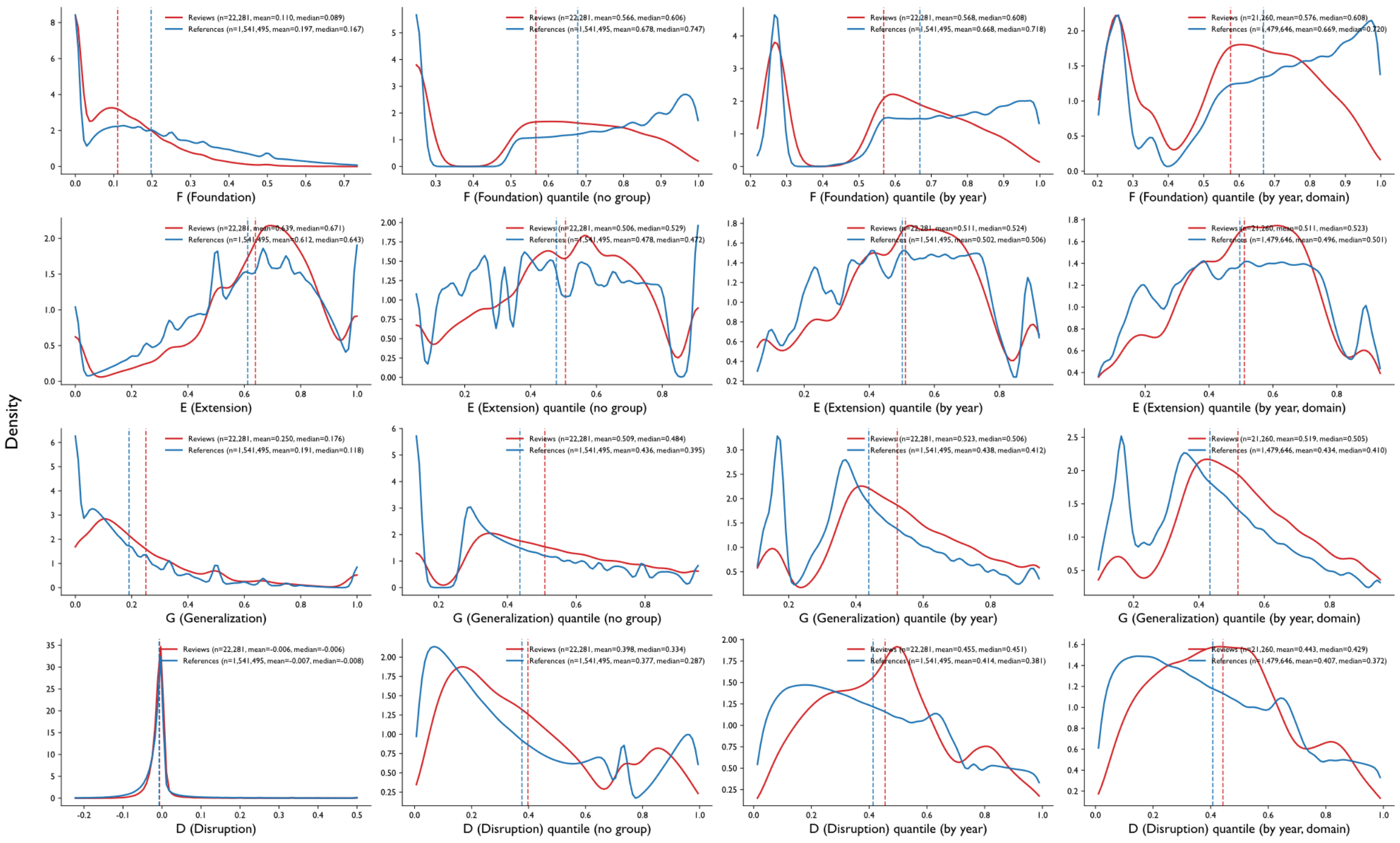

**Fig. S47 | Raw values and quantiles of the disruption, foundation, extension, and generalization indices for review articles and their referenced works, with review articles identified based on publication venue names in the WoS dataset.** The first column presents the raw values of each index, while the second through fourth columns present the corresponding quantiles under different normalization schemes. All metrics are computed for WoS publications that contain at least one reference, receive at least one citation within five years of publication, and were published no later than 2019. Review articles are identified based on publication venues, following the approach of Wu et al. [29]. Quantile measures are calculated within the corresponding grouping schemes shown in each column.

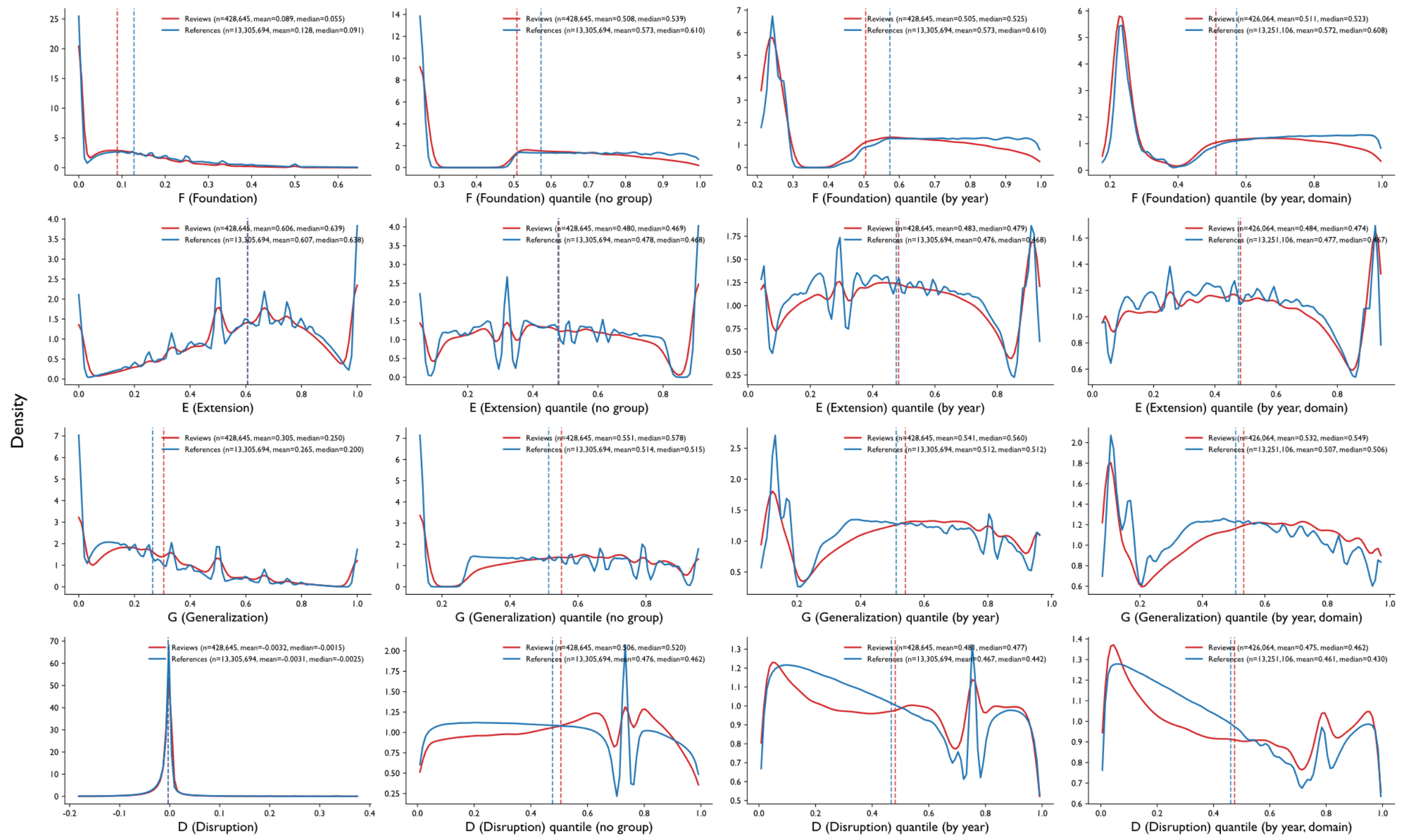

**Fig. S48 | Raw values and quantiles of the disruption, foundation, extension, and generalization indices for review articles and their referenced works, with review articles identified based on paper titles in the WoS dataset.** The first column presents the raw values of each index, while the second through fourth columns present the corresponding quantiles under different normalization schemes. All metrics are computed for WoS publications that contain at least one reference, receive at least one citation within five years of publication, and were published no later than 2019. Review articles are identified based on the appearance of the term "*review*" in the paper's title. Quantile measures are calculated within the corresponding grouping schemes shown in each column.

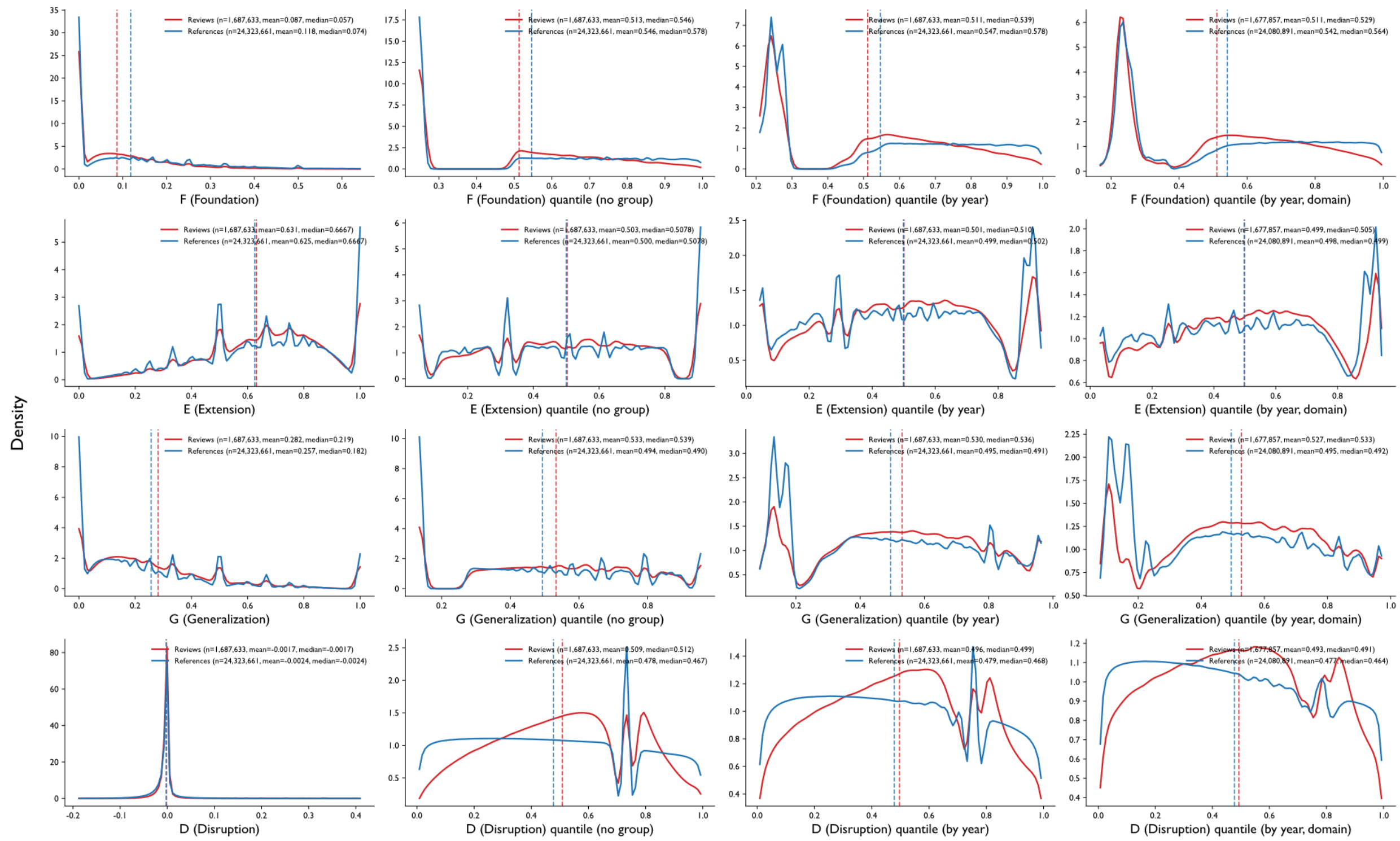

**Fig. S49 | Raw values and quantiles of the disruption, foundation, extension, and generalization indices for review articles and their referenced works, with review articles identified based on document type in the WoS dataset.** The first column presents the raw values of each index, while the second through fourth columns present the corresponding quantiles under different normalization schemes. All metrics are computed for WoS publications that contain at least one reference, receive at least one citation within five years of publication, and were published no later than 2019. Review articles are identified based on the labelled document type from the dataset. Quantile measures are calculated within the corresponding grouping schemes shown in each column.

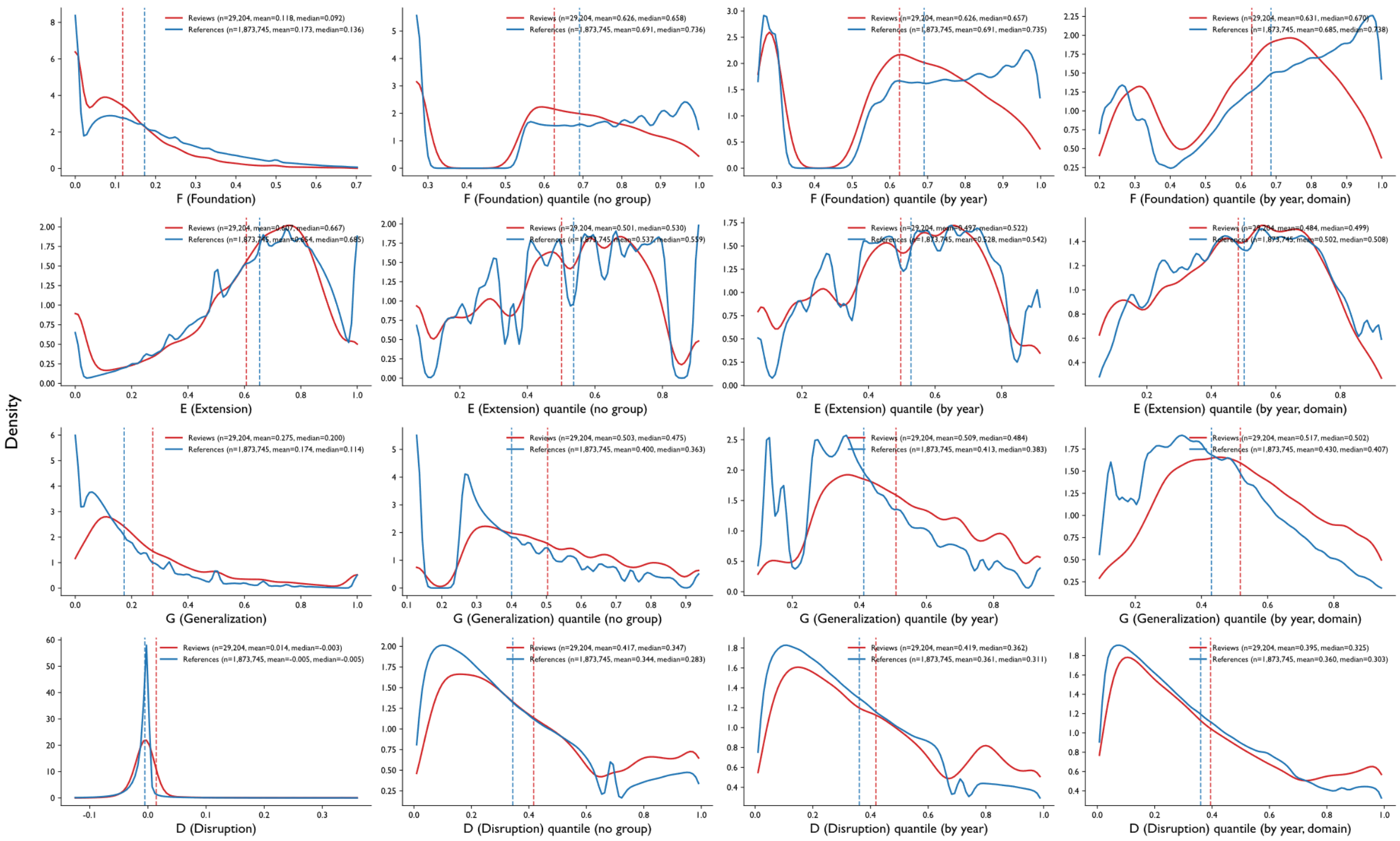

**Fig. S50 | Raw values and quantiles of the disruption, foundation, extension, and generalization indices for review articles and their referenced works, with review articles identified based on publication venue names in the OpenAlex dataset.** The first column presents the raw values of each index, while the second through fourth columns present the corresponding quantiles under different normalization schemes. All metrics are computed for OpenAlex publications that contain at least one reference, receive at least one citation within five years of publication, and were published no later than 2019. Review articles are identified based on publication venues, following the approach of Wu et al. [29]. Quantile measures are calculated within the corresponding grouping schemes shown in each column.

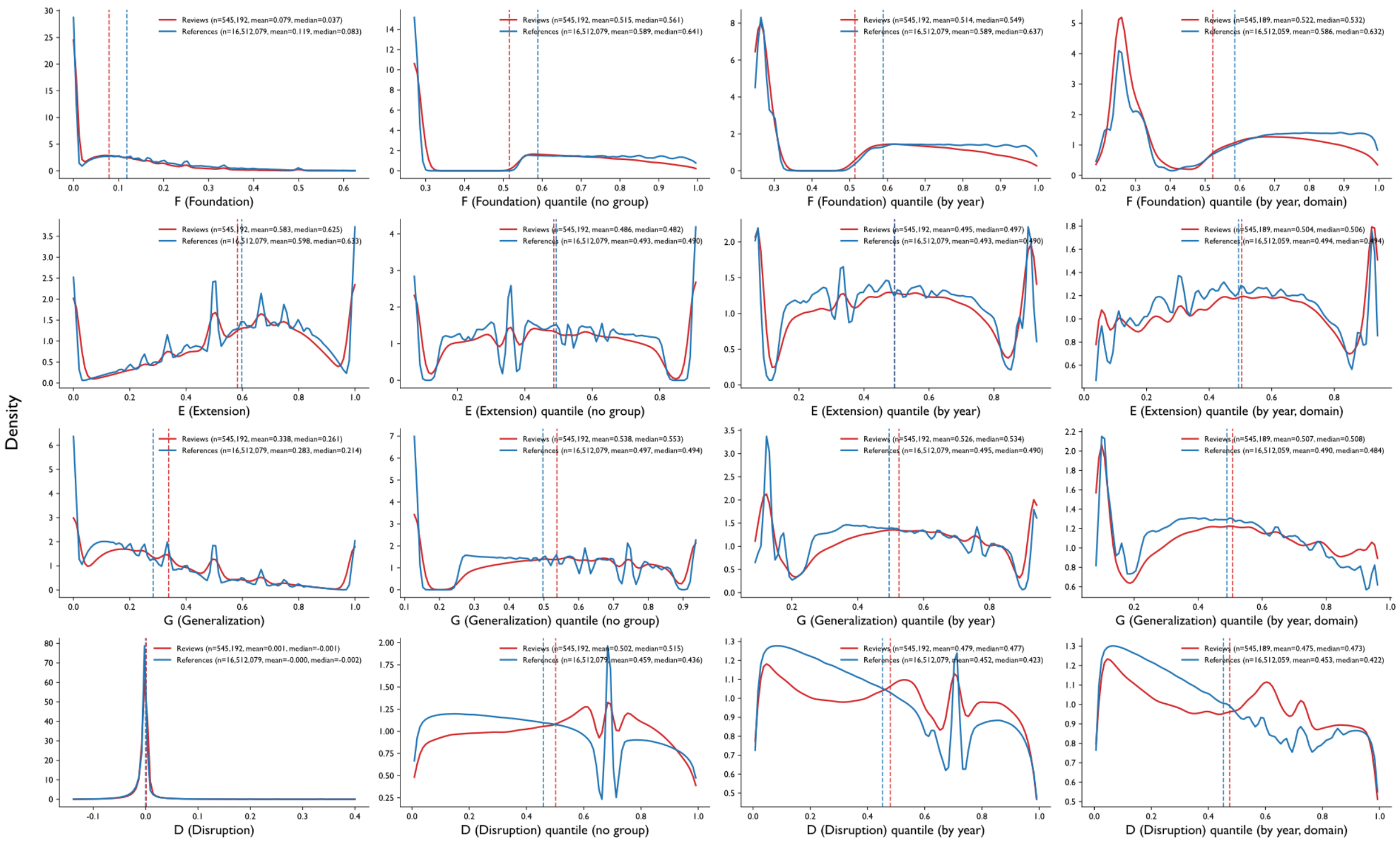

**Fig. S51 | Raw values and quantiles of the disruption, foundation, extension, and generalization indices for review articles and their referenced works, with review articles identified based on paper titles in the OpenAlex dataset.** The first column presents the raw values of each index, while the second through fourth columns present the corresponding quantiles under different normalization schemes. All metrics are computed for OpenAlex publications that contain at least one reference, receive at least one citation within five years of publication, and were published no later than 2019. Review articles are identified based on the appearance of the term “*review*” in the paper’s title. Quantile measures are calculated within the corresponding grouping schemes shown in each column.

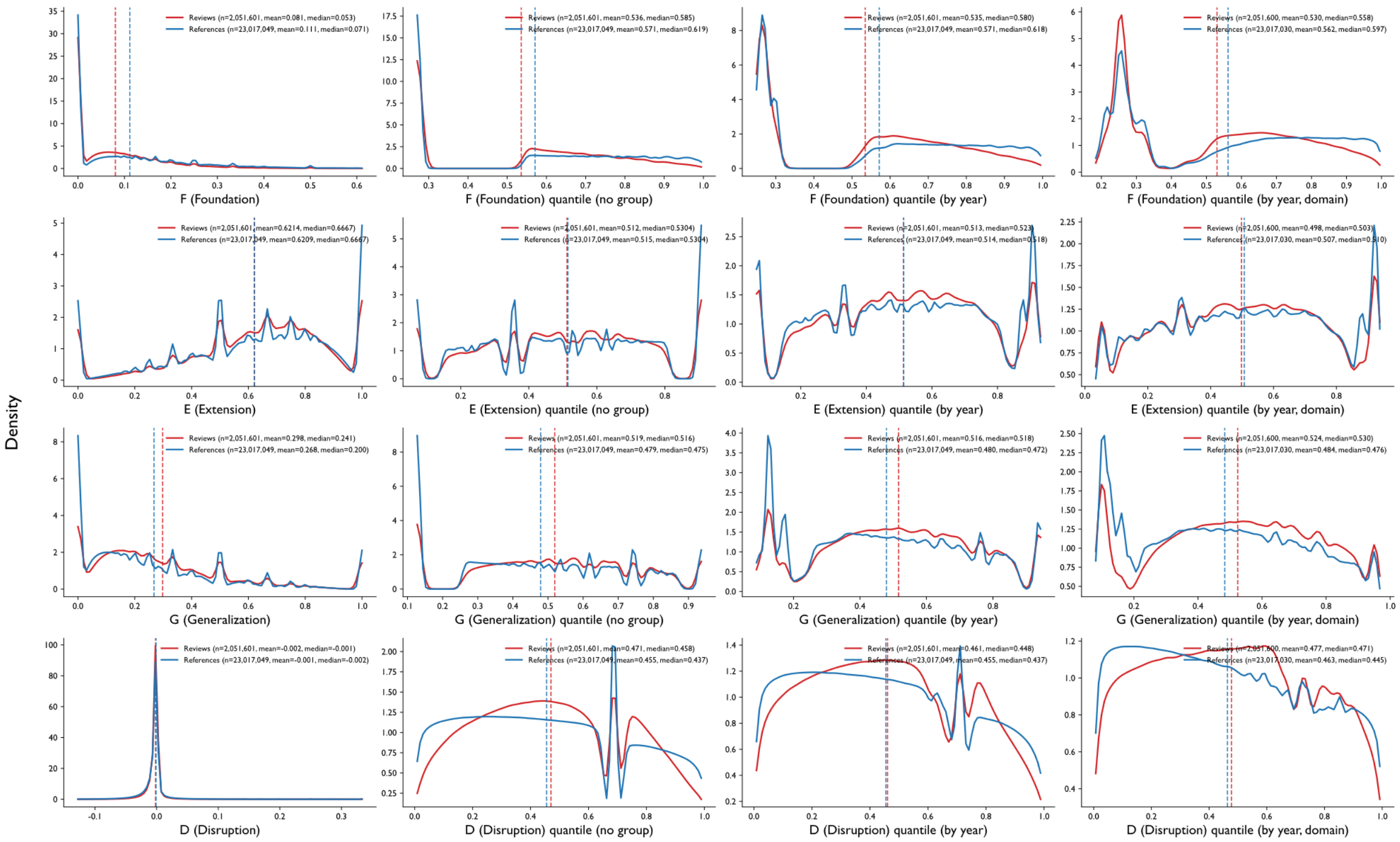

**Fig. S52 | Raw values and quantiles of the disruption, foundation, extension, and generalization indices for review articles and their referenced works, with review articles identified based on document types in the OpenAlex dataset.** The first column presents the raw values of each index, while the second through fourth columns present the corresponding quantiles under different normalization schemes. All metrics are computed for OpenAlex publications that contain at least one reference, receive at least one citation within five years of publication, and were published no later than 2019. Review articles are identified based on the labelled document type from the dataset. Quantile measures are calculated within the corresponding grouping schemes shown in each column.

**The decline in the Disruption Index.** We argue that the decline in disruption reported by Park et al. [25] is driven largely by the disproportionate increase in K relative to changes in I and J. As shown in Fig. S53a, the ratio $\frac{I+J}{K}$ declines from 0.60 in 1945 to 0.03 in 2019, indicating that K increased substantially faster than I+J (or C, the total number of citations received by the average paper). When K is excluded from the original definition of the disruption index, the resulting variants, $\frac{I}{I+J}$ and $\frac{I-J}{I+J}$, exhibit longitudinal patterns similar to that of the generalization index reported in Fig. 3 (Fig. S53b), consistent with the strong association between the disruption and generalization indices. Thus, our findings do not necessarily contradict the mechanism underlying the temporal decline documented by Park et al.[25]; rather, they reflect a different analytical focus by separating changes in citation structure captured by I and J from the pronounced growth in K. A similar pattern has also been reported and discussed by Petersen et al. [82].

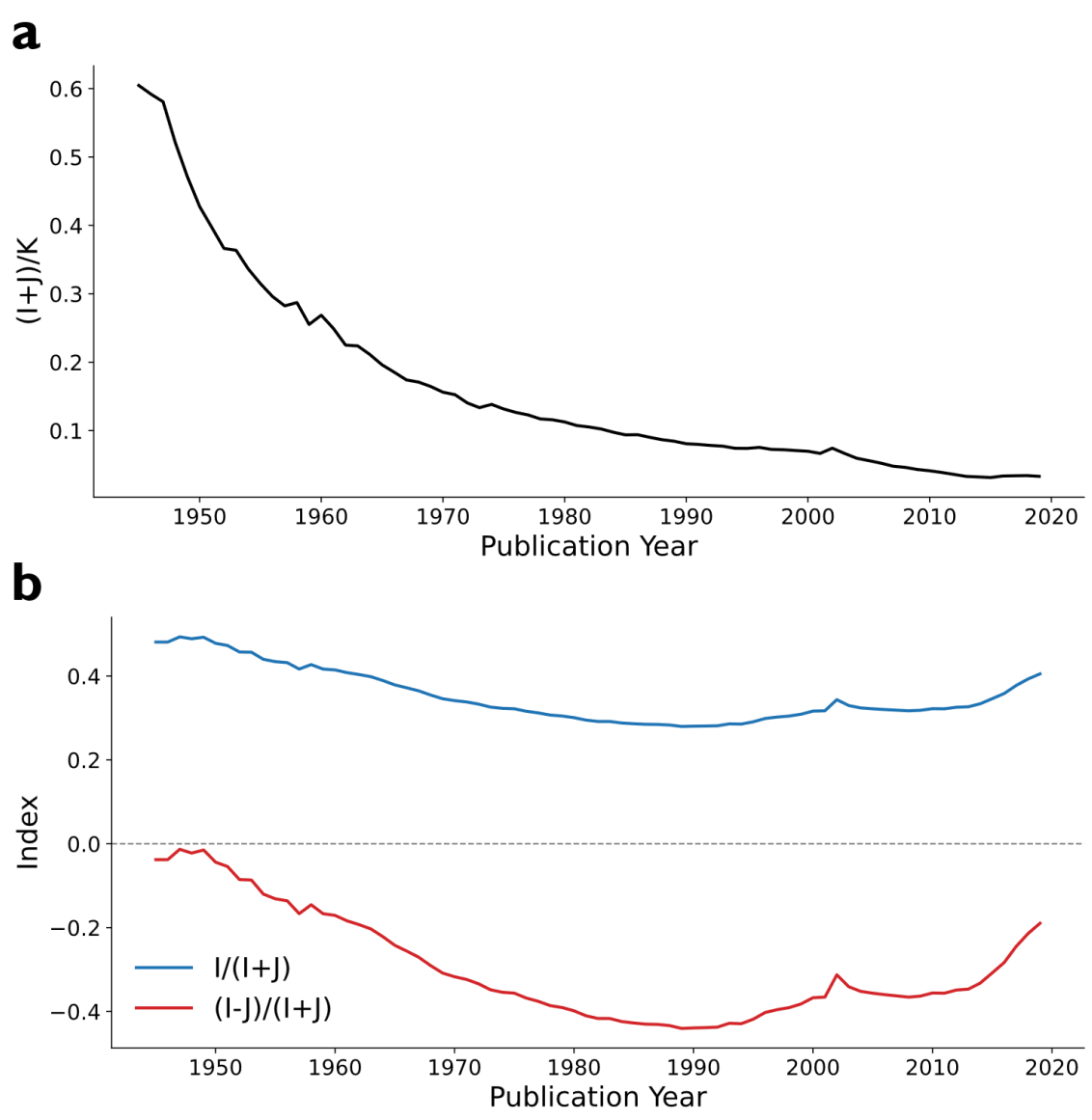

**Fig. S53 | Longitudinal change of $\frac{I+J}{K}$, $\frac{I}{I+J}$, and $\frac{I-J}{I+J}$.** The analysis includes 23,448,431 papers in OpenAlex published between 1945 and 2019, each with at least one reference and at least five citations within five years of publication. All indices are computed using citations accumulated within five years of publication. **a,** Longitudinal change in $\frac{I+J}{K}$. **b,** Longitudinal changes in $\frac{I}{I+J}$ and $\frac{I-J}{I+J}$.

**Decreased Atypicality over Time**. A less discussed trend in Park et al's *Nature* paper[85] is the reported decline in the atypicality of scientific publications over time, which serves as a key piece of evidence supporting their argument that scientific creativity has decreased. In contrast, our results shown in Fig. S21 reveal the opposite pattern: the average atypicality of papers increases over time.

To investigate the source of this discrepancy, Panel **a** of Fig. S5 presents our calculation of atypicality across publication years, while Panel **b** represents the corresponding measure from Park et al. (the details of their computation are not reported in the paper, and we infer their approach from their reported results). A critical methodological difference lies in the treatment of venue-level z-scores. In our approach, the venue-level z-score is held constant across periods when calculating paper-level atypicality, whereas Park et al. [85] recompute the venue-level z-score separately for different time periods.

Because our venue-level z-score benchmark remains fixed, a higher average atypicality in a given period indicates that papers published during that period are more likely to cite venue pairs that would have been considered distant according to historical citation patterns. As shown in Fig. S21, the distribution of atypicality values (measured as the median z-score among referenced venue pairs) for more recent papers, such as those published in the 2010s, is shifted toward lower values relative to papers published in earlier decades, such as the 1980s. This pattern implies an increase in atypicality, or equivalently, greater distance among referenced knowledge sources.

Although we are able to largely replicate the empirical pattern reported by Park et al. [85] in Fig. S54, the comparison between our result and theirs suggests that their conclusion of declining atypicality is driven not solely by changes in citation behavior, but also by temporal variation in the underlying venue-level z-scores used to construct the measure. Consequently, the observed decline in atypicality in their paper reflects a combination of changing reference patterns and shifts in the normalization benchmark, rather than an unambiguous decrease in the novelty or distance of cited knowledge combinations.

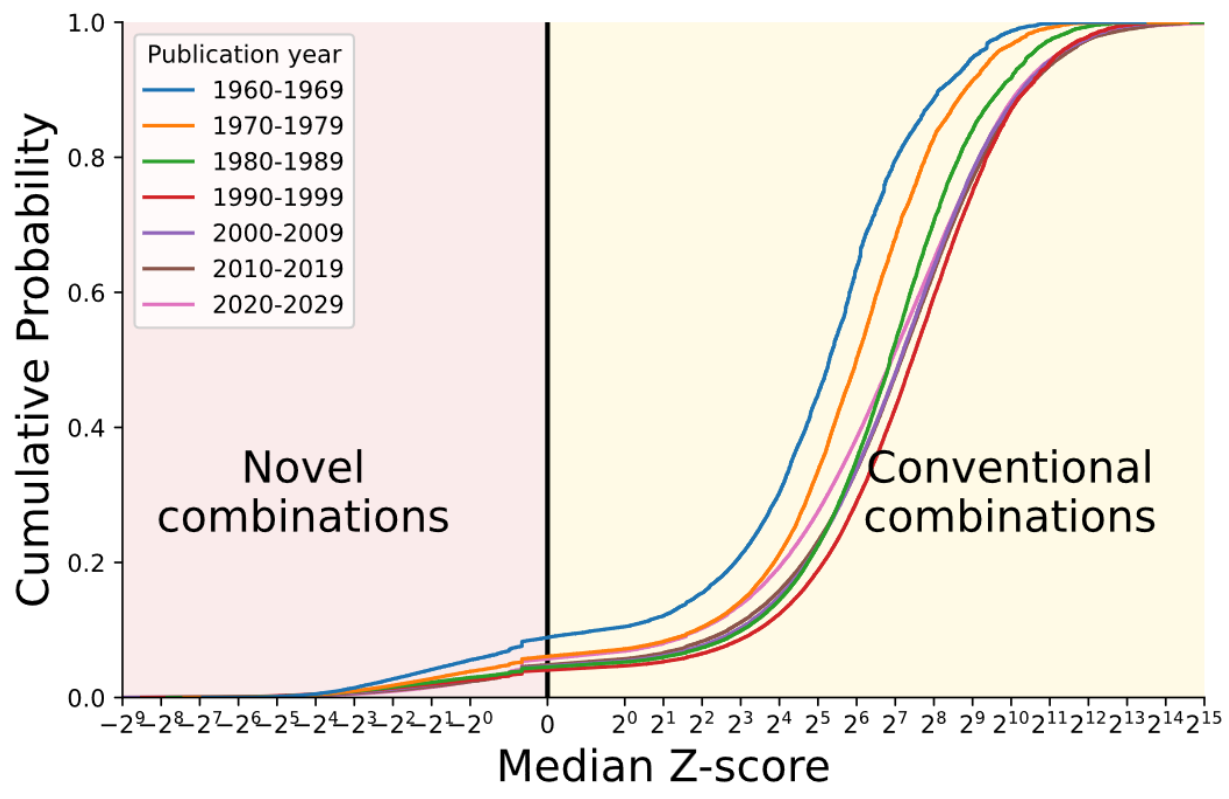

**Fig. S54 | Changes in atypicality under time-varying venue-level z-score computation.** Atypicality is computed separately for papers published in different periods, with both the venue-level z-score computation and the paper-level atypicality measures derived from period-specific sets of publications. The resulting patterns largely replicate those reported by Park et al.[25].

## 22. *Other Technical Details and Validations*

**Deleting Citations towards Earlier Publications.** In *OpenAlex*, some publications contain references to articles that were published after the citing paper, resulting in temporal inconsistencies and citation cycles within the network. Because such anomalies may distort the computation of the F, E, and G indices, we removed all citations that violate publication-year ordering from our analyses.

After removing 17,596,686 temporally inconsistent citation links (1.05% of all citation links), the resulting citation network contains 1,651,960,324 citation links. Given the small proportion of removed links, we do not expect this filtering procedure to materially affect the results reported in this study.

As a robustness check, we replicated our main analysis—the longitudinal trajectories of the F, E, and G indices—using the original citation network without removing citations based on publication year. The results remained substantively unchanged, confirming that our findings are robust to the treatment of these anomalous citation links.

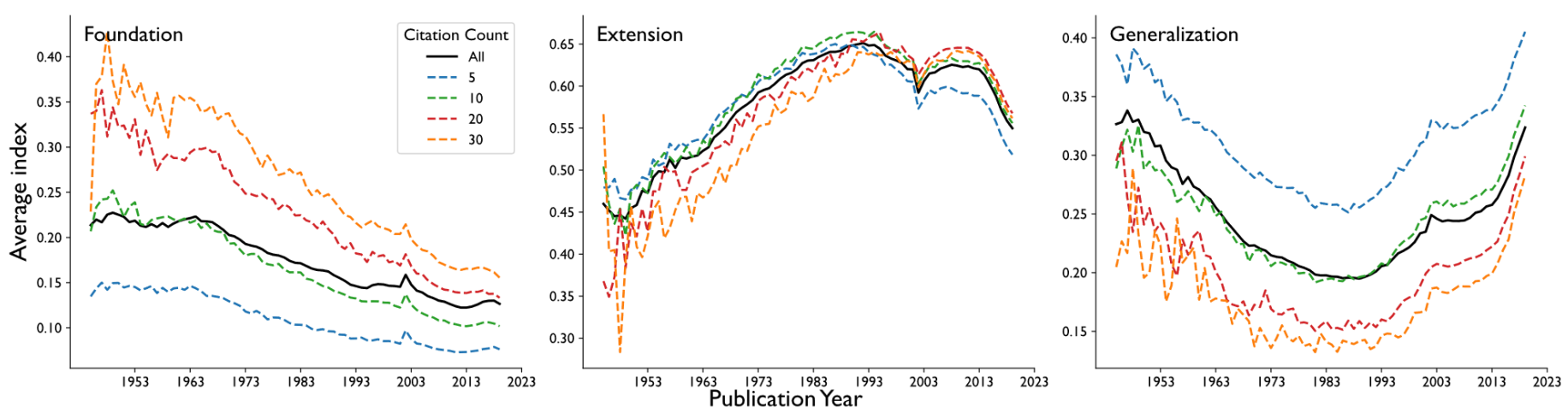

**Fig. S55 | Longitudinal changes in the F, E, and G indices computed from the OpenAlex citation network recorded in the original dataset, without removing citations to papers published after the citing paper.** The analysis is restricted to 23,911,215 papers that contain at least one reference, received at least five citations within five years of publication, and were published no later than 2019.

**The Approximation in Network Reshuffling**. In our analysis, we construct a null model through network reshuffling. Ideally, the reshuffled network should preserve the indegree and outdegree distributions of the original network. Achieving this requires maintaining, throughout the reshuffling process, the set of articles cited by each paper.

Fig. S56 illustrates the issue. Suppose the citation from article X to A is swapped with the citation from article Y to B. If article X already cites article B, then after the swap, article X would contain two citation edges pointing to the same paper. Because only unique citation relationships are retained, the duplicate edge is removed. Consequently, the outdegree of article X and the indegree of article B are each reduced by one relative to their values before reshuffling.

Preventing such degree changes would require dynamically tracking and updating the citation lists of all papers during the reshuffling procedure to avoid creating duplicate citation links. However, this additional constraint substantially increases the computational complexity of the algorithm and renders the reshuffling process nearly intractable for large networks. As a practical approximation, we allow the situations illustrated in Fig. S56 to occur. We show that this approximation affects only a very small fraction of articles and introduces negligible changes to the network structure. In the observed network, after excluding papers without valid venue information, there are 145,412,322 papers connected by 1,396,986,658 citation links. Following reshuffling, the number of citation links decreases to 1,391,839,498, corresponding to a reduction of only 0.37%. At the article level, only 2,470,581 papers experience a change in indegree (1.70% of all papers), while 3,989,856 papers experience a change in outdegree (2.74% of all papers). These results indicate that the reshuffling procedure preserves the vast majority of citation relationships and degree distributions, with only minimal deviations from the observed network.

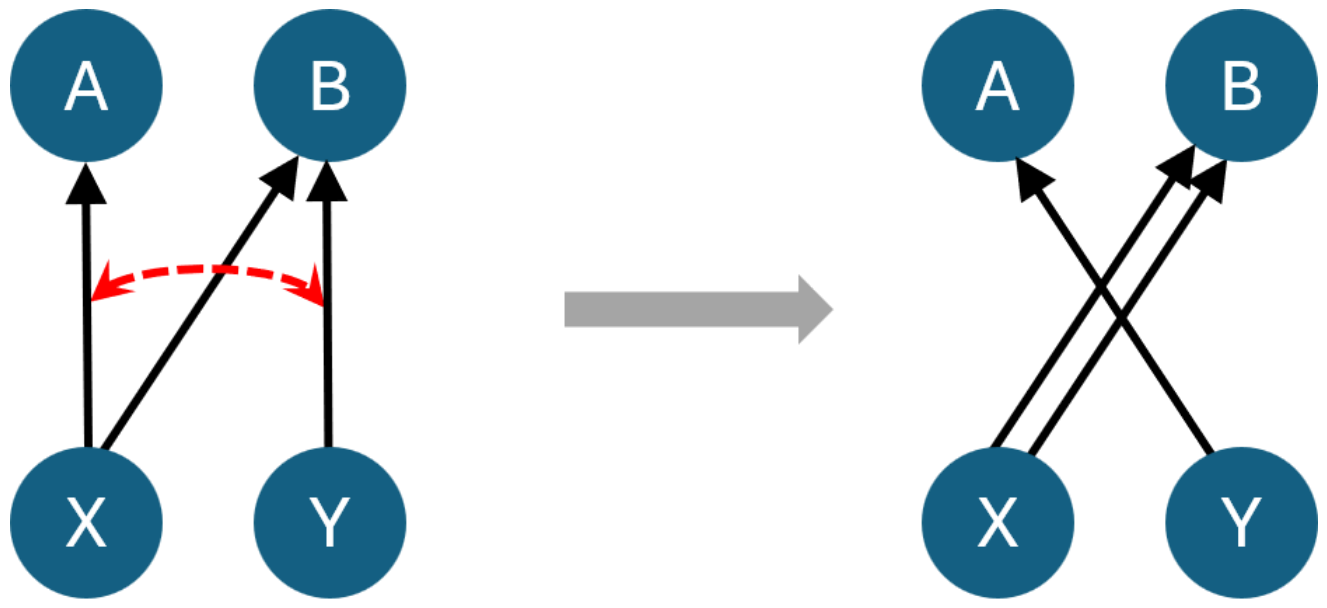

**Fig. S56 | Illustration of the issue encountered during the random reshuffling process.**

**The Influence of Papers with Missing References.** In OpenAlex, more than 60% of all papers have zero reference recorded, and this dataset coverage problem may have a major impact on the robustness of our finding, given that the computation of foundation, extension, and generalization depend on the references and citations of the focal work. Therefore, we provide several robustness checks.

At the citation level, if a paper cites another paper with no reference recorded, this citation link can only be generalizational, or foundational per our measurement, even if it can be extensional in reality. Therefore, we replicate our analysis on the property of citation links by only keeping citations whose cited paper has at least one reference recorded, and the replication results are shown in Figs. S57-S62; all are consistent with the original results reported when excluding papers with zero reference.

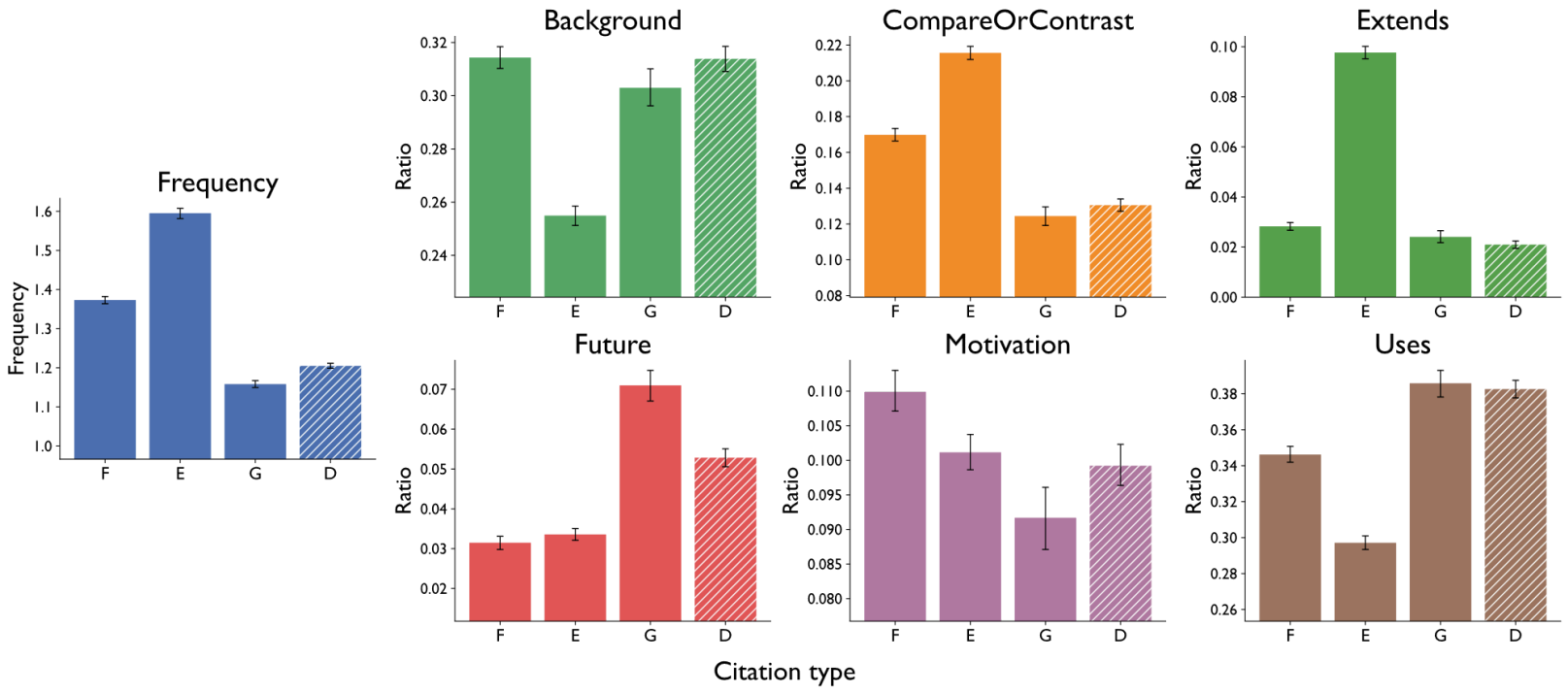

**Fig. S57 | Citation purposes associated with foundational, extensional, generalizational, and disruptive citations in the NLP citation purpose dataset.** This analysis replicates that shown in Fig. S17, while restricting the sample to citation links for which the cited paper has at least one reference. A total of 78,193 citation links and all of their 111,583 citation instances were included in the analysis.

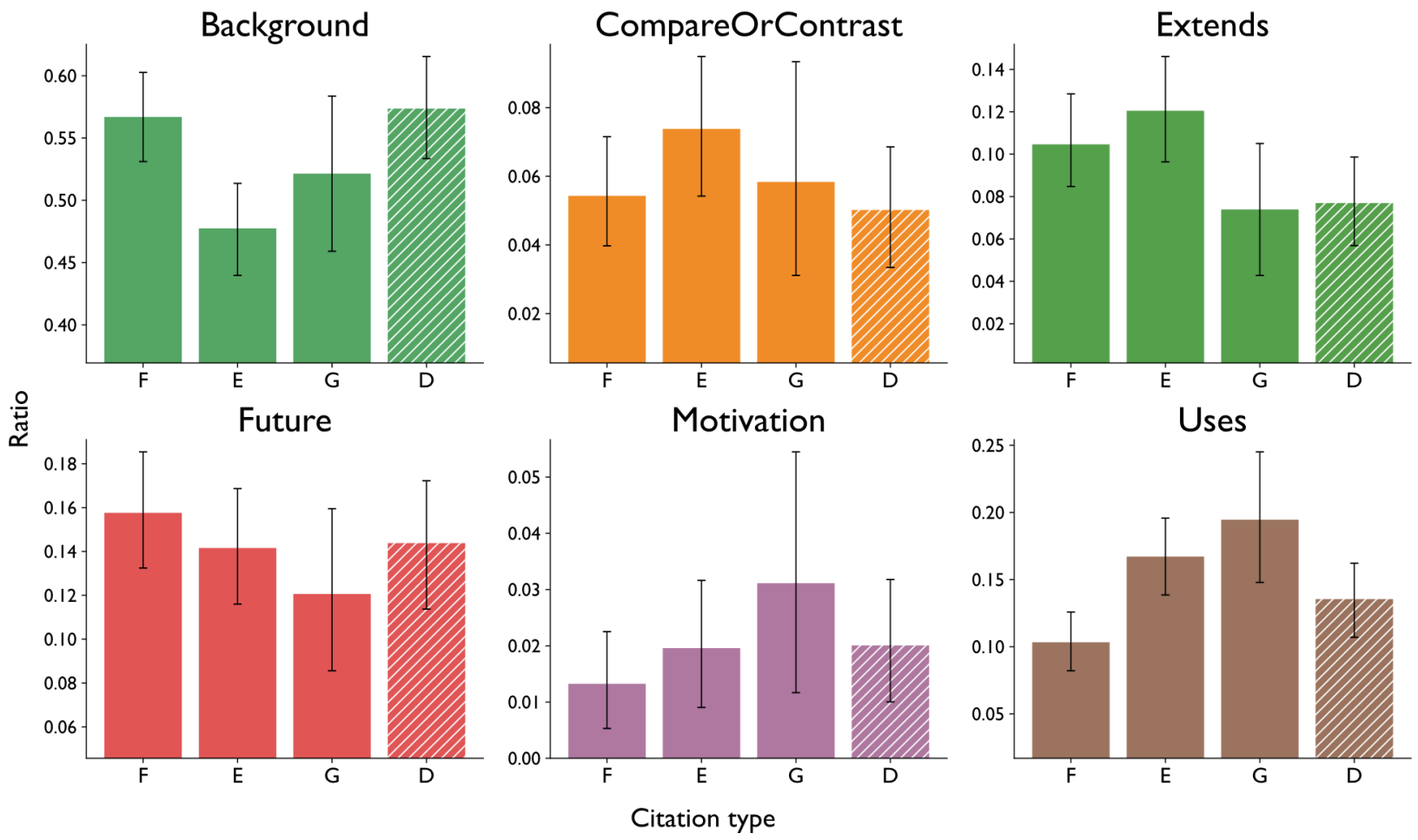

**Fig. S58 | Citation purposes associated with foundational, extensional, generalizational, and disruptive citations in the ACT2 dataset.** This analysis replicates the result shown in Fig. S18, while restricting the sample to citation links for which the cited paper has at least one reference. A total of 1,641 citation links and 1,676 citation instances were included in the analysis.

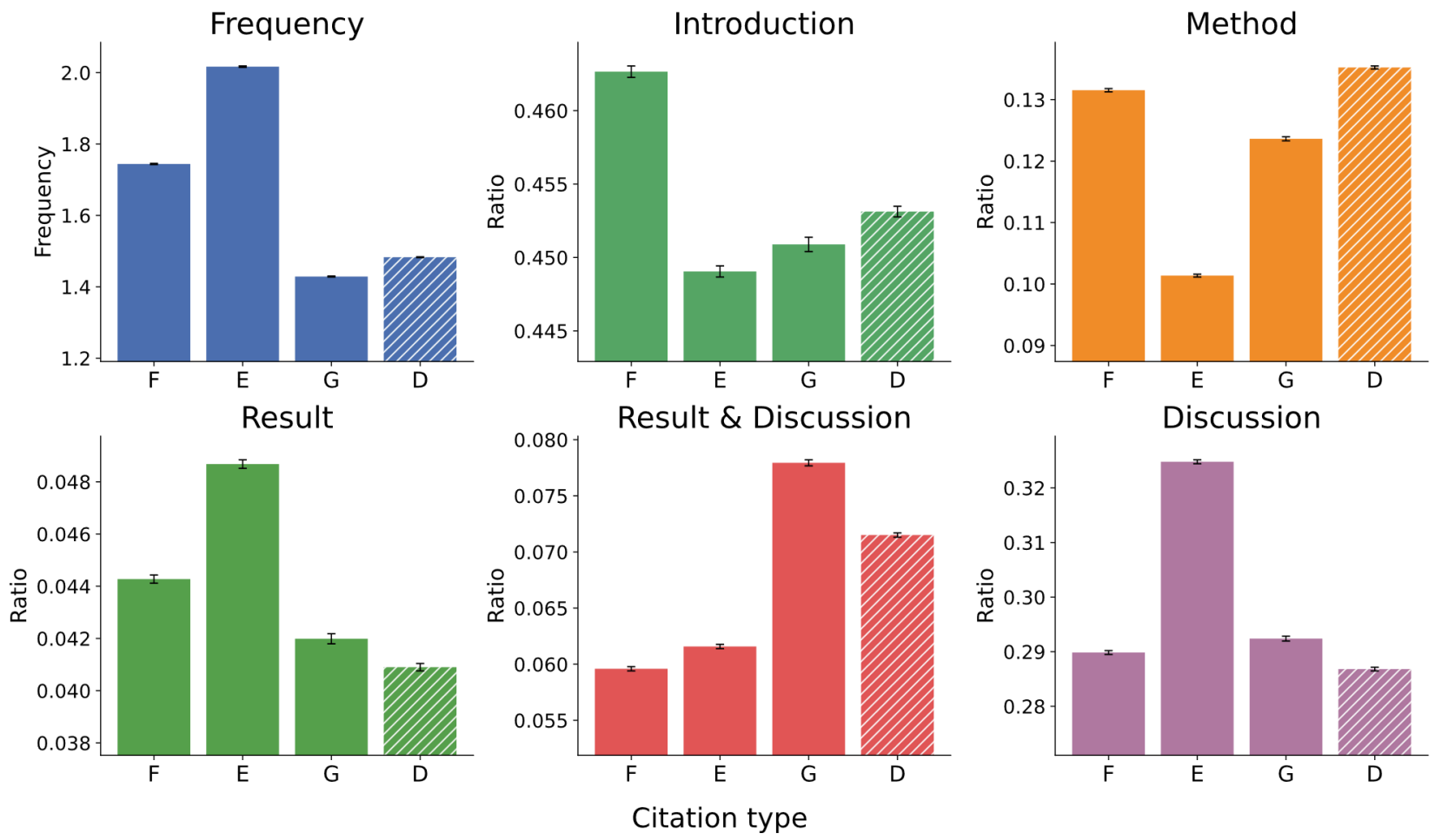

**Fig. S59 | The location of reference associated with foundational, extensional, generalizational, and disruptive citations in the *Semantic Scholar* full-text dataset.** This analysis replicates that shown in Fig. 2g, while restricting the sample to citation links for which the cited paper has at least one reference. A total of 14,769,210 citation links

and their 26,087,658 citation instances were included in the analysis.

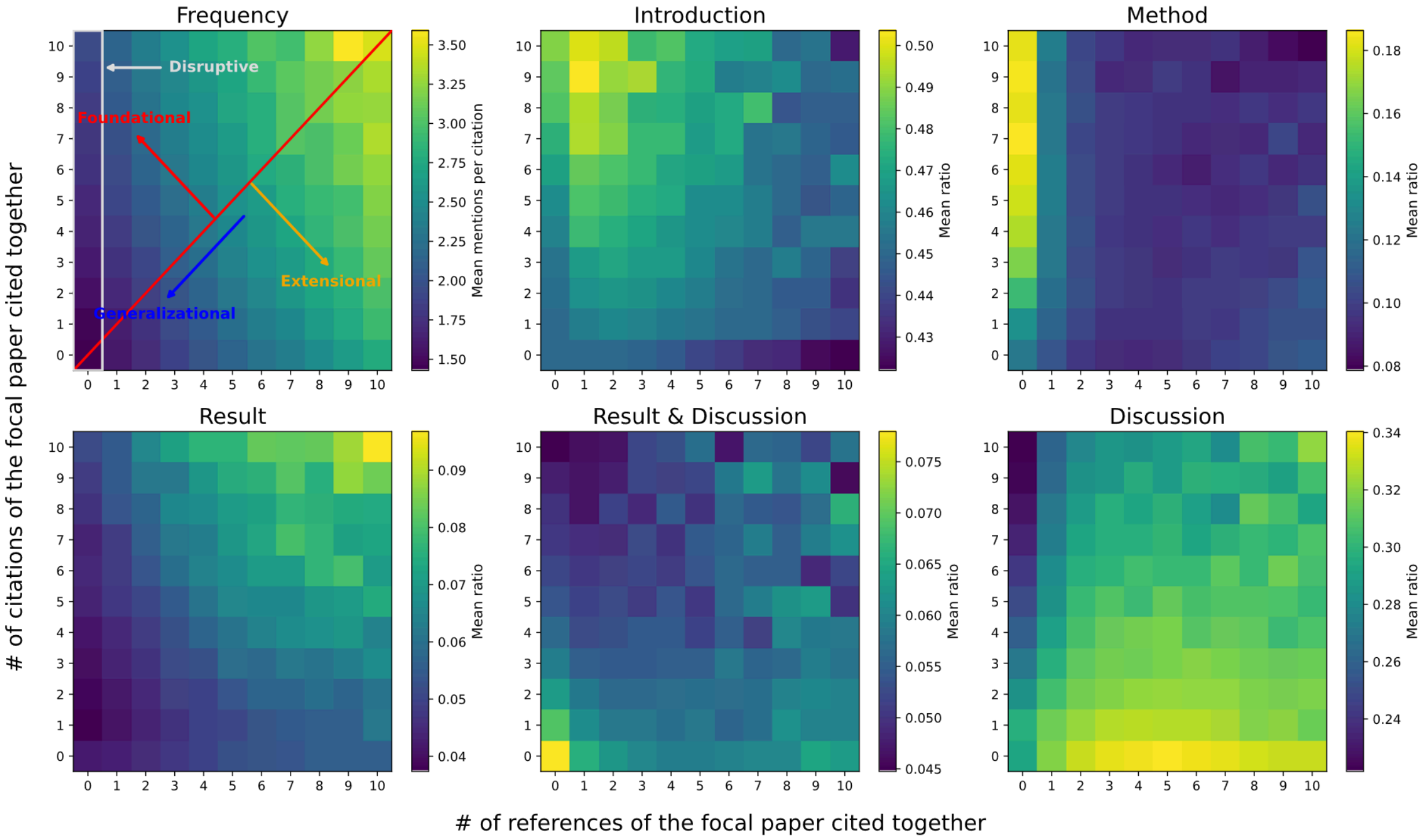

**Fig. S60 | Distribution of reference locations by the number of co-cited references and citations received by the focal paper.** Replication of the analysis presented in Extended Data Fig. 3 using the subset of Semantic Scholar citation links for which reference location information is available. The analysis includes 14,217,144 citation links corresponding to 24,372,805 citation instances with all cited papers containing at least one reference.

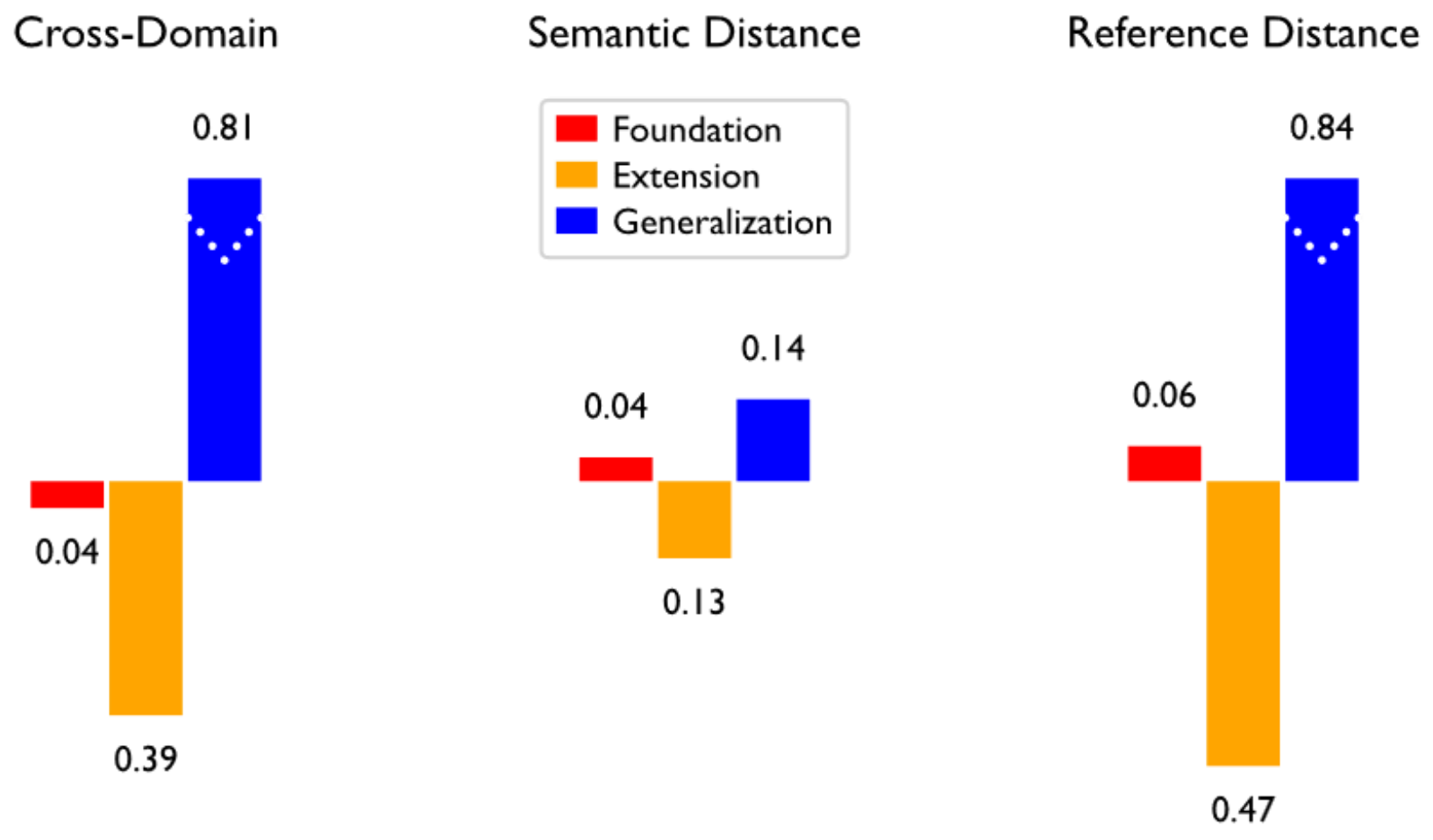

**Fig. S61 | Relationship between citation types and alternative measures of interdisciplinarity.** Replication of the analysis presented in Fig. 2e using alternative measures of interdisciplinarity. The analysis is restricted to citations among 57,887,123 OpenAlex publications published between 1951 and 2019 that contain at least one reference. Citation types are classified as foundational, extensional, or generalizational.

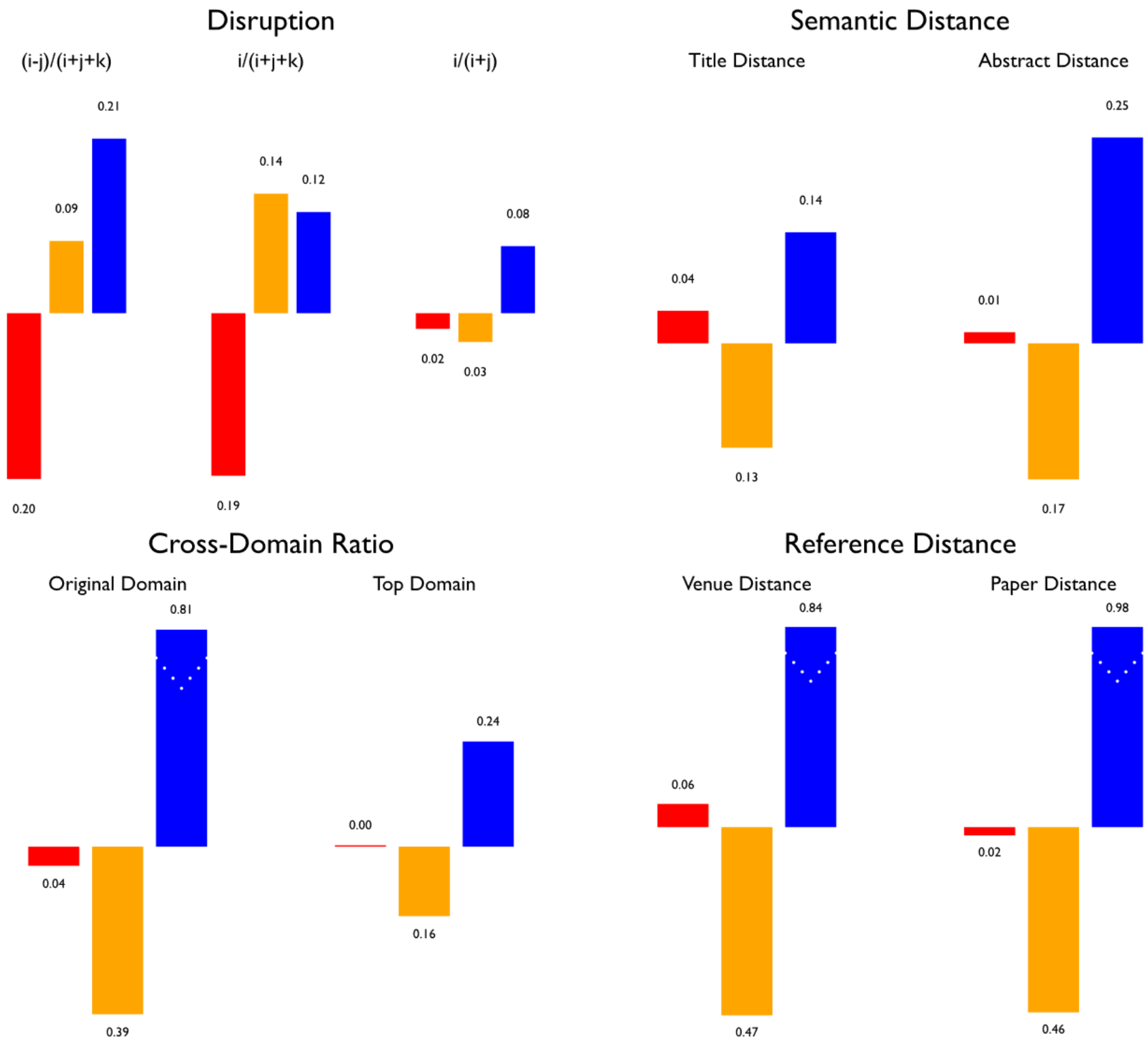

**Fig. S62 | Relationship between citation types and alternative measures of interdisciplinarity.** Replication of the analysis presented in Fig. S12 using alternative measures of interdisciplinarity. The analysis is restricted to citations among 57,887,123 OpenAlex publications published between 1951 and 2019 that contain at least one reference. Citation types are classified as foundational, extensional, or generalizational.

At the paper level, we examined the longitudinal trends of the F, E, and G indices while including papers with zero references. The overall patterns remained qualitatively consistent with the main analysis: the F and E indices decreased over time, whereas the G index increased, but the temporal transition occurred substantially later than the 1990s reported in Fig. 3a. For example, the increase in the G index began around 2012 (Fig. S63). Despite this, the longitudinal changes in the F, E, and G indices continue to exhibit patterns that differ from those observed in the randomized network, even after including papers with zero references (Fig. S64). This result further confirms that the observed temporal changes in the F, E, and G indices are not solely driven by changes in publication volume or citation practice.

Given that including papers with zero references does not qualitatively alter our main findings, we further report the proportion of such papers across disciplines and publication years to better characterize the extent of missing reference data. Our analysis reveals substantial temporal and disciplinary variation in the prevalence of papers with zero references. Across all domains, the proportion of papers lacking references declined from 86.7% in 1945 to 31.4% in 2022, indicating progressively improved citation coverage over time. This proportion subsequently increased to 41.0% in 2024, likely reflecting incomplete citation coverage for the most recent publications. Citation coverage also varied considerably across disciplines. Materials science exhibited the highest overall coverage, with only 37.1% of publications lacking references, whereas the arts showed the lowest coverage, with 79.3% of publications containing zero references (Fig. S65).

These patterns highlight the need for caution when comparing citation-based indices across publication years or research domains, as differences in citation coverage may introduce systematic biases. Nevertheless, these coverage differences do not alter the main conclusions of this study, as all analyses yielded qualitatively similar trends. Moreover, validation using the WoS dataset, which has a substantially lower proportion of papers with zero references (10.2%), produced comparable results, further supporting the robustness of our findings.

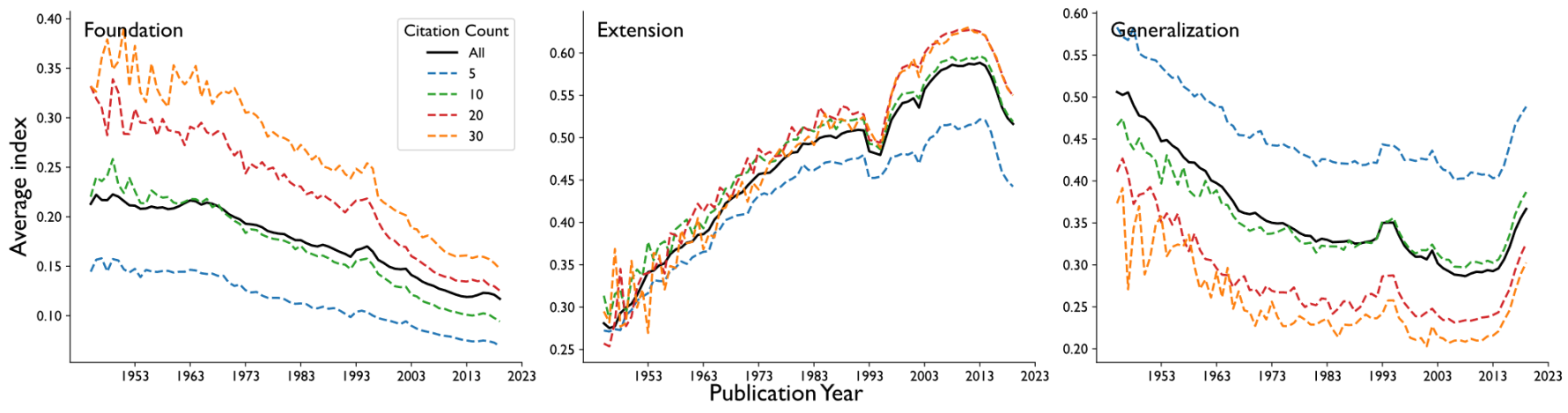

**Fig. S63 | The longitudinal trend for F, E and G indices including papers with zero references.** It is computed among the 26,068,520 publications in OpenAlex that received at least five citations within five years of publication, and published between 1945 and 2019.

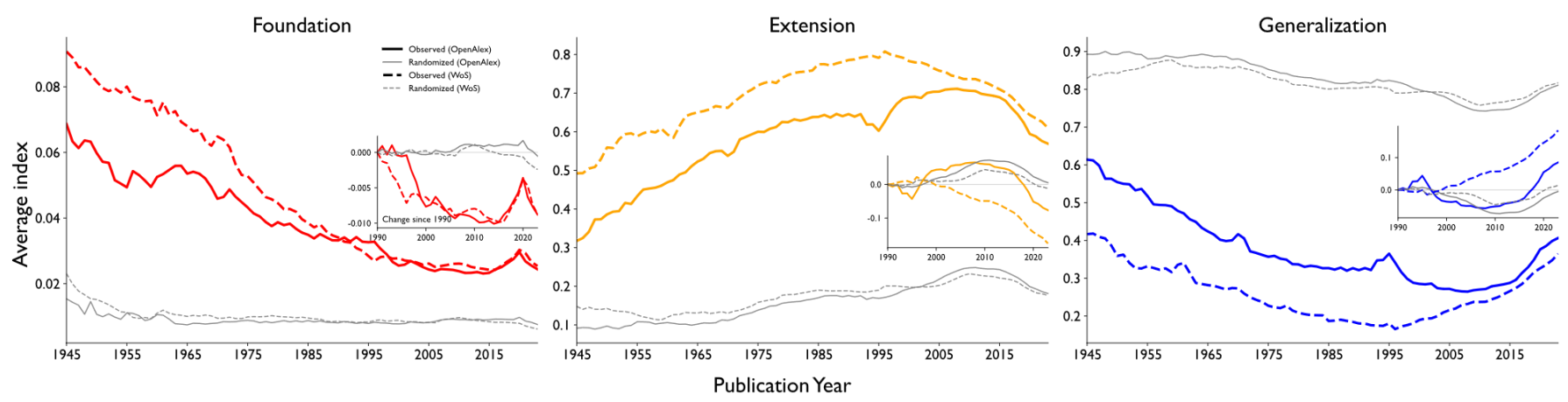

**Fig. S64 | Longitudinal changes in the F, E, and G indices in the observed and randomized citation networks, including papers with zero references.**

Replication of the analysis presented in Fig. 3b. The F, E, and G indices were computed using only citations accrued within one year of publication and were restricted to papers published between 1945 and 2024 that received at least two citations within this one-year window, yielding 20,965,168 papers in OpenAlex and 17,079,812 papers in WoS.

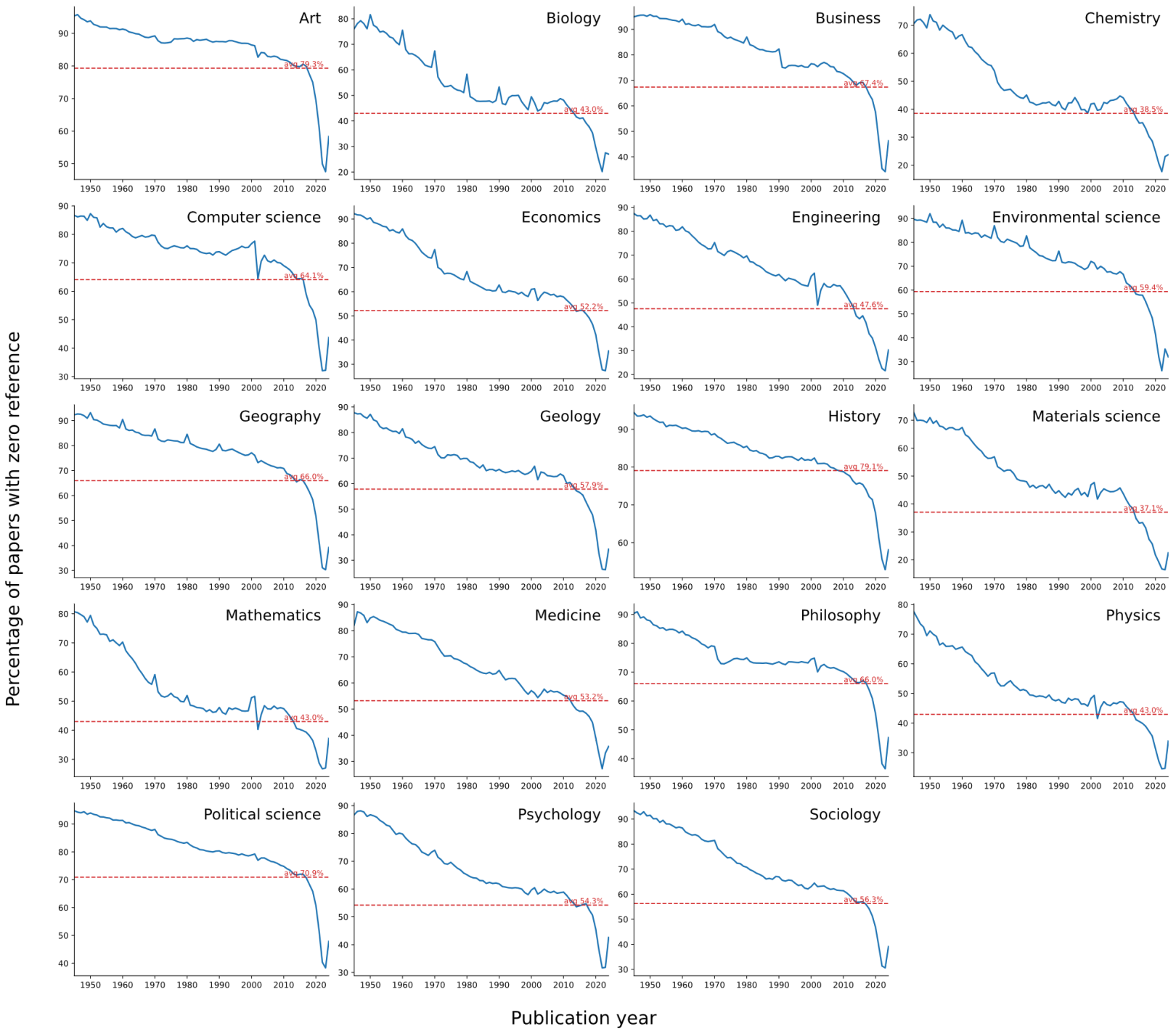

**Fig. S65 | Longitudinal and disciplinary variation in the proportion of publications with zero references.** It is computed with 191,457,232 publications indexed in OpenAlex between 1945 and 2024. The horizontal line indicates the average proportion of zero-reference publications within each discipline across the entire study period.

**Computation of the disruption index.** In computing the disruption index ($D = \frac{I-J}{I+J+K}$), we define $K$ as the number of citations received by all references cited by the focal paper after the publication of the focal paper and within the subsequent five-year citation window. An alternative implementation defines $K$ as the total number of citations received by the focal paper's references within five years after the publication of the focal paper, including citations accrued before the focal paper itself was published. Both definitions are consistent with the conceptual formulation of the disruption index [32]. Because the disruption index is used solely for comparative purposes in our analyses, we do not further evaluate the sensitivity of our results to this alternative implementation.

**Supplementary References**